%% file: main.tex
\documentclass[fleqn,usenatbib,useAMS]{mnras}

\usepackage{graphicx}	% Including figure files
\usepackage{amsmath}	% Advanced maths commands
\usepackage{amssymb}	% Extra maths symbols
\usepackage{multicol}        % Multi-column entries in tables
\usepackage{bm}		% Bold maths symbols, including upright Greek
\usepackage{pdflscape}	% Landscape pages
\usepackage{rotating}
\usepackage{lscape}
\usepackage[usenames,dvipsnames]{xcolor}
\usepackage{hyperref}
\usepackage{cancel}
\usepackage{pifont}

\newcommand{\aas}{AAS}

\usepackage[T1]{fontenc}
\usepackage{ae,aecompl}

\usepackage{newtxtext,newtxmath}
\newcommand{\pp}{\arcsec}

\newcommand{\sigT}{\mbox{$\sigma_{\mbox{\tiny T}}$}}
\newcommand{\Tcmb}{\mbox{$T_{\mbox{\tiny CMB}}$}}

\newcommand{\kB}{\mbox{$k_{\mbox{\tiny B}}$}}
\newcommand{\Ne}{{n_{\rm e}}}

\newcommand{\vek}[1]{{\bm #1}}
\newcommand{\expf}[1]{{{\rm e}^{#1}}}
\newcommand{\id}{{{\rm d}}}

\title[An ACA 1mm Survey of HzRGs]{An ACA 1mm survey of HzRGs in the ELAIS-S1: survey description and first results}

\author[H. Messias]{Hugo G. Messias$^{1,2}$\thanks{Contact e-mail: \href{mailto:hugo.messias@alma.cl}{hugo.messias@alma.cl}}\thanks{Present address: Joint ALMA Observatory, Alonso de C\'ordova 3107, Vitacura 763-0355, Santiago, Chile}
	Evanthia Hatziminaoglou$^{3}$,
	Pascale Hibon$^{2}$,
	Tony Mroczkowski$^{3}$,\newauthor
	Israel Matute$^{4,5}$,
	Mark Lacy$^6$,
	Brian Mason$^6$,
	Sergio Mart\'in$^{1,2}$,
	Jos\'e~M. Afonso$^{4,5}$,\newauthor
	Edward Fomalont$^{1,6}$,
	Stergios Amarantidis$^{4,5}$,
	Sonia Ant\'on$^{7,8}$,
	Ricardo Demarco$^{9}$,\newauthor
	Marie-Lou Gendron-Marsolais$^{1,2}$,
	Andrew M.~Hopkins$^{10}$,
	R\"udiger Kneissl$^{1,2}$,\newauthor
	Cristian Lopez$^{1}$,
	David Rebolledo$^{1,6}$,
	Chentao Yang$^{1,2}$
	\\
	$^{1}$Joint ALMA Observatory, Alonso de C\'ordova 3107, Vitacura 763-0355, Santiago, Chile \\
	$^{2}$European Southern Observatory, Alonso de C\'ordova 3107, Vitacura, Casilla 19001, Santiago de Chile, Chile \\
	$^{3}$European Southern Observatory, Karl-Schwarzschild-Str. 2, 85748 Garching bei M\"unchen, Germany \\
	$^{4}$Instituto de Astrof\'isica e Ci\^encias do Espa\c{c}o, Tapada da Ajuda - Edif\'icio Leste - 2$^o$ Piso 1349-018 Lisboa, Portugal \\
	$^{5}$Departamento de F\'isica, Faculdade de Ci\^encias da Universidade de Lisboa, Edif\'icio C8, Campo Grande, PT1749-016 Lisboa, Portugal\\
	$^{6}$National Radio Astronomy Observatory, 520 Edgemont Road, Charlottesville, VA 22903, USA\\
	$^{7}$Instituto de Telecomunica\c{c}\~oes, Campus Universit\'ario de Santiago, 3810-193 Aveiro, Portugal\\
    $^{8}$CIDMA, Departamento de F\'isica, Universidade de Aveiro, Campus Universit\'ario de Santiago, 3810-193 Aveiro, Portugal\\
	$^{9}$ Departamento de Astronom\'ia, Facultad de Ciencias F\'isicas y Matem\'aticas, Universidad de Concepci\'on, Concepci\'on, Chile\\
	$^{10}$ Australian Astronomical Optics, Macquarie University, 105 Delhi Rd, North Ryde, NSW 2113, Australia
	}

\date{Last updated ; in original form }

\pubyear{2020}

\begin{document}
	\label{firstpage}
	\pagerange{\pageref{firstpage}--\pageref{lastpage}}
	\maketitle
	
	\begin{abstract}
	\input{sections/abstract}
	\end{abstract}
	
	
	\begin{keywords}
	    %https://academic.oup.com/DocumentLibrary/mnras/keywords.pdf
	    cosmic background radiation;
	    %cosmology: observations;
		galaxies: active;
		galaxies: high-redshift;
		%galaxies: ISM;
		%galaxies: jets;
		%galaxies: starburst;
		%galaxies: statistics;
		ISM: jets and outflows;
		%ISM: molecules;
		radio continuum: galaxies;
		%submillimetre: ISM;
		submillimetre: galaxies
	\end{keywords}
	%%%%%%%%%%%%%%%%%%%%%%%%%%%%%%%%%%%%%%%%%%%%%%%%%%
	
	%%%%%%%%%%%%%%%%% BODY OF PAPER %%%%%%%%%%%%%%%%%%
	
	% The MNRAS class isn't designed to include a table of contents, but for this document one is useful.
	% I therefore have to do some kludging to make it work without masses of blank space.
%	\begingroup
%	\let\clearpage\relax
	%\tableofcontents
%	\endgroup
%	\newpage

	\section{Introduction}
	\input{sections/intro}

	\section{Radio data and sample selection} \label{sec:samp}
	
	\subsection{The choice of the ELAIS-S1 field} \label{sec:es1}
	\input{sections/es1field}
	
	\subsection{Radio photometry} \label{sec:radphot}
    \input{sections/radphot}
    
    \subsection{HzRGs selection criteria} \label{sec:selectioncriteria}
	\input{sections/selectionCriteria}
    
	\subsection{The HzRG candidates sample}
	\input{sections/hzrgSample}

	\section{ACA observations} \label{sec:acaobs}
	\input{sections/acaobs}

	\subsection{Continuum fluxes} \label{sec:b6flx}
	\input{sections/contflux}

    \section{The radio spectrum of HzRGs}\label{sec:radspec}
    \input{sections/radspec}

 	\section{Serendipitous line emitters} \label{sec:specline}
 	\input{sections/specline}

 	\section{Significance and possible implications of negative peaks}
 	\label{sec:sz}
	\input{sections/szeffect}

 	\section{Discussion} \label{sec:disc}
 	
 	\subsection{The selection criteria} \label{sec:discselec}
    \input{sections/disc-selecCrit} 

 	\subsection{Millimetre continuum detections}
 	\input{sections/disc-mmdetec}
 	
 	\subsection{Line detections}
 	\input{sections/disc-linedetec}
 	
 	\subsection{Flux decrements around radio galaxies}
 	\input{sections/disc-sze}
 	
 	\section{Conclusions} \label{sec:conc}
 	\input{sections/conclusions}

	\section*{Acknowledgements}
	% Entry for the table of contents, for this guide only
	\addcontentsline{toc}{section}{Acknowledgements}
	This paper makes use of the following ALMA data: ADS/JAO.ALMA\#2018.A.00046.S. ALMA is a partnership of ESO (representing its member states), NSF (USA) and NINS (Japan), together with NRC (Canada), MOST and ASIAA (Taiwan), and KASI (Republic of Korea), in cooperation with the Republic of Chile. The Joint ALMA Observatory is operated by ESO, AUI/NRAO and NAOJ.
	
	This research made use of iPython \citep{perez07}, Numpy \citep{vanderwalt11}, Matplotlib \citep{hunter07}, SciPy \citep{jones01}, Astropy \citep[a community-developed core Python package for Astronomy,][]{astropy13}, APLpy \citep[an open-source plotting package for Python,][]{robitaille12}, and Topcat \citep{taylor05}.
	
	EH and HM acknowledge support from the Joint ALMA Observatory Visitor Program.
	S.A. acknowledges support by FCT, through CIDMA, within project UIDB/04106/2020, and ENGAGE-SKA through POCI-01-0145-FEDER022217, funded by COMPETE 2020 and FCT, Portugal.
	R.D. gratefully acknowledges support from the Chilean Centro de Excelencia en Astrof\'isica y Tecnolog\'ias Afines (CATA) BASAL grant AFB-170002. JA, IM and SA gratefully acknowledge support from the Science and Technology Foundation (FCT, Portugal) through the research grants PTDC/FIS-AST/29245/2017, UIDB/04434/2020 and UIDP/04434/2020.

	\section*{Data availability}
	
	The data underlying this article are publicly available in ALMA science archive at \href{https://almascience.eso.org/asax/}{almascience.eso.org/asax}, and can be accessed with the project code identifier 2018.A.00046.S. Other supporting publicly available data used in this article is properly cited throughout the manuscript.
	
	%%%%%%%%%%%%%%%%%%%%%%%%%%%%%%%%%%%%%%%%%%%%%%%%%%
	
	
	%\bibliographystyle{mnras}

	%%%%%%%%%%%%%%%%%%%%%%%%%%%%%%%%%%%%%%%%%%%%%%%%%%
	
	%%%%%%%%%%%%%%%%% APPENDICES %%%%%%%%%%%%%%%%%%%%%
	
	\appendix
	
	\section{Input coordinates in ACA observations}\label{app:offset}
	\input{sections/app-offset}

	\section{ACA maps for the remainder sources}\label{app:extraimgs}
	\input{sections/app-extraimgs}

	\section{Other observed sources}\label{app:sampdrop}
	\input{sections/app-sampdrop}

	\section{The master table}\label{app:masttab}
	\input{sections/app-mastertable}
	
	%%%%%%%%%%%%%%%%%%%%%%%%%%%%%%%%%%%%%%%%%%%%%%%%%%

	% Don't change these lines
	\bsp	% typesetting comment
	\label{lastpage}
\end{document}

%% file: sections/abstract.tex
% limit of 250words
%https://wordcounter.net/ -> currently 248words

% context
Radio-emitting jets might be one of the main ingredients shaping the evolution of massive galaxies in the Universe since early cosmic times. However, identifying early radio active galactic nuclei (AGN) and confirming this scenario has been hard to accomplish, with studies of samples of radio AGN hosts at $z>2$ becoming routinely possible only recently.
% the survey
With the above in mind, we have carried out a survey with the Atacama Compact Array (ACA, or Morita Array) at 1.3 mm (rms=0.15 mJy) of 36 high-redshift radio AGN candidates found within 3.9\,deg$^2$ in the ELAIS-S1 field. The work presented here describes the survey and showcases a preliminary set of results. The selection of the sample was based on three criteria making use of infrared (IR) and radio fluxes only. 
% results
The criterion providing the highest selection rate of high-redshift sources (86\% at $z>0.8$) is one combining an IR colour cut and radio flux cut ($S_{\rm 5.8\mu m} / S_{\rm 3.6\mu m}>1.3$ and $S_{\rm 1.4\,GHz}>1\,$mJy). Among the sample of 36 sources, 16 show a millimetre (mm) detection. In eight of these cases, the emission has a non-thermal origin. A $z_{sp}=1.58$ object, with a mm detection of non-thermal origin, shows a clear spatial offset between the jet-dominated mm continuum emission and that of the host's molecular gas, as traced by serendipitously detected CO(5-4) emission. Among the objects with serendipitous line detections there is a source with a narrow jet-like region, as revealed by CS(6-5) emission stretching 20\,kpc out of the host galaxy.

%% file: sections/intro.tex
	Most galaxies are believed to harbour a super-massive black hole (SMBH) at their nuclear regions \citep[][and references therein]{kormendyho13}. Accretion of matter onto these SMBHs is the source of the emission of enormous amounts of radiation, driving powerful jets, winds or outflows (\citealt{lynden-bell69}; \citealt{rees84}) feeding back energy to the environment surrounding them, thus playing a significant role in the evolution of their hosts \citep[e.g.,][but see discussion in \citealt{kormendyho13} and \citealt{pitchford16}]{heckman14}. An early study by \cite{benson03} has shown the possible impact of 
	mechanisms quenching star formation in galaxies, driving the shape of the bright end of the galaxy luminosity function. Ever since, feedback from active galactic nuclei (AGN) has positioned itself as the main contributor to star formation quenching.
	
	AGN feedback comes in two flavours, namely {\it quasar} or {\it radiative} mode and {\it radio} or {\it kinetic} or {\it mechanical} mode. Quasar mode feedback occurs when most of the energy from the AGN is deposited to the interstellar medium (ISM) in the form of radiation, which can heat most of the gas in the ISM (and even expel it from the system), eventually quenching star formation in the AGN host. It is thought to occur mainly in young AGN with the peak of their activity at redshifts around 2 to 3, when the SMBH accretes via a geometrically thin, optically thick accretion disk, at rates higher than 1\% of the Eddington limit \citep[][and references therein]{shakura73,heckman14}.
	
	In the mechanical mode, on the other hand, the kinetic energy is deposited by means of winds and relativistic jets launched by the AGN, as they collide with the ISM. The jets eventually disturb the gas beyond the ideal turbulence point for gas clumps to form stars (e.g., \citealt{leroy15} versus \citealt{alatalo15}), or expel completely the gas from the galactic gravitational potential \citep{heckman14}, thereby quenching star formation at the high-mass end of the galaxy population. These processes are linked with the highest mass SMBHs, found in older galaxies with morphologies consistent with classical bulges and massive ellipticals \citep{hickox09,griffith10}. The gas content of these galaxies is low and they undergo little star formation. Mechanical feedback is believed to regulate gas cooling in a wide range of scales \citep[e.g.,][]{alatalo15,cotton20} and, thus, star formation in the most massive galaxies residing in the centres of galaxy groups and clusters. Yet, a robust physical understanding of the processes involved is still lacking.
	
	This mechanical mode is thought to be more common since the peak of activity in the Universe \citep[$z<2$, e.g.][]{croton06,amarantidis19}. Indeed, in the second half of the cosmic history, galaxies hosting radio AGN are highly clustered massive systems, exhibiting old stellar populations and little to no star formation activity \citep{hickox09,griffith10}. Nevertheless, the presence of high redshift radio galaxies (HzRGs; $z \gtrsim 1$, L$_{1.4 \,\, \rm GHz} > 10^{24.5}$ W Hz$^{-1}$; \citealt{afonso05,norris11}), accompanied at times by a double-lobed radio morphology, are hinting towards mechanical feedback in action at earlier epochs. HzRGs are found to be among the most massive and most luminous galaxies at any redshift and are thought to be the progenitors of the most massive galaxies in the local universe \citep[e.g.,][]{seymour07, debreuck10}. These findings thus make the study of this population at redshifts near cosmic noon a necessary step towards constraining the SMBH growth models as well as that of their hosts.
	
	Recent observations associate HzRGs with highly star-forming systems with significant gaseous content \citep{papadopoulos00,dannerbauer14, drouart14, emonts14, emonts15a, emonts15b, gullberg16}. The (sub)mm emission of HzRGs is not typical of that of the population of (sub)mm galaxies (SMGs) in that it is often extended on spatial scales of or above 100 kpc \citep[e.g.,][]{stevens03} and does not necessarily coincide spatially with the radio or optical position of the AGN or the centre of the host galaxy \citep{ivison08}. If indeed HzRGs are associated with actively star-forming systems, (sub)mm observations, sensitive to cold dust that is re-radiating emission from young stars, can probe their star formation history by revealing thermal dust emission from the host galaxies and neighbouring sources.
	
	The study of HzRGs is therefore of paramount importance to our understanding of structure formation, of the star formation processes in the most massive galaxies and of SMBH growth models. Nevertheless, only a handful of systematic studies of samples of HzRGs in (sub)mm wavelengths have taken place in the past two decades, looking for dust continuum emission \citep[e.g.,][]{archibald01, reuland04,falkendal19} or for CO transitions \citep[e.g.,][]{emonts14}. On the other hand, blind wide-field (sub)mm surveys only now begin to cover large enough areas to include significant numbers of HzRGs, but since these are conducted with single-dish facilities, source-confusion imposes a survey sensitivity limit usually around the 1\,mJy level (\citealt{bertoldi07,scott08,geach17}, but see \citealt{staguhn14,magnelli19}). As a result, only the most extreme sources can be picked up at high redshift. For instance, at a redshift of $\sim$2 an M82 or an Arp220-like spectral energy distribution (SED)\footnote{Here we use those from \citet{polletta07}.} detected with such sensitivity at 0.85 or 1.2\,mm would imply star formation rates of $\gtrsim200\,$M$_\odot$/yr or $\gtrsim600\,$M$_\odot$/yr, respectively.
	
	In order to contribute to the general understanding of HzRGs and in an effort to address the issues raised above, we present the first results of a survey targeting HzRG candidates selected within a sky-area of 3.9\,deg$^2$. This was carried out with the Atacama Compact Array \citep[ACA, or \textit{Morita} Array,][]{iguchi09} at 1.3\,mm (Band 6 at 233 GHz) as an Observatory Filler Programme (OFP). The paper is structured as follows: Section~\ref{sec:samp} explains the choice of field and describes the sample selection. Section~\ref{sec:acaobs} presents the ACA observations and derived Band 6 continuum fluxes. The multi-frequency radio SEDs of the sample are discussed in Sec.~\ref{sec:radspec}, while Sec.~\ref{sec:specline} presents emission lines found serendipitously in the ACA cubes. The possible significance of negative peaks are debated in Sec.~\ref{sec:sz}. The paper concludes with a discussion of the first set of findings of the survey (Sec.~\ref{sec:disc}) and some concluding remarks (Sec.~\ref{sec:conc}).

%% file: sections/es1field.tex
The radio spectrum is generally deprived of clear or bright features that can be used to identify high-redshift sources. As a result, radio information combined with other wavelengths has been the most effective way to identify high-redshift radio sources. 

In this work, we will adopt selection criteria for high-$z$ radio AGN to target two types of radio sources, namely the ultra-steep spectrum (USS; \citealt{blumenthal79}, \citealt{tielens79}) sources and the infrared (IR) faint radio sources \citep[IFRS;][]{norris06}. IFRSs are characterised by faint or absent near-infrared (NIR) counterparts and, consequently, extreme radio-to-infrared flux density ratios of up to several thousand. One interesting scenario inducing such characteristics is when a bright radio AGN is hosted by a very red high-redshift galaxy \citep{filho11,norris11}. USSs, on the other hand, are selected based on the assumption that a redshifted GHz-peaked radio spectrum will result in steep MHz-to-GHz spectral slopes. 

This approach requires a good coverage of the NIR, mid-infrared (MIR) and radio photometric information. As OFPs are meant to fill the ACA observing queue in under-subscribed Local Sidereal Time (LST) ranges, the targets had to be selected in a field in the adequate RA range with, at the same time, available multi-wavelength ancillary data. 

The European Large Area ISO Survey South \citep[ELAIS-S1;][]{oliver00} field fulfils these requirements. Thanks to its high galactic latitude (-73$^{\circ}$ 18$^{\prime}$ 46\arcsec) ELAIS-S1 ($\alpha_{2000}$ = 00$^h$ 34$^m$ 44.4$^s$, $\delta_{2000}$ = -43$^{\circ}$ 28$^{\prime}$ 12\arcsec) has a low level of galactic and Zodiacal-light foreground contamination, particularly important both for mid- to far-infrared observations conducted in the past (e.g. {\it Spitzer} Space Telescope, \citealt{werner04}; {\it Herschel} Space Observatory, \citealt{pilbratt10}) and for observations with upcoming or future facilities such as the \textit{James Webb} Space Telescope \citep[JWST;][]{gardner06}.

The wealth of deep multi-wavelength (from X-rays to radio) data available in ELAIS-S1 makes it an ideal target for follow up observations of selected targets. In the radio spectral-range, ELAIS-S1 has been observed at multiple radio frequencies, which is key for the USS criterion. The Australia Telescope Large Area Survey (ATLAS) covered the field at 1.4\,GHz (at 10\arcsec\ $\times$ 7\arcsec\ spatial resolution) and 2.3\,GHz (at 33\arcsec\ $\times$ 20\arcsec\ spatial resolution; \citealt{norris06}, \citealt{middelberg08}, \citealt{zinn12}, \citealt{hales14}, \citealt{franzen15}; Section~\ref{sec:radphot}) with the Australia Telescope Compact Array \citep[ATCA,][]{wilson11}. Molonglo Observatory Synthesis Telescope \citep[MOST,][]{mills81} observations in the same field provided 843\,MHz counterparts (at 62\arcsec $\times$43\arcsec spatial resolution) for about 10\% of the 1.4 GHz catalogue \citep{randall12}. Giant Metrewave Radio Telescope \citep[GMRT, ][]{swarup90} 610\,MHz data (at 12\arcsec\ $\times$ 12\arcsec\ resolution, Intema et al.~in prep) are also available for part of the ELAIS-S1 field. Finally, ELAIS-S1 is covered by the GaLactic and Extragalactic All-Sky MWA survey (GLEAM; \citealt{wayth15}, \citealt{hurleyWalker17}) in the frequency range between 72 and 231 MHz (FWHM$\sim$2.2\arcmin), sub-divided into 20 7.68 MHz-wide sub-channels, as well as the alternative data release of the TIFR GMRT Sky Survey \citep[TGSS,][]{intema17} at 150\,MHz (FWHM$\sim$25\arcsec \,\footnote{Formally the resolution of the TGSS at declinations south of $\delta=+19^{\circ}$ is $25\arcsec \times 25\arcsec$ / cos($\delta - 19^{\circ}$).}).

%% file: sections/radphot.tex
    The ATLAS 2.3\,GHz, MOST 843\,MHz and GMRT 610\,MHz catalogues were matched with the 1.4\,GHz positions with a 9\arcsec, 25\arcsec\, and 6\arcsec\ radius, respectively. These numbers are within the 1$\sigma$ range of the combined beam region of each radio-frequency map.
    
\begin{figure}
	\includegraphics[width=0.5\textwidth]{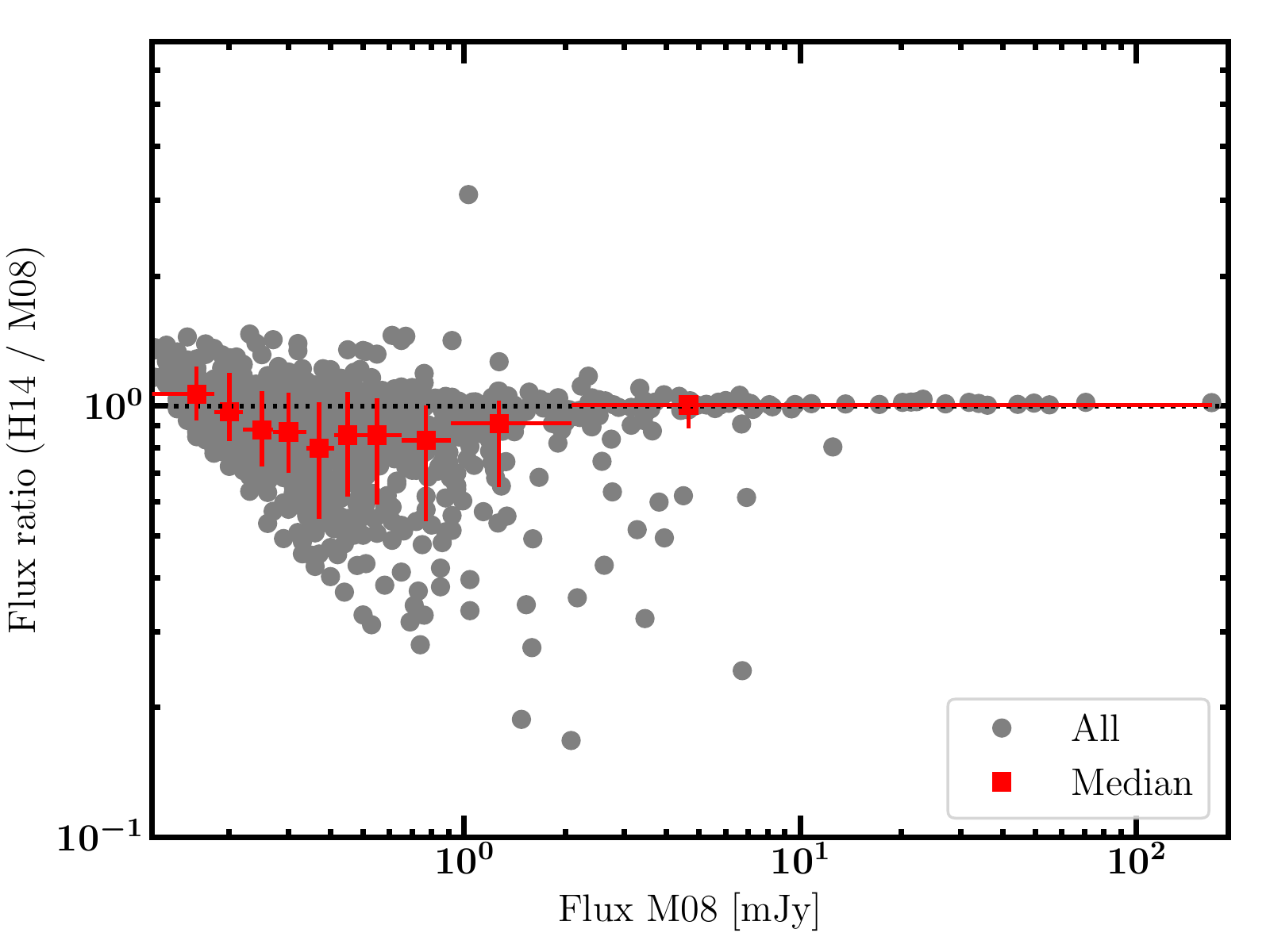}
	\includegraphics[width=0.5\textwidth]{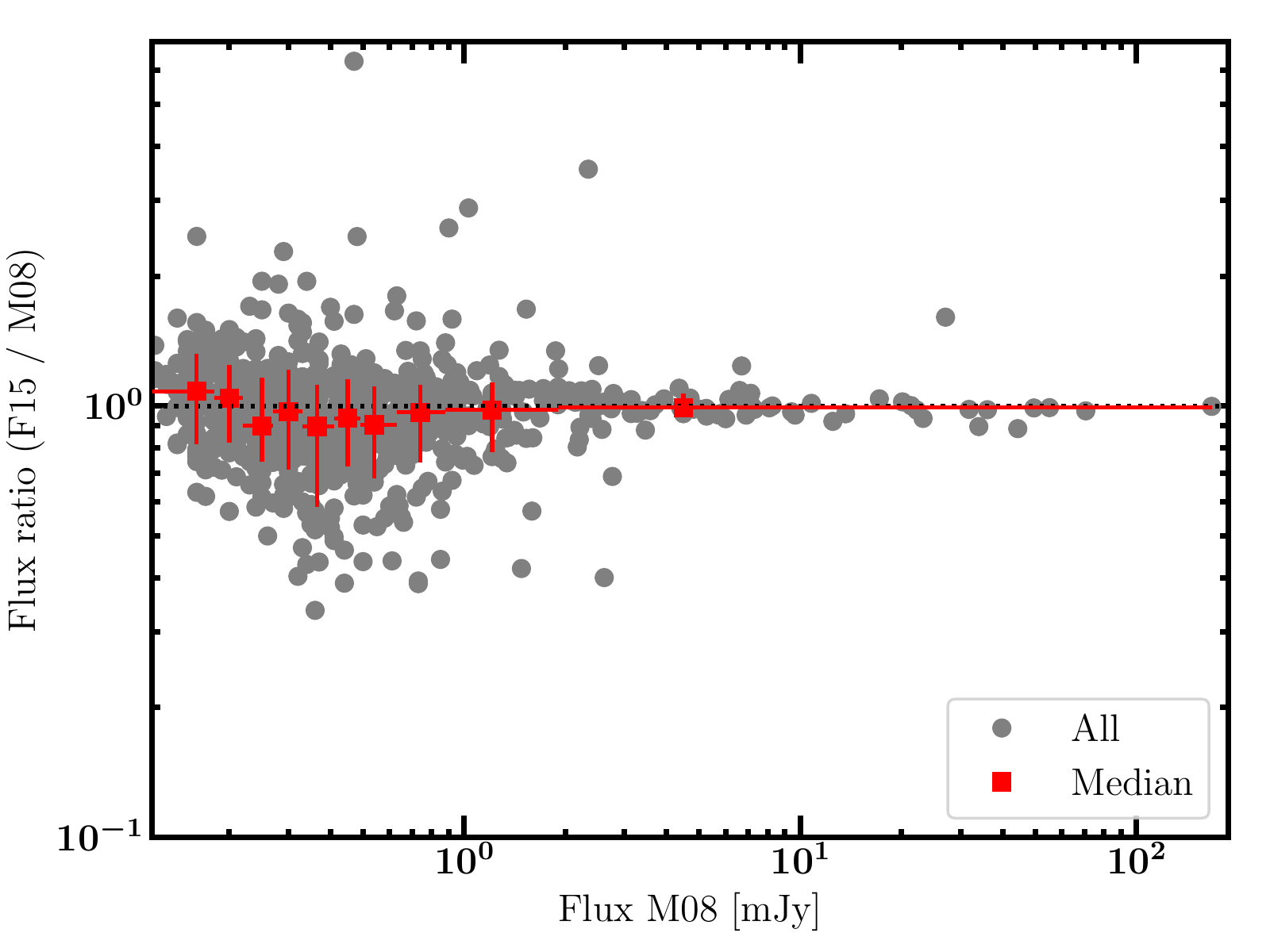}
	\caption{Comparing the 1.4\,GHz photometry between different data releases. The comparison is done only for isolated sources. The catalogued sources in M08, H14 and F15 are shown as grey dots, while the red squares show the median values of the population in ten different quantiles. The error bars show the difference from the median to the 16$^{\rm th}$ and 84$^{\rm th}$ percentiles.}
	\label{fig:photcomp}
	\end{figure}

    The 1.4~GHz catalogue has seen multiple releases \citep{norris06, middelberg08, zinn12, hales14, franzen15}. The catalogue in M08 encompasses a larger area than that included in later data releases. Figure~\ref{fig:photcomp} shows the 20\,cm flux difference between \cite[][hereafter M08]{middelberg08} and those reported in \citealt{hales14} (hereafter H14) and \citealt{franzen15} (hereafter F15), for sources that are not part of core+lobe(s) systems. The ratios of H14 and F15 to M08 were found to be $0.9\pm0.2$ and $1.0\pm0.2$, respectively. The difference between M08 and H14 (greater at $<1\,$mJy) owes to the latter adopting a flux-deboosting approach. Such a correction will be greater for fainter point-like sources detected at low significance levels. Down to 1\,mJy there is no systematic deviation between M08 and the (deeper) F15 data release, despite the different imaging approach pursued. Given the consistency between the two catalogues, we chose not to use F15 as their analysis does not consider the full radio-coverage and does not group multiple radio lobes into single sources, as it is done in M08, increasing the possibility for erroneous cross-matchings.
    
%% file: sections/selectionCriteria.tex
The HzRG candidate selection relies on radio information at the four available frequencies of 2.3~GHz, 1.4~GHz, 843~MHz and 610~MHz, as well as MIR information from the \textit{Spitzer} Enhanced Imaging Products (SEIP\footnote{https://irsa.ipac.caltech.edu/data/SPITZER/Enhanced/SEIP/overview.html}) Cryogenic Release version 3.0 (DR3). The latter has since been updated to DR4, that we now adopt throughout this work for consistency. The cross-match between the radio and NIR catalogues was done adopting a 7\arcsec\ radius nearest-neighbour cone search between the 1.4\,GHz coordinates reported in M08 and those in SEIP. The 1.4~GHz fluxes used were those reported in M08, for the reasons discussed in the previous section. For the selection of IFRSs and USSs we adopt the corresponding criteria from \cite{norris06} and \cite{afonso11} and references therein, respectively, shown in Tab. \ref{tab:selectioncriteria}. 

In order to target $z\gtrsim1.5$ galaxies and in conjunction with the radio criteria, we adopted NIR colour cuts to select either the redshifted 1.6\,$\mu$m stellar-bump \citep{pope06} or an AGN-dominated spectrum \citep{lacy07}. The colour cut applied to the USS sub-sample is more conservative since the radio selection itself is known to pick a significant number of sources at $z<1$ \citep{afonso11}. When only one radio-frequency is available --- preventing the USS assessment --- we consider the same NIR colour cut together with a cut in radio flux density of $S_{\rm 1.4\,GHz}>1$\,mJy, to select luminous radio sources (LRS, Tab. \ref{tab:selectioncriteria}). The cut in radio flux ensures an AGN-like radio emitter ($L_{\rm 1.4\,GHz}>10^{24.5}\,$W Hz$^{-1}$) at $z\gtrsim1.5$ \citep{afonso05}. Note that red $S_{5.8}/S_{3.6}\gtrsim1$ flux ratios are not expected in LRSs in the local Universe, since these are mostly found in massive evolved galaxies with a blue NIR spectrum dominated by old stellar populations \citep{hickox09,griffith10}. Possible contamination by $z\lesssim0.1$ starbursts with strong 6$\,\mu$m emission features are excluded as none of the selected sources appear extended in IR imaging from  \textit{Spitzer} and the {\it Wide-field Infrared Survey Explorer} \citep[WISE,][]{wright10}. 

To select the USS HzRG candidates we use the ATLAS, MOST and GMRT catalogues. For the radio sources we adopt the 1.4\,GHz coordinates resulting from the grouping presented in Tab. 5 in M08.

\begin{table}
\begin{tabular}{lll}
\hline
Type of candidates & $S_{\rm 5.8\mu m} / S_{\rm 3.6\mu m}$ & Radio criteria\\
\hline
 USS & $>1.3$ & $\alpha_{\rm LF}^{\rm HF}<-1.3$ \\
 LRS & $>1.3$ & $S_{\rm 1.4\,GHz}>1\,$mJy \\
 IFRS & $>0.79$ & $S_{\rm 1.4\,GHz} / S_{\rm (3.6~or~4.5\mu m)} > 100$ \\
\hline 
\end{tabular}
\caption{MIR and radio criteria applied for the selection of the HzRGs of the present study. $S\propto\nu^\alpha$, where $\alpha$ is computed in pairs with the lower and higher frequencies (LF and HF, respectively) taking their values among [1.4, 0.84, 0.61\,GHz] and [2.3, 1.4, 0.84\,GHz], respectively. An object is classified as a USS if any of the $\alpha_{\rm LF}^{\rm HF}$ values is below $-1.3$.}
\label{tab:selectioncriteria}
\end{table}

%% file: sections/hzrgSample.tex
The coordinates and redshift, HzRG class, MIR and radio photometry, radio spectral indices and MIR-to-radio flux-ratios of the 36 HzRG candidates selected based on the criteria shown in Tab.~\ref{tab:selectioncriteria} are listed in Tabs. 
~\ref{tab:samp}, \ref{tab:phot}, and \ref{tab:ratios} (a master table, described in Appendix~\ref{app:masttab}, is also made available). The IR photometry is that from SEIP, retrieved using a 3.8\arcsec\ aperture diameter from DR4 already corrected to account for the total fluxes.

\input{tables/tab-samp}
\input{tables/tab-phot}
\input{tables/tab-ratios}

For 15 out of the 36 sources in the sample, shown in Tab. \ref{tab:samp}, there are spectroscopic redshifts available from \citealt{mao12} (indicated by M12 in Tab. \ref{tab:samp}) and \citealt{vaccari15} (MVF). Another 15 objects have photometric redshift estimates from \citealt{rowanrobinson13} (RR13), the \textit{Herschel} Extragalactic Legacy Project \cite[HELP;][]{duncan18}, the Dark Energy Survey \cite[DES;][]{hoyle18}, and the photometric redshifts catalogue of the \textit{Spitzer} Extragalactic Representative Volume Survey \cite[SERVS;][]{pforr19}. Column~8 ($z$) in Tab. \ref{tab:samp} shows the redshift of the HzRG candidates. The values with errors or those marked by $P$ indicate a photometric redshift. The reported values were obtained by cross-matching with a 2\arcsec\ matching radius the radio and the redshift catalogues reported in column~9 ($z_{ref}$). Six out of the 36 objects do not have any distance measurements. In a forthcoming work, we will revise the redshift estimates by using a common tool.

\begin{figure*}
\includegraphics[width=0.45\textwidth]{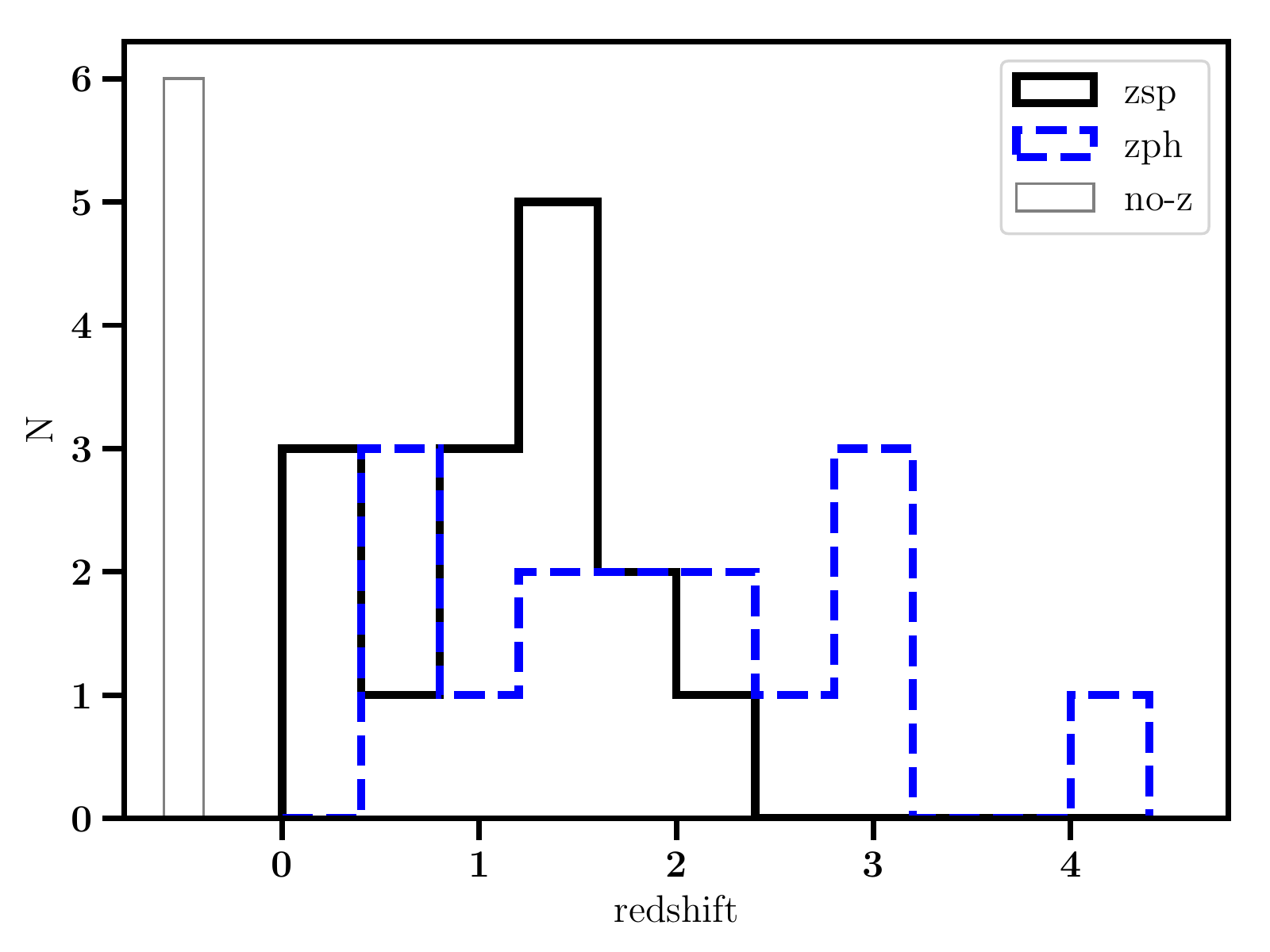}
\includegraphics[width=0.45\textwidth]{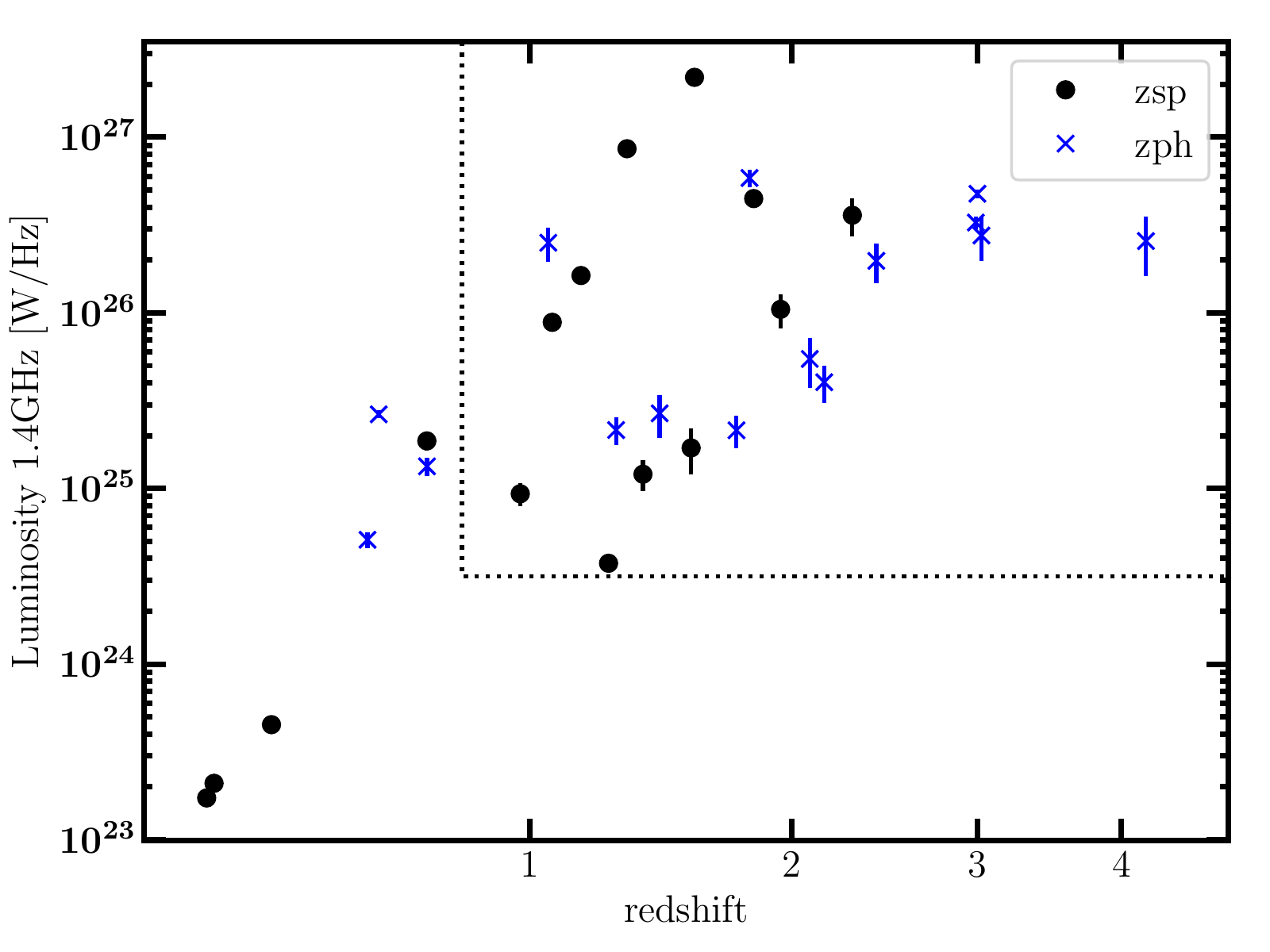}
\caption{The redshift (left) and the redshift-luminosity (right) distributions of the observed sample. Sources with spectroscopic and photometric-type measurements are displayed separately, respectively, as solid black and dashed blue histograms (left) or black circles and blue crosses (right). The six sources with no redshift measurements are shown as a single column centred at $z=-0.5$ in the left panel. In the right panel only flux and spectral-index errors are considered (Sec.~\ref{sec:radphot}), while redshift uncertainties are not shown, both for clarity purposes and because of their absence in some cases. The upper right region delimited by the dotted line is populated by galaxies in the first half of cosmic history whose radio emission is AGN-dominated \citep{afonso05}.
}
\label{fig:redshift}
\end{figure*}

Note that at the time the sample was selected the redshift information in ELAIS-S1 (spectroscopic and photometric combined) was not available for all sources. As a result, no cut in redshift was imposed. The selection relied, instead, exclusively on the MIR and radio criteria shown in Tab.~\ref{tab:selectioncriteria}. We find that 77\% of our sample with redshift measurements lie in the earliest half of cosmic history ($z>0.8$; see Tab. \ref{tab:samp}). Figure~\ref{fig:redshift} shows the redshift distribution separately for those with a spectroscopic and photometric measurements. The overall sample median is $z=1.4$ with 68\% of the population being in the redshift range between 0.7 and 2.3. Despite the incidence of low-redshift sources in the sample of HzRG candidates, we chose to present the full sample, as observations of the lower redshift objects also carry relevant information not only for assessment purposes of the selection criteria reliability, but also individually. Appendix~\ref{app:sampdrop} describes six sources that were observed as part of our OFP, but do not comply with the selection criteria presented in Tab. \ref{tab:selectioncriteria}. Except for Sec.~\ref{sec:sz}, these sources were not considered in the analysis.

% redshift distribution

Deeper IRAC 3.6--4.5\,$\mu$m imaging from SERVS \citep{mauduit12} in post-cryogenic cycles are available for the eastern half of the field covered by the radio observations. Using this SEIP/SERVS overlap region, we find the false-positive\footnote{By false-positive it is meant that a radio source has no reliable MIR counterpart due to either non-detections or clustered fields.} matching-probability, defined as the fraction of the radio sources whose best SEIP counterpart position differs by more than 1.7\arcsec\ to that in SERVS, to be 2\%. We note that IRAC short-wavelength bands spatial resolution is 1.7--2.0\arcsec\ (where the upper limit is from the warm mission\footnote{Section~2.2.2 from the IRAC Instrument Handbook.}). In our sample of 36 sources, we therefore expect one false-positive counterpart association. Finally, 3.8\% and 5.1\% of the radio sources in M08 and F15, respectively, have no SEIP counterparts, but this value is not corrected for flagged regions in the IR imaging.

%% file: tables/tab-samp.tex
\begin{table*}
\begin{tabular}{ccrrrrrrcccc}
ID & ID$_{M08}$ & RA$_{rad}$ & Dec$_{rad}$ & RA$_{IR}$ & Dec$_{IR}$ & sep & $z$ & $z_{ref}$ & USS & LRS & IFRS \\
 & & [deg] & [deg] & [deg] & [deg] & [arcsec] & & & & & \\
\hline
1 & S14 & 9.40915 & -44.64373 & 9.40926 & -44.64400 & 1.0 & 1.3825 & MVF &  & $\checkmark$ &  \\
2 & S15 & 8.65602 & -44.64474 & 8.65603 & -44.64474 & 0.0 & 1.29$~^p$ & HELP  & $\checkmark$ & $\checkmark$ &  \\
3 & S38 & 9.57743 & -44.53606 & 9.57749 & -44.53604 & 0.2 & 3.00$_{-1.95}^{+3.00}$ & SERVS &  & $\checkmark$ & $\checkmark$ \\
4 & S45 & 9.32451 & -44.50390 & 9.32457 & -44.50393 & 0.2 & 1.3243 & MVF &  & $\checkmark$ & $\checkmark$ \\
5 & S48 & 8.34848 & -44.50004 & 8.34850 & -44.50001 & 0.1 & 0.70$_{-0.09}^{+0.14}$ & SERVS &  &  & $\checkmark$ \\
6 & S51 & 7.79623 & -44.48275 & 7.79675 & -44.48279 & 1.3 & 0.58$~^p$ & HELP  &  &  & $\checkmark$ \\
7 & S66 & 9.92688 & -44.45383 & 9.92671 & -44.45381 & 0.5 & 1.81$_{-0.23}^{+0.17}$ & SERVS &  & $\checkmark$ & $\checkmark$ \\
8 & S100 & 7.41331 & -44.38871 & 7.41384 & -44.38857 & 1.5 & 1.06$_{-0.95}^{+-0.95}$ & DES &  & $\checkmark$ & $\checkmark$ \\
9 & S120 & 8.20546 & -44.36402 & 8.20551 & -44.36404 & 0.1 & 1.8284 & M12 &  & $\checkmark$ & $\checkmark$ \\
10 & S178 & 8.00136 & -44.28361 & 8.00154 & -44.28361 & 0.5 &  &  &  & $\checkmark$ &  \\
11 & S267 & 9.47662 & -44.18484 & 9.47685 & -44.18531 & 1.8 & 0.2120 & MVF & $\checkmark$ & $\checkmark$ &  \\
12 & S296 & 9.20793 & -44.15058 & 9.20753 & -44.15088 & 1.5 & 0.7047 & M12 &  &  & $\checkmark$ \\
13 & S331 & 8.90657 & -44.12183 & 8.90496 & -44.12177 & 4.2 & 0.56$_{-0.06}^{+0.05}$ & SERVS & $\checkmark$ & $\checkmark$ & $\checkmark$ \\
14 & S384 & 8.98189 & -44.04370 & 8.98190 & -44.04367 & 0.1 & 1.45$_{-0.22}^{+1.35}$ & SERVS & $\checkmark$ & $\checkmark$ &  \\
15 & S390 & 9.33108 & -44.02906 & 9.33094 & -44.02873 & 1.2 & 2.42$_{-0.08}^{+0.10}$ & SERVS &  & $\checkmark$ &  \\
16 & S467 & 9.01788 & -43.92944 & 9.01785 & -43.92938 & 0.2 & 3.02$_{-0.03}^{+0.05}$ & SERVS & $\checkmark$ & $\checkmark$ &  \\
17 & S500 & 9.66772 & -43.88792 & 9.66742 & -43.88793 & 0.8 & 0.3400 & M12 &  & $\checkmark$ &  \\
18 & S543 & 9.76442 & -43.82975 & 9.76443 & -43.82970 & 0.2 & 1.1644 & HELP &  & $\checkmark$ & $\checkmark$ \\
19 & S598 & 7.46450 & -43.75780 & 7.46446 & -43.75775 & 0.2 &  &  &  &  & $\checkmark$ \\
20 & S612 & 9.62472 & -43.74835 & 9.62472 & -43.74847 & 0.4 & 1.5670 & MVF & $\checkmark$ & $\checkmark$ &  \\
21 & S653 & 9.33771 & -43.71134 & 9.33784 & -43.71111 & 0.9 & 0.2259 & M12 &  & $\checkmark$ &  \\
22 & S671 & 7.87021 & -43.68915 & 7.87028 & -43.68915 & 0.2 &  &  &  &  & $\checkmark$ \\
23 & S769 & 8.63547 & -43.56550 & 8.63567 & -43.56545 & 0.5 & 2.2953 & MVF &  & $\checkmark$ & $\checkmark$ \\
24 & S915 & 8.18569 & -43.41299 & 8.18577 & -43.41292 & 0.3 & 1.9487 & M12 &  & $\checkmark$ &  \\
25 & S928 & 9.28583 & -43.39804 & 9.28589 & -43.39807 & 0.2 & 1.2588 & MVF &  & $\checkmark$ &  \\
26 & S943 & 7.44019 & -43.36375 & 7.44027 & -43.36371 & 0.2 & 1.0702 & MVF &  & $\checkmark$ & $\checkmark$ \\
27 & S977 & 9.44165 & -43.32158 & 9.44158 & -43.32182 & 0.9 & 2.99$_{-0.02}^{+0.02}$ & SERVS &  & $\checkmark$ & $\checkmark$ \\
28 & S1002 & 9.94689 & -43.29181 & 9.94744 & -43.29182 & 1.4 & 2.08$_{-0.89}^{+1.51}$ & SERVS &  & $\checkmark$ &  \\
29 & S1021 & 8.23139 & -43.27421 & 8.23093 & -43.27406 & 1.3 &  &  &  & $\checkmark$ & $\checkmark$ \\
30 & S1073 & 8.62464 & -43.22416 & 8.62464 & -43.22421 & 0.2 & 2.15$_{-0.70}^{+0.05}$ & SERVS &  & $\checkmark$ &  \\
31 & S1122 & 7.79730 & -43.15464 & 7.79738 & -43.15462 & 0.2 &  &  &  &  & $\checkmark$ \\
32 & S1127 & 9.06962 & -43.15960 & 9.06966 & -43.15969 & 0.3 & 1.5805 & MVF &  &  & $\checkmark$ \\
33 & S1187 & 7.96913 & -43.04566 & 7.96916 & -43.04568 & 0.1 & 1.75$~^p$ & RR13  &  & $\checkmark$ &  \\
34 & S1232 & 8.49895 & -42.96400 & 8.49900 & -42.96406 & 0.2 & 0.9700 & MVF & $\checkmark$ & $\checkmark$ &  \\
35 & S1256 & 7.72196 & -42.87093 & 7.72220 & -42.87091 & 0.6 &  &  & $\checkmark$ & $\checkmark$ & $\checkmark$ \\
36 & S1257 & 9.74207 & -42.86718 & 9.74139 & -42.86787 & 3.1 & 4.19$~^p$ & HELP  &  & $\checkmark$ & $\checkmark$ \\
\hline
\end{tabular}
\caption{The sample considered in this study. Here, we list the radio (1.4\,GHz; \citealt[][M08]{middelberg08}) and IR (from SEIP) coordinates, the separation between the two, and the redshift of each source. Spectroscopic redshifts are reported when available, while photometric redshifts are those followed by 1$\sigma$ or 68\% confidence levels or, when these are not reported, by a $^P$. The $z_{ref}$ column indicates the catalogue from which the values were retrieved (for details, see the text).
The last three columns indicate which selection criteria each source complies with.
}
\label{tab:samp}
\end{table*}

%% file: tables/tab-phot.tex
\begin{table*}
\begin{tabular}{crrrrrrr}
ID & S$_{3.6}$ & S$_{4.5}$ & S$_{5.8}$ & S$_{2.3}$ & S$_{1.4}$ & S$_{0.8}$ & S$_{0.6}$  \\
& [$\mu$Jy] & [$\mu$Jy] & [$\mu$Jy] & [mJy] & [mJy] & [mJy] & [mJy] \\
\hline
1 & 148$\pm$1 & 248$\pm$1 & 382$\pm$6 & \ldots & 1.4$\pm$0.2 & \ldots & 2.9$\pm$0.3 \\
2 & 48.8$\pm$0.7 & 61$\pm$1 & 98$\pm$3 & \ldots & 2.9$\pm$0.3 & 5$\pm$1 & 7.1$\pm$0.7 \\
3 & 21.0$\pm$0.6 & 19.4$\pm$0.9 & 40$\pm$3 & 4.2$\pm$0.5 & 7.2$\pm$0.8 & 11$\pm$2 & 14$\pm$1 \\
4 & 128.9$\pm$0.9 & 170$\pm$1 & 250$\pm$4 & \ldots & 100$\pm$10 & 160$\pm$20 & 190$\pm$20 \\
5 & 36.3$\pm$0.6 & 40$\pm$1 & 29$\pm$4 & 5.2$\pm$0.6 & 7.1$\pm$0.8 & 11$\pm$2 & 12$\pm$1 \\
6 & 124$\pm$1 & 96$\pm$1 & 99$\pm$5 & 16$\pm$2 & 21$\pm$2 & 32$\pm$4 & 38$\pm$4 \\
7 & 17.4$\pm$0.6 & 23$\pm$10 & 49$\pm$4 & \ldots & 34$\pm$4 & 61$\pm$7 & \ldots \\
8 & 144$\pm$1 & 172$\pm$1 & 246$\pm$7 & \ldots & 46$\pm$5 & 42$\pm$5 & \ldots \\
9 & 345$\pm$1 & 565$\pm$2 & 932$\pm$6 & 34$\pm$4 & 44$\pm$5 & 39$\pm$4 & 47$\pm$5 \\
10 & 20.9$\pm$0.6 & 30.8$\pm$0.8 & 60$\pm$4 & 1.3$\pm$0.2 & 1.6$\pm$0.2 & \ldots & 1.0$\pm$1.0 \\
11 & 215$\pm$10 & 265$\pm$1 & 395$\pm$5 & 0.56$\pm$0.09 & 1.3$\pm$0.2 & \ldots & \ldots \\
12 & 73.0$\pm$0.7 & 58$\pm$1 & 76$\pm$5 & 6.7$\pm$0.8 & 10$\pm$1 & 14$\pm$2 & 16$\pm$2 \\
13 & 25.2$\pm$0.5 & 20.2$\pm$0.8 & 38$\pm$3 & 1.1$\pm$0.1 & 4.7$\pm$0.5 & \ldots & 8.2$\pm$0.8 \\
14 & 35.0$\pm$0.5 & 44$\pm$1 & 59$\pm$3 & 1.2$\pm$0.2 & 2.1$\pm$0.2 & 4$\pm$1 & 3.4$\pm$0.3 \\
15 & 81.8$\pm$0.7 & 149$\pm$1 & 262$\pm$4 & \ldots & 6.7$\pm$0.8 & \ldots & 12$\pm$1 \\
16 & 63.9$\pm$0.7 & 74$\pm$1 & 120$\pm$5 & 2.0$\pm$0.2 & 3.3$\pm$0.4 & 7$\pm$1 & 6.8$\pm$0.7 \\
17 & 306$\pm$1 & 397$\pm$2 & 560$\pm$6 & 0.8$\pm$0.1 & 1.2$\pm$0.2 & \ldots & 1.7$\pm$0.2 \\
18 & 156.8$\pm$0.8 & 198$\pm$1 & 272$\pm$5 & \ldots & 23$\pm$3 & 35$\pm$4 & 38$\pm$4 \\
19 & 53.5$\pm$0.7 & 58$\pm$1 & 49$\pm$4 & \ldots & 5.4$\pm$0.6 & 8$\pm$2 & 9.2$\pm$0.9 \\
20 & 711$\pm$2 & 1187$\pm$2 & 1993$\pm$7 & 0.40$\pm$0.08 & 1.2$\pm$0.2 & \ldots & 1.2$\pm$0.1 \\
21 & 568$\pm$2 & 761$\pm$2 & 1175$\pm$6 & 0.7$\pm$0.1 & 1.3$\pm$0.2 & \ldots & 2.4$\pm$0.2 \\
22 & 45.0$\pm$0.6 & 44$\pm$10 & 36$\pm$4 & \ldots & 220$\pm$20 & 300$\pm$30 & 380$\pm$40 \\
23 & 31.2$\pm$0.4 & 34.5$\pm$0.6 & 57$\pm$4 & 9$\pm$1 & 14$\pm$2 & 22$\pm$3 & 26$\pm$3 \\
24 & 65.4$\pm$0.6 & 90$\pm$1 & 135$\pm$4 & 5.8$\pm$0.7 & 5.7$\pm$0.6 & 6$\pm$2 & 5.2$\pm$0.5 \\
25 & 134.2$\pm$0.9 & 173$\pm$1 & 207$\pm$3 & 5.9$\pm$0.7 & 2.5$\pm$0.3 & \ldots & 1.0$\pm$0.1 \\
26 & 94.7$\pm$0.7 & 107$\pm$1 & 130$\pm$4 & \ldots & 17$\pm$2 & 24$\pm$3 & 33$\pm$3 \\
27 & 56.6$\pm$0.7 & 70.0$\pm$0.9 & 94$\pm$4 & 5.1$\pm$0.6 & 6.7$\pm$0.7 & 10$\pm$2 & 12$\pm$1 \\
28 & 33.0$\pm$0.6 & 41$\pm$1 & 66$\pm$4 & \ldots & 2.5$\pm$0.6 & \ldots & \ldots \\
29 & 28.0$\pm$0.5 & 35.9$\pm$0.7 & 42$\pm$3 & 11$\pm$1 & 16$\pm$2 & 25$\pm$3 & 30$\pm$3 \\
30 & 30.2$\pm$0.3 & 31.8$\pm$0.6 & 42$\pm$2 & 1.0$\pm$0.1 & 1.8$\pm$0.2 & \ldots & 2.6$\pm$0.3 \\
31 & 95.1$\pm$0.8 & 115$\pm$10 & 109$\pm$4 & 19$\pm$2 & 29$\pm$3 & 38$\pm$4 & \ldots \\
32 & 100.2$\pm$0.7 & 116.0$\pm$0.6 & 107$\pm$3 & 110$\pm$10 & 170$\pm$20 & 250$\pm$30 & 330$\pm$30 \\
33 & 59.2$\pm$0.6 & 168$\pm$1 & 406$\pm$4 & 0.9$\pm$0.1 & 1.5$\pm$0.2 & \ldots & 2.6$\pm$0.3 \\
34 & 73.5$\pm$0.6 & 78.3$\pm$0.6 & 126$\pm$3 & \ldots & 2.4$\pm$0.3 & 6$\pm$2 & \ldots \\
35 & 16.2$\pm$0.5 & 19.6$\pm$0.9 & 25$\pm$3 & \ldots & 1.6$\pm$0.2 & 4$\pm$2 & \ldots \\
36 & 12.9$\pm$0.6 & 18$\pm$10 & 30$\pm$4 & \ldots & 2.8$\pm$0.5 & \ldots & \ldots \\
\hline
\end{tabular}
\caption{The sample's radio and IR photometry relevant for the selection criteria shown in Tab. \ref{tab:samp}. The numbers 2.3, 1.4, 0.8, and 0.6 denote frequencies in GHz, while 3.6, 4.5, and 5.8 denote wavelength in $\mu$m. The 1.4\,GHz fluxes are taken from M08. When a value is not reported, that means that either there is no mapping coverage or there is no catalogue entry.}
\label{tab:phot}
\end{table*}

%% file: tables/tab-ratios.tex
%\begin{landscape}
%\begin{table}
\begin{table*}
\begin{tabular}{crrrrrrrrr}
ID & $\alpha^{2.3}_{1.4}$ & $\alpha^{2.3}_{0.8}$ & $\alpha^{2.3}_{0.6}$ & $\alpha^{1.4}_{0.8}$ & $\alpha^{1.4}_{0.6}$ & $\alpha^{0.8}_{0.6}$ & S$_{5.8}$/S$_{3.6}$ & S$_{1.4}$/S$_{3.6}$ & S$_{1.4}$/S$_{4.5}$\\
\hline
1 & \ldots & \ldots & \ldots & \ldots & -0.9$\pm$0.2 & \ldots & 2.58$\pm$0.04 & 9$\pm$1 & 5.6$\pm$0.8 \\
2 & \ldots & \ldots & \ldots & -0.9$\pm$0.5 & -1.1$\pm$0.2 & -1.4$\pm$0.8 & 2.01$\pm$0.07 & 60$\pm$7 & 48$\pm$6 \\
3 & -1.1$\pm$0.3 & -0.9$\pm$0.2 & -0.9$\pm$0.1 & -0.7$\pm$0.4 & -0.8$\pm$0.2 & -0.9$\pm$0.6 & 1.9$\pm$0.2 & 340$\pm$40 & 370$\pm$50 \\
4 & \ldots & \ldots & \ldots & -0.9$\pm$0.3 & -0.7$\pm$0.2 & -0.5$\pm$0.5 & 1.94$\pm$0.03 & 790$\pm$90 & 600$\pm$70 \\
5 & -0.6$\pm$0.3 & -0.8$\pm$0.2 & -0.6$\pm$0.1 & -0.9$\pm$0.4 & -0.7$\pm$0.2 & -0.3$\pm$0.5 & 0.8$\pm$0.1 & 190$\pm$20 & 180$\pm$20 \\
6 & -0.6$\pm$0.3 & -0.7$\pm$0.2 & -0.7$\pm$0.1 & -0.8$\pm$0.3 & -0.7$\pm$0.2 & -0.5$\pm$0.5 & 0.80$\pm$0.04 & 170$\pm$20 & 220$\pm$30 \\
7 & \ldots & \ldots & \ldots & -1.2$\pm$0.3 & \ldots & \ldots & 2.8$\pm$0.2 & 1900$\pm$200 & 1500$\pm$700 \\
8 & \ldots & \ldots & \ldots & 0.1$\pm$0.3 & \ldots & \ldots & 1.71$\pm$0.05 & 320$\pm$40 & 270$\pm$30 \\
9 & -0.5$\pm$0.3 & -0.1$\pm$0.2 & -0.2$\pm$0.1 & 0.3$\pm$0.3 & -0.1$\pm$0.2 & -0.6$\pm$0.5 & 2.70$\pm$0.02 & 130$\pm$10 & 79$\pm$9 \\
10 & -0.3$\pm$0.3 & \ldots & 0.3$\pm$0.1 & \ldots & 0.6$\pm$0.2 & \ldots & 2.9$\pm$0.2 & 75$\pm$9 & 51$\pm$6 \\
11 & -1.8$\pm$0.4 & \ldots & \ldots & \ldots & \ldots & \ldots & 1.84$\pm$0.09 & 6.2$\pm$0.9 & 5.1$\pm$0.7 \\
12 & -0.8$\pm$0.3 & -0.8$\pm$0.2 & -0.7$\pm$0.1 & -0.8$\pm$0.3 & -0.6$\pm$0.2 & -0.3$\pm$0.5 & 1.04$\pm$0.07 & 140$\pm$20 & 170$\pm$20 \\
13 & -2.9$\pm$0.3 & \ldots & -1.5$\pm$0.1 & \ldots & -0.7$\pm$0.2 & \ldots & 1.5$\pm$0.1 & 180$\pm$20 & 230$\pm$30 \\
14 & -1.1$\pm$0.3 & -1.3$\pm$0.3 & -0.7$\pm$0.1 & -1.4$\pm$0.6 & -0.5$\pm$0.2 & 0.8$\pm$0.9 & 1.69$\pm$0.09 & 61$\pm$7 & 49$\pm$6 \\
15 & \ldots & \ldots & \ldots & \ldots & -0.7$\pm$0.2 & \ldots & 3.20$\pm$0.06 & 82$\pm$9 & 45$\pm$5 \\
16 & -1.0$\pm$0.3 & -1.2$\pm$0.2 & -0.9$\pm$0.1 & -1.4$\pm$0.4 & -0.9$\pm$0.2 & -0.0$\pm$0.7 & 1.88$\pm$0.08 & 52$\pm$6 & 45$\pm$5 \\
17 & -1.0$\pm$0.4 & \ldots & -0.6$\pm$0.1 & \ldots & -0.4$\pm$0.2 & \ldots & 1.83$\pm$0.02 & 4.1$\pm$0.6 & 3.1$\pm$0.4 \\
18 & \ldots & \ldots & \ldots & -0.8$\pm$0.3 & -0.6$\pm$0.2 & -0.2$\pm$0.5 & 1.73$\pm$0.03 & 150$\pm$20 & 120$\pm$10 \\
19 & \ldots & \ldots & \ldots & -0.8$\pm$0.4 & -0.6$\pm$0.2 & -0.3$\pm$0.6 & 0.92$\pm$0.08 & 100$\pm$10 & 90$\pm$10 \\
20 & -2.3$\pm$0.5 & \ldots & -0.8$\pm$0.2 & \ldots & 0.1$\pm$0.2 & \ldots & 2.80$\pm$0.01 & 1.7$\pm$0.2 & 1.0$\pm$0.1 \\
21 & -1.3$\pm$0.4 & \ldots & -0.9$\pm$0.1 & \ldots & -0.7$\pm$0.2 & \ldots & 2.07$\pm$0.01 & 2.3$\pm$0.3 & 1.7$\pm$0.2 \\
22 & \ldots & \ldots & \ldots & -0.6$\pm$0.3 & -0.7$\pm$0.2 & -0.7$\pm$0.5 & 0.80$\pm$0.09 & 4900$\pm$500 & 5000$\pm$1000 \\
23 & -0.8$\pm$0.3 & -0.9$\pm$0.2 & -0.8$\pm$0.1 & -1.0$\pm$0.3 & -0.8$\pm$0.2 & -0.5$\pm$0.5 & 1.8$\pm$0.1 & 440$\pm$50 & 400$\pm$40 \\
24 & 0.1$\pm$0.3 & -0.1$\pm$0.3 & 0.1$\pm$0.1 & -0.2$\pm$0.5 & 0.1$\pm$0.2 & 0.6$\pm$0.8 & 2.06$\pm$0.06 & 90$\pm$100 & 63$\pm$7 \\
25 & 1.7$\pm$0.3 & \ldots & 1.3$\pm$0.1 & \ldots & 1.1$\pm$0.2 & \ldots & 1.54$\pm$0.02 & 19$\pm$2 & 15$\pm$2 \\
26 & \ldots & \ldots & \ldots & -0.7$\pm$0.3 & -0.8$\pm$0.2 & -0.9$\pm$0.5 & 1.37$\pm$0.04 & 180$\pm$20 & 160$\pm$20 \\
27 & -0.5$\pm$0.3 & -0.7$\pm$0.2 & -0.6$\pm$0.1 & -0.9$\pm$0.4 & -0.7$\pm$0.2 & -0.3$\pm$0.6 & 1.66$\pm$0.07 & 120$\pm$10 & 100$\pm$10 \\
28 & \ldots & \ldots & \ldots & \ldots & \ldots & \ldots & 2.0$\pm$0.1 & 80$\pm$20 & 60$\pm$20 \\
29 & -0.8$\pm$0.3 & -0.8$\pm$0.2 & -0.8$\pm$0.1 & -0.9$\pm$0.3 & -0.8$\pm$0.2 & -0.6$\pm$0.5 & 1.5$\pm$0.1 & 580$\pm$70 & 450$\pm$50 \\
30 & -1.2$\pm$0.4 & \ldots & -0.7$\pm$0.1 & \ldots & -0.5$\pm$0.2 & \ldots & 1.39$\pm$0.07 & 58$\pm$7 & 55$\pm$6 \\
31 & -0.9$\pm$0.3 & -0.7$\pm$0.2 & \ldots & -0.5$\pm$0.3 & \ldots & \ldots & 1.15$\pm$0.04 & 300$\pm$30 & 250$\pm$40 \\
32 & -0.9$\pm$0.3 & -0.9$\pm$0.2 & -0.9$\pm$0.1 & -0.8$\pm$0.3 & -0.8$\pm$0.2 & -0.8$\pm$0.5 & 1.07$\pm$0.03 & 1700$\pm$200 & 1400$\pm$200 \\
33 & -0.9$\pm$0.4 & \ldots & -0.8$\pm$0.1 & \ldots & -0.7$\pm$0.2 & \ldots & 6.9$\pm$1.0 & 25$\pm$3 & 9$\pm$10 \\
34 & \ldots & \ldots & \ldots & -1.8$\pm$0.7 & \ldots & \ldots & 1.71$\pm$0.04 & 33$\pm$4 & 31$\pm$3 \\
35 & \ldots & \ldots & \ldots & -2.0$\pm$0.8 & \ldots & \ldots & 1.5$\pm$0.2 & 100$\pm$10 & 80$\pm$10 \\
36 & \ldots & \ldots & \ldots & \ldots & \ldots & \ldots & 2.3$\pm$0.3 & 210$\pm$40 & 150$\pm$90 \\
\hline
\end{tabular}
\caption{The two-point radio spectral indices and flux density ratios relevant for the selection criteria adopted in this work. The super/subscript coding as in Table\,\ref{tab:phot}. Note that the 1.4\,GHz photometry used for the IFRS was that of M08 which covers a larger area than \citet{hales14} and \citet{franzen15}. When a value is not reported, that means that either there is no mapping coverage or there is no catalogue entry}
\label{tab:ratios}
\end{table*}
%\end{table}
%\end{landscape}

%% file: sections/acaobs.tex
The sample\footnote{The initially observed sample comprised 42 sources, six of which were later on found not to comply with the criteria in Tab.~\ref{tab:selectioncriteria} after opting for the use of the M08 and SEIP DR4 fluxes, as described in Secs. \ref{sec:radphot} and \ref{sec:selectioncriteria}.} These extra six sources are presented in App.~\ref{app:sampdrop}. was observed under the ACA OFP 2018.A.00046.S (PI Messias). Each source was observed for 35--40~minutes and targeted with a single pointing, which provides a field of view (FoV) of $\sim$41\,arcsec (as reported by the ALMA Observing Tool). This value translates to $\sim$250\,kpc at $z=0.5$, or $\sim$315--350\,kpc in the range $z=1-3$. The reference frequency was set to 233\,GHz, with four 1.875\,GHz wide spectral windows centred at 242, 240, 226, and 224\,GHz, with a 15.625\,MHz channel width (corresponding to $\sim$20\,km/s). The data was calibrated with the Cycle~6 ALMA Science Pipeline (version 5.4.0-70, r42254). The requested sensitivity of 0.2\,mJy was set to match that of $\sim$1.2\,mJy at 850\,$\mu$m provided by SCUBA2 maps \citep{geach17} assuming a thermal-like spectral slope of $\alpha=3.9$ (with $S_\nu\propto\nu^\alpha$). The achieved sensitivity reached $\sim$0.15\,mJy by imaging with a Briggs weighting and a robust parameter of 1.5 (a robust parameter of 0.5 would have resulted in an {\sc rms}$\sim$0.23\,mJy). The weighting scheme resulted in a synthesised beam of $7.5\arcsec \times 4.7\arcsec$ (sampled with a 1\arcsec pixel size). The baselines for ACA range from 9 to 47\,m (implying a maximum recoverable large angular scale of $\sim$30\arcsec). The imaging was done using the {\sc tclean} task of {\sc casa} \citep{mcmullin07}.

\begin{figure*}
    \centering
	\includegraphics[width=0.24\textwidth]{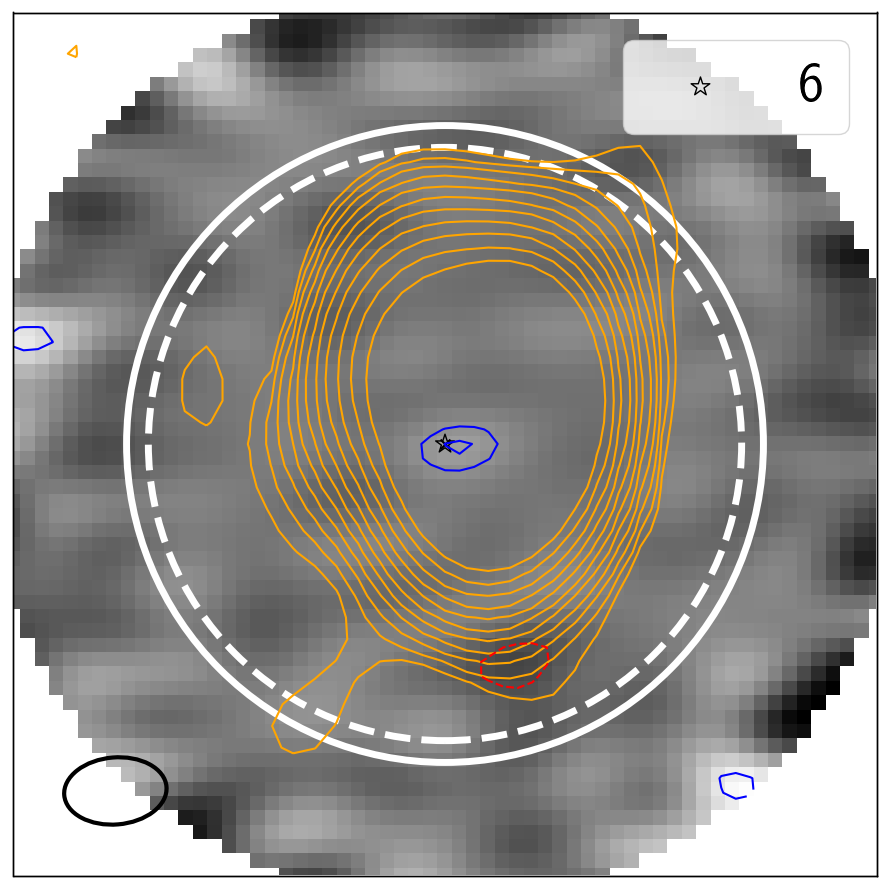}
	\includegraphics[width=0.24\textwidth]{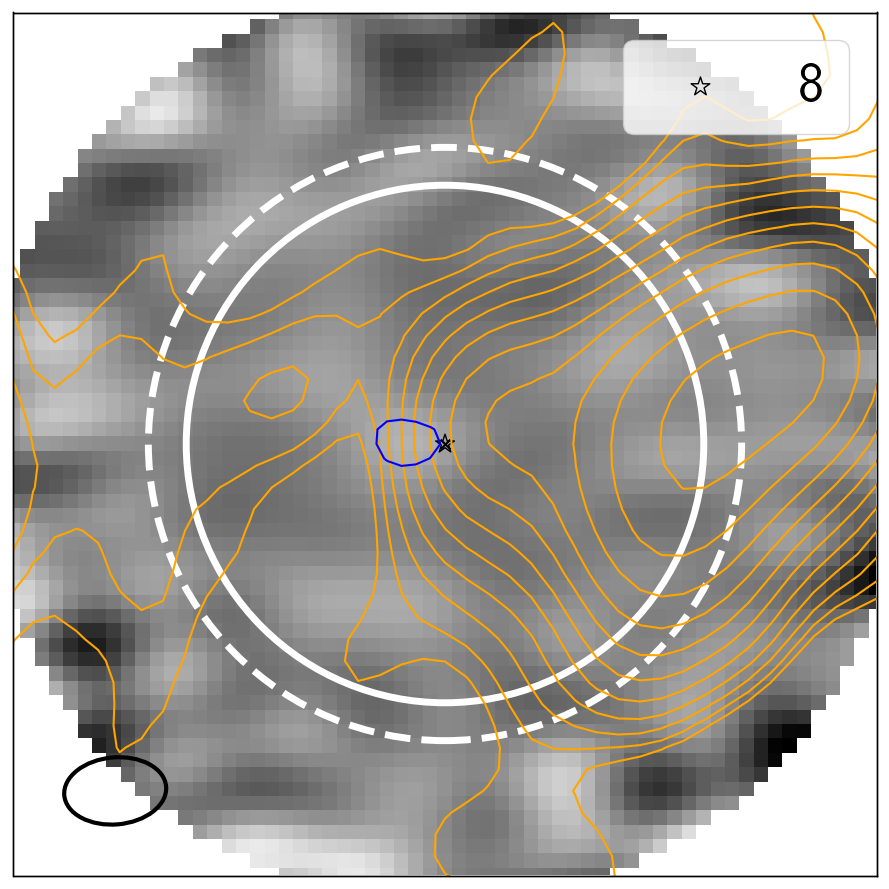}
	\includegraphics[width=0.24\textwidth]{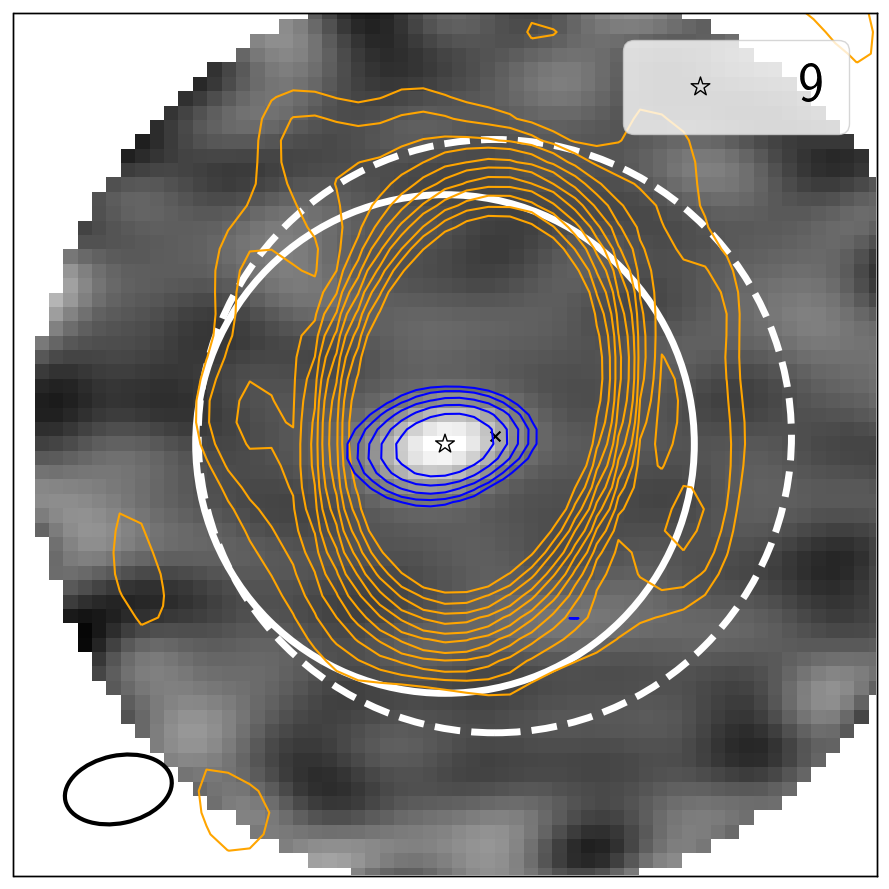}
	\includegraphics[width=0.24\textwidth]{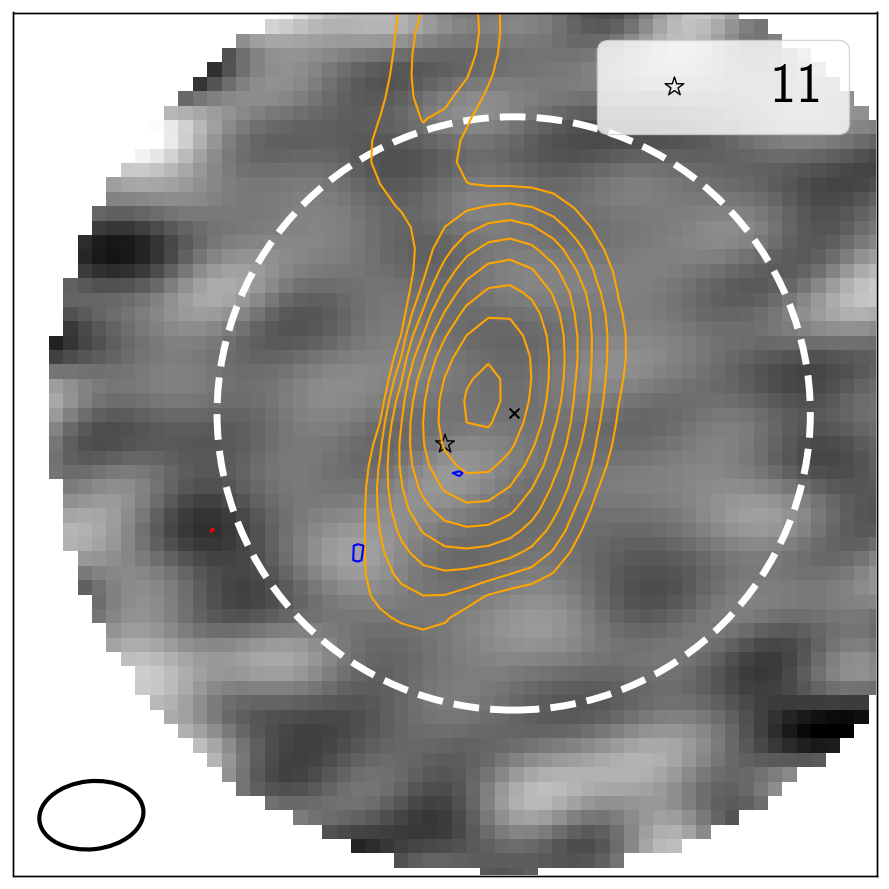}\\
	\includegraphics[width=0.24\textwidth]{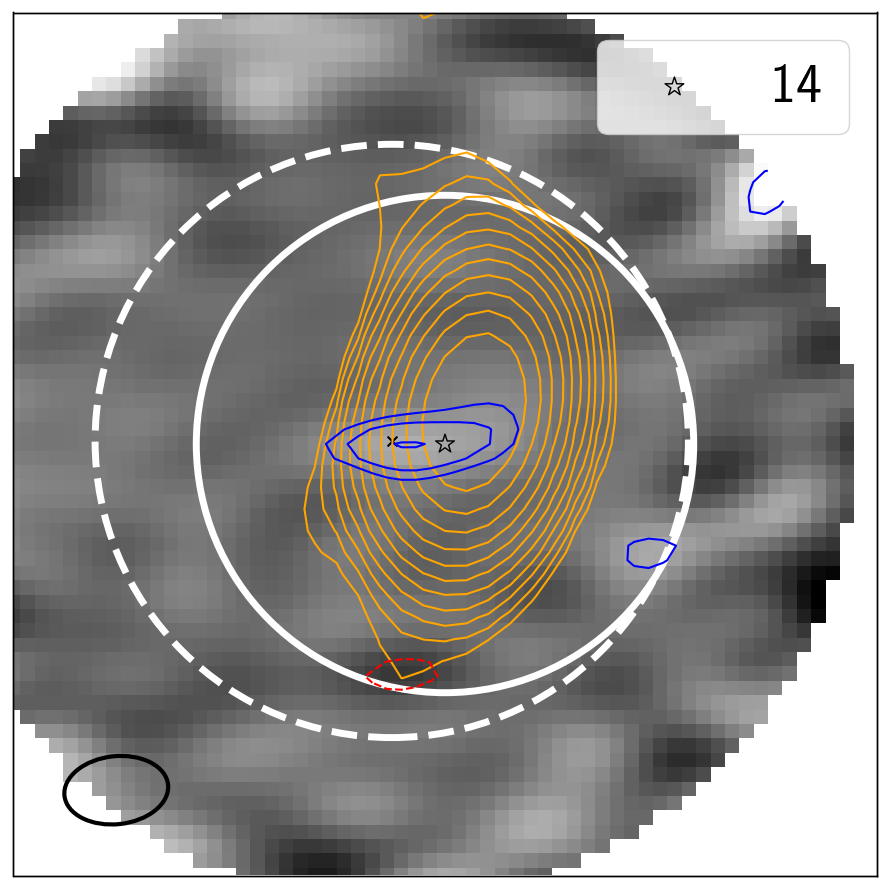}
	\includegraphics[width=0.24\textwidth]{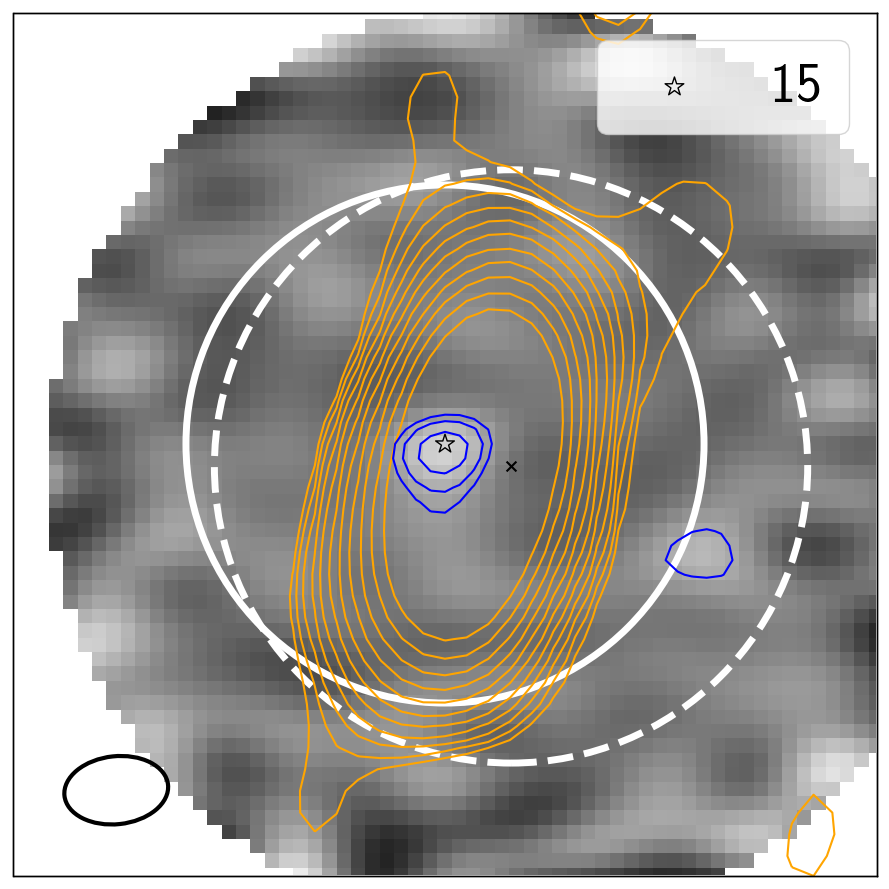}
	\includegraphics[width=0.24\textwidth]{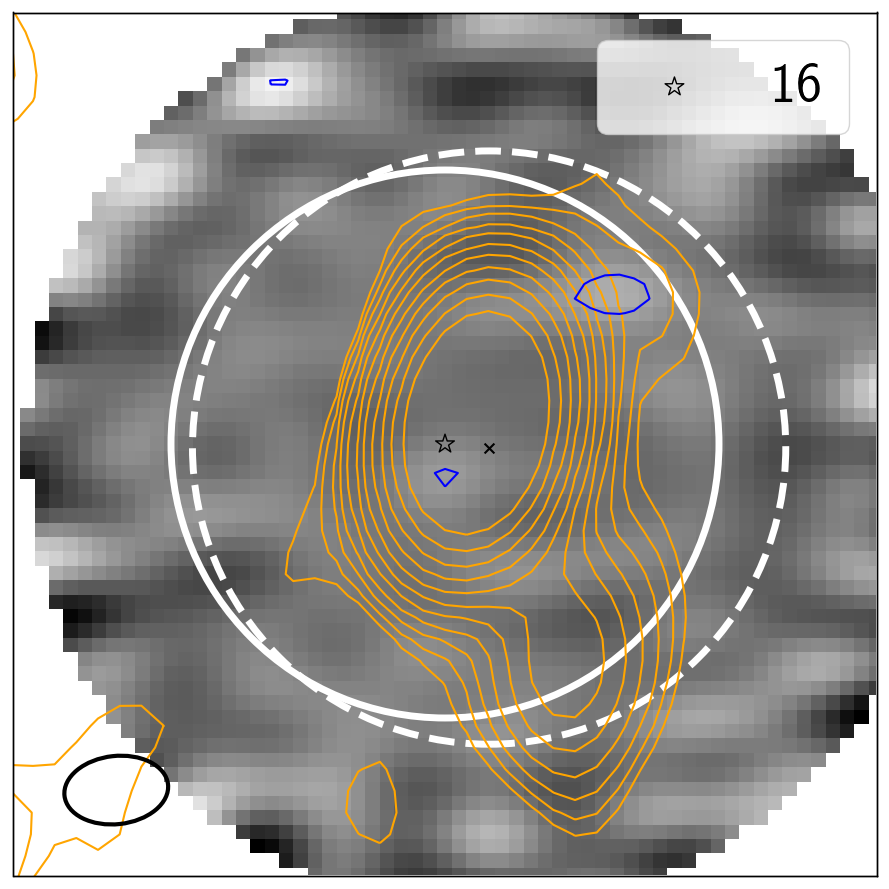}
	\includegraphics[width=0.24\textwidth]{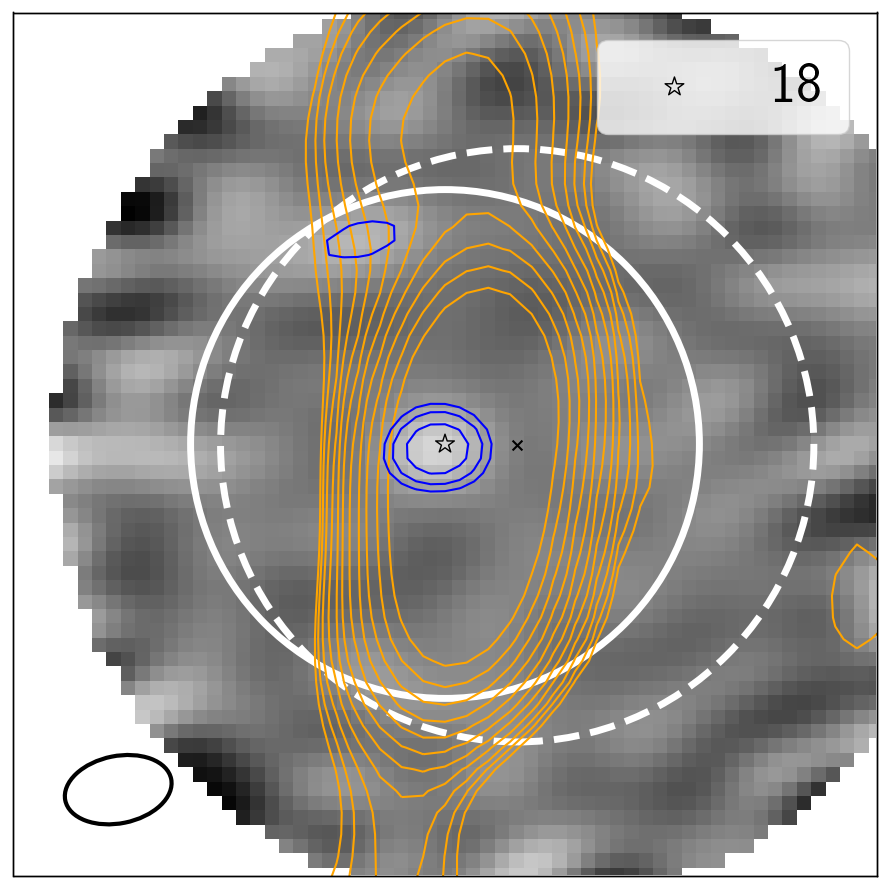}\\
	\includegraphics[width=0.24\textwidth]{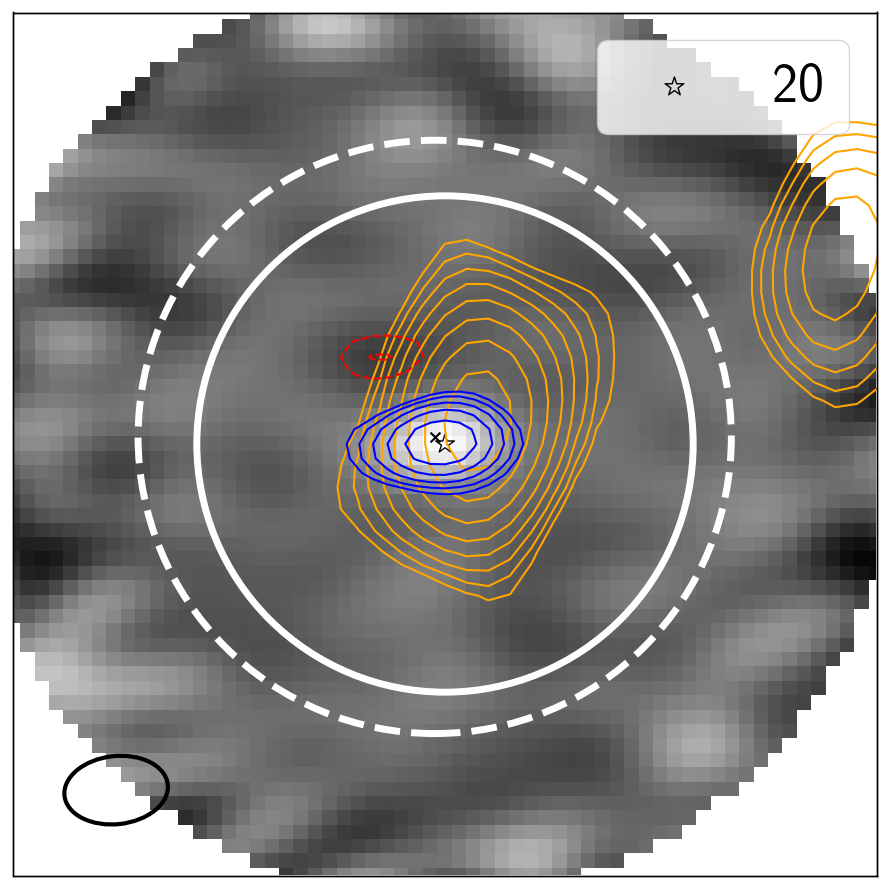}
	\includegraphics[width=0.24\textwidth]{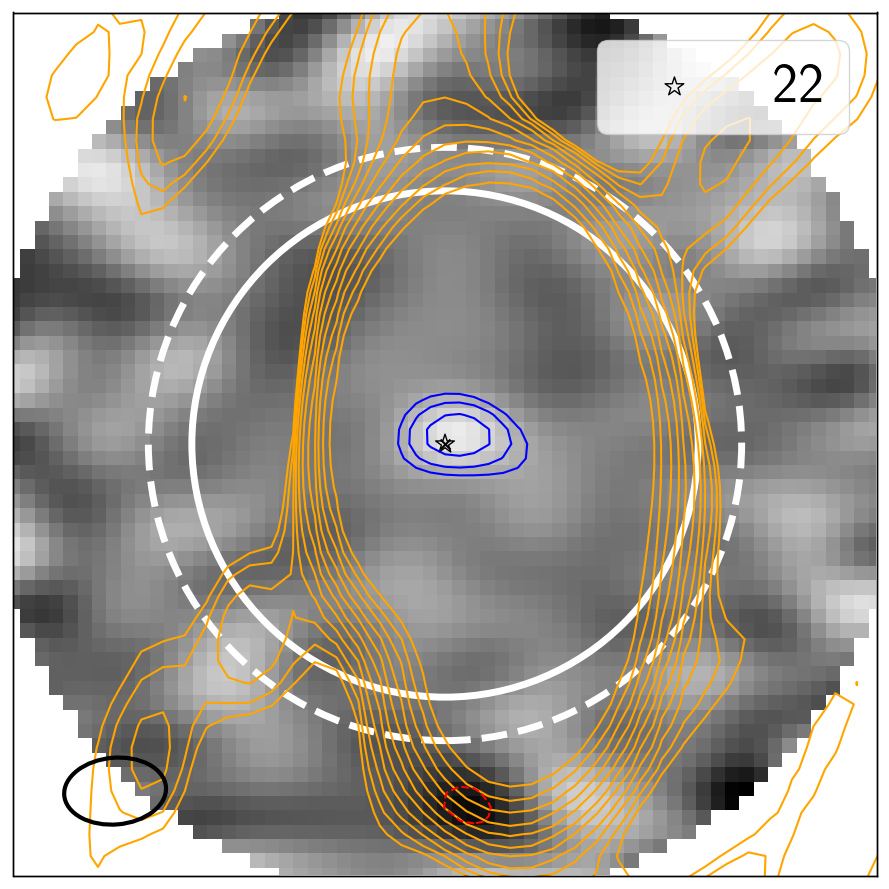}
	\includegraphics[width=0.24\textwidth]{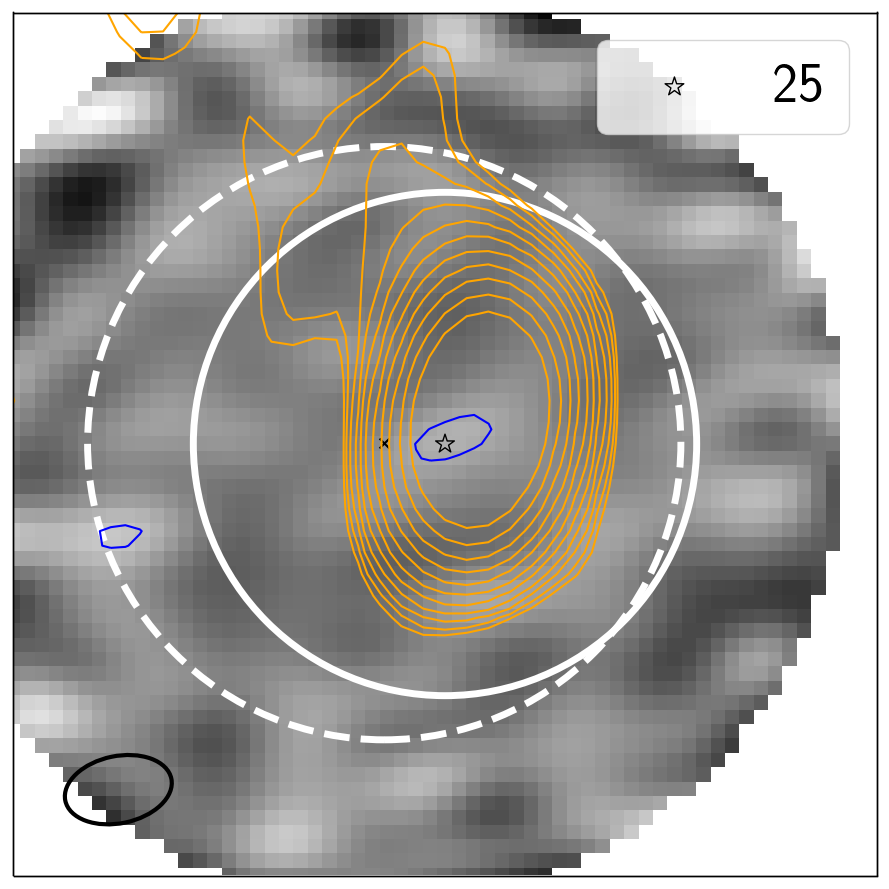}
	\includegraphics[width=0.24\textwidth]{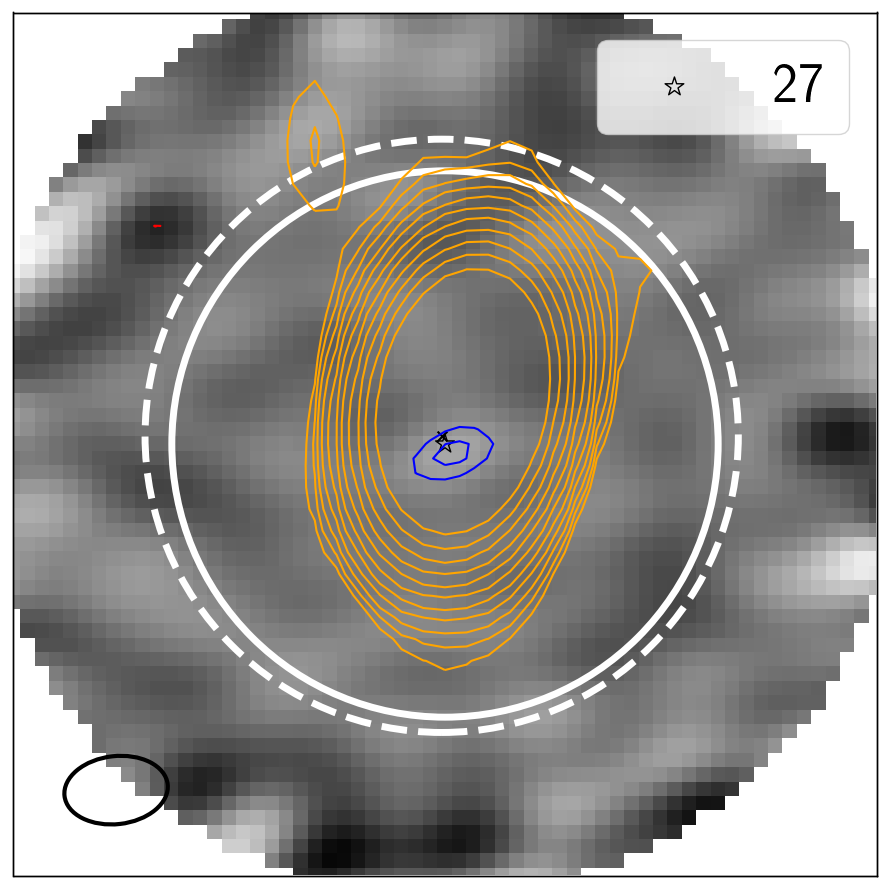}\\
	\includegraphics[width=0.24\textwidth]{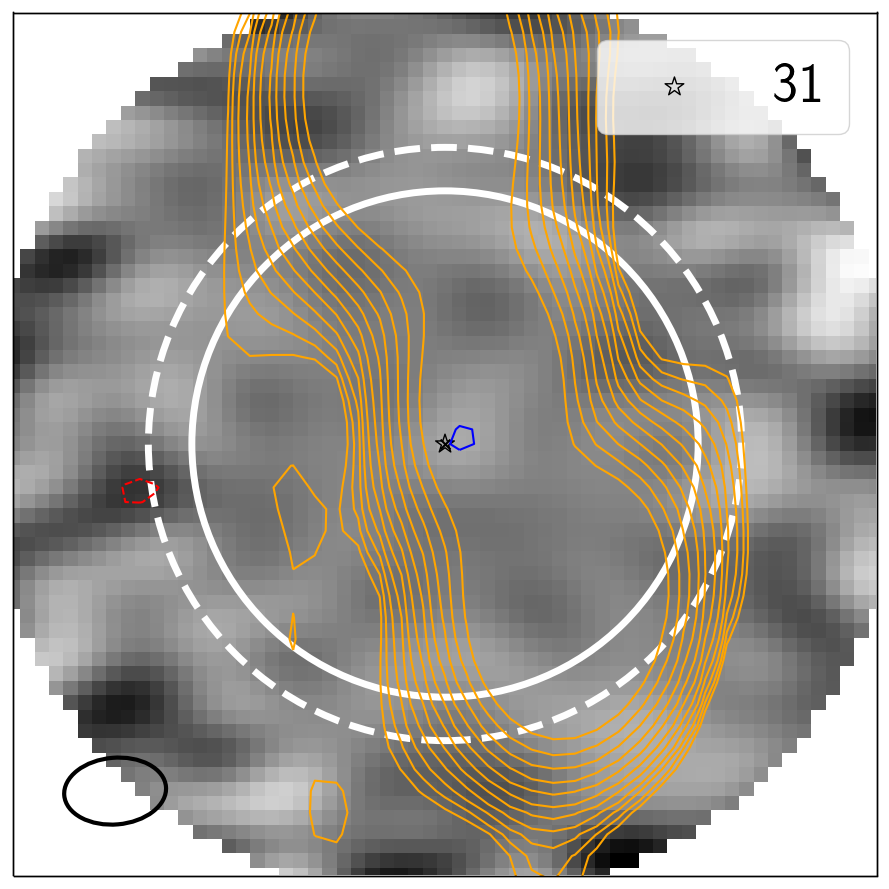}
	\includegraphics[width=0.24\textwidth]{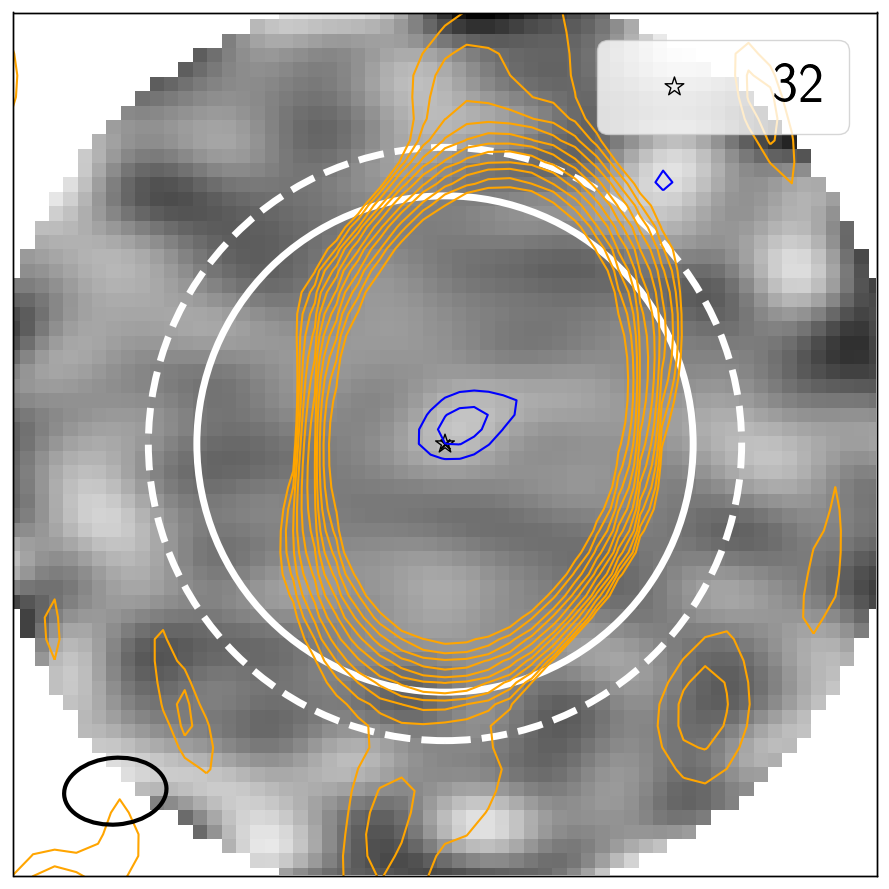}
	\includegraphics[width=0.24\textwidth]{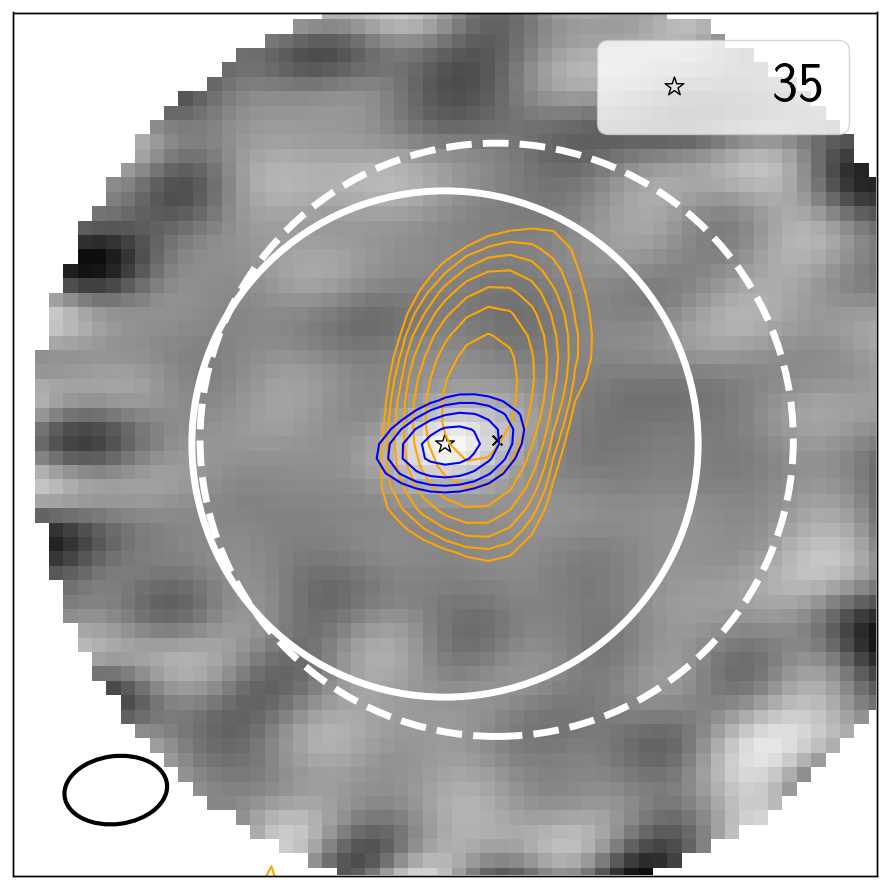}
	\includegraphics[width=0.24\textwidth]{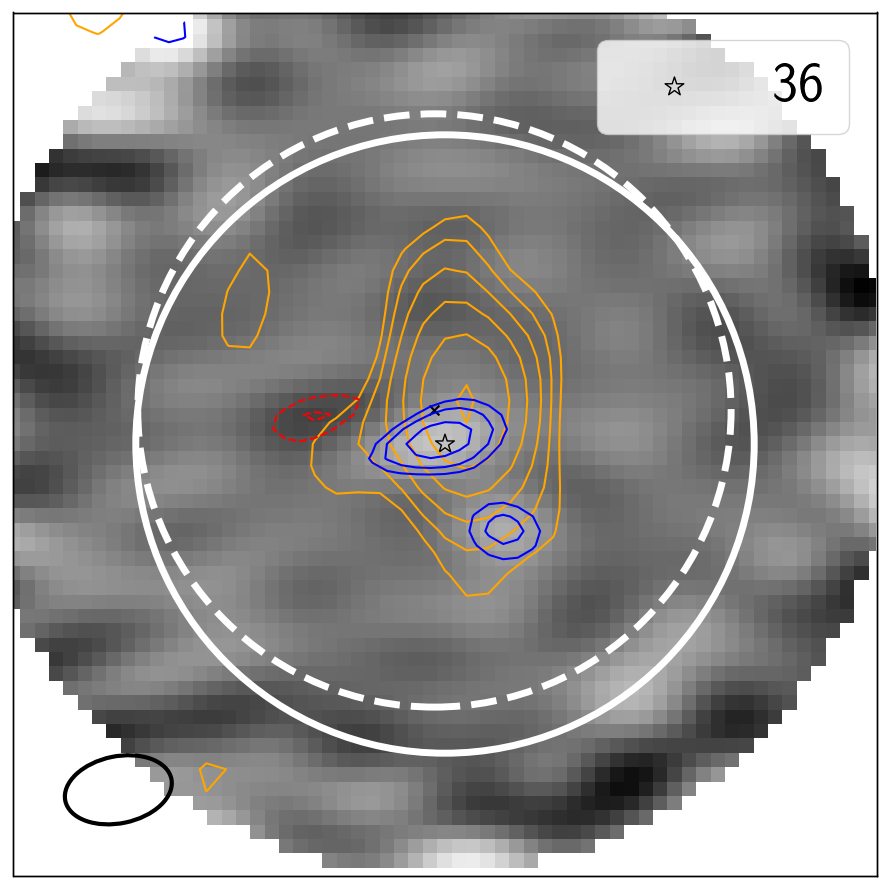}
	\caption{Primary-beam-corrected ACA 1.3\,mm maps (grey linear scale) for sources where a central emission has been detected at $>4.2\sigma$ or at $>3\sigma$ and is closer than 3.5\arcsec\ from the IR counterpart (Sec.~\ref{sec:b6flx}). Each panel is 1\arcmin\ wide with a North-up~/~East-right orientation. Emission at $\sigma=3$, 4.2, 6, 8.5, 12 is shown with blue contours, while that at $\sigma=-3$ and -4.2 is highlighted with red contours. The ACA FOV is overlaid as a dashed circumference centred in the cross, while the solid circumference --- centred at the coordinates of the IR counterpart (star) --- marks the 150\,kpc radius at each source's redshift (see Appendix\,\ref{app:offset} on why they are offset from each other). The 1.4\,GHz emission is overlaid as orange contours (levels at 3 to 32$\sigma$ with respect to the local median absolute deviation with multiplying steps of $\sqrt{2}$). The source identification number is shown at the upper right corner in each panel. The ACA synthesised beam of $7.5\arcsec \times 4.7\arcsec$ is shown at the lower left corner. The maps of the remainder undetected sources are displayed in Figure~\ref{afig:aca3}.}\label{fig:aca1}
\end{figure*}

Figure~\ref{fig:aca1} shows the primary-beam-corrected ACA maps for HzRG candidates with an ACA detection corresponding to the central source (throughout the manuscript we will refer to the radio-source IR counterpart as ``central source''). Positive significant emission is shown with blue contours (at $\sigma$ levels of 3, 4.2, 6, 8.5, 12), while negative emission is highlighted with red contours ($\sigma=-3$ and $-4.2$). The ACA FoV is overlaid as a dashed circle, while the solid circle marks the 150\,kpc radius at each source's redshift (for an explanation of the offsets see Appendix\,\ref{app:offset}). Figure~\ref{fig:aca2} shows sources where, although the central source was not detected, a significant detection was found in the field.

% central detections

% neighbour detections
\begin{figure*}
	\includegraphics[width=0.24\textwidth]{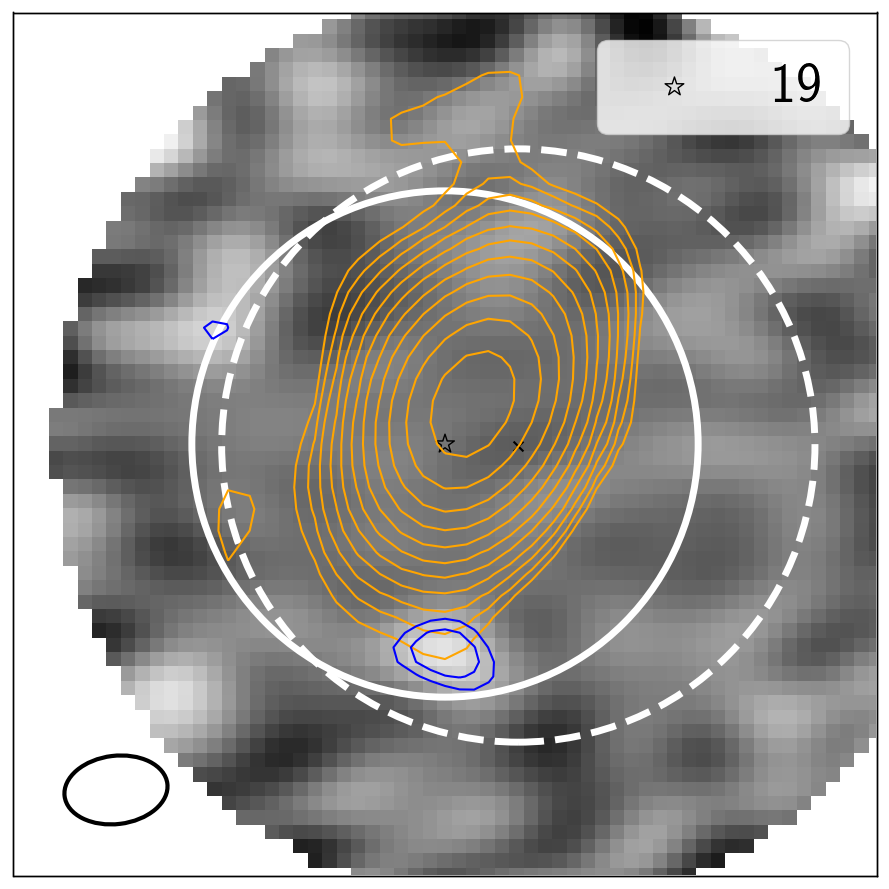}
	\includegraphics[width=0.24\textwidth]{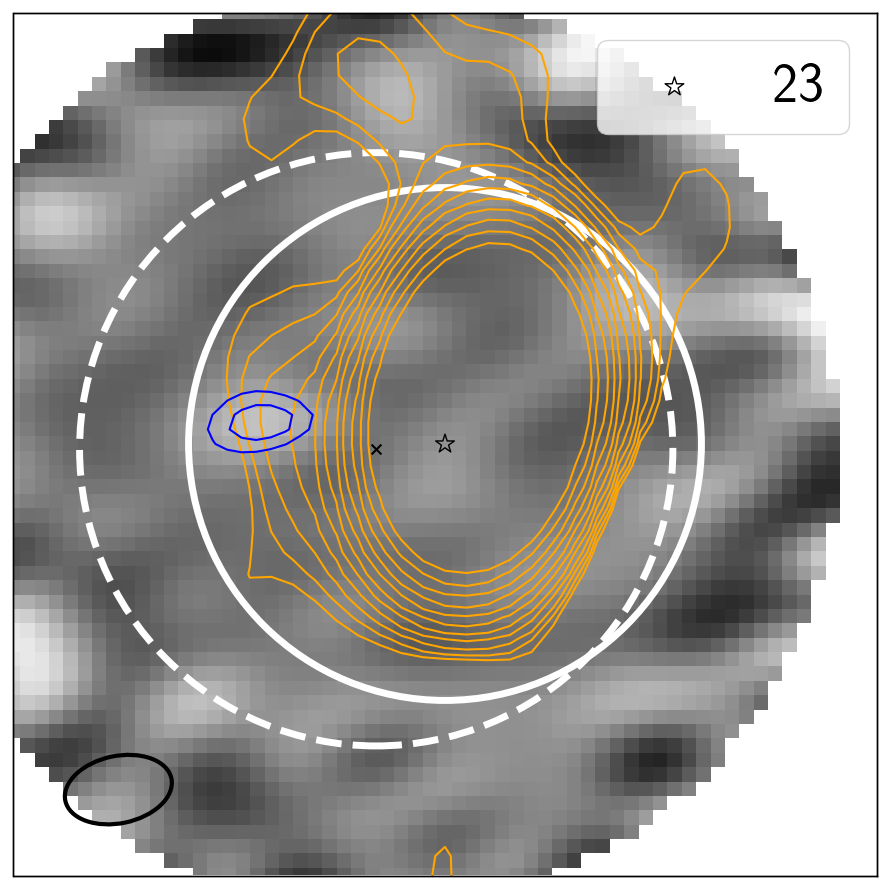}
	\includegraphics[width=0.24\textwidth]{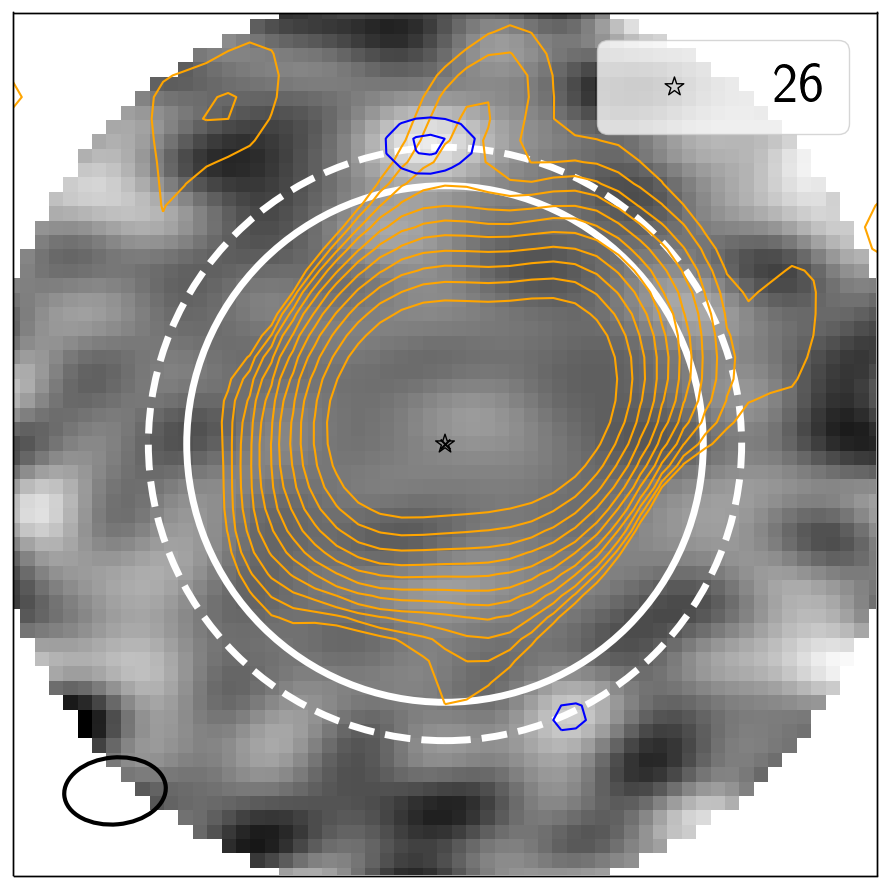}
	\caption{The same as in Figure~\ref{fig:aca1}, but for sources where a neighbouring emission has been detected at $>4.2\sigma$.}\label{fig:aca2}
\end{figure*}

% IR imaging
 \begin{figure*}
\includegraphics[width=0.21\textwidth]{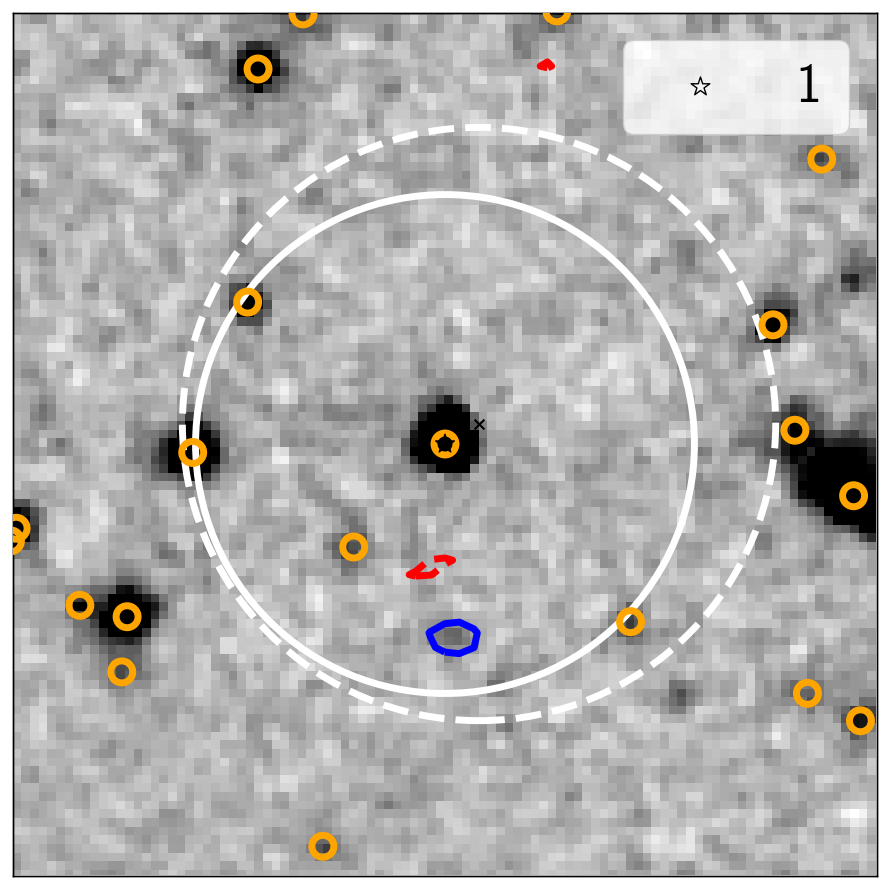}
\includegraphics[width=0.21\textwidth]{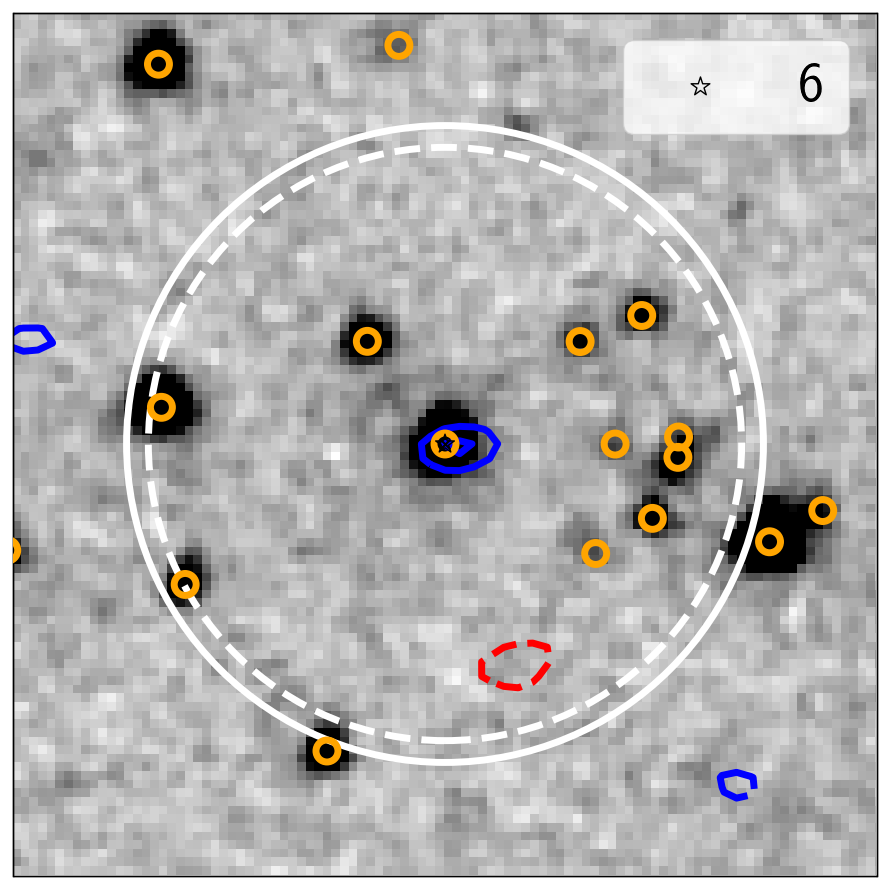}
 \includegraphics[width=0.21\textwidth]{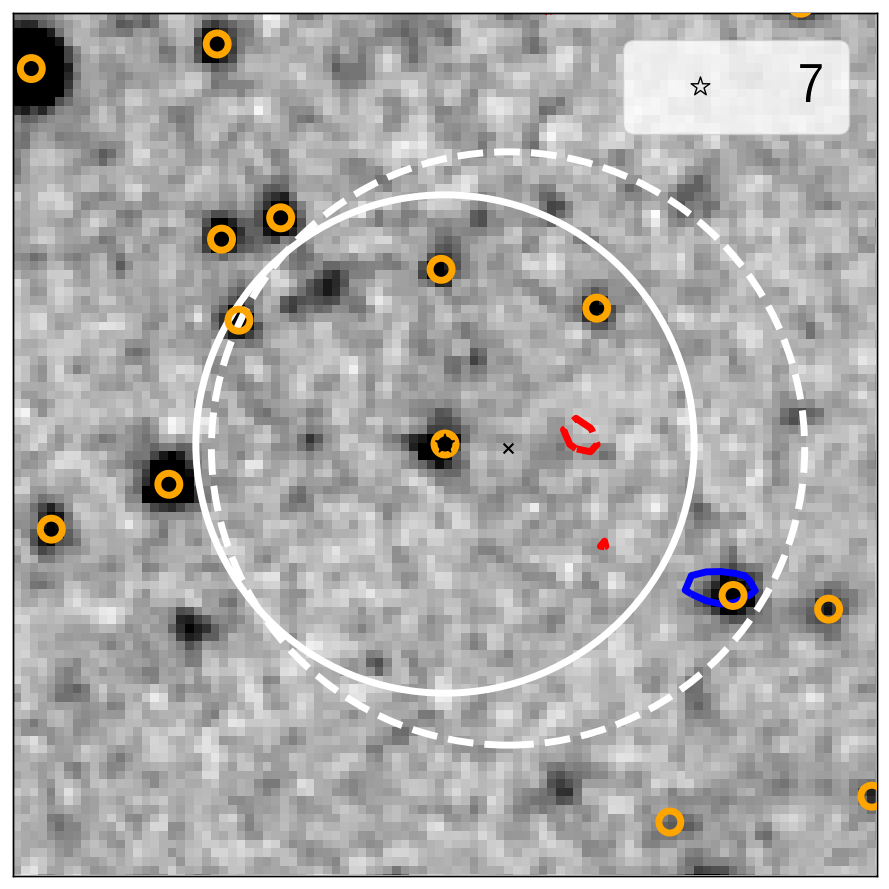}
\includegraphics[width=0.21\textwidth]{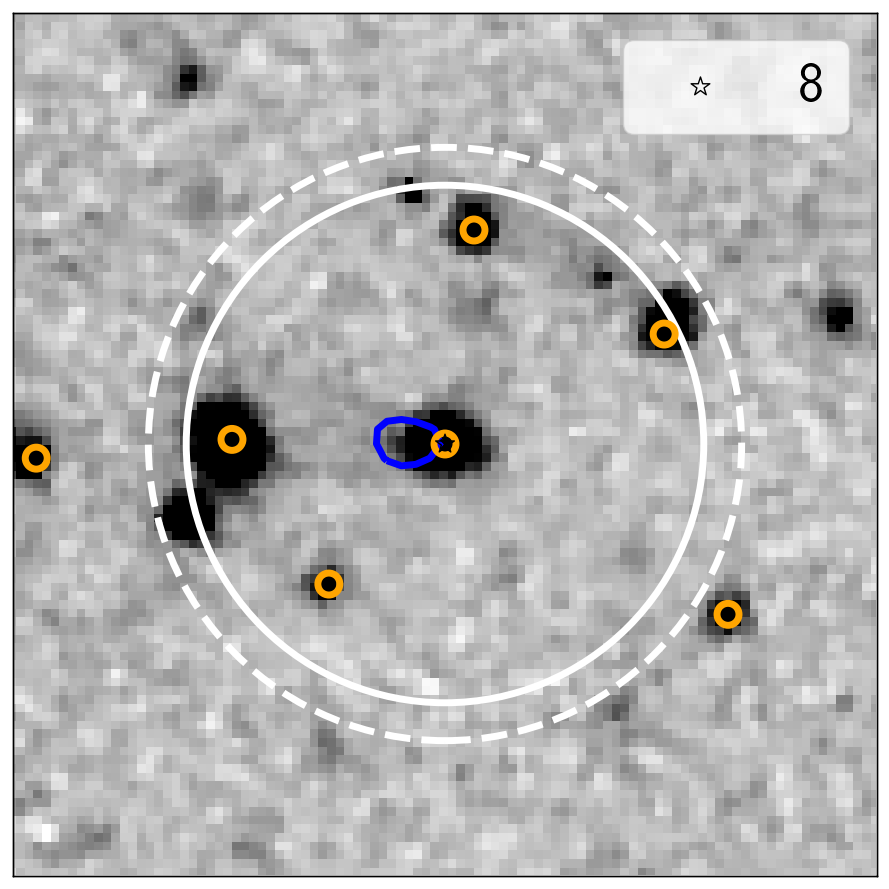}
\includegraphics[width=0.21\textwidth]{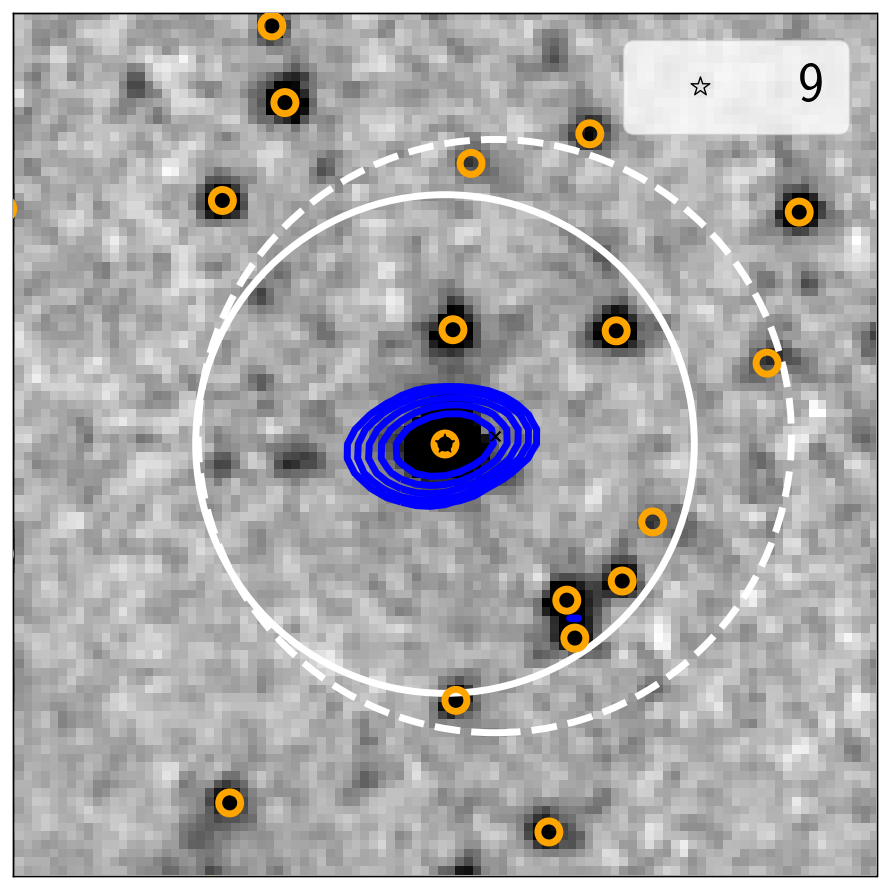}
\includegraphics[width=0.21\textwidth]{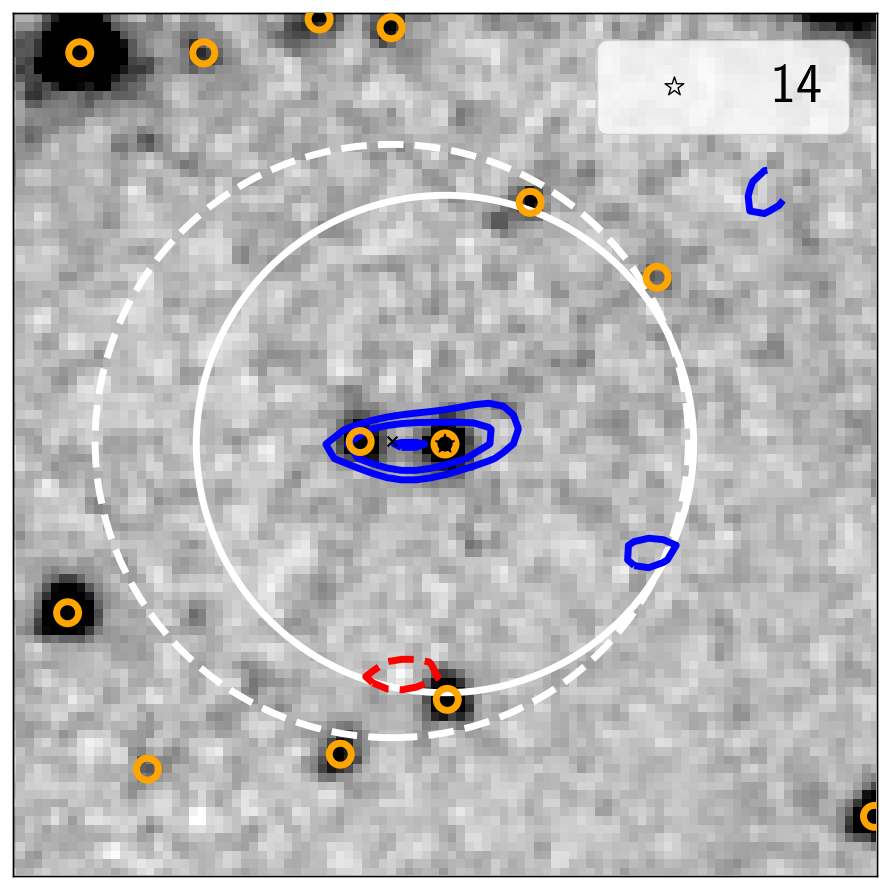}
\includegraphics[width=0.21\textwidth]{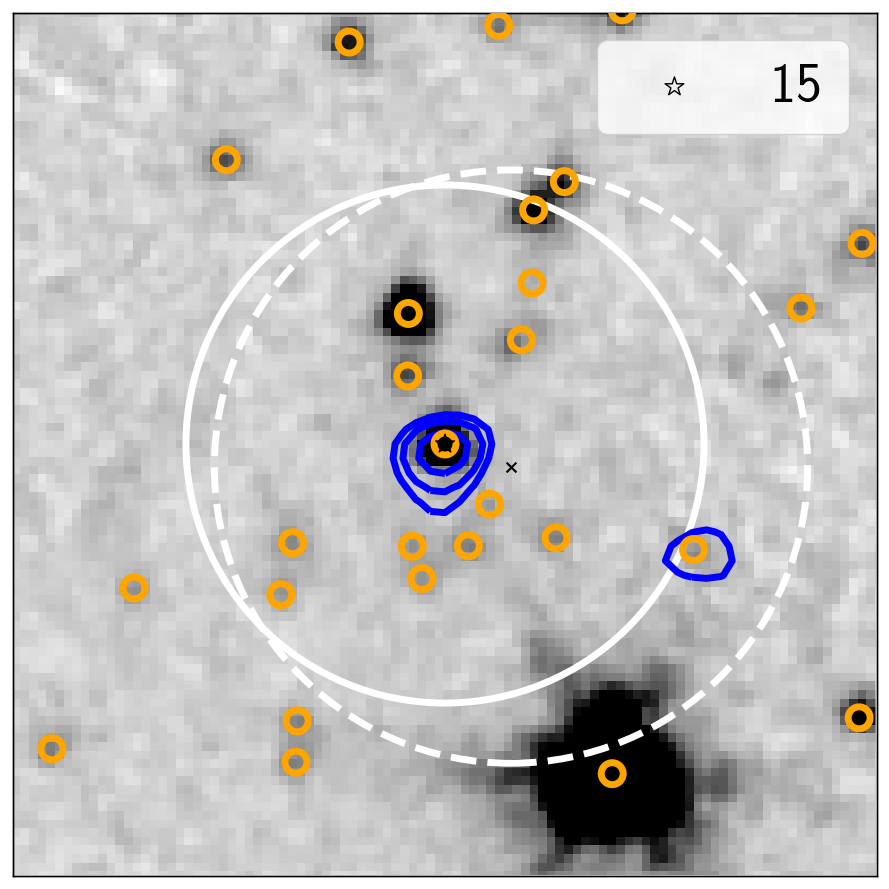}
\includegraphics[width=0.21\textwidth]{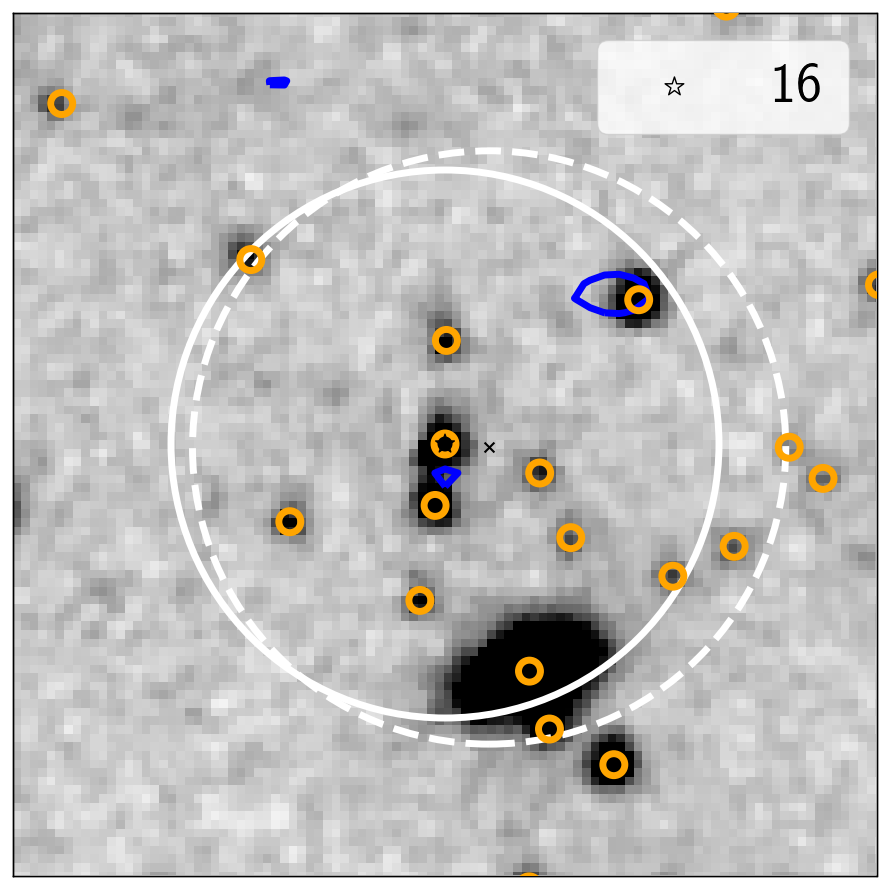}
\includegraphics[width=0.21\textwidth]{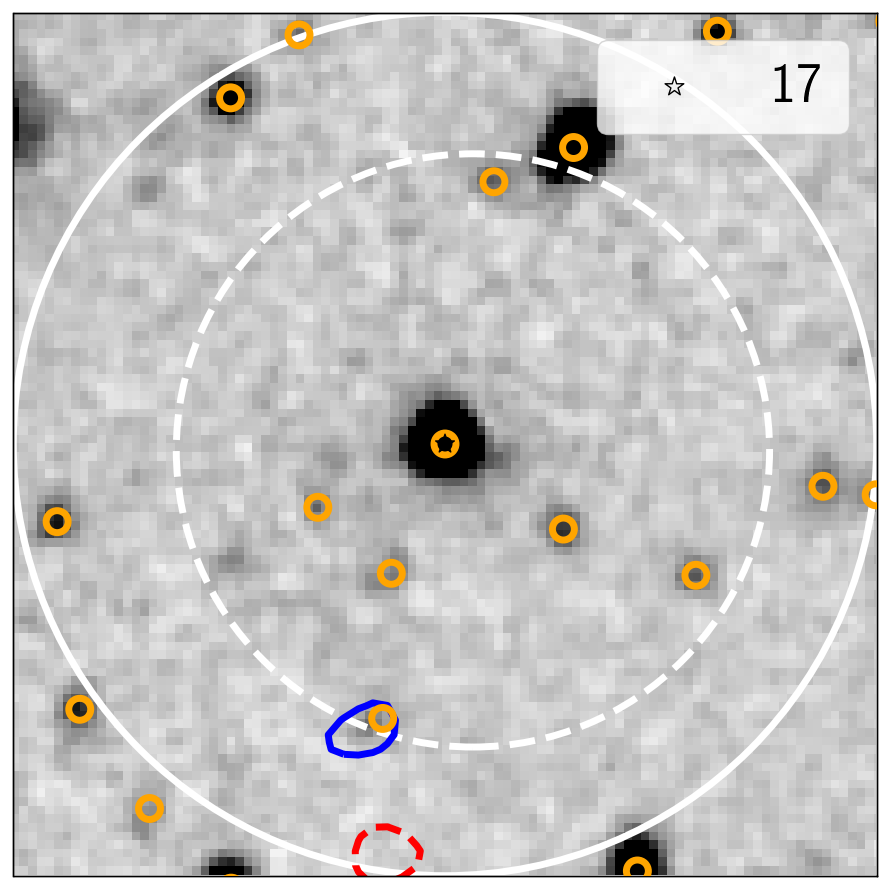}
\includegraphics[width=0.21\textwidth]{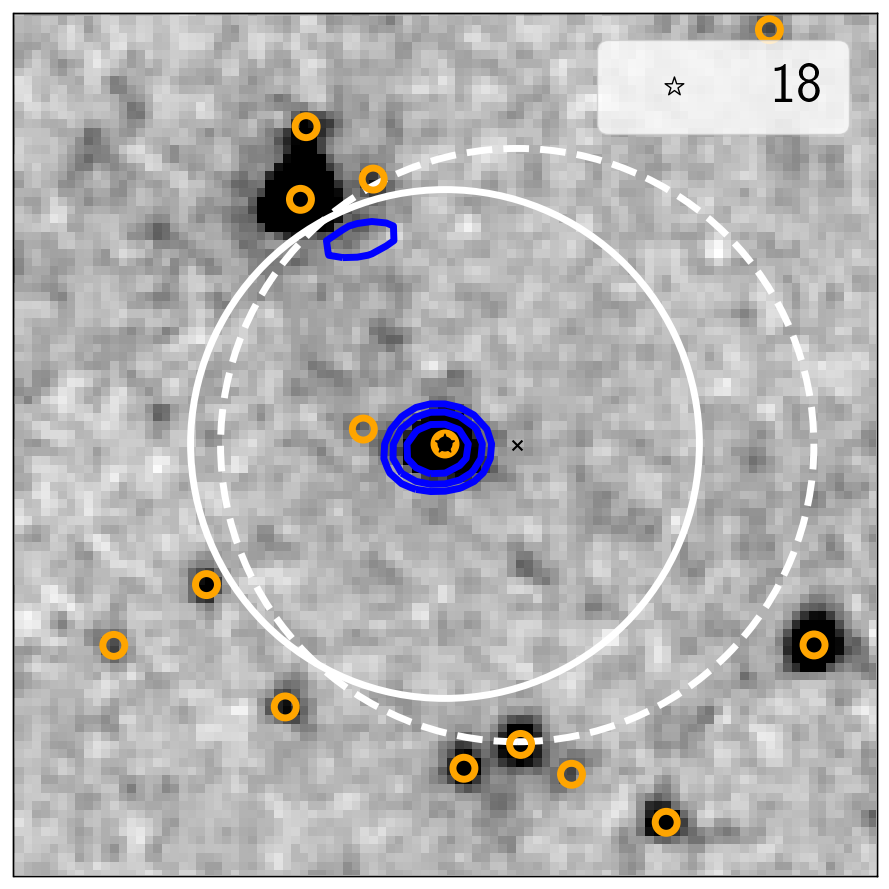}
\includegraphics[width=0.21\textwidth]{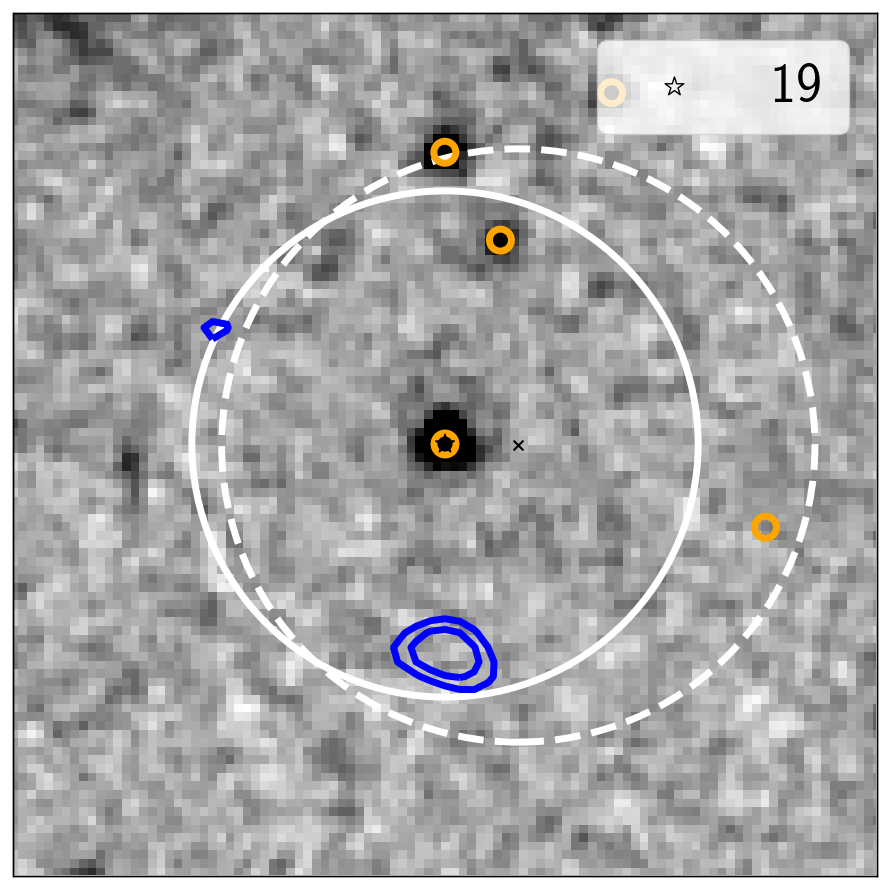}
\includegraphics[width=0.21\textwidth]{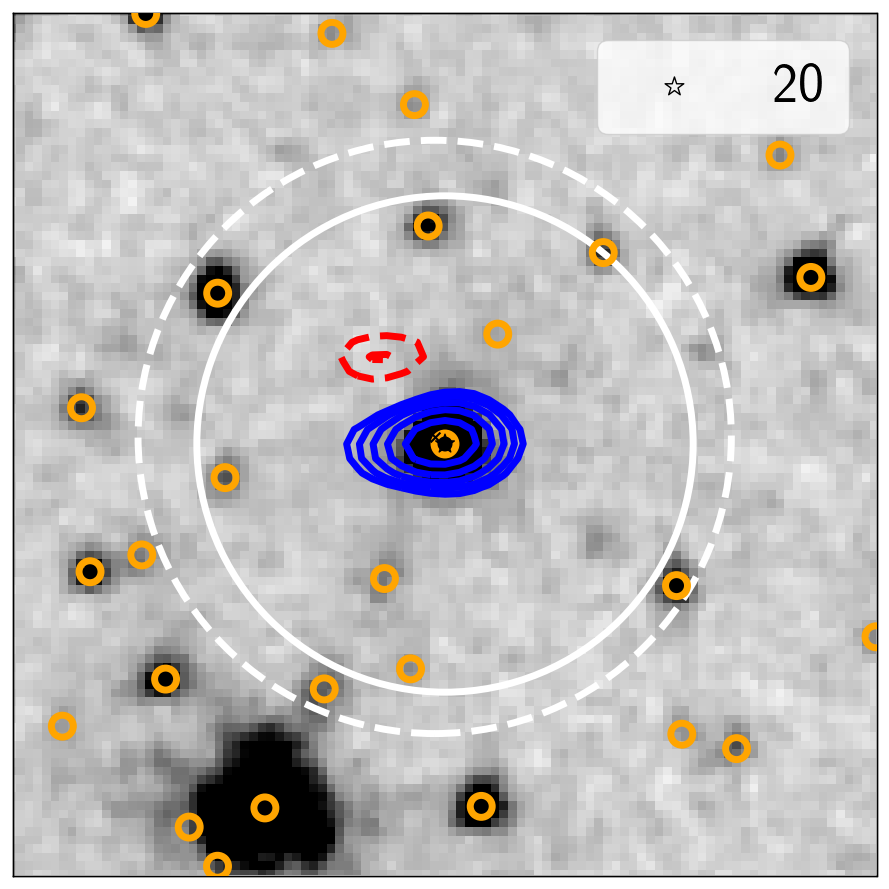}
\includegraphics[width=0.21\textwidth]{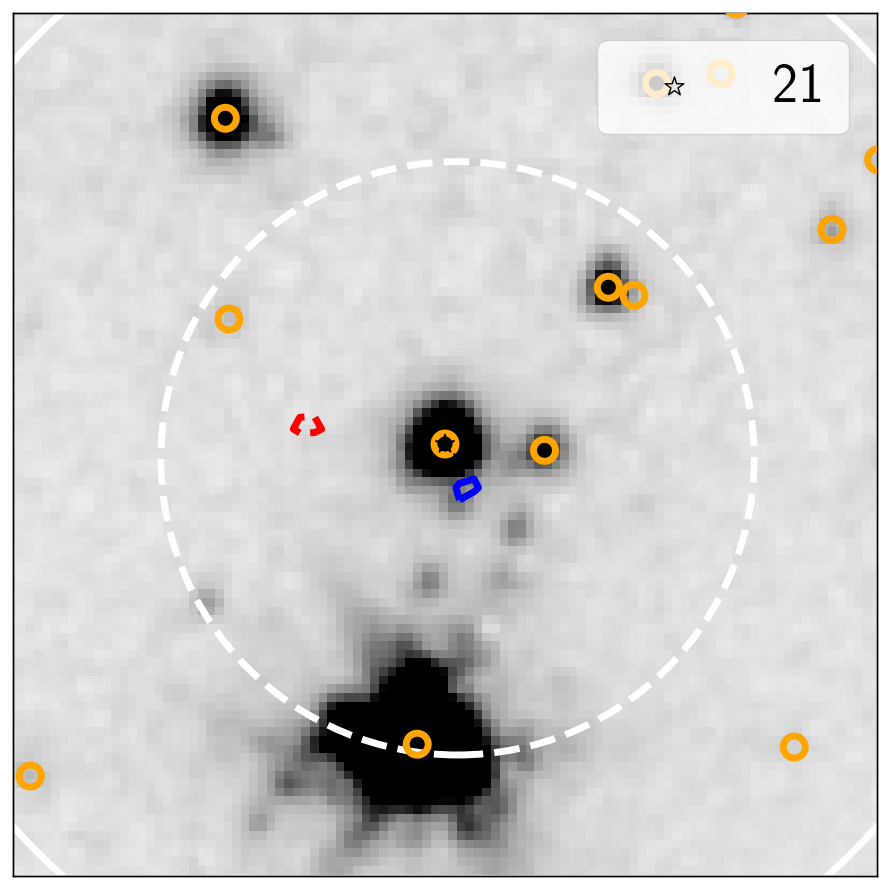}
\includegraphics[width=0.21\textwidth]{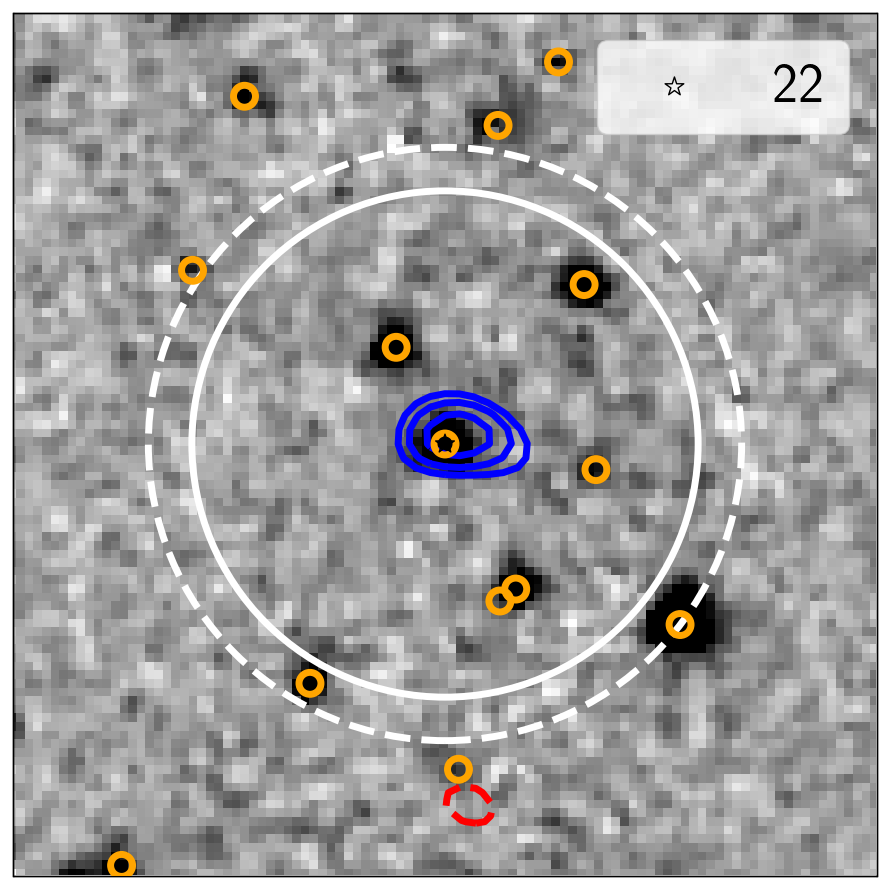}
\includegraphics[width=0.21\textwidth]{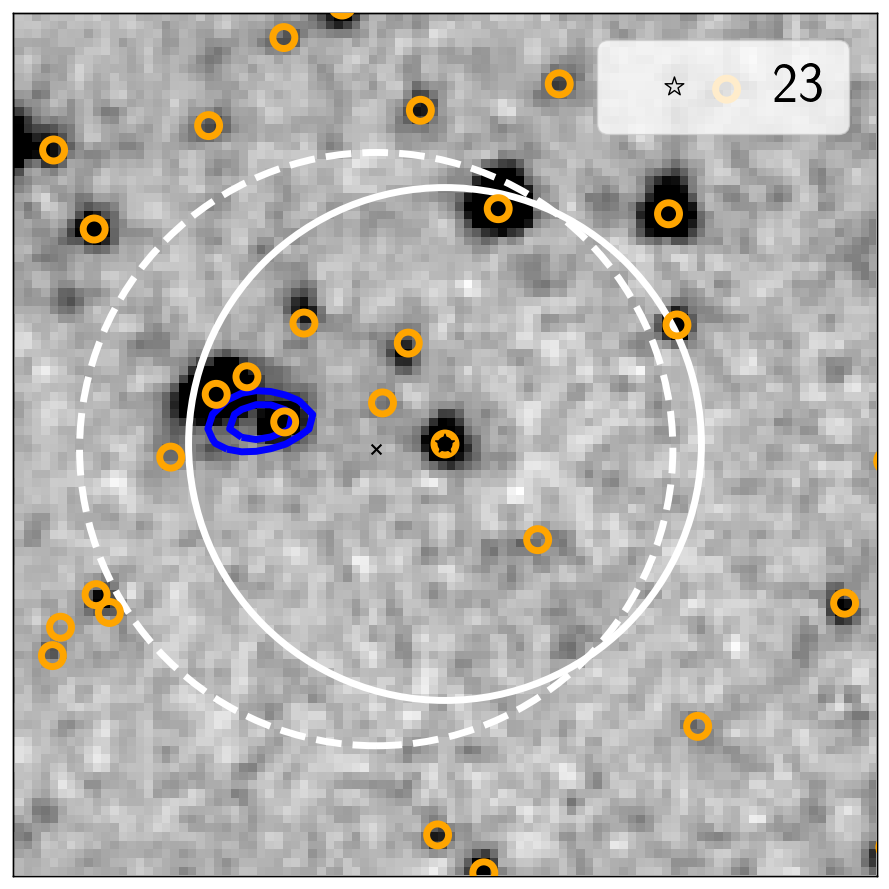}
\includegraphics[width=0.21\textwidth]{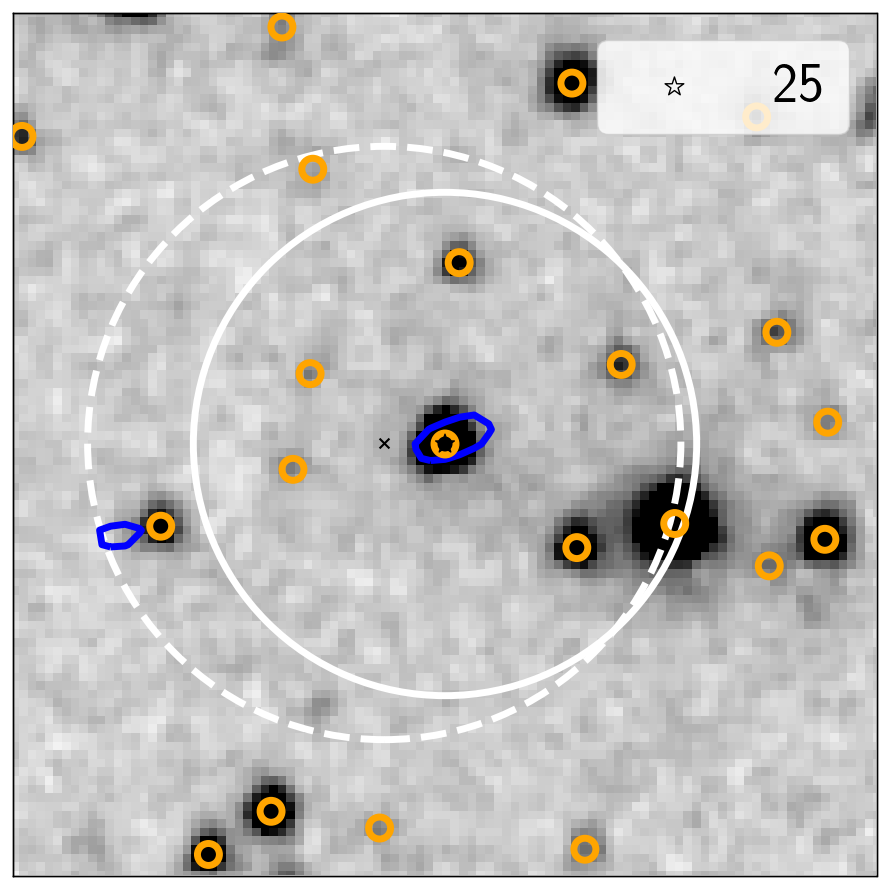}
\includegraphics[width=0.21\textwidth]{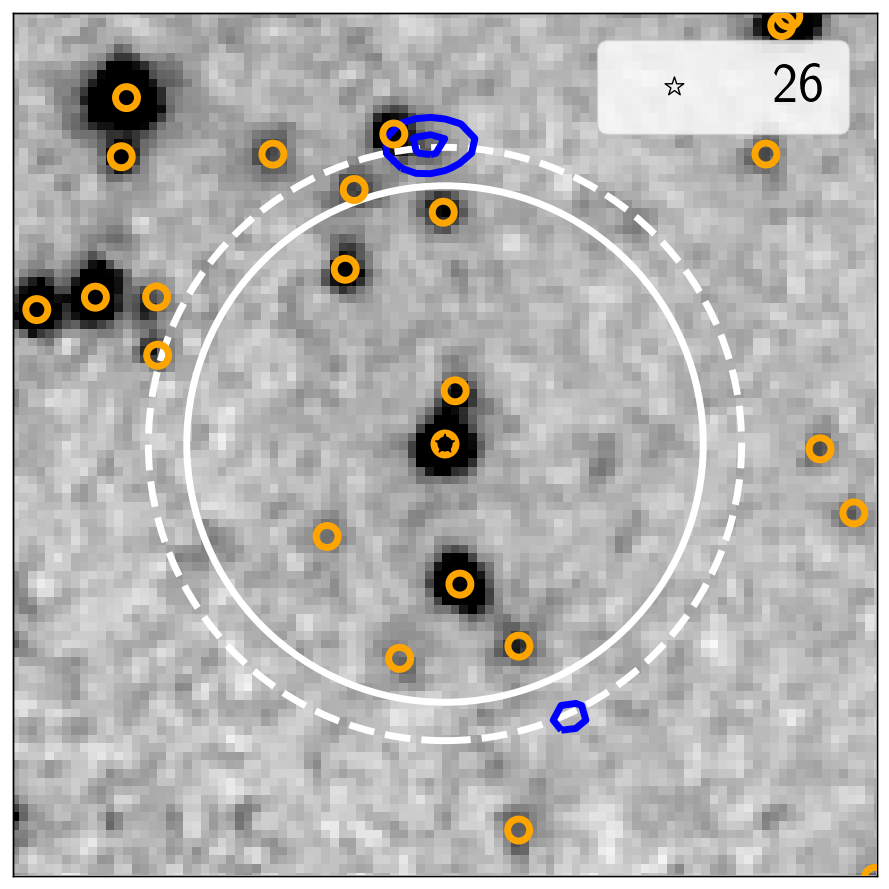}
\includegraphics[width=0.21\textwidth]{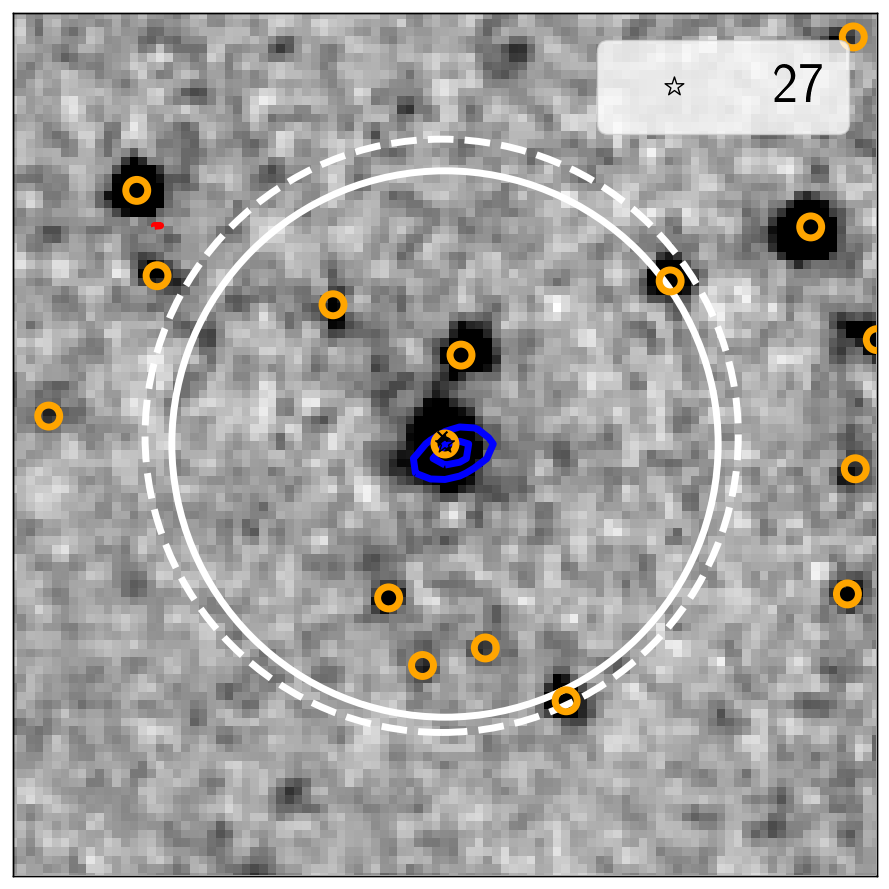}
\includegraphics[width=0.21\textwidth]{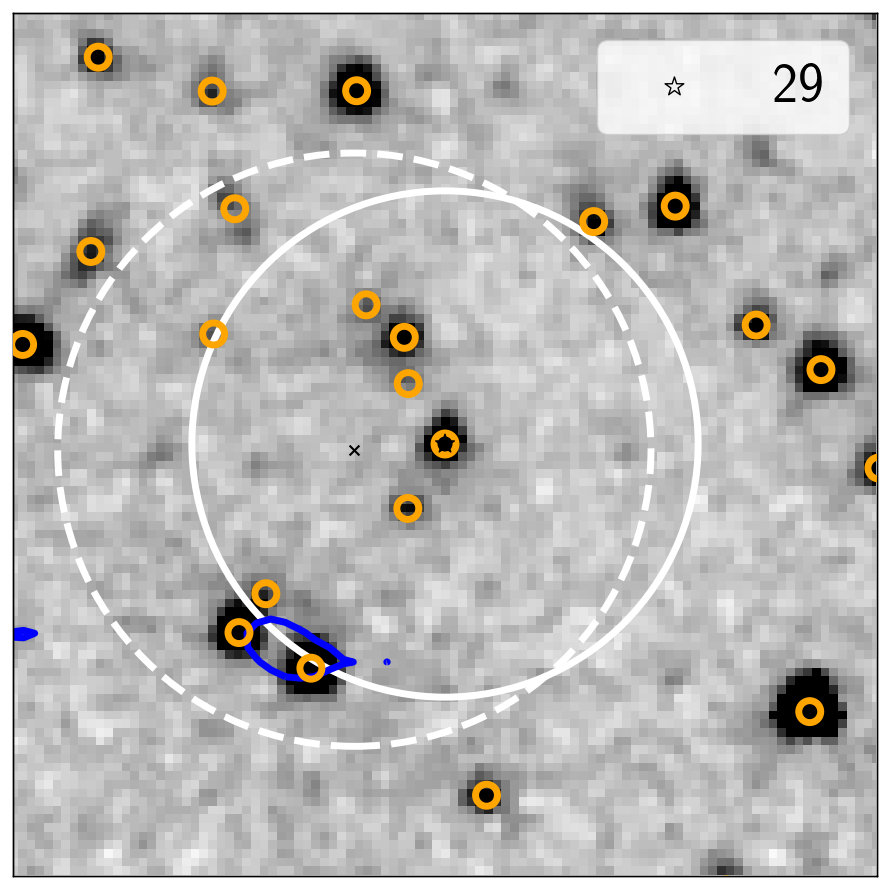}
\includegraphics[width=0.21\textwidth]{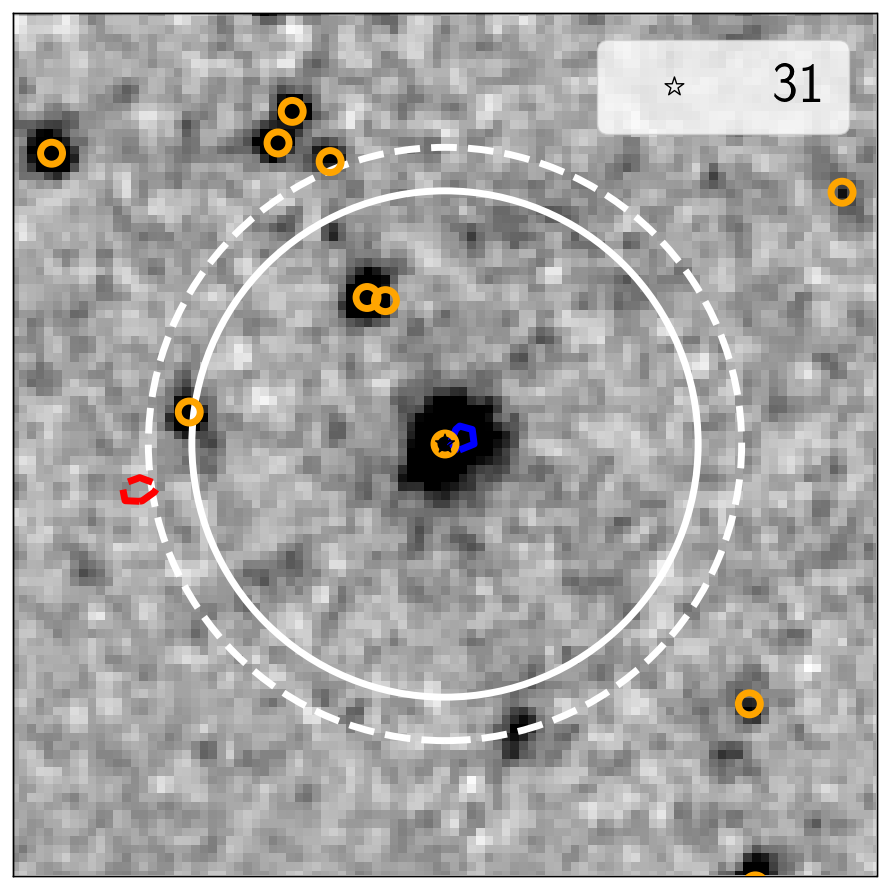}
\includegraphics[width=0.21\textwidth]{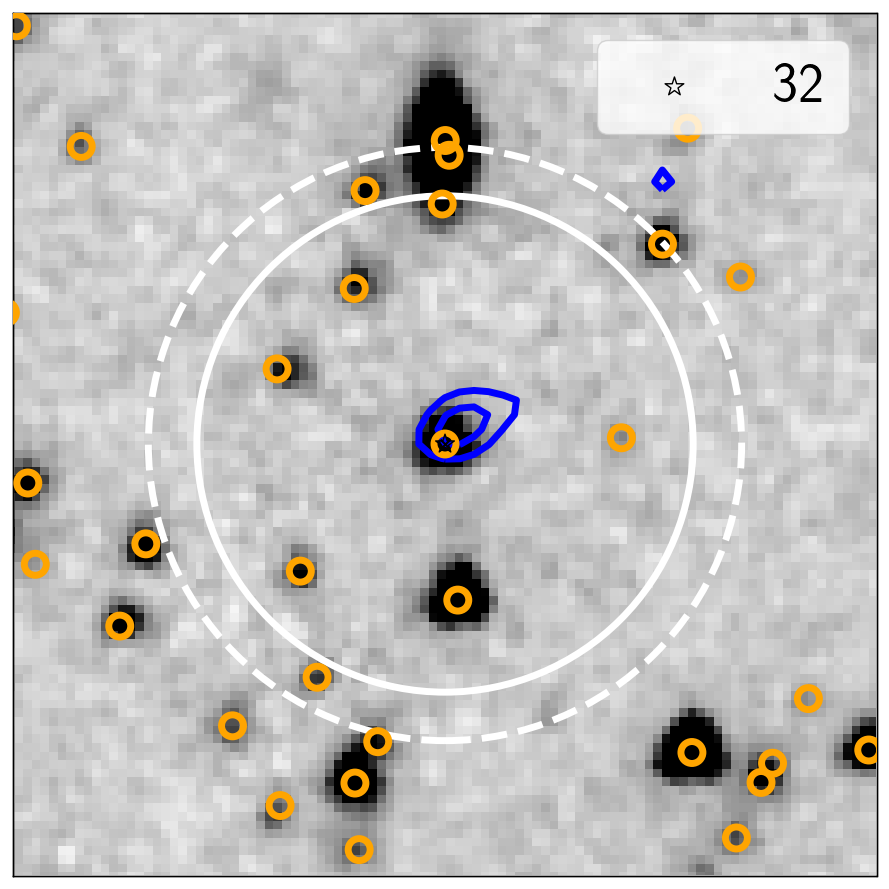}
\includegraphics[width=0.21\textwidth]{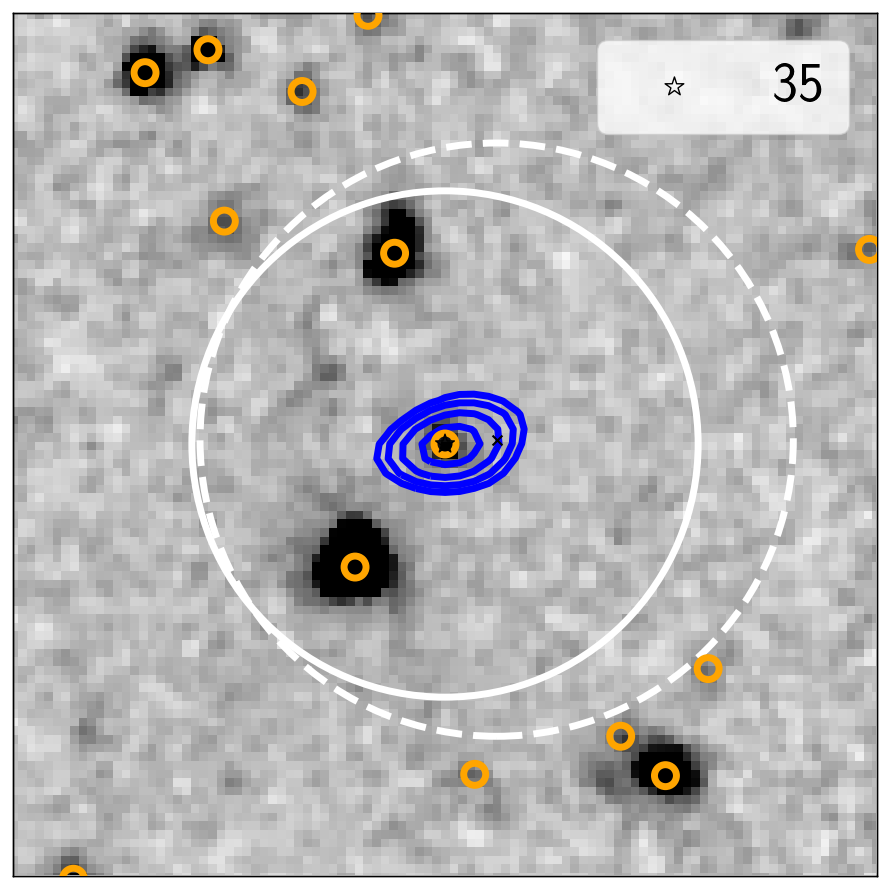}
\includegraphics[width=0.21\textwidth]{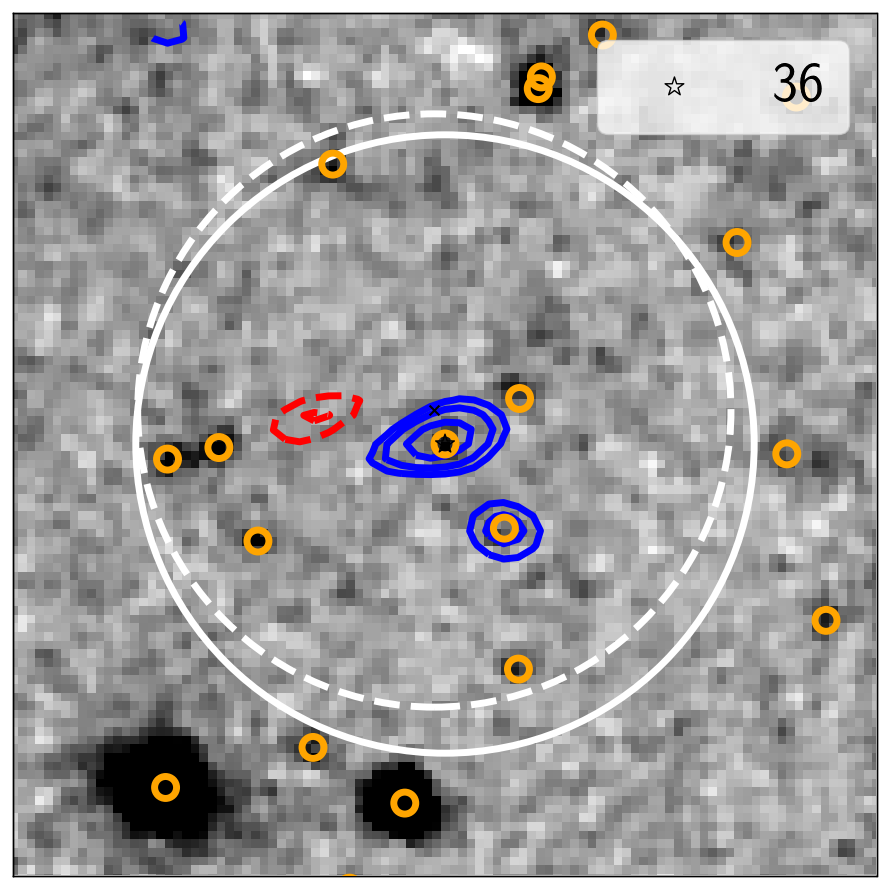}
 \caption{IRAC 3.6~$\mu$m maps around the targeted sources where emission is detected at $>4.2\sigma$ or where $>3\sigma$ emission matches the location of an IRAC source (orange circles as SEIP catalogued sources). Contour levels as in Fig.~\ref{fig:aca1}.}\label{fig:irmm}
 \end{figure*}

%% file: sections/contflux.tex
Using the {\sc imfit} task from {\sc casa}, which fits elliptical Gaussian components to specified regions, we retrieved the fluxes for all detections above $4.2\sigma$ as well as for detections above $3\sigma$ matching with a known IRAC source (Figs.~\ref{fig:aca1}, \ref{fig:aca2}, and \ref{fig:irmm}). The results are listed in Tab.~\ref{tab:b6phot}. The `a' suffix indicates a central source detected at or above 3$\sigma$ within 3.5\arcsec\ from the radio IR counterpart coordinates. The `b' suffix indicates sources in the ACA FoV detected at or above 3$\sigma$ further away than 3.5\arcsec\ from the radio source (and therefore not associated to it), but co-located with an IRAC detection above 3$\sigma$ to within 2\arcsec. The 3.5\arcsec\ is about $1\sigma$ of the ACA synthesised beam.

Of the 36 HzRG candidates, 14 have ACA detections corresponding to the radio source (Tab. \ref{tab:b6phot}). Twelve additional sources have been detected in the fields of 12 out of the 36 objects. Note that sources 9a, 21b, and 26b are reported as resolved by {\sc imfit}, and source 14a (with a peak-to-integrated flux ratio of $1.7\pm0.4$) is most likely the counterpart of an IRAC blend of two sources (Fig. \ref{fig:irmm}). Sources 1b and 21b were associated to neighbouring uncatalogued sources after visual inspection of IRAC imaging. The significant negative detections observed in a few sources are discussed in Sec.~\ref{sec:sz}.

The simulations presented in Sec.~\ref{sec:sz} are also used to derive the reliability of the detections at a signal-to-noise ratio (SNR) in the range $3\sigma<$~SNR~$<4.2\sigma$. The simulations show that there is a 63\% probability that an ACA map observed under the same conditions may show a flux maximum in that SNR range. In other words, it is expected that 23 of the maps in the sample contain such spurious detections. In total, we find 39 detections in the $3\sigma<$~SNR~$<4.2\sigma$ range, where 26 are discarded as spurious, since no IRAC sources are detected at those positions. We also note that we find 27 negative features with significances between $3\sigma$ and $4.2\sigma$. None of them matches in position with an IRAC source. This implies that of order 27 of the positive detections in this signal-to-noise range would be expected to be spurious, consistent with the 26 we discard. Consequently, the remaining low signal-to-noise detections, listed in Tab.~\ref{tab:b6phot}, are expected to be real.

\input{tables/tab-b6phot}

\begin{figure}
\includegraphics[width=0.45\textwidth]{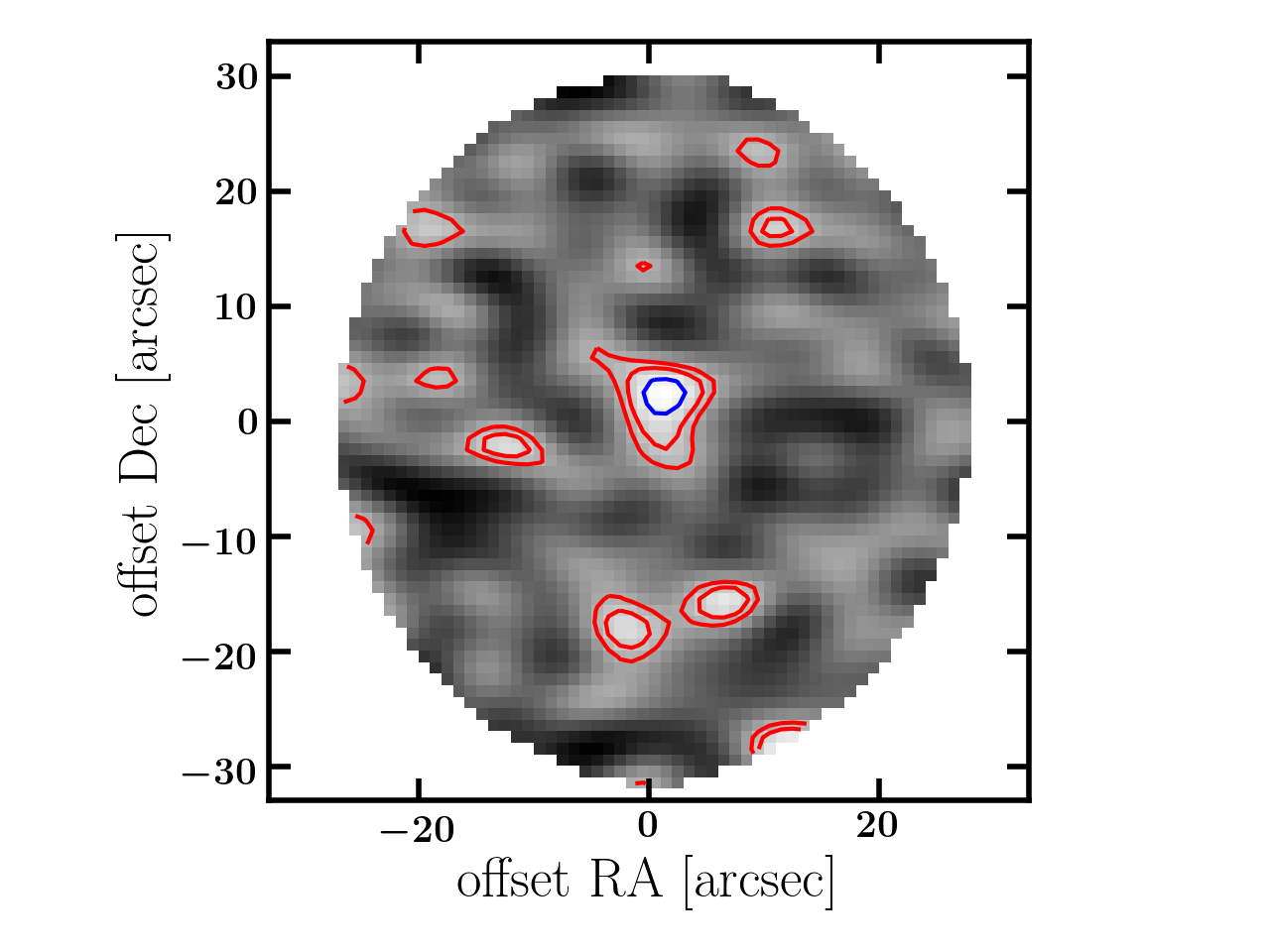}
\caption{Stacking of the maps of the 11 undetected sources at $z>0.8$. Only positive isocontours are shown at 1.5, $1.5\times\sqrt(2)$, and 3$\sigma$, where the first two levels are in red, and the last in blue.}
\label{fig:stack}
\end{figure}

In order to assess the flux of the undetected sources in the sample, we have pursued a stacking analysis considering only those sources at $z>0.8$ and devoid of $>3\sigma$ signal in the central region. We note that, despite the large redshift range adopted, the negative K-correction of the thermal SEDs implies that galaxies similar in nature would present comparable flux levels. This assumes the thermal SED dominates in less radio-luminous galaxies, which is supported by our findings (Tab.~\ref{tab:radiosed}). The 11 sources complying with these cuts yield a stacked signal of $0.16\pm0.05\,$mJy (3.5$\sigma$, Fig.~\ref{fig:stack}) which shows that deeper continuum surveys are required to directly detect these fainter sources. Out of the 8 sources at $z>2$, the 3 undetected ones yield a $3\sigma$ upper-limit of 0.28\,mJy, when stacked.

% figure with the stacked signal

%% file: tables/tab-b6phot.tex
	\begin{table}
	\begin{tabular}{rrrrrr}
    ID & S$_p$ & S$_i$ & RA & Dec & Dist \\
    & [mJy] & [mJy] & [deg~(\pp)] & [deg~(\pp)] & [\pp] \\
    \hline
    \multicolumn{5}{c}{SNR~$>4.2\sigma$}\\
      6a & 0.7$\pm$0.2 & 0.5$\pm$0.2 & 7.7964~(0.7) & -44.48289~(0.2) & 1.0 \\
     9a & 3.5$\pm$0.2 & 3.7$\pm$0.3 & 8.20557~(0.2) & -44.36406~(0.08) & 0.2 \\
     14a & 0.9$\pm$0.1 & 1.5$\pm$0.3 & 8.9825~(1) & -44.04366~(0.2) & 1.5\\
     15a & 1.3$\pm$0.2 & 1.1$\pm$0.3 & 9.33102~(0.3) & -44.02902~(0.3) & 1.0 \\
     18a & 1.2$\pm$0.2 & 1.1$\pm$0.2 & 9.76465~(0.3) & -43.82978~(0.2) & 0.6 \\
     19b & 1.2$\pm$0.2 & 1.2$\pm$0.3 & 7.46443~(0.4) & -43.76181~(0.2) & 18 \\
     20a & 2.4$\pm$0.1 & 2.4$\pm$0.2 & 9.62482~(0.2) & -43.74845~(0.08) & 0.3 \\
     22a & 1.3$\pm$0.2 & 1.5$\pm$0.3 & 7.86991~(0.5) & -43.68901~(0.2) & 1.1 \\
     23b & 0.9$\pm$0.2 & 1.0$\pm$0.3 & 8.6407~(0.7) & -43.56502~(0.2) & 13 \\
    26b & 1.5$\pm$0.3 & 2.0$\pm$0.4 & 7.44068~(0.5) & -43.35799~(0.3) & 21 \\
    27a & 0.8$\pm$0.2 & 0.6$\pm$0.2 & 9.4414~(0.6) & -43.32202~(0.3) & 0.8 \\
    32a & 0.8$\pm$0.2 & 1.0$\pm$0.3 & 9.0691~(0.8) & -43.1593~(0.4) & 2.0 \\
    35a & 1.8$\pm$0.2 & 2.0$\pm$0.3 & 7.7220~(0.4) & -42.87094~(0.2) & 0.5 \\
    36a & 1.0$\pm$0.1 & 1.0$\pm$0.2 & 9.7414~(0.5) & -42.86779~(0.2) & 0.3 \\
    36b & 0.7$\pm$0.2 & 0.6$\pm$0.2 & 9.7399~(0.4) & -42.8696~(0.5) & 7.4 \\
    \hline
    \multicolumn{5}{c}{$3\sigma<$~SNR~$<4.2\sigma$}\\
      1b & 0.7$\pm$0.2 & 0.4$\pm$0.1 & 9.40893~(0.3) & -44.64772~(0.1) & 13 \\
      7b & 0.7$\pm$0.2 & 0.7$\pm$0.2 & 9.9191~(0.8) & -44.45656~(0.2) & 22 \\
      8a & 0.6$\pm$0.2 & 0.7$\pm$0.3 & 7.4151~(0.8) & -44.3885~(0.4) & 3.3 \\
     9b & 0.6$\pm$0.2 & 0.9$\pm$0.3 & 8.2019~(2) & -44.36739~(0.2) & 15 \\
      11a & 0.5$\pm$0.2 & 0.5$\pm$0.2 & 9.4768~(1) & -44.1861~(0.7) & 2.8\\
     15b & 0.9$\pm$0.2 & 0.8$\pm$0.3 & 9.32398~(0.6) & -44.03087~(0.3) & 20 \\
     16a & 0.5$\pm$0.2 & 0.3$\pm$0.1 & 9.01782~(0.6) & -43.93005~(0.3) & 2.4 \\
     16b & 0.7$\pm$0.2 & 0.7$\pm$0.2 & 9.0132~(0.8) & -43.92650~(0.2) & 16 \\
     17b & 1.2$\pm$0.3 & 1.2$\pm$0.3 & 9.66972~(0.5) & -43.89345~(0.3) & 21 \\
     21b & 0.5$\pm$0.2 & 1.1$\pm$0.5 & 9.3375~(1) & -43.7121~(1) & 3.7 \\
     25a & 0.5$\pm$0.1 & 0.7$\pm$0.3 & 9.2856~(1) & -43.3980~(0.7) & 0.8 \\
    29b & 0.8$\pm$0.2 & 1.2$\pm$0.4 & 8.2350~(1) & -43.2780~(0.4) & 18 \\
    31a & 0.5$\pm$0.2 & 0.5$\pm$0.2 & 7.7968~(0.7) & -43.1544~(0.6) & 1.7 \\
    \hline
    \end{tabular}
	\caption{ACA Band~6 detections at $>4.2\sigma$ (top part) and between 3 and 4.2 $\sigma$ that positionally match with an IRAC source (bottom part). The letter `a' in the ID indicates a detection of the central source while `b' refers to other sources detected in the FoV of the HzRG candidates, not coinciding with the target source. The peak flux densities (S$_p$) are those measure in the image, while the integrated fluxes (S$_i$) are those measured by {\sc casa} task {\sc imfit}, which also provides an estimate for the source coordinates (the values in parenthesis refer to the errors in sky arcsec). The last column indicates the distance between the radio IR counterpart and the ACA source in arcsec.}
	\label{tab:b6phot}
	\end{table}

%% file: sections/radspec.tex
\begin{figure*}
	\includegraphics[width=0.3\textwidth]{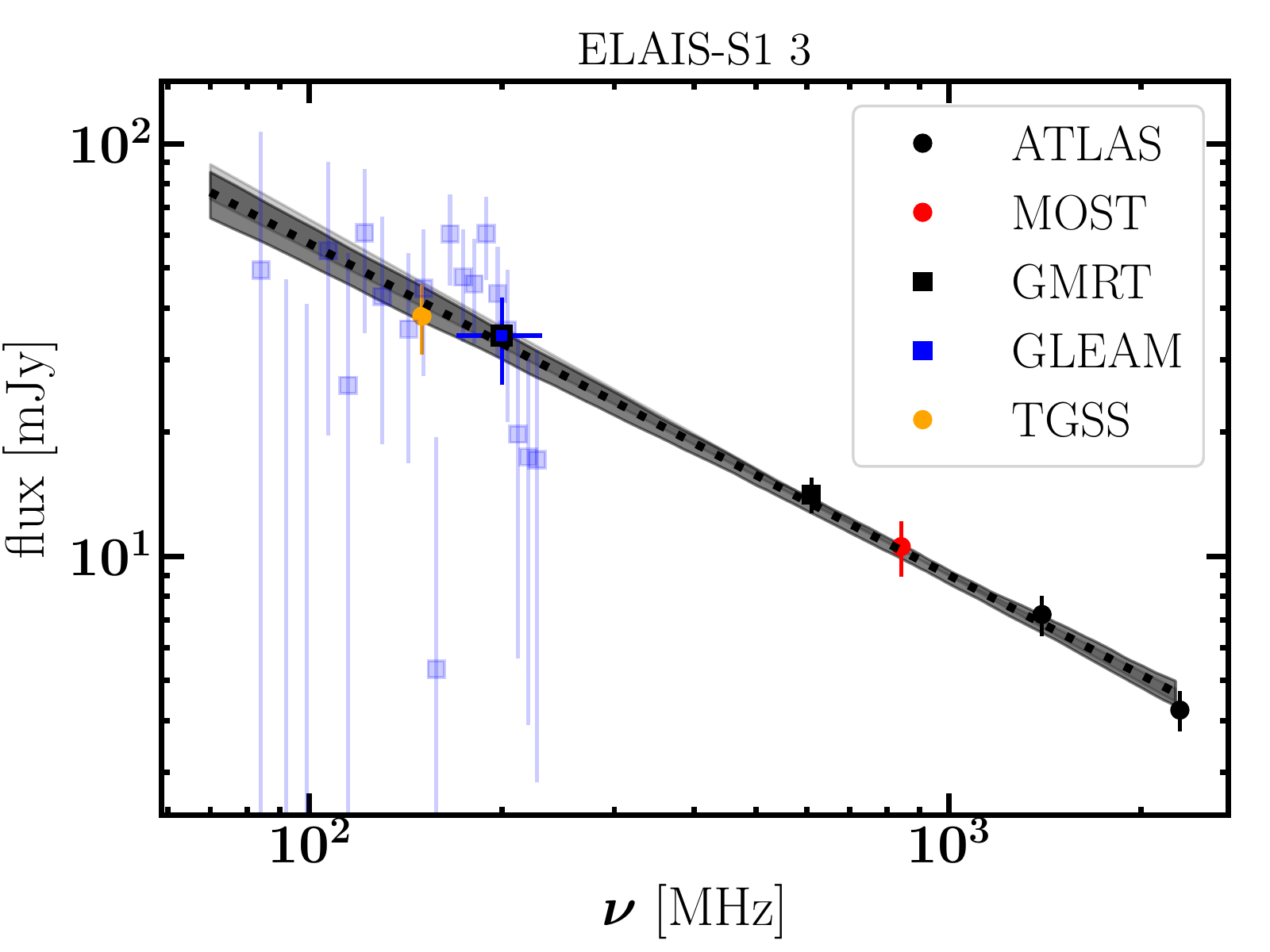}
	\includegraphics[width=0.3\textwidth]{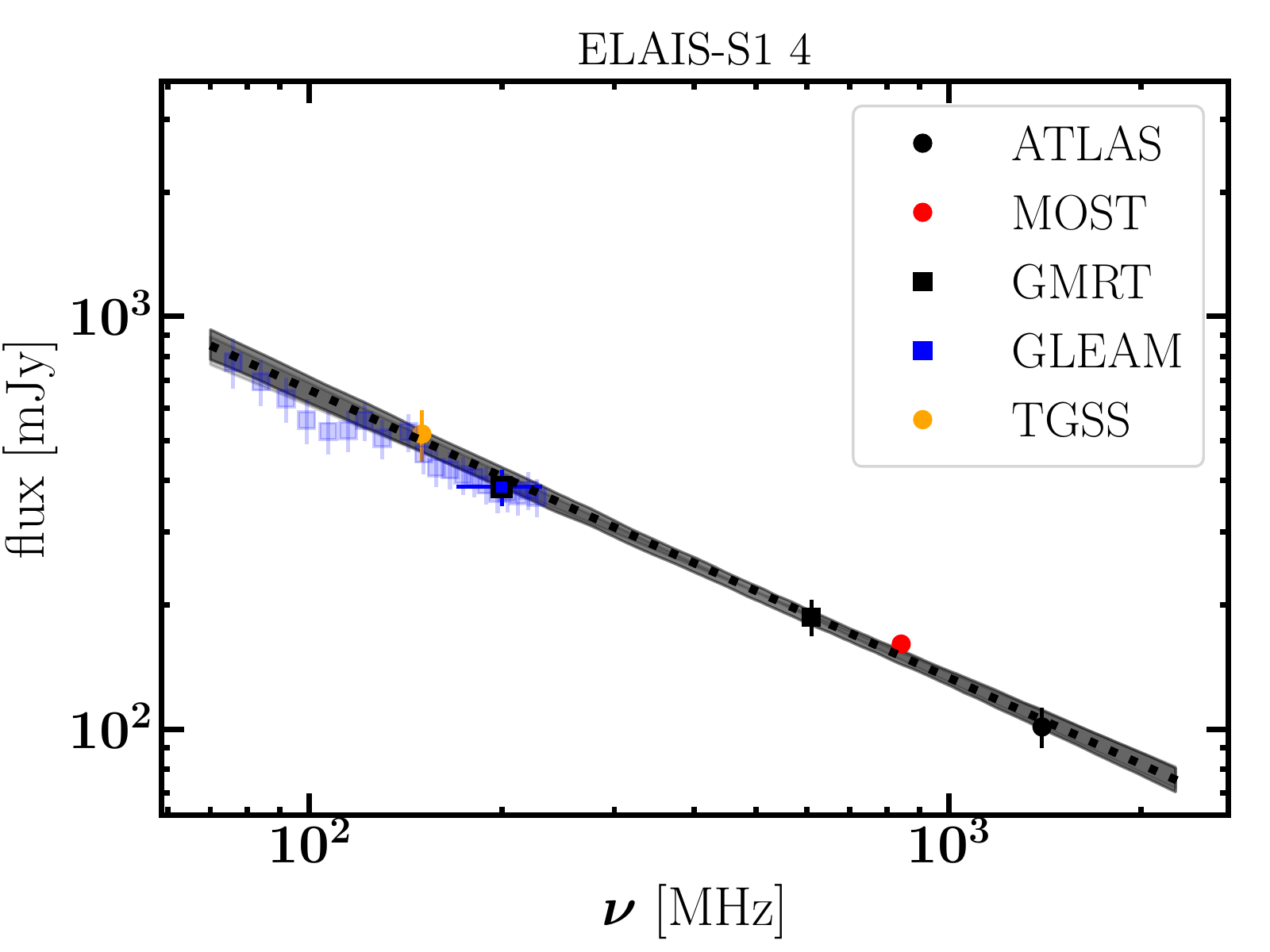}
	\includegraphics[width=0.3\textwidth]{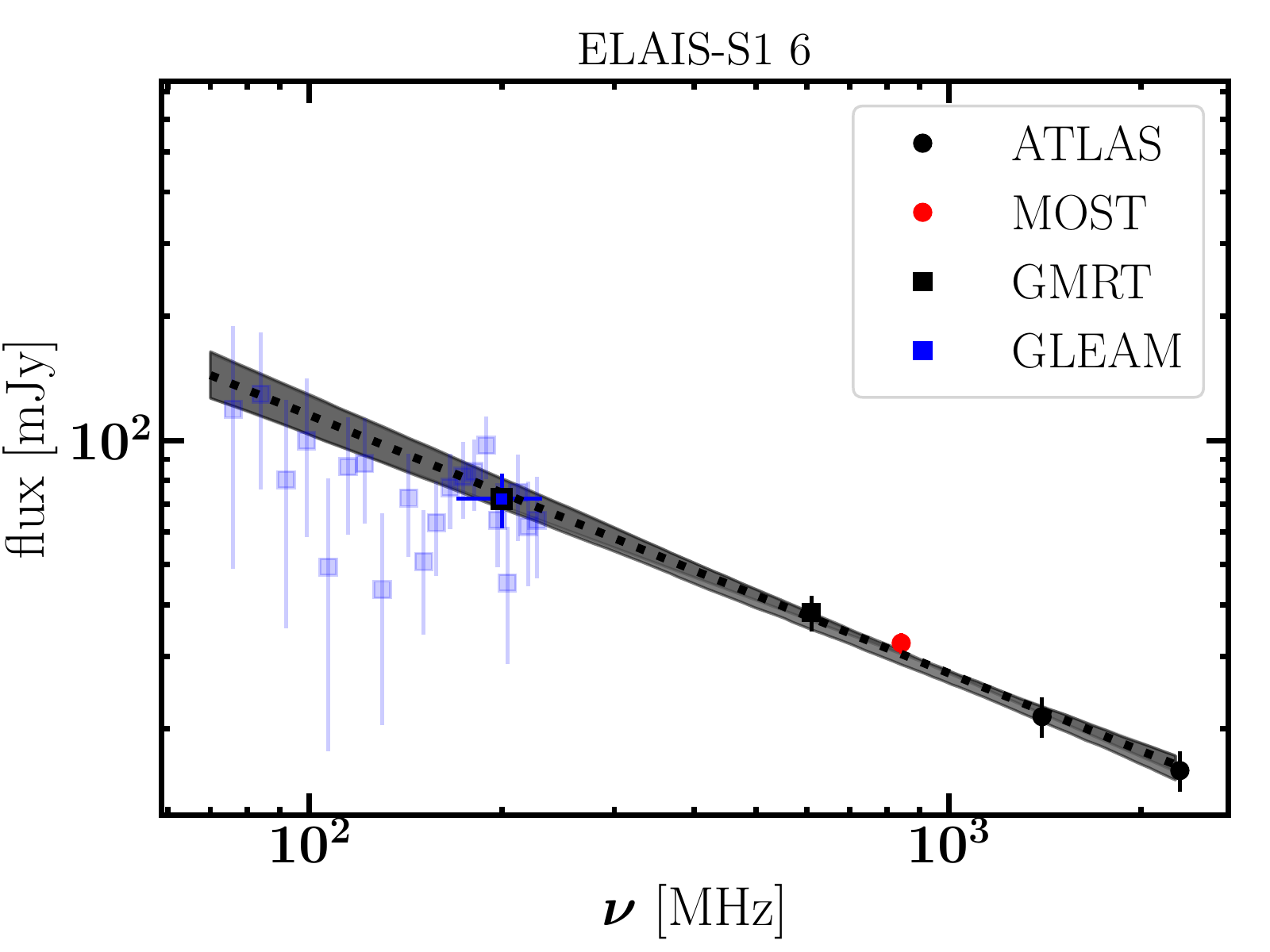}
	\includegraphics[width=0.3\textwidth]{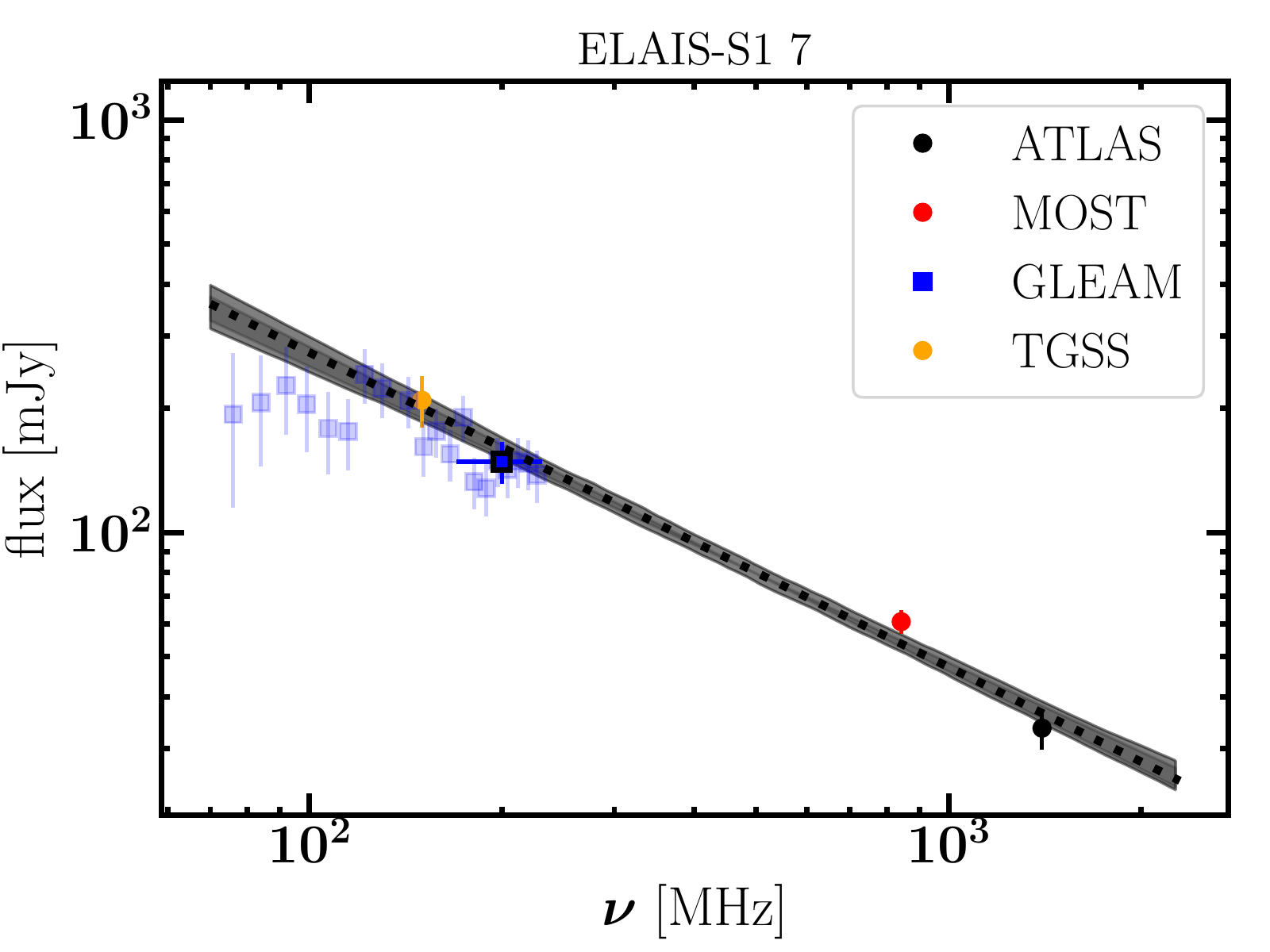}
	\includegraphics[width=0.3\textwidth]{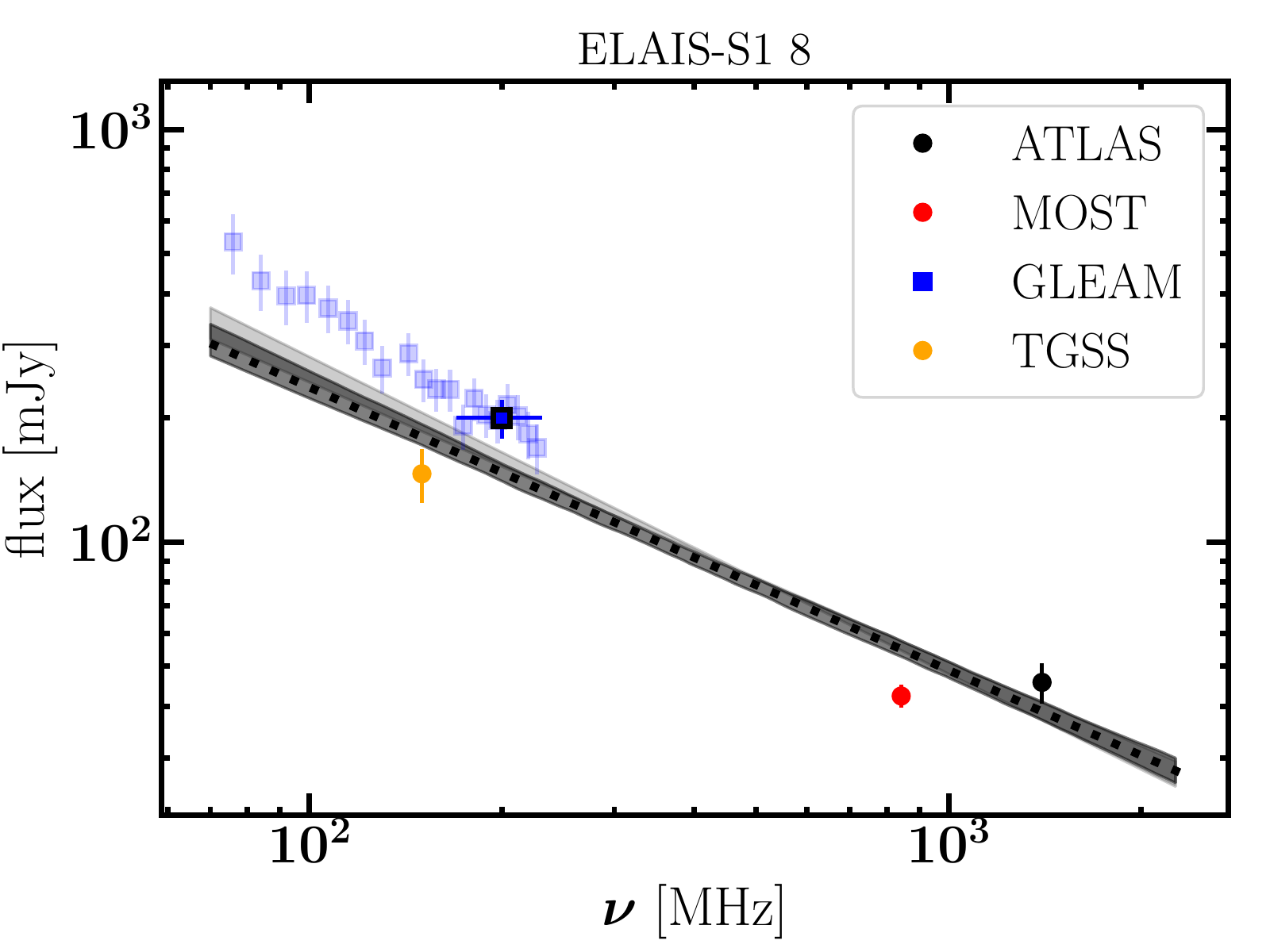}
	\includegraphics[width=0.3\textwidth]{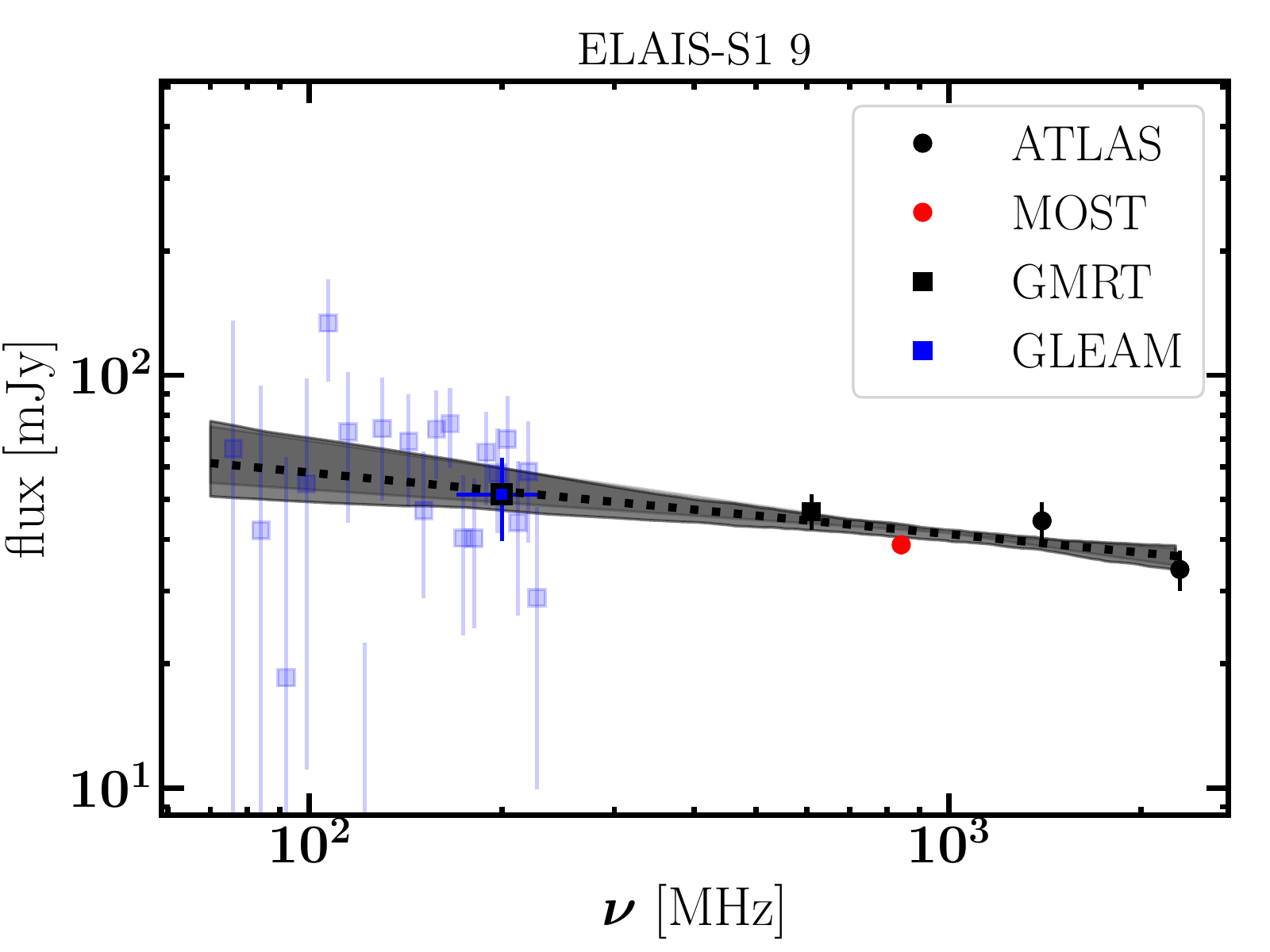}
	\includegraphics[width=0.3\textwidth]{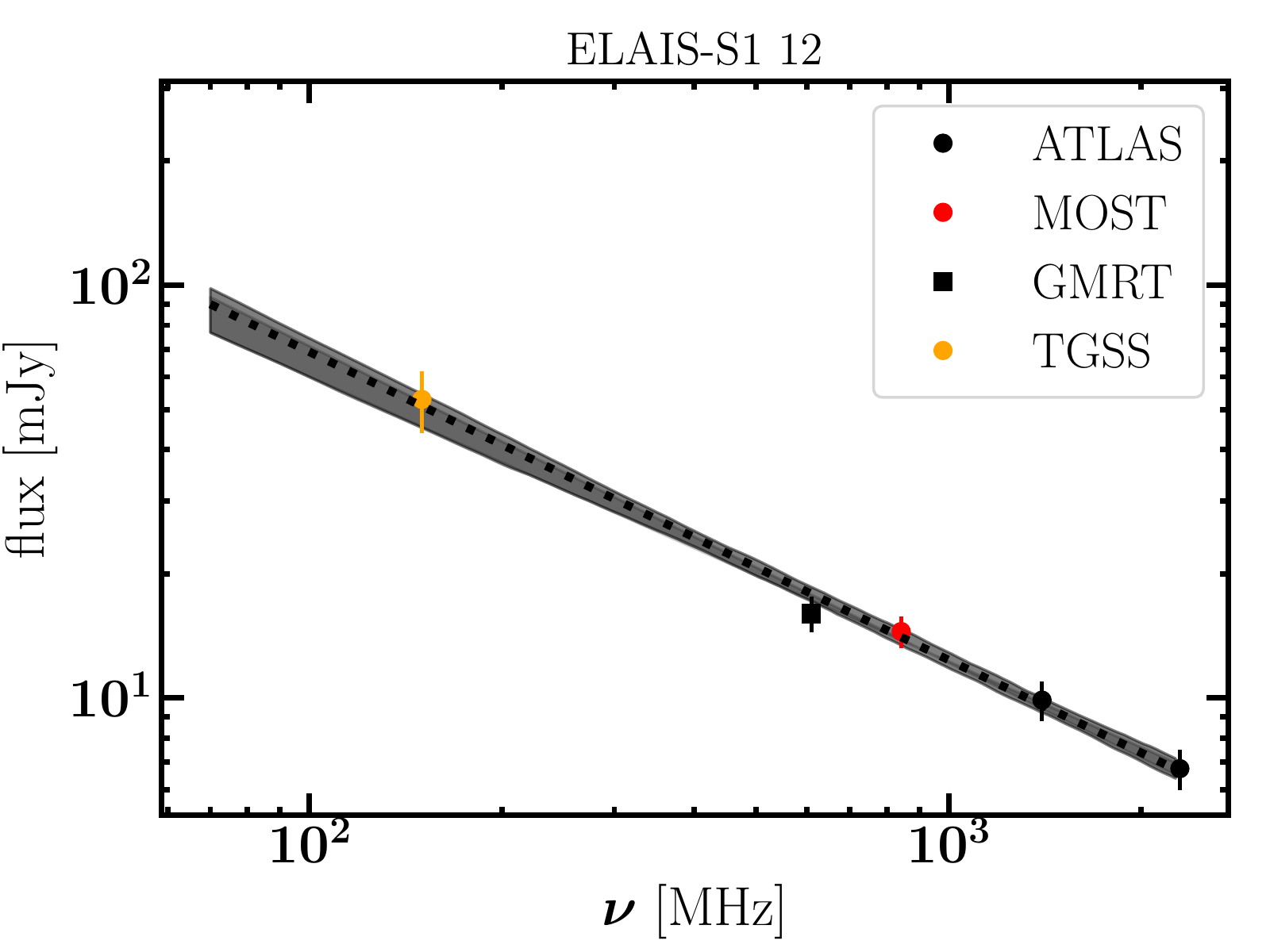}
	\includegraphics[width=0.3\textwidth]{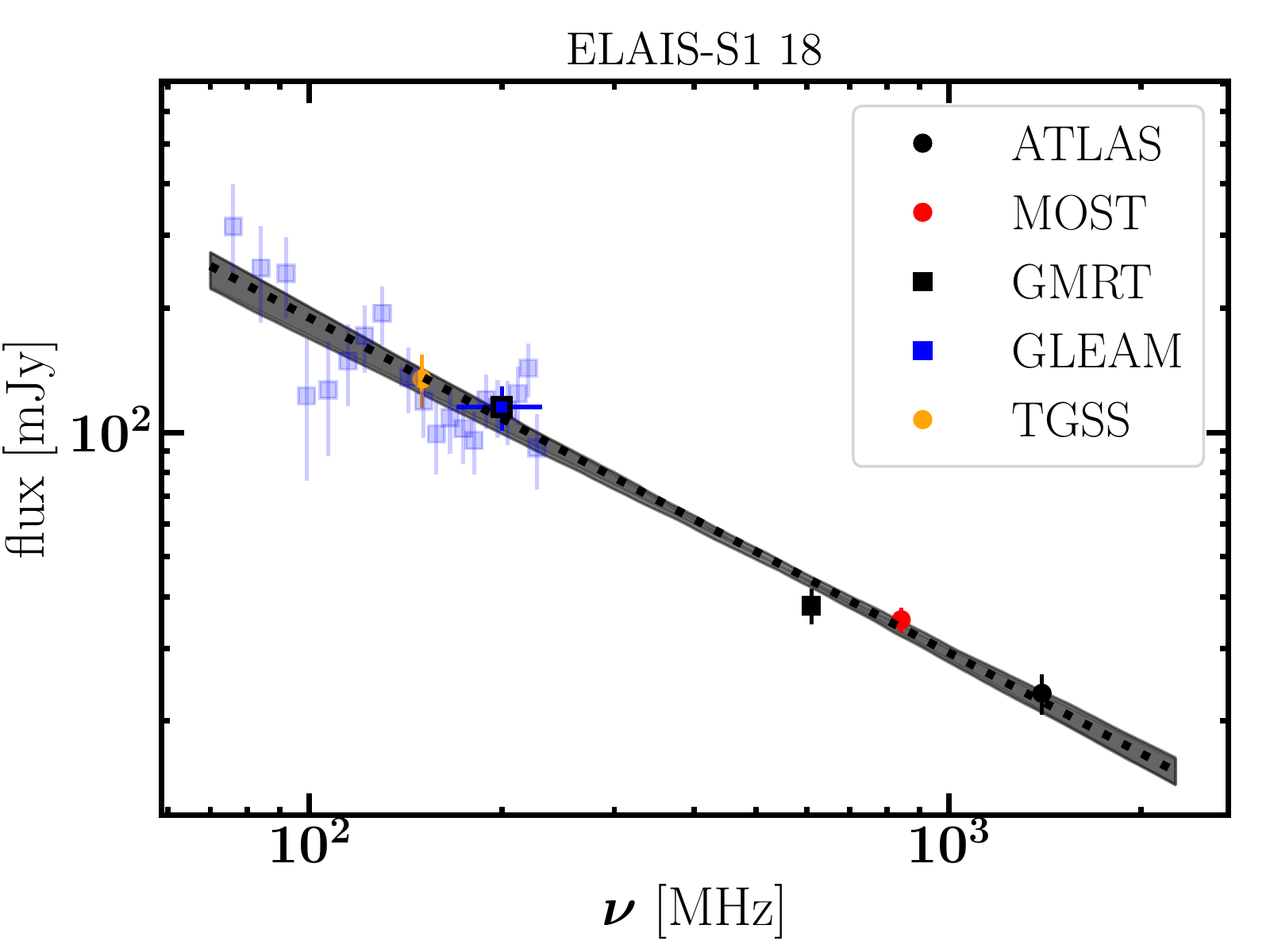}
	\includegraphics[width=0.3\textwidth]{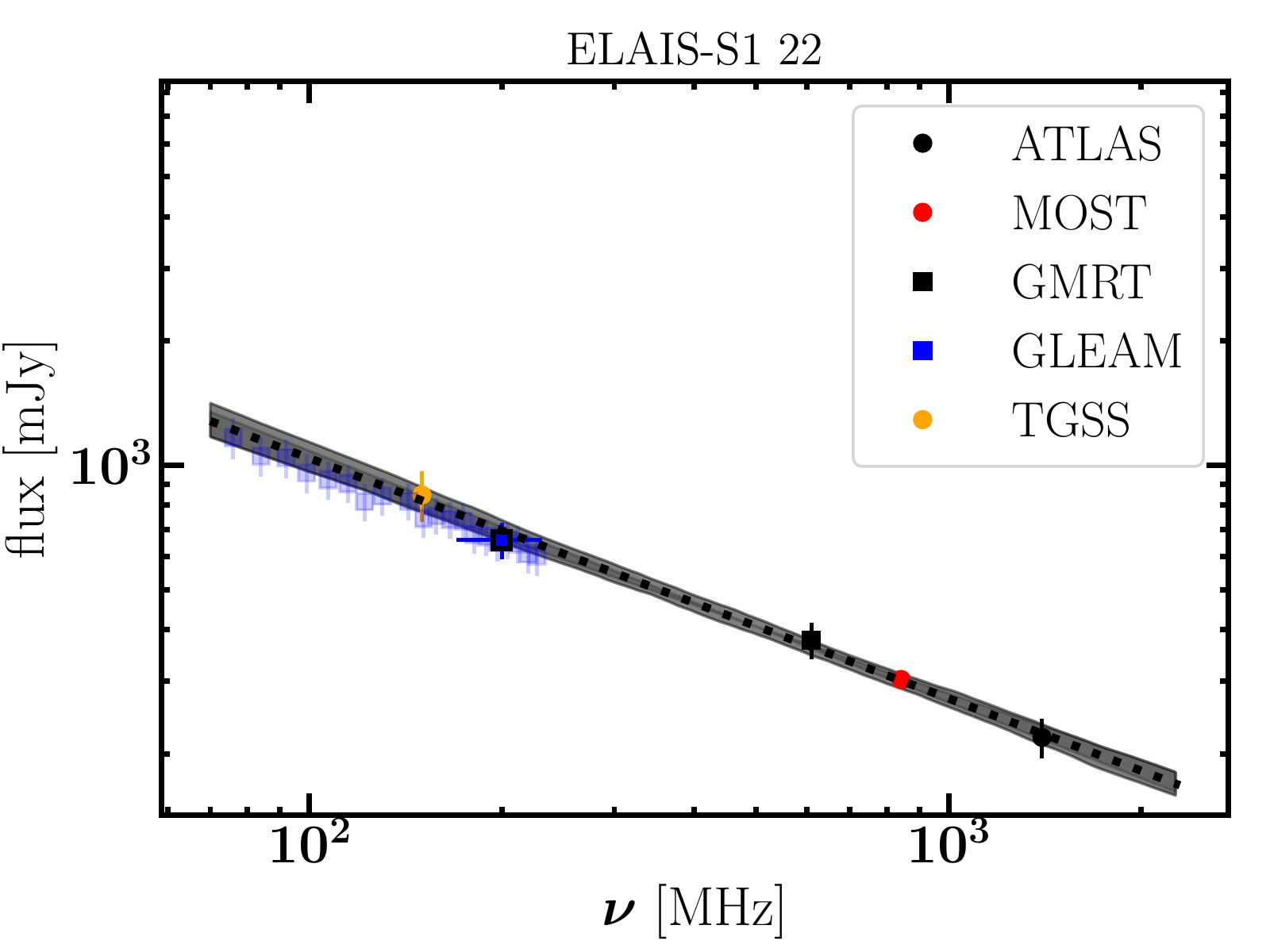}
	\includegraphics[width=0.3\textwidth]{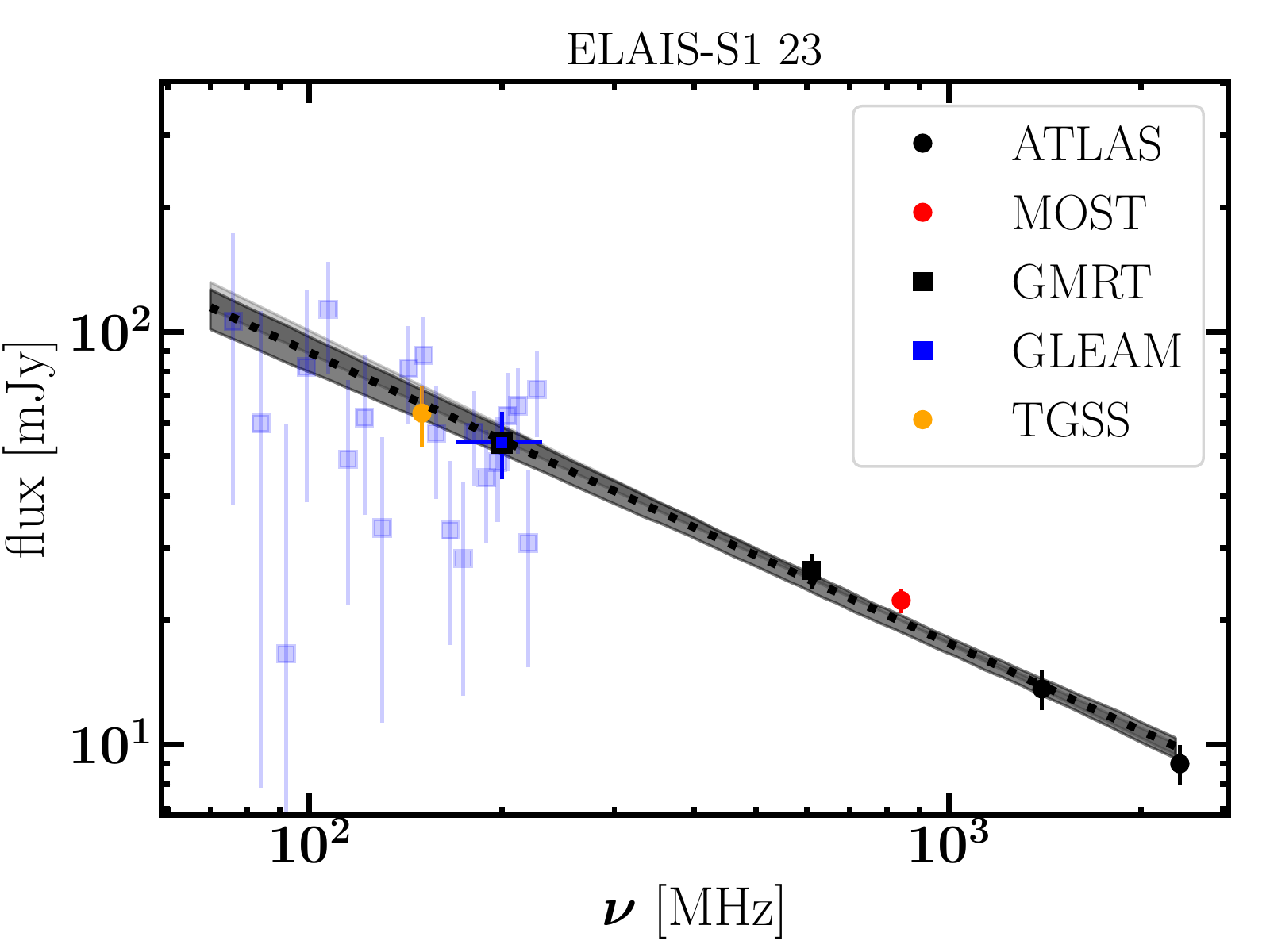}
	\includegraphics[width=0.3\textwidth]{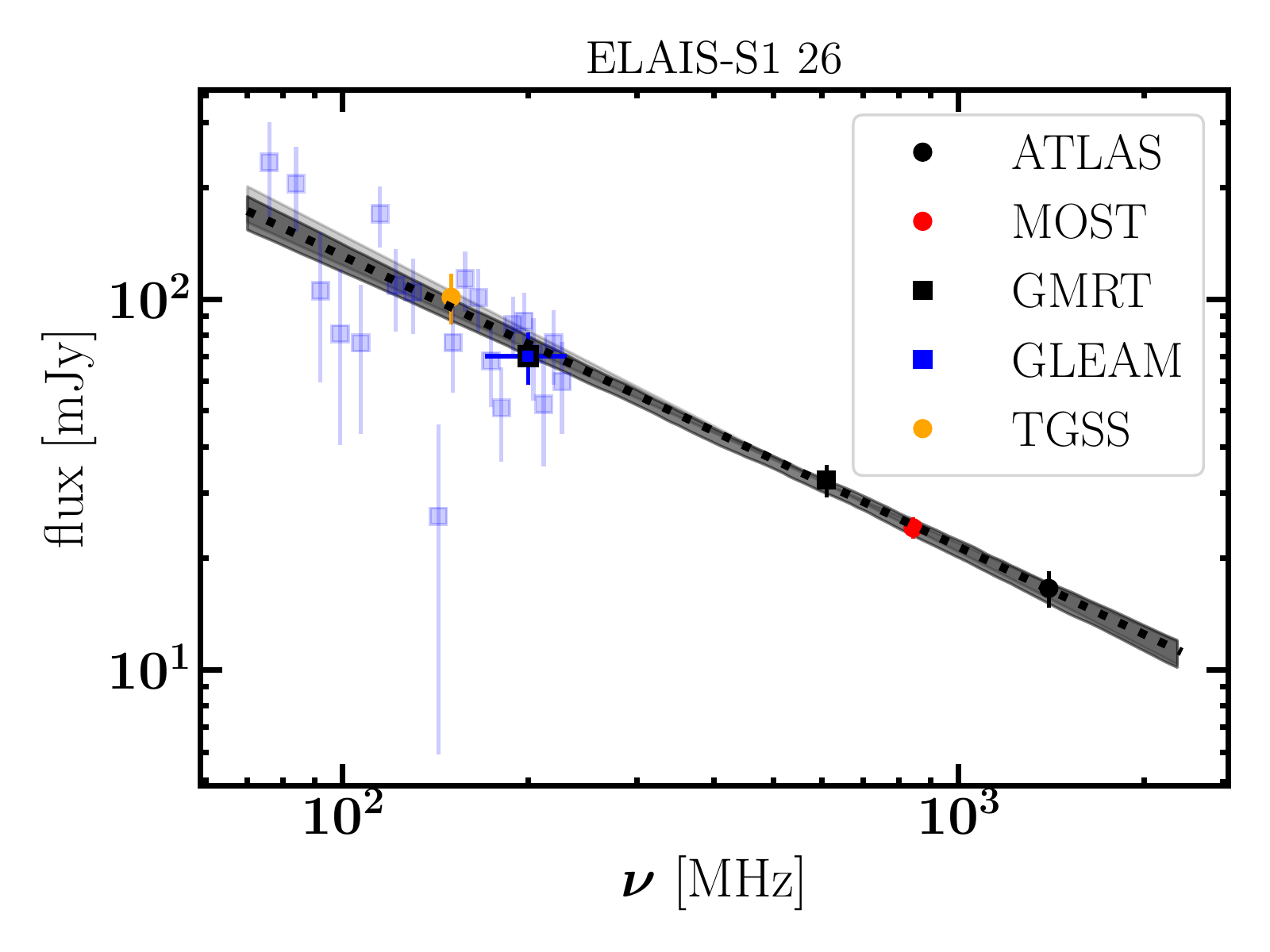}
	\includegraphics[width=0.3\textwidth]{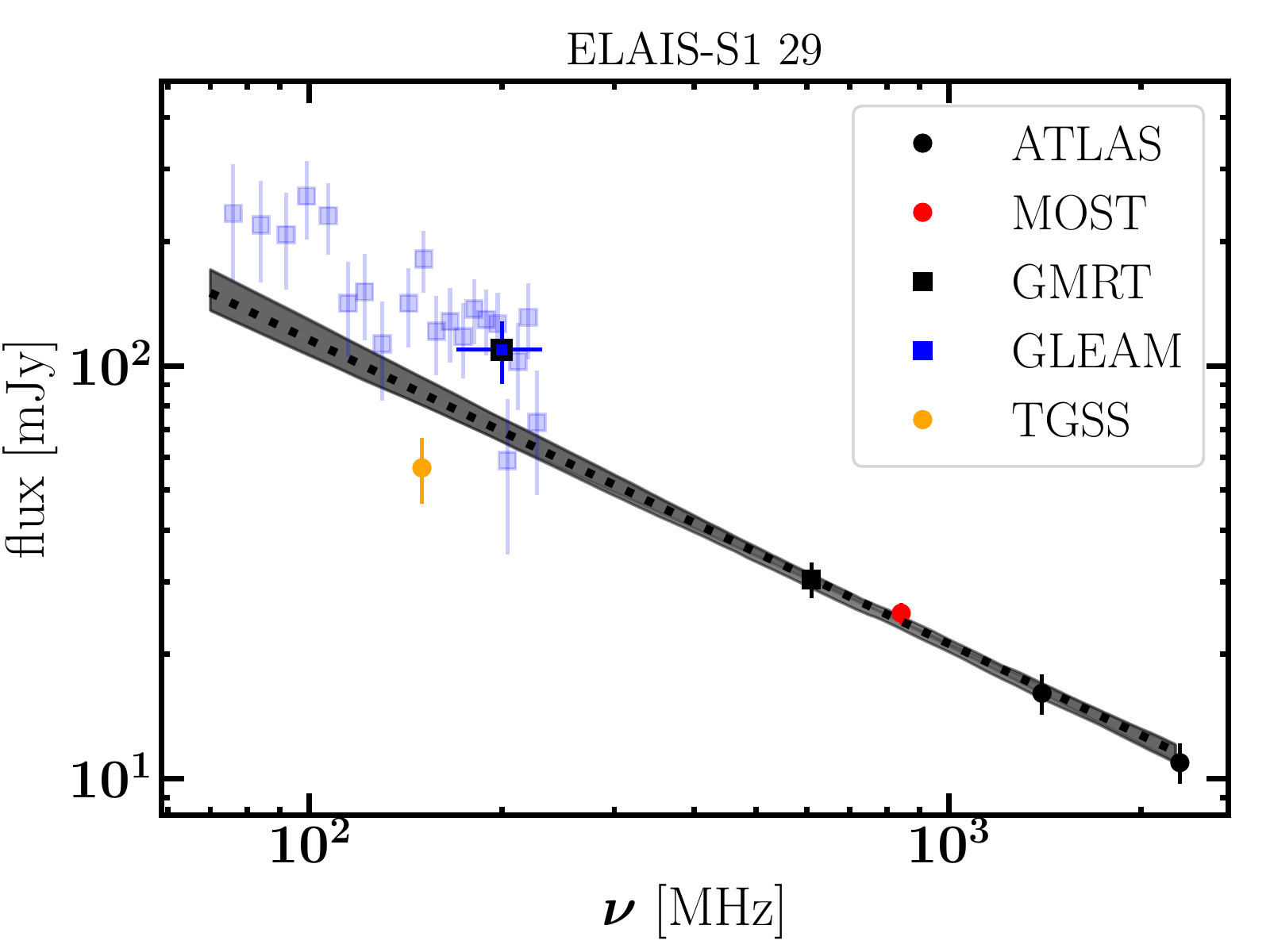}
	\includegraphics[width=0.3\textwidth]{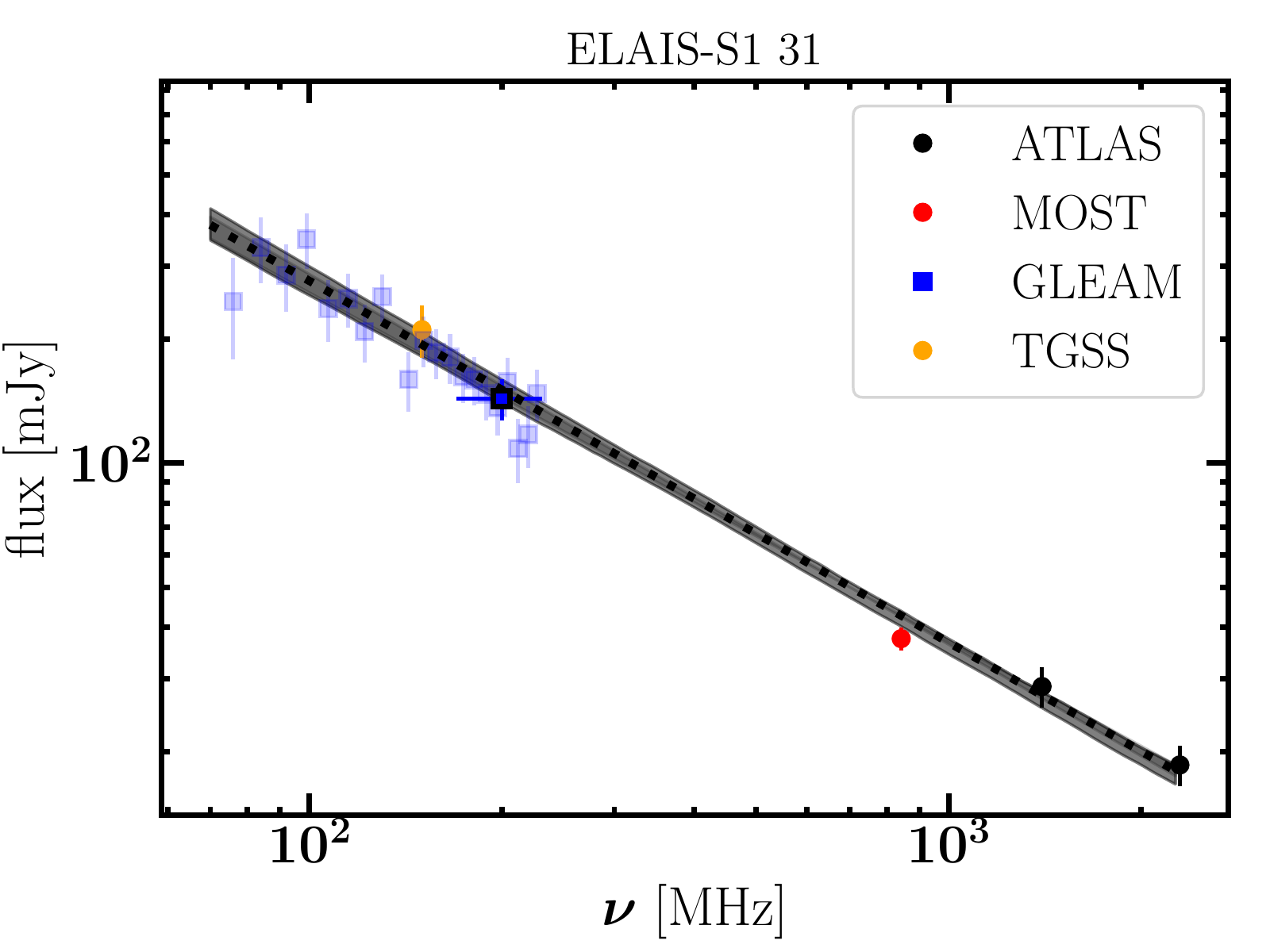}
	\includegraphics[width=0.3\textwidth]{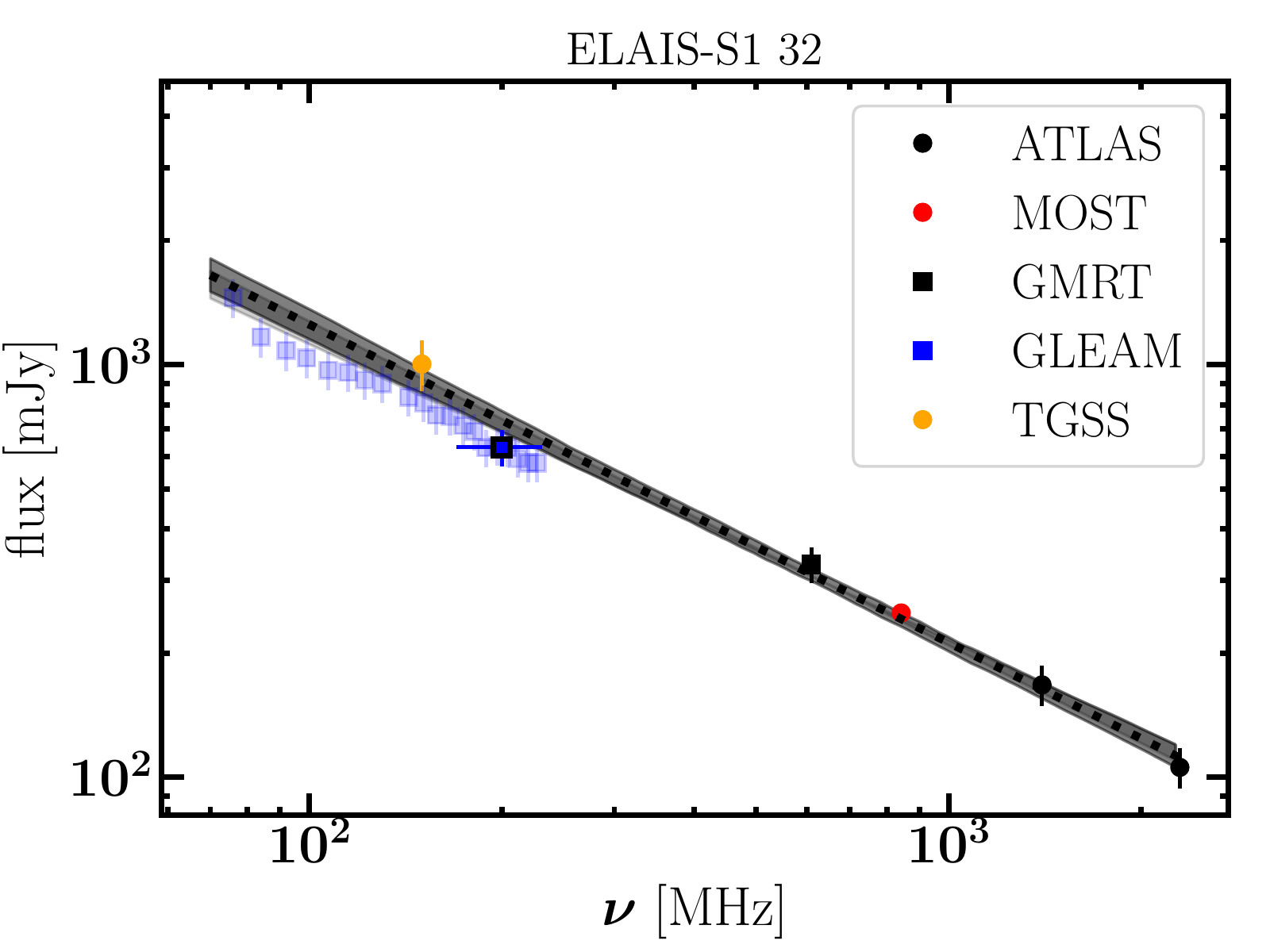}
	\caption{Radio SEDs for the sources detected by GLEAM or TGSS. The dotted line shows the best fit (assuming $S_\nu\propto\nu^{\alpha}$) to the photometry data in the 0.61--2.3\,GHz spectral regime, the GLEAM 200\,MHz-wide band, and the 150\,MHz flux from TGSS (the shaded region represents the $\pm1\sigma$ range, see text for details). The spectral index, $\alpha$ and the extrapolated 233\,GHz flux ($S_{233}$) are reported in Tab.~\ref{tab:radiosed}.}
	\label{fig:radiosed}
	\end{figure*}

	To extend the radio SEDs to lower frequencies, we matched the 36 HzRG candidates with the two low frequency catalogues GLEAM and TGSS described in Sec. \ref{sec:es1}. For this we adopted 65\arcsec\ and 15\arcsec\  matching radii with GLEAM and TGSS, respectively, equivalent to a 1$\sigma$ matching region (see Sec. \ref{sec:es1}), as we have done with the matching of the other radio catalogues. The radio SEDs of the sources with a GLEAM and/or TGSS counterparts are shown in Fig.~\ref{fig:radiosed}. They include all the GLEAM sub-channels and TGSS observations, as well as the ATCA points at 1.4 and 2.3 GHz, the MOST points at 843 MHz and the GMRT points at 610 MHz. We fit the radio SEDs with a power law with a single spectral index  ($S_\nu\propto\nu^\alpha$), indicated by the dotted line. To account for the error budget in a realistic manner, we used the {\tt SciPy} function {\tt curve\_fit} \citep{jones01} that uses a non-linear least-square fit and a bootstrapping alternative. The latter was pursued by refitting the radio SED after randomly drawing new flux values (N$=$100) assuming a normal distribution characterised by the flux uncertainty around the mean value. The darker grey region marks the 1~$\sigma$ scatter from the bootstrapping technique, while the lighter grey area indicate the uncertainties derived from the use of the {\tt curve\_fit} function.
	In the GLEAM spectral range, only the 200 MHz-wide band is used (170--231\,MHz), not the fluxes in the individual sub-bands (shown in light blue in Fig. \ref{fig:radiosed}). The good alignment between the lower and higher frequency data points, all following one single power-law spectrum, indicates that the low-frequency counterparts found for the HzRGs candidates as a result of the matching were the correct ones, reflecting the adequacy of the choice of the matching radii. The extrapolation of the SED fit to the reference frequency of our ACA observations at 233\,GHz, assuming the same single power law, allows for an estimate of the flux due to synchrotron emission at that frequency.
	
	Table~\ref{tab:radiosed} reports the observed and SED-fitting-based extrapolated 233\,GHz ACA fluxes (S$^{obs}_{233}$ and S$^{fit}_{233}$, respectively) and the slope, $\alpha$, of the single power law. The last column indicates whether we consider the observed flux at this frequency to be possibly mostly due to synchrotron emission (marked with a check mark). This is the case when the predicted flux at 233~GHz, S$^{fit}_{233}$, is greater than 50\% of the observed flux within 1~$\sigma$. On the other hand, if S$^{fit}_{233}$ falls below 50\% of the observed flux by more than 3~$\sigma$, we consider that thermal emission dominates at these frequencies (marked with a cross). An S$^{obs}_{233}$ value lying between these two limits (${\rm log_{10}(S^{fit}_{233})+1\sigma < log_{10}(S^{obs}_{233}/2) < log_{10}(S^{fit}_{233})+3\sigma}$) should be considered to be the result of comparable contribution from synchrotron and thermal emissions (marked as empty column in Table~\ref{tab:radiosed}). In case only upper limits are available at 233\,GHz, and if the predicted flux, S$^{fit}_{233}$, falls below half that limit by more than 1~$\sigma$, nothing can be said about the origin of the Band 6 emission for such objects. The upper part of the table lists the objects with GLEAM or TGSS counterparts. In the middle and lower parts of the table we report the extrapolated fluxes at 233\,GHz for objects without GLEAM or TGSS detections but with data in the 0.61--2.3\,GHz spectral range. 
 
    % radio SEDs properties
    \input{tables/tab-radiosed}
   	
\begin{figure}
	\includegraphics[width=0.45\textwidth]{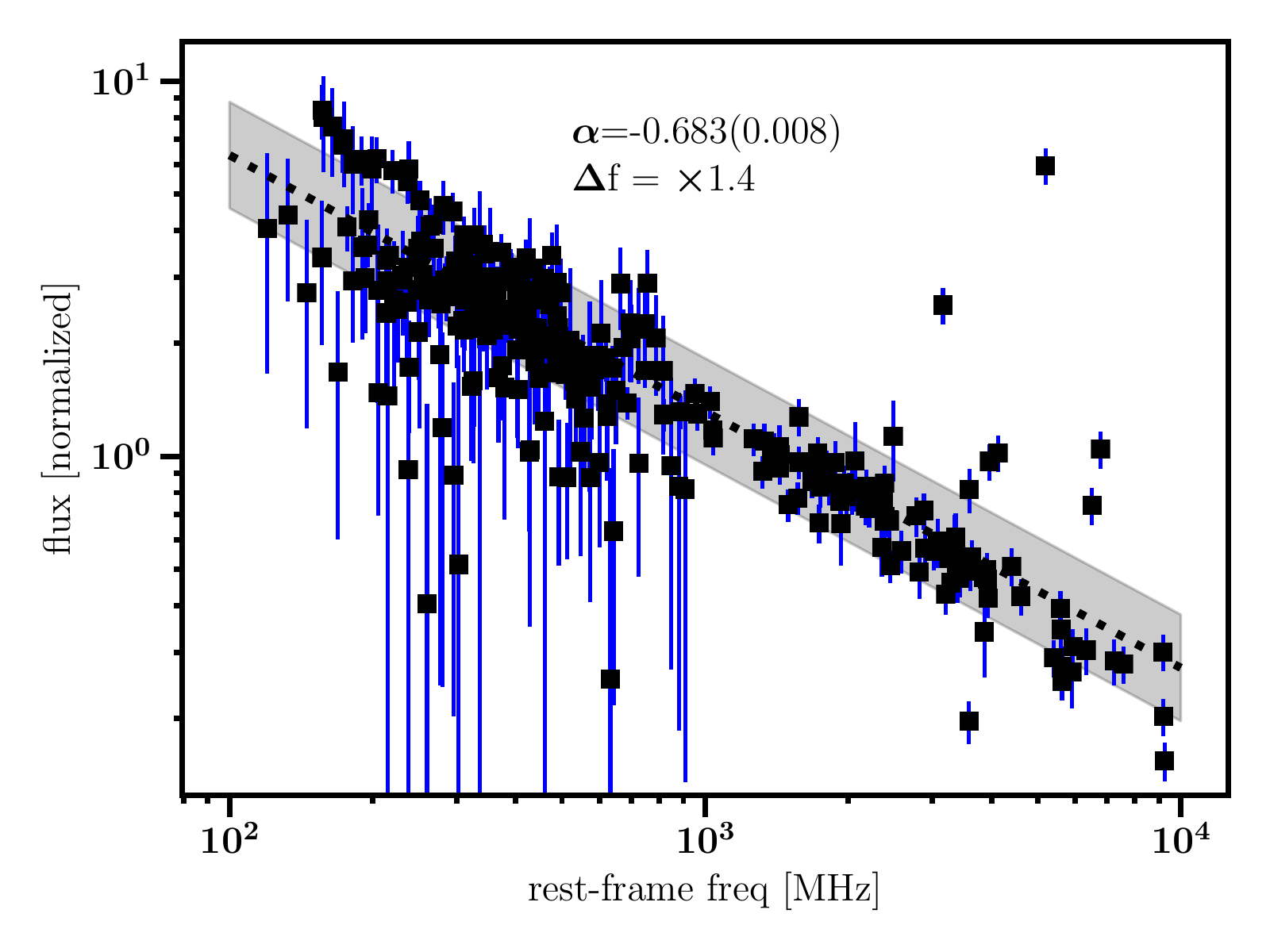}
	\caption{Rest-frame radio SED of the HzRG candidates with available spectroscopic or photometric redshift measurements. The SED is normalised at 1.4\,GHz. The best-fit (dotted line) presents a spectral index of $-0.683\pm0.008$, with an overall population dispersion of a factor of 1.4 (grey region).}
	\label{fig:radiosedrf}
	\end{figure}

	Source 25 ($z_{sp}=1.24882$) shows a rising SED at radio wavelengths ($\alpha=1.3\pm0.1$), hinting at a GHz-peaked spectrum source (peak at $\nu_{\rm rest}>5\,$GHz) revealing the early evolutionary stages of a jet or a ``frustrated'' one \citep[see][and references therein]{callingham17}. Since the SED is unconstrained, it is unclear whether the emission by ACA is synchrotron or not. Adopting the flux observed at 2.3\,GHz and assuming a spectral index of $-0.683$ (Fig.~\ref{fig:radiosedrf}), the synchrotron emission at 233\,GHz should be at least 0.25\,mJy. Since this estimate has an uncertainty of a factor of 1.4, it is $<3\sigma$ away from the observed 233\,GHz flux. It is therefore considered that this source also shows significant contribution from synchrotron emission. The same reasoning can be applied to the flat-spectrum sources 10 ($\alpha=0.39\pm0.08$) and 24 ($\alpha=0.05\pm0.07$), where the synchrotron emission at 233\,GHz should be at least 0.05 and 0.24\,mJy, respectively, assuming a spectral index of $-0.683$ extrapolated from the 2.3\,GHz flux. As a result, we consider that the 233\,GHz emission of source 24 is likely non-thermal in nature even though it is not detected by ACA.

    % figure with radio SEDs including GLEAM data
	
	% figure with rest-frame radio SEDs

 	Figures~\ref{fig:radiosed} and \ref{fig:radiosedrf} and Tab.~\ref{tab:radiosed} show that, within the uncertainties, most of the sources with GLEAM detections show a clean power-law shaped spectrum spanning more than one order of magnitude in frequency, with spectral indices ranging from $-0.1$ to $-1$. The flatness of some of these radio SEDs compared to the classic synchrotron-like shapes of others together with the high luminosities of most of the sample (Fig.~\ref{fig:redshift}) clearly show that the emission at radio frequencies of the majority of the sample reported here is powered by an AGN. Such a mechanism is also shown to dominate (i.e., $>50\%$ flux contribution) down to millimetre wavelengths in seventeen of the sources.

%% file: tables/tab-radiosed.tex
\begin{table}%[!htb]
\centering
\begin{tabular}{rrrrr}
ID & S$^{obs}_{233}$ & S$^{fit}_{233}$ & $\alpha$ & Synchr. \\
& [mJy] & [mJy] & & \\
\hline\hline
\multicolumn{4}{c}{with GLEAM or TGSS}\\
\hline
3 & $<$0.46 & 0.11~(1.3) & -0.81$\pm$0.04 &   \\
4 & $<$0.48 & 3.07~(1.3) & -0.69$\pm$0.04 & \ding{52} \\
6 & 0.7$\pm$0.2 & 0.88~(1.3) & -0.63$\pm$0.05 & \ding{52} \\
7 & $<$0.39 & 0.80~(1.3) & -0.75$\pm$0.04 & \ding{52} \\
8 & 0.6$\pm$0.2 & 1.06~(1.3) & -0.70$\pm$0.04 & \ding{52} \\
9 & 3.5$\pm$0.2 & 17.2~(1.4) & -0.16$\pm$0.07 & \ding{52} \\
12 & $<$0.45 & 0.23~(1.3) & -0.73$\pm$0.04 & \ding{52} \\
18 & 1.2$\pm$0.2 & 0.36~(1.3) & -0.81$\pm$0.05 &   \\
22 & 1.3$\pm$0.2 & 11.6~(1.3) & -0.58$\pm$0.04 & \ding{52} \\
23 & $<$0.52 & 0.35~(1.3) & -0.72$\pm$0.04 & \ding{52} \\
26 & $<$0.57 & 0.28~(1.4) & -0.80$\pm$0.05 & \ding{52} \\
29 & $<$0.48 & 0.37~(1.3) & -0.74$\pm$0.04 & \ding{52} \\
31 & 0.5$\pm$0.2 & 0.31~(1.2) & -0.88$\pm$0.03 & \ding{52} \\
32 & 0.8$\pm$0.2 & 3.32~(1.2) & -0.76$\pm$0.04 & \ding{52} \\
\hline\hline
\multicolumn{4}{c}{without GLEAM/TGSS, with ACA}\\
\hline
11 & 0.5$\pm$0.2 & 0.040~(1.4) & (-0.68) & \ding{55} \\
14 & 0.9$\pm$0.1 & 0.04~(1.5) & -0.78$\pm$0.08 & \ding{55} \\
15 & 1.2$\pm$0.2 & 0.20~(1.4) & (-0.68) & \ding{55} \\
16 & 0.5$\pm$0.2 & 0.03~(1.6) & -0.95$\pm$0.08 & \ding{55} \\
20 & 2.4$\pm$0.1 & 0.04~(1.9) & -0.6$\pm$0.1 & \ding{55} \\
25 & 0.5$\pm$0.1 & 2135.42~(1.5) & 1.29$\pm$0.08 & \ding{52} \\
27 & 0.8$\pm$0.2 & 0.3~(1.6) & -0.63$\pm$0.08 & \ding{52} \\
35 & 1.8$\pm$0.2 & 0.048~(1.4) & (-0.68) & \ding{55} \\
36 & 1.0$\pm$0.1 & 0.081~(1.4) & (-0.68) & \ding{55} \\
\hline\hline
\multicolumn{4}{c}{without GLEAM, TGSS or ACA}\\
\hline
1 & $<$0.46 & 0.041~(1.4) & (-0.68) &   \\
2 & $<$0.50 & 0.012~(1.9) & -1.1$\pm$0.1 &   \\
5 & $<$0.46 & 0.25~(1.5) & -0.66$\pm$0.07 & \ding{52} \\
10 & $<$0.49 & 6~(1.6) & 0.29$\pm$0.08 &  \\
13 & $<$0.42 & 0.003~(1.6) & -1.38$\pm$0.09 &   \\
17 & $<$0.48 & 0.07~(1.7) & -0.5$\pm$1.0 &   \\
19 & $<$0.50 & 0.2~(2.1) & -0.7$\pm$0.1 & \ding{52} \\
21 & $<$0.51 & 0.013~(1.7) & -0.9$\pm$1.0 &   \\
24 & $<$0.46 & 7~(1.5) & 0.05$\pm$0.07 & \ding{52} \\
28 & $<$0.53 & 0.076~(1.4) & (-0.68) &   \\
30 & $<$0.43 & 0.04~(1.6) & -0.69$\pm$0.09 &   \\
33 & $<$0.48 & 0.03~(1.6) & -0.78$\pm$0.09 &   \\
34 & $<$0.55 & 0.071~(1.4) & (-0.68) &   \\
\hline
\end{tabular}
\vspace{-0.3cm}
\caption{Observed 233\,GHz fluxes (S$^{obs}_{233}$) and fluxes extrapolated from the radio SED fit at the same frequency (S$^{fit}_{233}$) assuming a spectral shape of the form $S_\nu\propto\nu^\alpha$. The observed upper-limits correspond to the $3\sigma$ level measured on the images. The value in parenthesis in the third column indicates the error (see text for details). The spectral index, $\alpha$, reported in the fourth column in parenthesis indicates the availability of up to two radio data points only. In such cases, a value of $-0.68$ was adopted with population deviation factor of 1.4 (Fig.~\ref{fig:radiosedrf}). The right-most column indicates whether the observed signal is considered to be dominated by synchrotron emission, as described in the text. The top, middle and bottom groups list ACA sources with a GLEAM or TGSS counterparts, ACA sources with neither GLEAM nor TGSS counterparts and sources with neither ACA nor GLEAM nor TGSS detections.
}
\label{tab:radiosed}
\end{table}

%% file: sections/specline.tex
    The native spectral resolution enabled by the ACA observations described in Sec.~\ref{sec:acaobs} is $\sim$20\,km/s, enough to resolve typical widths of emission lines from the most abundant gas species in starburst or AGN host galaxies. With an instantaneous spectral coverage of 7.5\,GHz ($\sim10,000$\,km/s) it is likely that some of the 36 observed sources will show emission lines falling serendipitously into observed spectral range.
    
    In order to perform an unbiased identification of possible line emission features in the ACA cubes, we used the Source Finding Application (SoFiA
    % THIS URL MIGHT CAUSE A COMPILATION ERROR
    \footnote{https://github.com/SoFiA-Admin/SoFiA};
    \citealt{serra15}). SoFiA provides a selection of several algorithms for the search of emission (lines) in data (cubes) as well as flexibility in smoothing, spatially and spectrally, the three-dimensional cubes.
    
    In an effort to avoid biases introduced by averaging pixels in the image plane, and to check for consistency in feature detection, each ACA cube was fed into the code with three different spectral resolutions (20, 60 and 80~km~s$^{-1}$). We note that these were produced directly with the {\sc tclean} task in {\sc casa}. The cubes were also inverted (i.e., multiplied by $-1$) in a separate run as a means to assess the reliability of the positive features found in the original cubes (assuming all negative features are spurious). SoFiA was fed with the continuum-subtracted and primary-beam uncorrected cubes. The continuum was removed in the image plane assuming a point-like source at the centroid given by the continuum map. Two models, a constant flux density and a power-law, were independently fit to the continuum on all four spectral windows, showing no significant change in the results regarding line identification or flux over continuum estimate.
    
    The source finding adopted a circular Gaussian smoothing kernel with widths of 3, 6 and 9\arcsec\ in the spatial scale, and 60, 120, 180, 240, 300, and 360\,km/s in the spectral axis. The detection threshold on the frequency-integrated maps (moment 0) was set to 5, assuming a noise equal to the median absolute deviation (computed using negative and positive pixels). A noise normalisation was adopted along the spectral axis. Additionally, a line detection is only considered as such if it has a minimum size of 3\arcsec\ in the spatial scale, and a minimum width of 2 channels (i.e., 40, 120, and 160\,km/s, depending on the run). Detections in non-contiguous channels or pixels were merged into sources if within 4\arcsec\ and 3 channels.

\begin{figure*}
		\includegraphics[width=0.28\textwidth]{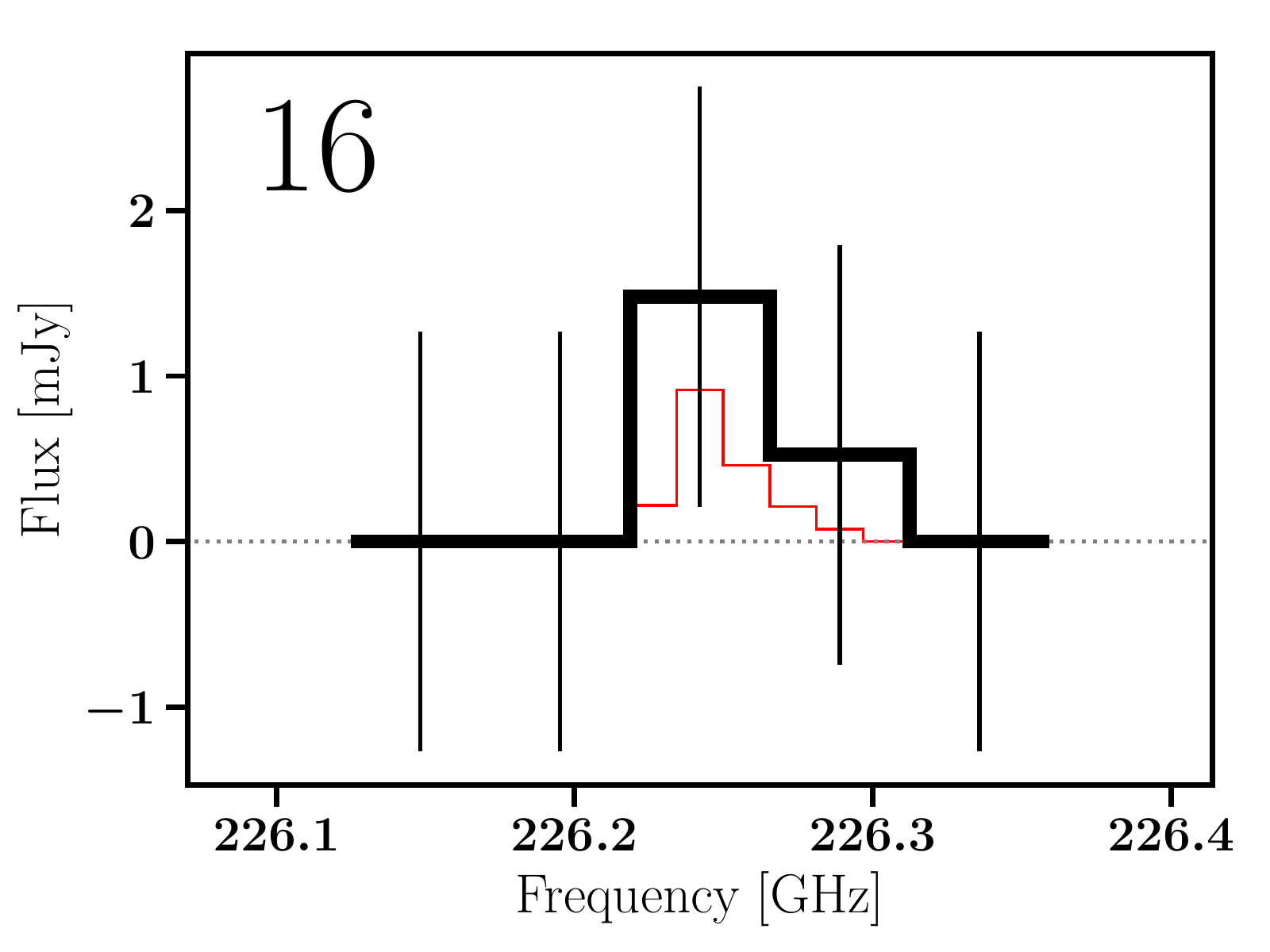}
		\includegraphics[width=0.21\textwidth]{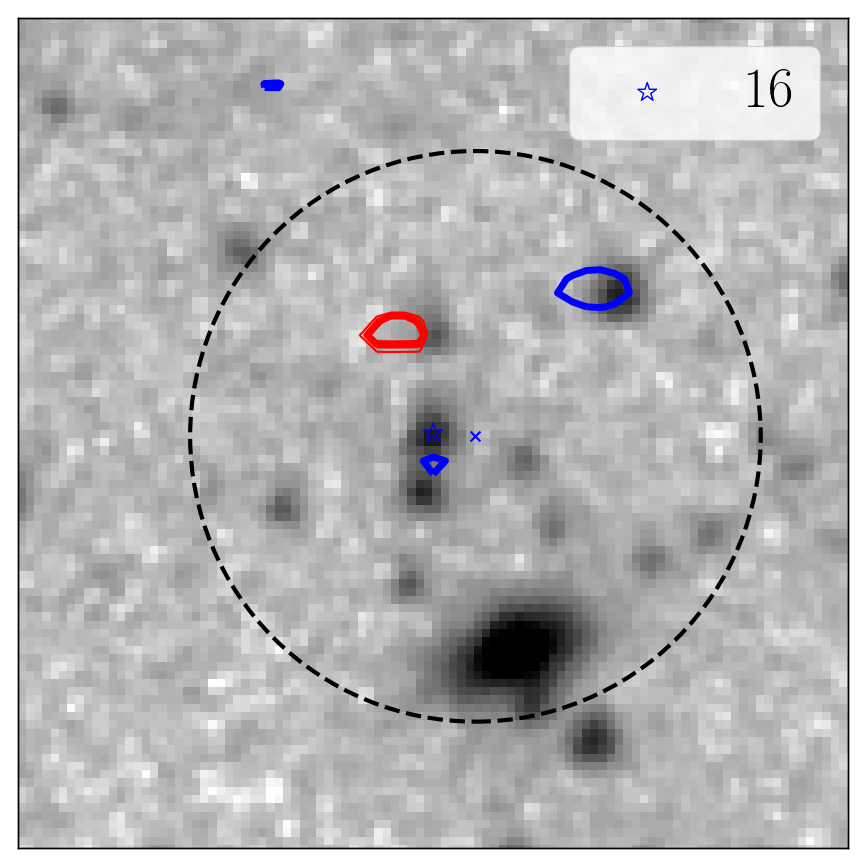}
		\hfill
		\includegraphics[width=0.28\textwidth]{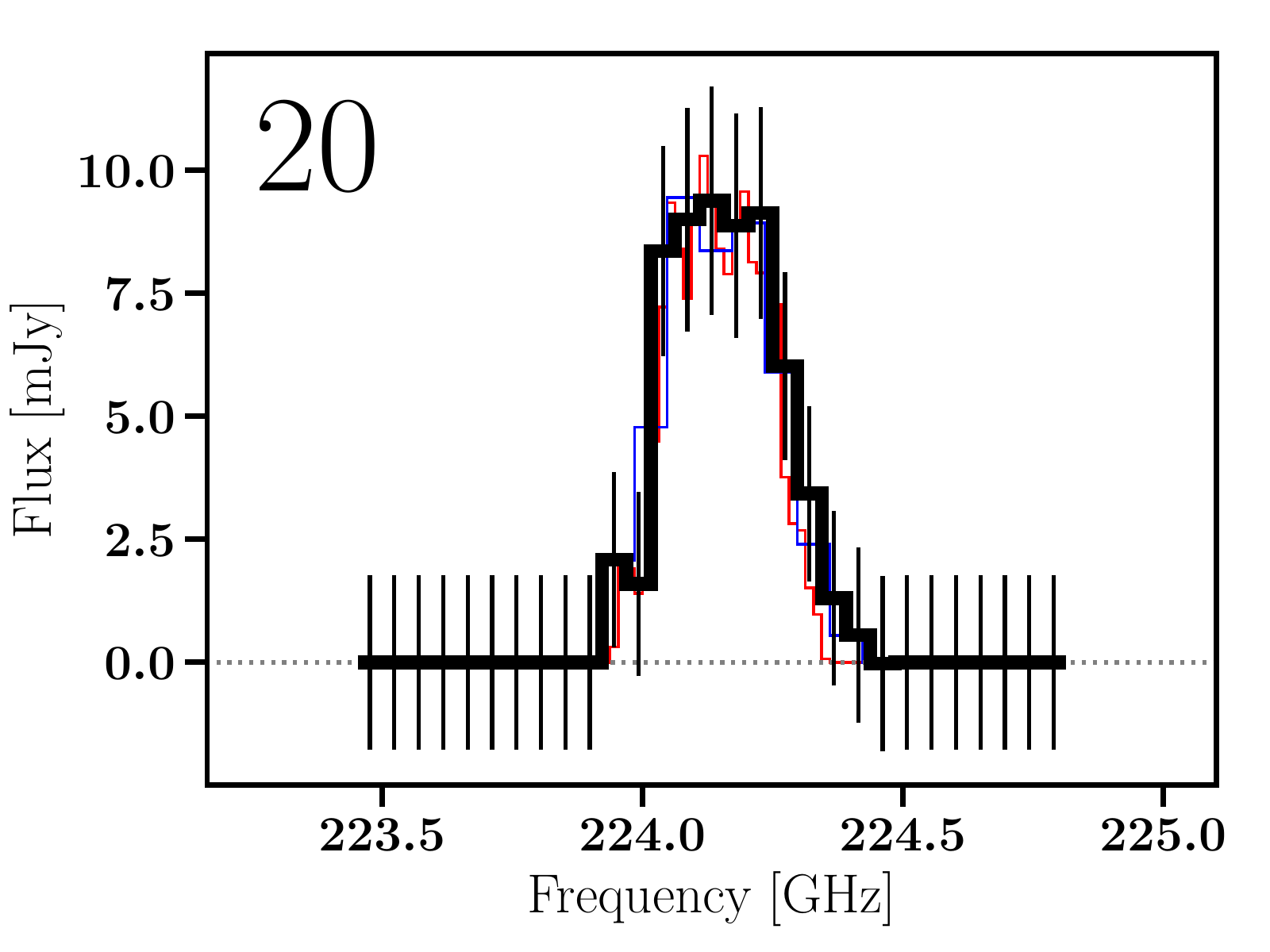}
		\includegraphics[width=0.21\textwidth]{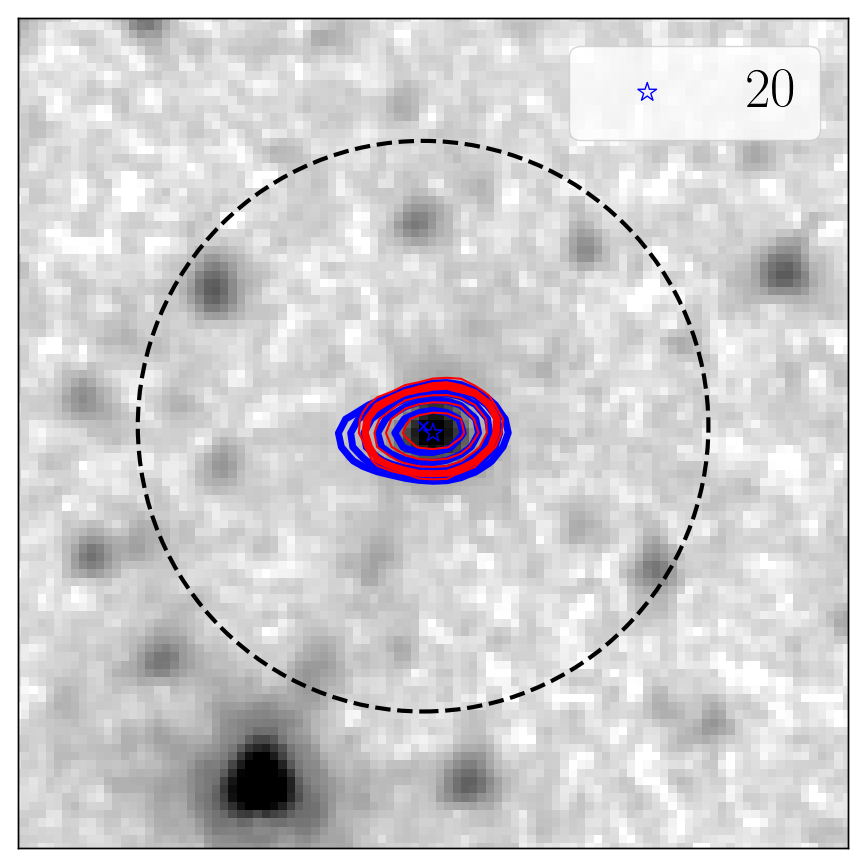}
		\hfill
		\includegraphics[width=0.28\textwidth]{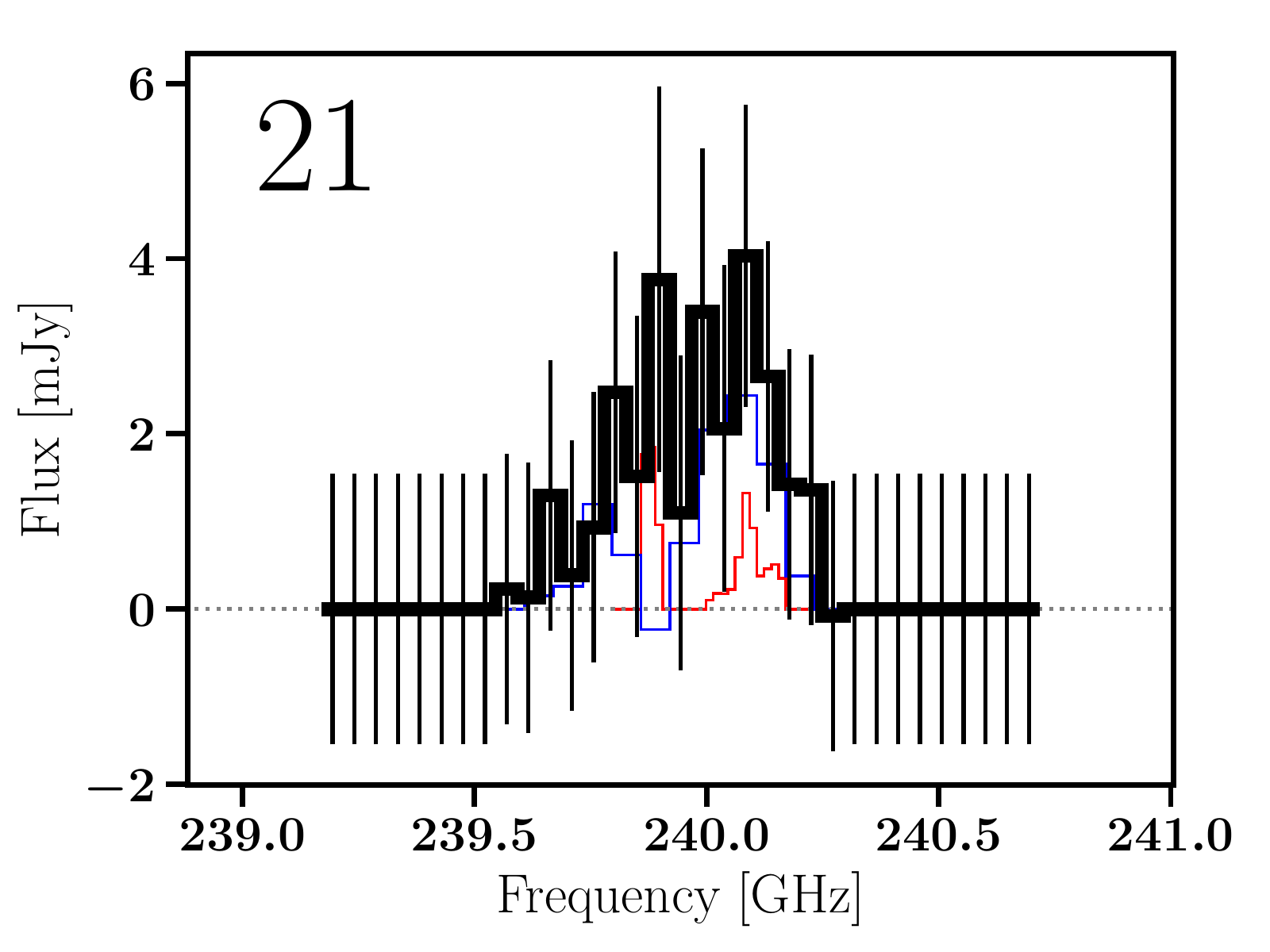}
		\includegraphics[width=0.21\textwidth]{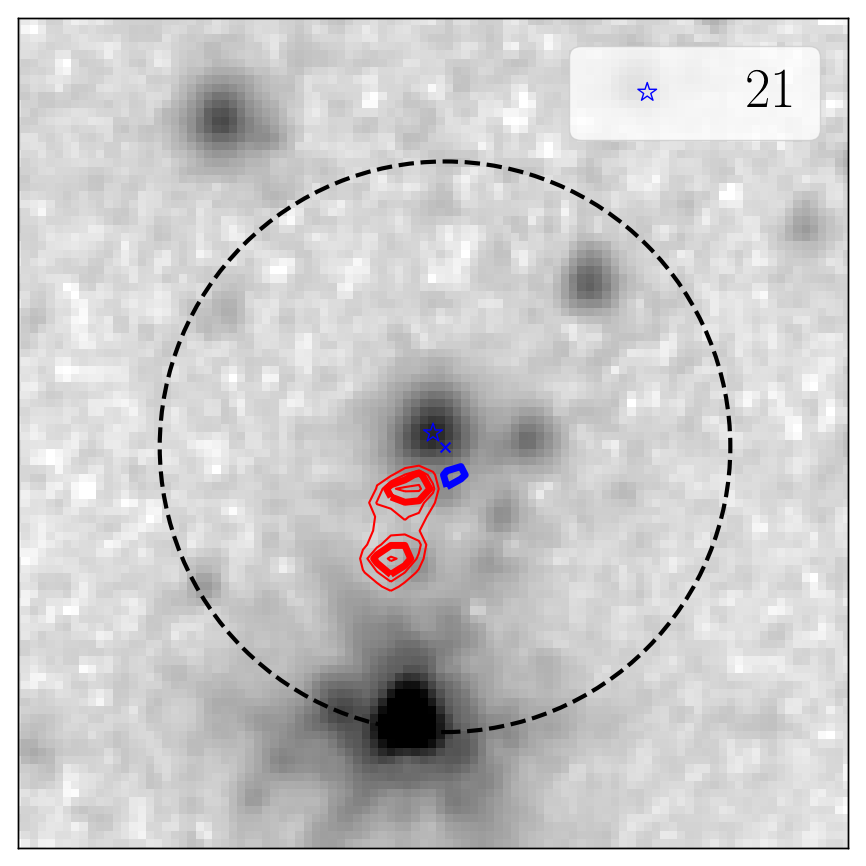}
		\hfill
		\includegraphics[width=0.28\textwidth]{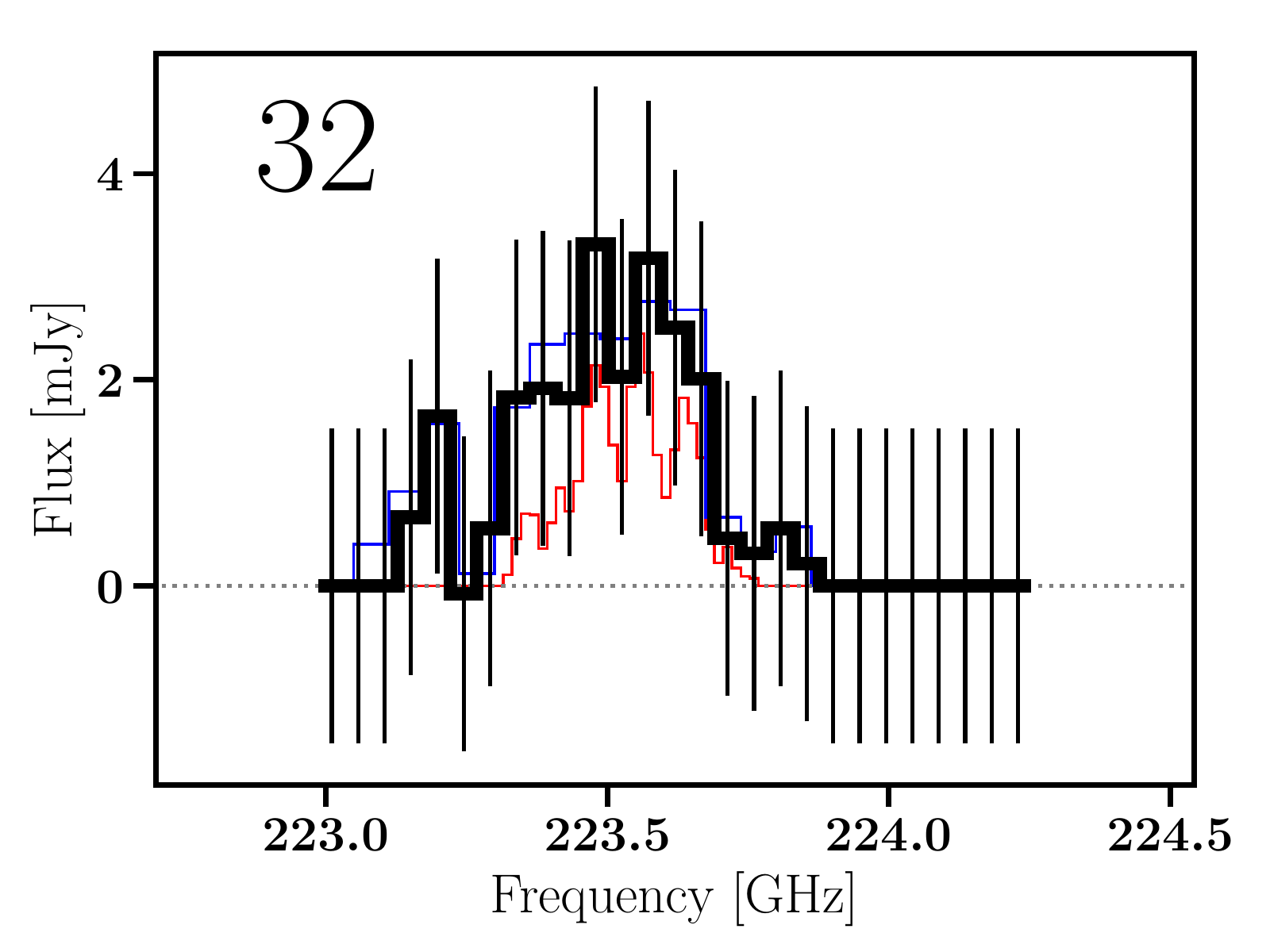}
		\includegraphics[width=0.21\textwidth]{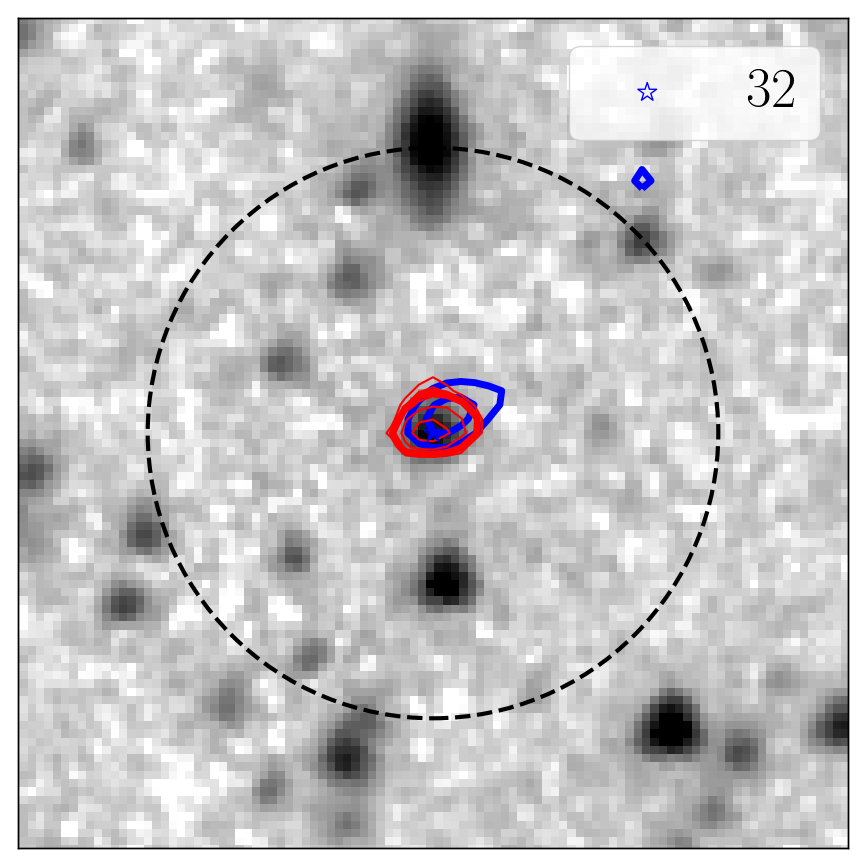}
		\caption{The spectral lines found in an unbiased SoFiA search. For each source, each (horizontal) pair of panels shows the identified spectrum on the left and the velocity integrated map (moment 0) on the right as red contours overlaid on the 3.6\,$\mu$m imaging. The latter also displays the Band~6 emission as blue contours (contour levels as in Fig.\ref{fig:aca1}). For consistency with the adopted SoFiA setup, the moment-0 contours are at 3, 4.2, 6, 8.5, 12 times the Median Absolute Deviation (MAD), with the thickest of the lines marking 5$\times$MAD. The error bars in the spectra show the error considering the channel's {\sc rms}. The spectra found in the 20~km~s$^{-1}$ raw spectral resolution (red line) or in the 80~km~s$^{-1}$ (blue) cubes are overlaid. The red-spectra show systematically fainter fluxes since the noise per 20~km~s$^{-1}$ channel is higher, resulting in more signal being diluted within the noise level, hence not recovered by SoFiA.}
		\label{fig:sofia}
	\end{figure*}

    Figure~\ref{fig:sofia} shows the four emission lines identified by SoFiA in the 30\,km/s cubes. The reported lines are those which, in a given spectral smoothing, showed a velocity integrated signal more significant than any of the negative spurious ones identified among all the inverted cubes at a given spectral smoothing. Comments on the individual line detections are given below:
    \begin{itemize}
        \item Source 16: A faint line is identified in the raw-resolution cube as well as in the 3-channel smoothed cube at 226.254$\pm$0.007\,GHz ($S_\nu\Delta\nu=0.1\pm0.1$). It is offset by $\sim1.5\arcsec$ from an IR source in the North-West quadrant, hence likely associated to it, and not the radio-counterpart. \citet{pforr19} reports a photometric redshift of $z_{ph}=1.21_{-0.06}^{+0.01}$ for this source. Within the $1\sigma$ intervals, the candidate species (and respective transitions) are atomic carbon ([CI]~${\rm ^3P_1-^3P_0}$, $\nu_{\rm RF}=492.161\,$GHz) at $z=1.17526\pm0.00007$ or carbon monosulfite, CS(10-9) ($\nu_{\rm RF}=489.751\,$GHz) at $z=1.16461\pm0.00007$. The latter transition has been detected in the past at $z=1$ even though in a gravitationally lensed galaxy \citep{messias14}. Another viable solution (2.9$\sigma$) is the detection of carbon monoxide, CO(4-3) ($\nu_{\rm RF}=461.041\,$GHz) at $z=1.03771\pm0.00006$.
        \item Source 20: the line detected at 224.156$\pm$0.003\,GHz is the brightest of the lines found serendipitously, with a velocity-integrated flux of 3.7$\pm$0.4~Jy~km/s and peaking at 9$\pm$2\,mJy (FWHM=380$\pm$10\,km/s). It is clearly co-located with the Band~6 continuum, which is associated to the dust emission in the host galaxy (Tab.\ref{tab:radiosed}). Based on the spectroscopic redshift of the source at $z_{\rm spec}=1.567$, this line is identified as the CO(5-4) transition ($\nu_{\rm RF}=576.268\,$GHz, $z_{\rm spec}=1.57083\pm0.00003$).
        \item Source 21: the feature is clearly extended and, despite not being co-located with the IR counterpart (nor the neighbouring mm-continuum detection), it might still be associated with it under the assumption of a jet-like or outflow component outside the host galaxy. This galaxy's spectroscopic redshift at 0.225 implies that the feature detected at 239.972$\pm$0.005\,GHz corresponds to the CS(6-5) transition ($\nu_{\rm RF}=293.912\,$GHz, $z_{\rm spec}=0.22478\pm0.00003$) with a velocity-integrated flux of 1.6$\pm$0.4~Jy~km/s and peaking at 4$\pm$2\,mJy (FWHM=460$\pm$40\,km/s). The morphology indicates that the emission extends about 20\,kpc out of the galaxy, yet SoFiA finds emission even co-located with the IR host, albeit at low SNR ($1-2\sigma$).
        \item Source 32: Given the redshift of the source at $z_{\rm spec}=1.58049$, the line detected at 223.490$\pm$0.005\,GHz corresponds to the CO(5-4) transition ($\nu_{\rm RF}=576.268\,$GHz, $z_{\rm spec}=1.57850\pm0.00006$) with a velocity-integrated flux of 1.4$\pm$0.4~Jy~km/s and peaking at 3$\pm$1\,mJy (FWHM=460$\pm$30\,km/s). We note that there is a significant offset between the centroid of the line and the continuum (1.5$\pm$0.8~arcsec), with the former being more aligned with the IR detection. This is compatible with the 233\,GHz detection being emission from the jet, in agreement with the results in Tab.~\ref{tab:radiosed}. We also note that this is the most radio-luminous source in the sample with $L_{\rm 1.4~GHz}=2.2\pm0.2\times10^{27}\,$W/Hz.
    \end{itemize}
    
    We do note that given the redshift of source~1 ($z_{sp}=1.3825$), a redshifted CO(5-4) transition (241.875\,GHz) is covered by the observed spectral coverage of one of the spectral-windows (241--243\,GHz). However, no significant continuum emission is detected (Fig.~\ref{fig:irmm}), nor does SoFiA report a line-feature detection toward this source.
    
    % specs and moments of SoFiA-identified lines

%% file: sections/szeffect.tex
\begin{figure*}
		\includegraphics[width=0.24\textwidth]{figs/acamaps/es1-020.png}  %  zsp=1.5670
		\includegraphics[width=0.24\textwidth]{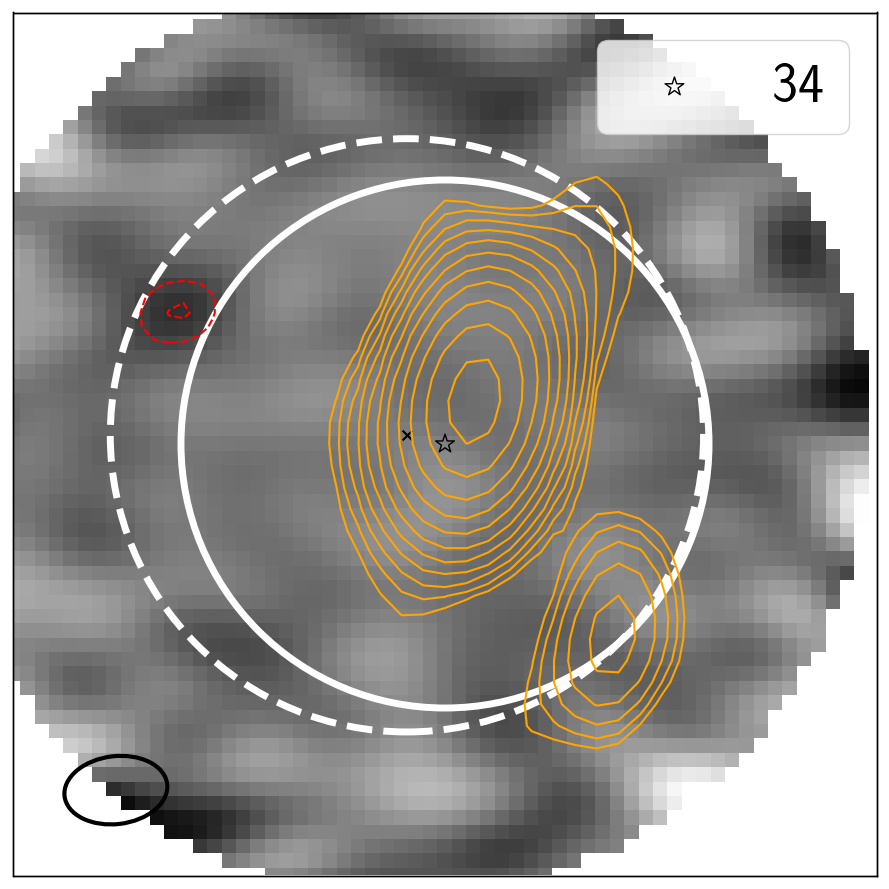}  %  zsp=0.9700
		\includegraphics[width=0.24\textwidth]{figs/acamaps/es1-036.png}  %  zph=4.19
		\includegraphics[width=0.24\textwidth]{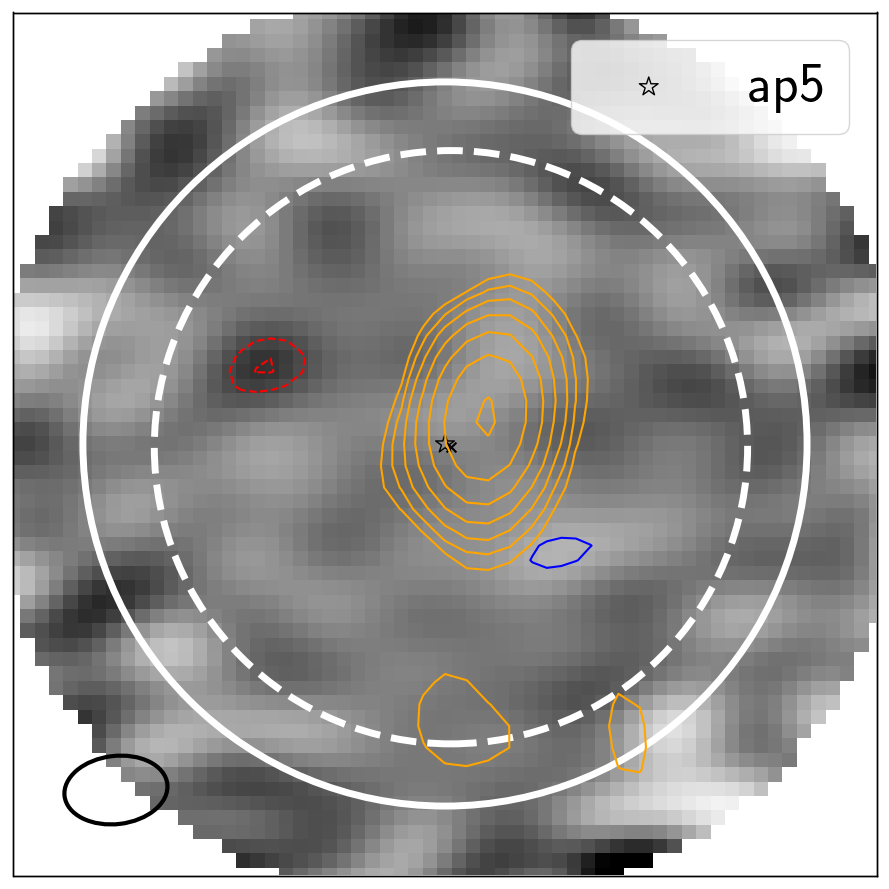}  %  zsp=0.4532
		\caption{The same as in Fig.~\ref{fig:aca1}, but for sources where a neighbouring negative emission has been detected at $>4.2\sigma$. Source ap5 is further introduced in Appendix~\ref{app:sampdrop}.}
		\label{fig:sz}
	\end{figure*}

	Sources with IDs 20, 34, 36, and ap5\footnote{See Appendix~\ref{app:sampdrop}} each exhibit a significant ($>4\sigma$) negative feature in their ACA continuum maps (Fig.~\ref{fig:sz}). Such features in interferometric maps are usually associated with missing short baselines that would otherwise recover large-scale emission, with bad calibration, or with strong side-lobes in the dirty beam. However, real negative features can occur e.g. as secondary imprints in a large-scale background. These are expected in the mm spectral range when photons from the Cosmic Microwave Background (CMB) interact with hot electron clouds, scattering a small percentage of the low-energy photons to higher energies \citep[the so-called Sunyaev-Zel'dovich effect, SZE; see][]{sunyaev72}. There are different components of the SZE that result from distinct velocity constituents of the scattering electrons. The thermal component (tSZE) is due to the random thermal velocities of the scattering electrons, and scales as the integrated pressure within the volume probed. The kinetic component (kSZE; \citealt{sunyaev80}) is due to the bulk velocity of the intra-cluster gas or outflows with respect to the restframe of the CMB \citep[for examples of observational signatures of the kSZE in individual observations, see][]{hand12,sayers13,adam17,lacy19,sayers19}.
	
	The tSZE is mostly seen associated with virialised galaxy clusters and is expected to induce a decrement in the CMB peaking at around $\approx$130\,GHz and an increment peaking at frequencies $\gtrsim$218\,GHz \citep[depending upon temperature; see e.g.][]{rephaeli95,Itoh1998,sazonov98,Chluba2012}. As a result, while observing at 233\,GHz, one should not expect a tSZE decrement. On the other hand, the intensity of the anisotropies caused by the kSZE is expected to peak in the 150--300\,GHz range. Ideally, in order to separate the contribution of both effects, one would need multi-frequency coverage. However, in this work, only a single frequency is available. Therefore, the observed negative features reported here, if real, might arise from a dominant contribution from kSZE.
	
	% figure with the ACA maps showing the negative dips
	
	\subsection{The significance of the negative peaks}
	
    In order to assess the likelihood that these features are real, we conducted a number of tests: (a) a re-imaging of the data after removing bright sources found in the field; (b) an assessment of the statistics directly from the visibilities; (c) simulation of the same observations bootstrapping the errors in each iteration (N$_{\rm iter}$=2000).
    
    Removing the central source in $uv$-space (option a), making use of three {\sc casa} tasks\footnote{The {\sc cl} task was used to produce a delta-function component model with the flux of the source of interest and at its location. This model was translated to visibilities with the {\sc ft} task and included in the {\sc model} column of the measurement set. The {\sc uvsub} task was then used to remove the source model from the {\sc data} column, and save the result into the {\sc corrected\_data} column, which was the one used for imaging.}, has no impact on the presence and significance of the negative feature in the fields of sources 20 and 36, indicating that these negative peaks are not related to negative side lobes. Separating the data either in time or frequency still reveals the features in a consistent manner. We note that these sources were observed in the same observing run as sources that show a clear central detection but without negative peaks in their vicinity (sources 15 and 35 were observed together with sources 20, 34, and ap5; source 9 together with source 36).
    
\begin{figure*}
	\includegraphics[width=0.8\textwidth]{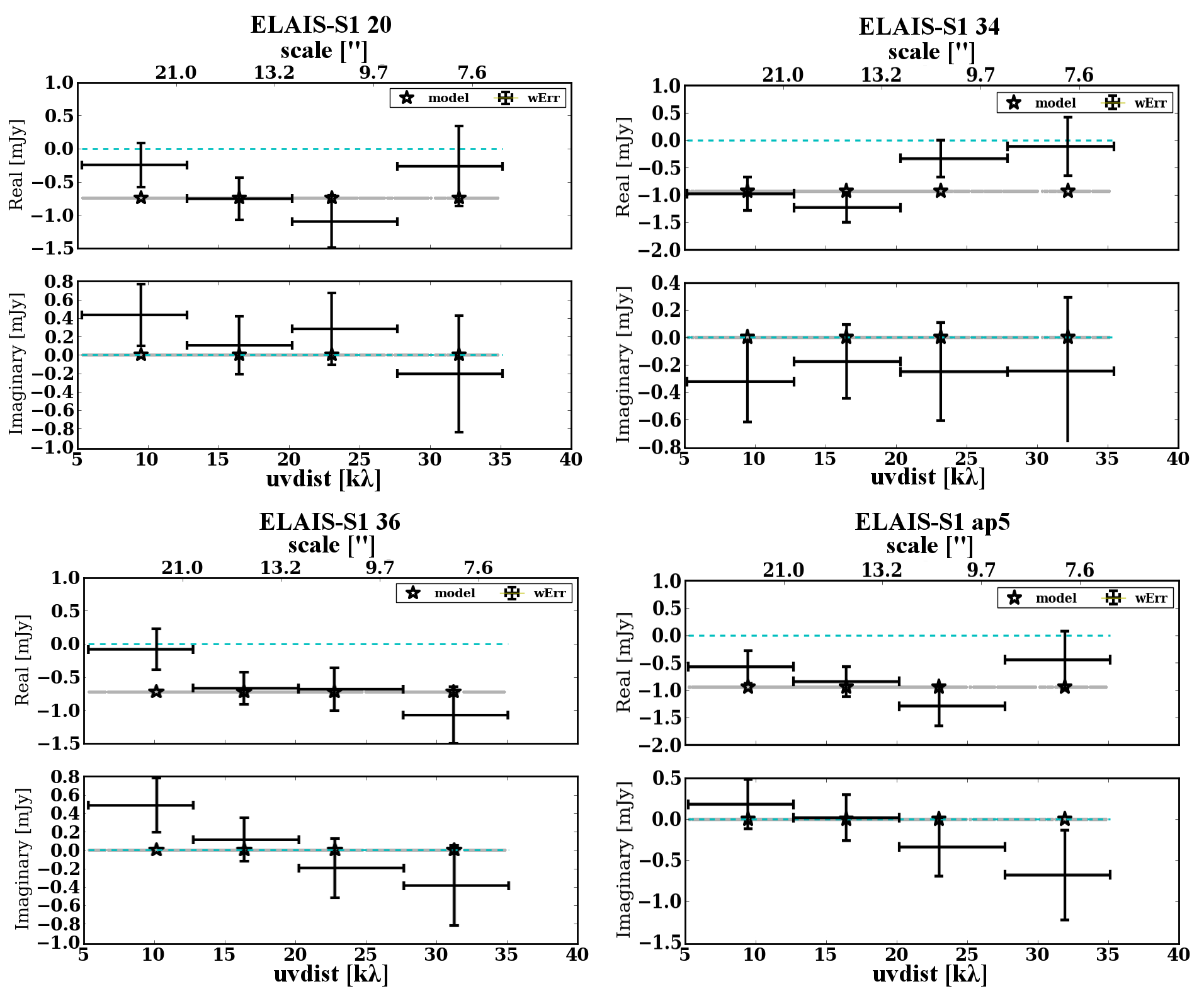}
	\caption{For each of the four sources showing a significant dip in the continuum map, the distribution of the visibilities' real and imaginary parts (upper and lower panels, respectively) are shown, with baseline length (in k$\lambda$, bottom axis, and sky arcseconds, top axis) averaged in four groups. The grey trends show the model for a point-like source at the dip position with a flux equal to that measured in the primary-beam-corrected continuum map. The star shows the average value of the model, which should be compared to the observed data (depicted as error bars). The dashed blue line corresponds to the zeroth level. For each source the visibilities were corrected to have the phase centre at the dip position. Bright sources, when present, were removed.
	}
	\label{fig:szvis}
	\end{figure*}

    Option (b) shows that there is indeed a significant decrement in the fields, that becomes clearer when visibilities are corrected to have the feature at the phase centre. This is shown in Fig.~\ref{fig:szvis}, where, in the case of the analysis done with the negative peak at the phase centre, the real part of the visibilities are significantly below the zeroth level. Given the significant amplitude drop beyond 20\,k$\lambda$ source 34 might, in fact, be resolved. We also made use of $uv$-{\sc multifit} \citep{martividal14} to fit the visibilities assuming point-like sources at the positions observed in the image plane. For sources 34 and ap5, the estimated fluxes range from $-0.8$ to $-1.1$\,mJy depending on whether we leave the upper flux bound free or limited to zero, respectively. Note that source 34 was also fit assuming an axisymmetric Gaussian morphology. This yields a flux of $-0.9\pm0.2$\,mJy and a size of 5\pp$\pm$4\pp, hence still consistent with being point-like. For the case of 20 and 36 where (positive) sources are found in the field, we fit all sources at once, and the negative peaks are found to have $-0.7$ and $-0.6$\,mJy, respectively. The uncertainties in these flux estimates are about 0.2\,mJy, but may increase if the provided initial guess is farther from the fluxes inferred from the image plane.
    
    % figure showing the Real/Imaginary parts with uv-dist
    
    Option (c) was addressed in several ways. We have adopted a Bootstrapping method to randomly draw from a Gaussian distribution new values for each observed $uv$ data point (i.e., visibility). We tested Gaussian randomisation of the visibility data of each source independently considering different visibility groups: all data (no grouping), per observation, per groups of equal-length baseline ranges, and per baseline range quantiles (i.e., ranges with equal number of integrations). The randomisation was done separately in the Real and Imaginary parts and by assuming the observed standard deviation of each group mentioned above, but we also tested in addition the consideration of the observed systematic deviations away from the expected average value of zero for a pure noise data set. Although such average deviations were not significant in each group, the location of the most significant spurious peaks became more spatially structured. We adopted the latter approach when using the observations targeting sources 34, 36 and ap5 (the brightness of source 20 prevented a proper test of the data standard deviation and systematics). The data was assumed to comprise pure noise, hence the features seen in the image were not removed from the visibilities.
    
\begin{figure}
	\includegraphics[width=0.5\textwidth]{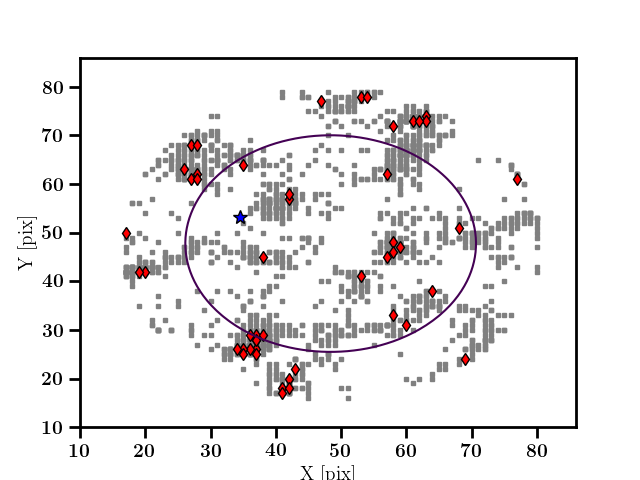}
	\includegraphics[width=0.5\textwidth]{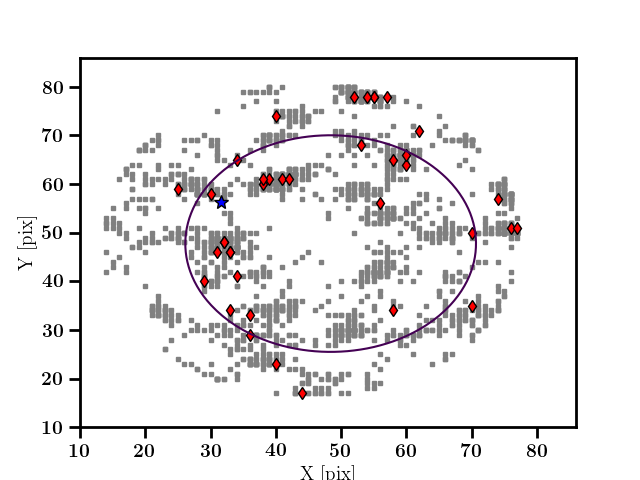}
	\includegraphics[width=0.5\textwidth]{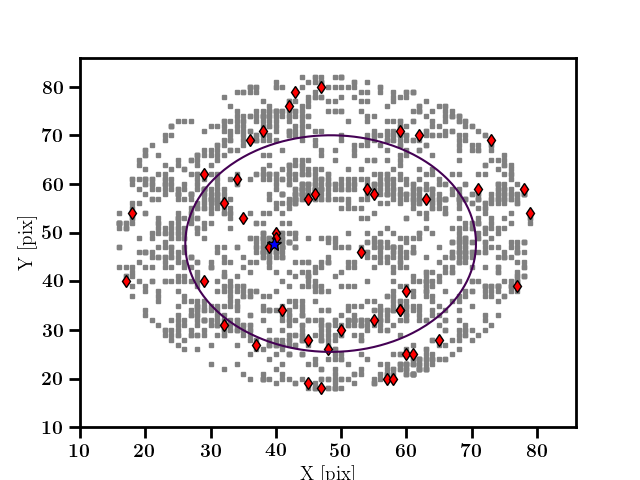}
	\caption{The distribution of the most significant negative dip in each of the 2000 simulated images (grey squares), while considering both data noise and systematics. The simulated features more significant than $-4\sigma$ are shown as red diamonds, while the location of the $-4.2\sigma$ dips reported in Fig.~\ref{fig:sz} are indicated with blue stars. The upper panel considers the data set on source ap5, while the middle panel that on 34, and the bottom on 36. The bright central source in the source~20 map, prevented us from testing systematics, hence its absence from this figure.}
	\label{fig:sims}
	\end{figure}

    Figure~\ref{fig:sims} shows that the minima found in the simulations tended to be clustered in specific regions and that the strongest of such fluctuations ($>4\sigma$; red diamonds) indeed fall in such regions. Nevertheless, it is clear that they do not fall only on the north-east region as seen in the four sources reported. Overall, the analysis shows that the probability of having negative features more significant than 4 or 4.2$\sigma$ in these maps is of 2.5\% or 0.8\%, respectively, regardless of the data set we use. Conservatively, one can consider these percentages as lower-limits (e.g., due to non-Gaussian effects) since the incidence of negative features at $3-4.2\sigma$ was already found to be 17\% higher when compared to what was predicted (23 versus 27 found; Sec.~\ref{sec:b6flx}). Hence, one should correct the above probabilities of having negative features more significant than 4 or 4.2$\sigma$ to 2.9\% and 0.9\%, respectively. The latter means that, in a total of 42 observed sources (Tabs.~\ref{tab:samp} and \ref{atab:samp}), we would expect at most one negative feature at $\geq4.2\sigma$ to be spurious at the 2.5$\sigma$ level. However, from the observed location of the detected features (Fig.~\ref{fig:sz}) with respect to the simulated ones, we acknowledge that 36 (and maybe 34) may indeed be spurious, while ap5 remains a good candidate. Regarding the case of 36, we re-ran the simulations, but removing in advance the candidate dip from the visibilities. This guarantees that the systematic deviations in the simulations are not tracing a real signal. The results still show that the dip location falls in a critical region in the image. Finally, regarding the possibility of non-Gaussian effects  inducing these extreme features, we do note that the statistics relative to the incidence of spurious negative and positive detections at 3 to 4$\sigma$ does point otherwise when comparing the results of the reported simulations and the observations as discussed in Sec.~\ref{sec:b6flx}.
    
    % figure showing the simulated dips location

	\subsection{The Sunyaev-Zel'dovich effect as a physical explanation}
	
	As mentioned above, the most probable physical explanation, at the observed frequencies, is the kSZE.  We have checked in Atacama Cosmology Telescope \citep[ACT;][]{Naess2020,Hilton2021} maps for evidence of large scale tSZE imprints on the CMB. However, we find no significant features toward the four mentioned galaxies. As a result, we discard the possibility that these features are due to bulk motion of the inter-cluster gas associated with any cluster or group host for the galaxies. We also find no evidence for co-location of the negative features and the AGN jets as traced by the radio maps (except maybe in source 36 at a lower significance level). We thus consider that these features, if real, originate from isolated outflows induced by feedback from radiative-mode accretion onto a SMBH.
	
	Classically, the kSZE effect can be computed \citep[see e.g.][]{mroczkowski19} as:
    \begin{equation}
    \label{eq:ksz_flux1}
    \Delta I_\nu =  - I_0 ~ \frac{x^4 \expf{x}}{(\expf{x}-1)^2} ~ \tau_e ~ \beta_z
    \end{equation}
	where $I_0 = 2 (\kB \Tcmb)^3/(h c)^2 = 270.06~\rm MJy/sr$ is a normalisation factor for the Planck spectrum of the CMB, $\kB$ is Boltzmann's constant, $h$ is Planck's constant, $\Tcmb = 2.7255~\rm K$ is the temperature of the CMB \citep{fixsen09}, $c$ is the speed of light, $x \equiv h \nu / \kB \Tcmb \approx \nu/[56.79~\rm GHz]$ is the dimensionless frequency, $\tau_e = \sigT \int \Ne(l) dl$ is the optical depth of the cloud along the line of sight $\ell$, $\sigT$ is the Thomson cross-section, $\Ne$ is the electron number density and $\beta_z = v_z/c$ is the dimensionless velocity (normalised by the speed of light $c$ and positive for motion along the redshift vector $\vek{z}$). We note, as discussed in {\it Lee \& Chluba in prep.} that for relativistic velocities and non-thermal electron distributions, the spectrum of both the thermal and kinetic SZ effects can vary significantly at 233~GHz.
	
    As an exercise, let us consider the negative peak in the field of source ap5, as the results of the simulations imply that its detection is the most reliable of the four. This source is located at $z_{sp}=0.4532$ and has an integrated flux decrement of $S_{233\,GHz} \sim -1$\,mJy at 233\,GHz. As previously discussed, we consider the kSZE to be the sole contributor to the 233\,GHz signal and assume that no relativistic corrections are required. From Eq. \eqref{eq:ksz_flux1}, the flux density due to the kSZE can be computed as:
    \begin{eqnarray}
    \label{eq:ksz_flux2}
    \Delta S_\nu &=& \int \Delta I_\nu d \Omega\\
        &=&  - I_0 \frac{x^4 \expf{x}}{(\expf{x}-1)^2} 
        {\sigT} \int\!\int  \Ne \beta_z \,\id \ell \,\id \Omega\\
        &=& - I_0 \frac{x^4 \expf{x}}{(\expf{x}-1)^2} \, \beta_z \frac{\sigT \langle \Ne \rangle V}{d_A^2}.
    \end{eqnarray}
    where $d_A$ is the angular diameter distance to the source, $V$ is the volume, $\langle \Ne \rangle$ is mean electron density, and all other variables are defined as above for Eq.~\ref{eq:ksz_flux1}.
    Assuming an electron weighting factor $\mu_e=1.17$ typical of low metallicity, fully ionised gas (see \citealt{andersGrevesse89}; for an example calculation of $\mu_e$, see \citealt{mroczkowski09} or \citealt{Adam2020}), the kSZ flux density scales simply as the product of gas mass ($M_{\rm gas} = \mu_e m_p \langle \Ne \rangle V$, where $m_p$ is the proton rest mass) times the line of sight velocity ($\beta_z$). 
    
    For the astrophysical properties of the putative kSZE source, let us assume similar properties as those reported in \citet{lacy19} toward the quasar HE~0515-4414 ($z=1.71$). Therein, the combined scenario of having both tSZE and kSZE at play in order to explain the observed decrement at 140\,GHz holds a cloud speed of -510\,km/s and $\langle \Ne \rangle \sim 0.16\,\rm cm^{-3}$, implying a total gas mass of $7\times10^{10}\rm M_\odot$ within a 66\,kpc feature. Here, we consider a source no larger than $5\arcsec \approx 30\,$kpc at $z_{sp}=0.4532$. Assuming the same cloud velocity and an observed flux of $\sim$1\,mJy toward ap5, the implied electron density is $\sim25\,\rm cm^{-3}$, more than two orders of magnitude higher than in HE~0515-4414. However, for the assumed cloud size, it implies a total gas mass comparable to $10^{13}\rm M_\odot$, which is highly implausible since such high masses are typical of galaxy clusters \cite[e.g][]{Pratt2019}. Inversely, assuming $\langle \Ne \rangle=0.16\,\rm cm^{-3}$ (implying a total mass for the gas of $\sim6\times10^{10}\rm M_\odot$), the computed velocity would have to be approximately 78,000\,km s$^{-1}$ ($0.26c$), which implies that relativistic corrections to the kSZ effect may need to be considered \citep[][{\it Lee \& Chluba in prep.}]{sazonov98}. Reducing the cloud size (which is supported by the fact that the feature appears unresolved) does not change the need for extreme conditions, since the effect scales with gas mass and $\beta_z$ (see Eq.\ \eqref{eq:ksz_flux2}). For instance, assuming a source size of 1\arcsec\ (5.8\,kpc at $z_{sp}=0.4532$), a cloud at a velocity of about -1700\,km s$^{-1}$ \citep[e.g.,][for the case of molecular clouds]{rupkeVeilleux11, sturm11, cicone14, gowardhan18} would produce a 1\,mJy decrement providing it had an electron density of approximately $1000\,\rm cm^{-3}$ (implying a total gas mass of $3\times10^{12}\rm M_\odot$), which is at the upper end of typical densities as estimated from optical spectra \citep[100--$1000\,\rm cm^{-3}$;][and references therein]{perna17,kakkad18}.
    
    At this point a proper characterisation of these features is required. On the one hand, a multi-frequency approach would allow one to constrain the contribution of the tSZE (the above assumption is that it is negligible at 233\,GHz) and the spectral shapes of both components. On the other hand, constraining the size of the cloud and/or its velocity and electron density would reduce the above uncertainties and assess whether relativistic corrections need to be considered.

%% file: sections/disc-selecCrit.tex
 Section~\ref{sec:selectioncriteria} describes the selection criteria listed in Tab.~\ref{tab:selectioncriteria}, while Tab.~\ref{tab:samp} indicates which of the criteria each source complies with. The colour selection criteria were chosen to identify HzRG candidates at $z>1.5$. Nevertheless, the adopted MIR flux-ratio cuts resulted in a significant fraction (23\%) of $z<0.8$ interlopers. This is effectively due to low-redshift dusty AGN with red MIR spectra mimicking the colours of high-redshift sources whose rest-frame spectrum at $<1.6\,\mu$m appears red due to dust absorption or old stellar population \citep{lacy07,messias12}. Overall, the LRS criterion is the least contaminated by $z<0.8$ objects in fractional terms (4 out of 29 sources or 14\% versus 20\% and 25\% for IFRS and USS samples, respectively). Interestingly this is the criterion that also provided the largest number of candidates to the sample (29 versus 20 and 8 from IFRS and USS samples, respectively). We note that a more conservative MIR flux-ratio cut of 1.3 applied to the IFRS sample would improve the efficiency by discarding three low-redshift interlopers, but one confirmed at $z_{sp}=1.58$ and three objects with unknown redshifts would have also been lost in the process. Finally, it is worth noting that all the adopted criteria require a detection at 5.8\,$\mu$m. The shallow nature of the underlying \textit{Spitzer} Wide-area IR Extragalactic survey \citep[SWIRE;][]{lonsdale03} in ELAIS-S1 and the lower sensitivity of the band with respect to the shorter-wavelength IRAC bands result in a bias towards the most luminous and rare sources, but also in an overall lower median redshift of the selected sample ($z_{\rm med}=1.4$; Fig.~\ref{fig:redshift}). 
 
 Adopting alternative colour cuts, such as the $K_s-[4.5]>0$ criterion proposed by \citet{messias12} that discards any non-AGN source at $z\lesssim1$, can improve the efficiency of the selection. With the extension of the near-to-mid-IR coverage of ELAIS-S1 beyond the surveyed regions by VIDEO \citep{jarvis13} and SERVS \citep{mauduit12} and by matching these surveys' depths via the DeepDrill \citep{lacy21} and VEILS \citep{honig17} programmes, such a NIR criterion will indeed be possible to apply to the radio-source sample in ELAIS-S1 in the future.
 
 We also highlight the fact that imposing a morphological criterion at radio wavelengths (e.g. point-like source selection) would most likely affect the fraction of high-redshift sources. Sources 18 at $z_{sp}=1.1644$ and 24 at $z_{sp}=1.9487$, for instance, have physical sizes beyond 400\,kpc. Furthermore, a preliminary $z_{ph}\sim1.3$ would also put source 22, a clear X-shaped source, in that category.
 
 Finally, we assess the selection effect of an IR flux cut such as that adopted by \citet{jarvis09}: $S_{\rm 3.6\mu m}<33\mu$Jy\footnote{We have correct up the flux cut of 30\,$\mu$m by a factor of 1.11 to reflect a total flux cut, since the initial cut was defined within a 1.9\arcsec\ aperture radius.}. Ten sources in our sample comply with this cut: three have no redshift estimate, one is at $z_{\rm ph}=0.56$, one at $z_{\rm ph}=1.81$ and the remaining five are found at $z\sim2-4$. Although this points to a contamination level of $\sim17\%$ by $z<0.8$ objects comparable to LRS, we note that sources such as 9, 15, 16, 24, or 27 at $z\sim2-3$ are brighter than the flux cut adopted by \citet{jarvis09}, tuned to $z>4$.

%% file: sections/disc-mmdetec.tex
 As reported in Tab.~\ref{tab:b6phot} (Sec.~\ref{sec:acaobs}), 11 (five) out of 36 central sources were detected at $>4.2\sigma$ ($3-4.2\sigma$) in the ACA maps, that reach a typical noise level of 0.15\,mJy. Out of the 16 central detections, 8 are of non-thermal origin (Sec.~\ref{sec:radphot}, Tab.~\ref{tab:radiosed}), leaving only 8 where ACA is detecting the dust emission from the host galaxy. This low detection fraction can be a result of higher dust temperatures, lower star-formation rates, lower dust masses, or a combination of the above. All three scenarios are consistent with the AGN nature of these sources, but a more careful analysis of their spectral energy distributions is required. However, this is beyond the scope of the present work.
 
 Adopting the relation between the mm continuum and the molecular gas content (M$_{\rm H_2}$) proposed by \citet[]{scoville16a,scoville16b} assuming a CO to H$_2$ conversion factor of $\alpha_{\rm CO} = 6.5\,{\rm M_\odot~/(K~km~s^{-1}~pc^2)}$, we find an average M$_{\rm H_2}=1.5-1.7\times10^{11} \,M_\odot$ for the nine detected sources with a redshift estimate (see Tabs.~\ref{tab:samp} and \ref{tab:radiosed}). The range in M$_{\rm H_2}$ reflects the range in dust temperatures between 25 and 35\,K used in the above study. The continuum flux values were corrected for the synchrotron emission contribution (Tab.~\ref{tab:radiosed}).
 
 We have also stacked the continuum maps of 11 undetected sources (no central detection above $3\sigma$) as described in Sec.\ref{sec:b6flx}, yielding a stacked signal of $0.16\pm0.05\,$mJy (3.5$\sigma$). This shows that deeper continuum surveys are required to directly detect these fainter sources. Note that we find a similar detection fraction ($\sim$50\%) among sources at $0.8<z<2$ and $z>2$ (7 out of 15, and 4 out of 8, respectively), indicating that increasing the efficiency of the selection criteria toward the identification of $z>2$ sources will still require deeper surveys ({\sc rms}$<$0.15\,mJy).
 
 We finally caution against the use of the mm spectral range to study the host galaxies of the most luminous radio-galaxies at least at $z<2$ (where most of our sample lies). At these redshifts the non-thermal component of the spectrum, typically decreasing with frequency, still dominates over the host's thermal emission which is increasing with frequency. This will gradually cease to apply with increasing redshifts and decreasing radio luminosities.
 
 In 12 of the maps we further report the detection of neighbouring emission (four at $>4.2\sigma$ and eight at $3-4.2\sigma$; Fig.~\ref{fig:aca1} and \ref{fig:aca2}, Tab.~\ref{tab:b6phot}). For a combined survey area of $36\times\pi(44\pp/2)^2=15\,$arcmin$^2$ and for a minimum source flux of 0.5\,mJy, one would expect between $10\pm1$ \citep{franco18} and $13\pm4$ \citep{umehata17} serendipitous detections of sources in our survey. This means that there is at most a $2\sigma$ excess in the neighbouring source-count in our sample with respect to the field at mm wavelengths. The clustering properties of the sample will be studied in detail in a forthcoming work.

%% file: sections/disc-linedetec.tex
 	In Sec.~\ref{sec:specline} we report emission features serendipitously detected in the data cubes. In three out of four cases we were able to identify the molecular species and the corresponding transition. 
 	%CO\,(5-4) - 605 at z=1.567 and 224.156$\pm$0.003\,GHz with a velocity-integrated flux of 3.7$\pm$0.4~Jy.km/s; 1119 at z=1.58049 and 223.490$\pm$0.005\,GHz with a velocity-integrated flux of 1.4$\pm$0.4~Jy.km/s
 	%lum=3.25e7*3.7*224.156**-2*11705.9**2*2.567**-3/1e10; err=lum*0.4/3.7
 	%lum=3.25e7*1.4*223.49**-2*11830.8**2*2.58049**-3/1e10; err=lum*0.4/1.4
 	
 	For the sources with IDs 20 and 32 toward which we detect CO(5-4) we estimate the molecular gas content in these systems. The measured fluxes imply line luminosities of (1.9$\pm$0.2) and (0.7$\pm$0.2)\,$\times10^{10}$~K.km/s.pc$^2$ for sources 20 and 32, respectively. Based on the quasar population average luminosity ratios between the J=1-0 and J=5-4 transitions of $L'_{J=5-4}/L'_{J=1-0}=0.3-0.7$ \citep{carilli13, kirkpatrick19}, and assuming a conservative range for the CO conversion factor of 0.5--1\,M$_\odot$/(K.km/s.pc$^2$) characteristic of luminous IR sources, we should expect molecular gas masses in the range of 1.4--6.5 and $0.5-2.4~\times10^{10}$\,M$_\odot$ for sources 20 and 32, respectively. The value for source 20 agrees with that derived by \citet{scoville16b} relation after correcting for different $\alpha_{\rm CO}$ values ($\sim7~\times10^{10}$\,M$_\odot$). These values place these two $z\sim1.5$ sources close to the knee of the molecular gas mass function at these redshifts \citep{decarli19}.
 	
 	With respect to source 21 towards which we detect CS, we note that this molecule is commonly assumed to be a dense gas tracer \citep[e.g.,][]{lada97,kelly15}. Specifically, the J=6-5 transition (${\rm %E_L=35~K,~~
 	E_U=49~K}$) has effective and critical densities of $\sim10^6$ and $\sim10^7$\,cm$^{-3}$, respectively \citep{wu10}. This is well within the cloud density regime ($n>10^{4-5}\,$cm$^{-3}$) which SF is restricted to \citep[][]{lada97,lada09,lada10}. The CS(6-5) emission toward source 21 extends to about 20\,kpc out of the galaxy in the South-East direction with a jet-like morphology. Due to its shape and location, it may resemble a case for molecular cloud formation induced by a jet. However, the radio imaging does not reveal a jet morphology down to its spatial resolution, even though low-SNR emission is seen in the opposite direction. An XMM-\textit{Newton} survey has covered this source (PI W.~Brandt) with a flux of $\sim2.5\times10^{-14}\,{\rm erg/s/cm}^2$ reported by the pipeline. We find no evidence for extended emission being detected, hence no cavity is visible to assess the jet-nature of the observed feature. Alternatively, the CS emission can result from excitation by shocks \citep{garciaBurillo14,viti14}, which would agree with the outflow scenario.

%% file: sections/disc-sze.tex
In Sec.~\ref{sec:sz} we discussed the reliability and possible physical interpretations of the $\sim 4\sigma$ detections of negative features around four observed sources.

Despite the fact that they all fall in the north-east quadrant, which hints at calibration artefacts, other cases in the sample observed in the same runs do not show such features, and the simulations carried out --- where we attempt to account for systematics --- indicate that at most one of these features would be spurious at the $2.5\sigma$ level. Purely based on location while comparing with the simulations results (Fig.~\ref{fig:sims}) it is shown that at least one source should be considered as highly reliable.

One tempting astrophysical explanation for the observed flux decrements is the kinetic Sunyaev-Zel'dovich effect (kSZE) caused by outflows moving along the line-of-sight away from the observer. With the lack of sufficient spectral information required to characterise these features, one can only discuss specific cases using reference values for the cloud velocity and electron density from the literature. However, only the most extreme of such values ($v>500$\,km~s$^{-1}$ and $\Ne>10\,$cm$^{-3}$) can produce the observed flux decrements, but still imply total gas masses of $10^{11}-10^{13}\,$M$_\odot$. These are considered to be implausible since they imply galactic- or cluster-like gas masses within galactic spatial scales. As a result, multi-frequency data are required, especially at longer wavelengths ($>2$\,mm), in order to separate the thermal and kinetic SZE components, and, in the process, confirm the contribution of the former at 233\,GHz (which was assumed negligible at these frequencies). The assessment of the cloud velocity will also reveal whether relativistic corrections are required.

%% file: sections/conclusions.tex
In this work we describe a survey conducted with the ACA at 1.3\,mm (Band 6 at 233\,GHz) targeting 36 HzRG candidates selected within 3.9\,deg$^2$ in the ELAIS-S1 field.

Three different selection criteria were considered (Sec.~\ref{sec:samp}), yielding a sample mostly falling in the $0.7<z<2.3$ range, showing rest-frame 1.4\,GHz luminosities of up to $10^{27}\,$W/Hz and with radio-emission sizes up to $\sim500\,$kpc. Given the usage of shallow $5.8\,\mu$m photometry in the criteria, we believe that the surveyed sample is biased towards lower redshifts and to the most extreme sources of the kind. We note that a simple IR colour-cut together with a cut in radio flux provides an 86\% reliability in the selection of sources in the first half of the Universe life ($z>0.8$). Among the latter, there is a 50\% detection rate by ACA down to {\sc rms}$\sim$0.15\,mJy (Sec.~\ref{sec:b6flx}). There is no significant change of this fraction with redshift, meaning that deeper mm surveys are required to directly detect a large fraction of the HzRG population ({\sc rms}<50\,$\mu$Jy). Note, however, that for a large fraction of the sources (8 out of 16 detected at $>0.5\,$mJy) the detected 1.3\,mm flux is non-thermal in nature, preventing the direct study of the host emission at these wavelengths (at least at $z\lesssim2$).

We show the potential of ACA as a survey machine of the continuum and gas properties of the most luminous HzRGs (Secs.~\ref{sec:specline} and  \ref{sec:disc}). This is exemplified by the serendipitous detection of CO(5-4) towards two galaxies at $z\sim1.6$, with derived gas masses $\sim10^{10}$\,M$_\odot$.  One of them shows evidence of a spatial offset (2$\sigma$) between the gas and the mm continuum emission, the latter resulting from the jet in the system.

Finally, we note the detection of four negative continuum features in the vicinity of four ACA maps (Sec.~\ref{sec:sz}). We discuss the nature of these features as induced by either calibration systematics or by gas clouds moving away from the observer, i.e., kinematic Sunyaev-Zel'dovich effect. While the former is plausible, we showed that the probability of occurrence in our survey is low. On the other hand, the kSZE on its own is found to require extreme gas conditions in order to induce these features. Future multi-frequency mm observations together with gas characterisation studies are required to provide a better understanding of these features, if they are real.

Forthcoming work making use of these data will focus on on-going and future multi-wavelength follow-up of some of the sources reported here in detail, as well as a revision of the photometric redshift estimates in a uniform manner accompanied by the analysis of the UV-to-mm SEDs, and finally an assessment of the galaxy environment in which the selected sources are found.

%% file: sections/app-offset.tex
Due to a mistake in the coordinates inserted in the ALMA Observing Tool, there is a noticeable offset (up to $\sim$1\,beam away from phase centre) between some of the targets and the corresponding observed phase centres (Tab.~\ref{tab:offset}). The imaging has been corrected to be centred at the source, but for record,  we report on the offset between the input phase-centre coordinates and those of the expected IRAC counterpart.

\begin{table*}
	\begin{tabular}{crrrrrccc}
		ID & RAir & DECir & RAin & DECin & dist & PA\\
		& [deg] & [deg] & [deg] & [deg] & [arcsec] & [deg]\\
		\hline
		1 & 9.409256 & -44.643998 & 9.408333 & -44.643611 & 2.7 & 301\\
		2 & 8.656027 & -44.644740 & 8.654167 & -44.644722 & 4.8 & 271\\
		3 & 9.577494 & -44.536039 & 9.579167 & -44.536111 & 4.3 & 93\\
		4 & 9.324575 & -44.503932 & 9.325000 & -44.503889 & 1.1 & 82\\
		5 & 8.348502 & -44.500012 & 8.348502 & -44.500012 & 0.0 & 0\\
		6 & 7.796753 & -44.482790 & 7.796753 & -44.482790 & 0.0 & 0\\
		7 & 9.926706 & -44.453807 & 9.925000 & -44.453889 & 4.4 & 266\\
		8 & 7.413837 & -44.388567 & 7.413837 & -44.388567 & 0.0 & 0\\
		9 & 8.205513 & -44.364041 & 8.204167 & -44.363889 & 3.5 & 279\\
		10 & 8.001535 & -44.283605 & 8.000000 & -44.283611 & 4.0 & 270\\
		11 & 9.476850 & -44.185312 & 9.475000 & -44.184722 & 5.2 & 294\\
		12 & 9.207525 & -44.150879 & 9.208333 & -44.150556 & 2.4 & 61\\
		13 & 8.904957 & -44.121771 & 8.908333 & -44.121944 & 8.7 & 94\\
		14 & 8.981902 & -44.043670 & 8.983333 & -44.043611 & 3.7 & 87\\
		15 & 9.330941 & -44.028734 & 9.329167 & -44.029167 & 4.8 & 251\\
		16 & 9.017850 & -43.929375 & 9.016667 & -43.929444 & 3.1 & 265\\
		17 & 9.667415 & -43.887932 & 9.666667 & -43.888056 & 2.0 & 257\\
		18 & 9.764433 & -43.829697 & 9.762500 & -43.829722 & 5.0 & 269\\
		19 & 7.464460 & -43.757749 & 7.462500 & -43.757778 & 5.1 & 269\\
		20 & 9.624725 & -43.748470 & 9.625000 & -43.748333 & 0.9 & 55\\
		21 & 9.337835 & -43.711113 & 9.337500 & -43.711389 & 1.3 & 221\\
		22 & 7.870282 & -43.689145 & 7.870282 & -43.689145 & 0.0 & 0\\
		23 & 8.635667 & -43.565454 & 8.637500 & -43.565556 & 4.8 & 94\\
		24 & 8.185767 & -43.412924 & 8.187500 & -43.413056 & 4.6 & 96\\
		25 & 9.285885 & -43.398072 & 9.287500 & -43.398056 & 4.2 & 89\\
		26 & 7.440267 & -43.363715 & 7.440267 & -43.363715 & 0.0 & 0\\
		27 & 9.441579 & -43.321825 & 9.441667 & -43.321667 & 0.6 & 22\\
		28 & 9.947444 & -43.291819 & 9.945833 & -43.291944 & 4.2 & 264\\
		29 & 8.230928 & -43.274059 & 8.233333 & -43.274167 & 6.3 & 94\\
		30 & 8.624639 & -43.224212 & 8.625000 & -43.224167 & 1.0 & 80\\
		31 & 7.797381 & -43.154622 & 7.797381 & -43.154622 & 0.0 & 0\\
		32 & 9.069656 & -43.159686 & 9.069656 & -43.159686 & 0.0 & 0\\
		33 & 7.969161 & -43.045682 & 7.970833 & -43.045556 & 4.4 & 84\\
		34 & 8.498995 & -42.964057 & 8.500000 & -42.963889 & 2.7 & 77\\
		35 & 7.722195 & -42.870914 & 7.720833 & -42.870833 & 3.6 & 275\\
		36 & 9.741389 & -42.867870 & 9.741667 & -42.867222 & 2.4 & 17\\
		\hline
		ap1 & 8.224183 & -44.490856 & 8.223996 & -44.490711 & 0.7 & 317\\
		ap2 & 9.633623 & -44.427915 & 9.633333 & -44.428056 & 0.9 & 236\\
		ap3 & 8.675558 & -44.224105 & 8.675000 & -44.224167 & 1.5 & 261\\
		ap4 & 9.324525 & -43.847568 & 9.324525 & -43.847568 & 0.0 & 0 \\
		ap5 & 9.643759 & -43.635349 & 9.643600 & -43.635411 & 0.5 & 242\\
		ap6 & 8.569710 & -43.513493 & 8.569710 & -43.513493 & 0.0 & 0\\
		\hline
	\end{tabular}
	\caption{The offset differences between input observed coordinates (RAin, DECin) and those of the expected IRAC counterpart (RAir, DECir). The position angle (PA) is measured in degrees clock-wise from North.}
	\label{tab:offset}
\end{table*}

%% file: sections/app-extraimgs.tex
\begin{figure*}
	\includegraphics[width=0.24\textwidth]{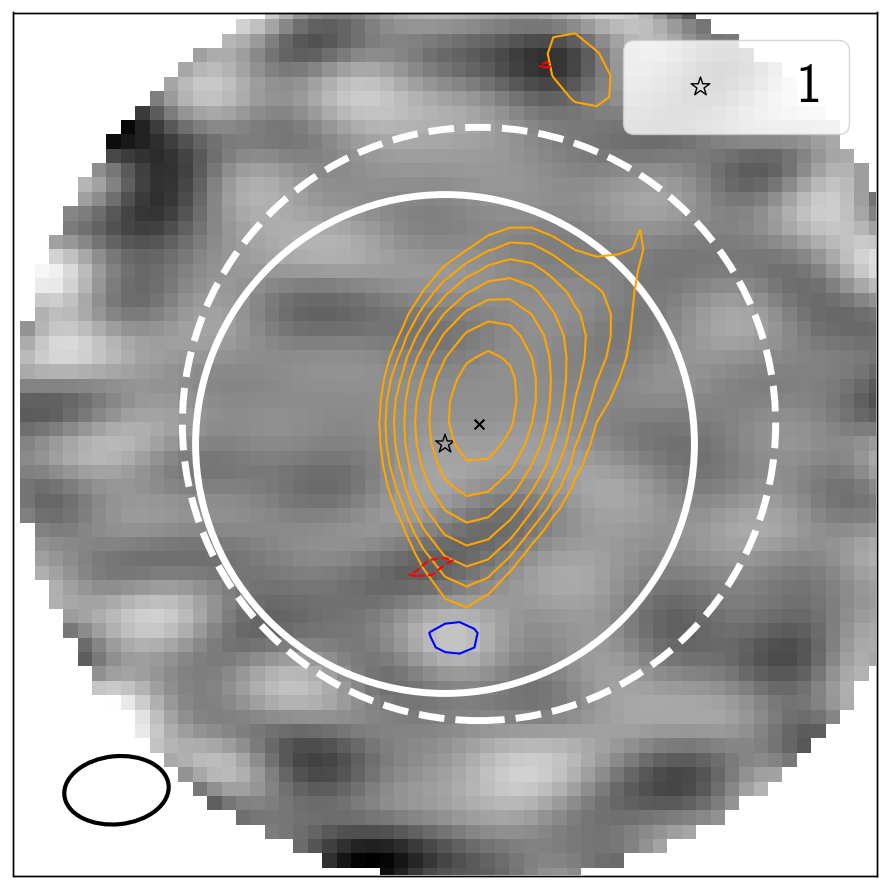}
	\includegraphics[width=0.24\textwidth]{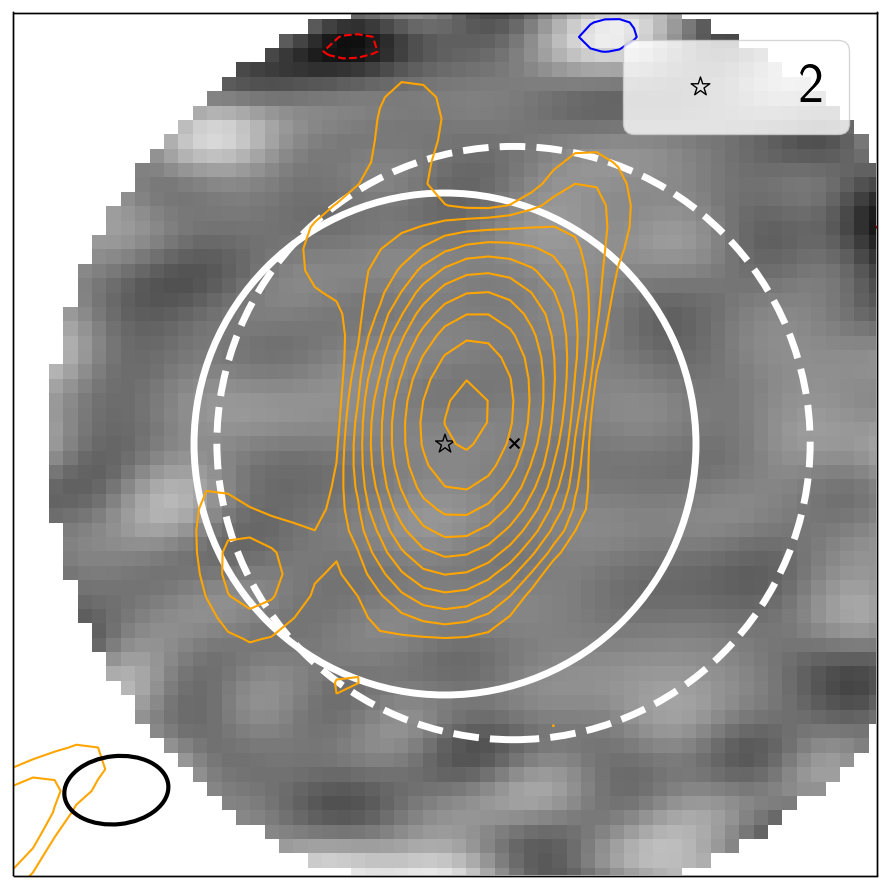}
	\includegraphics[width=0.24\textwidth]{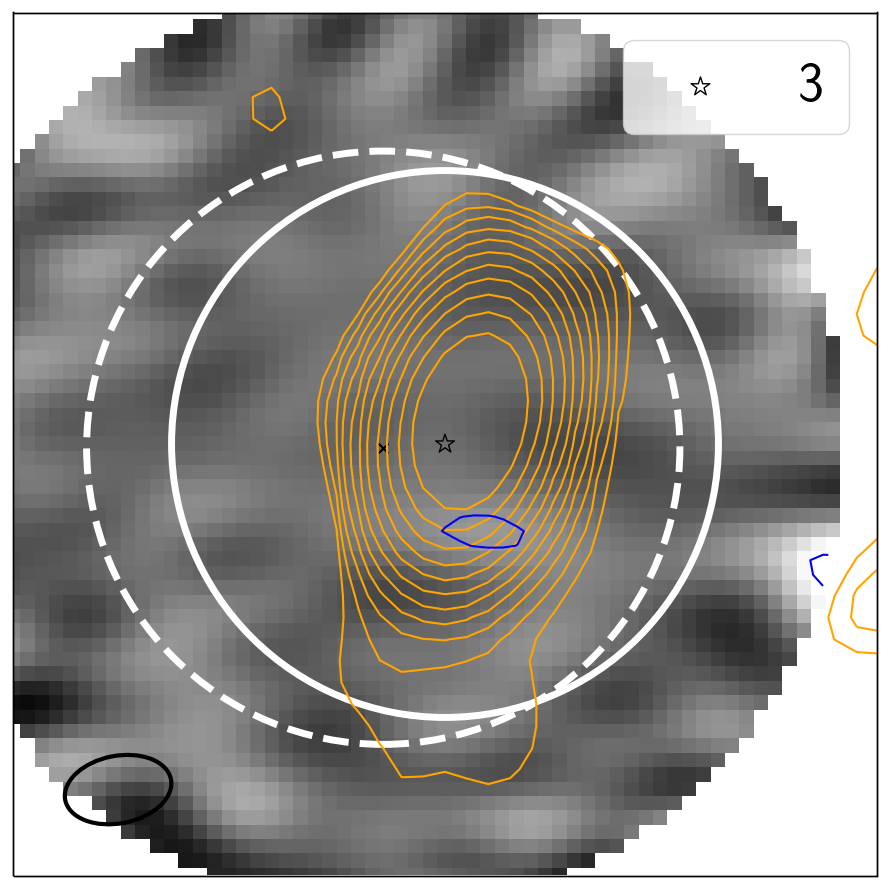}
	\includegraphics[width=0.24\textwidth]{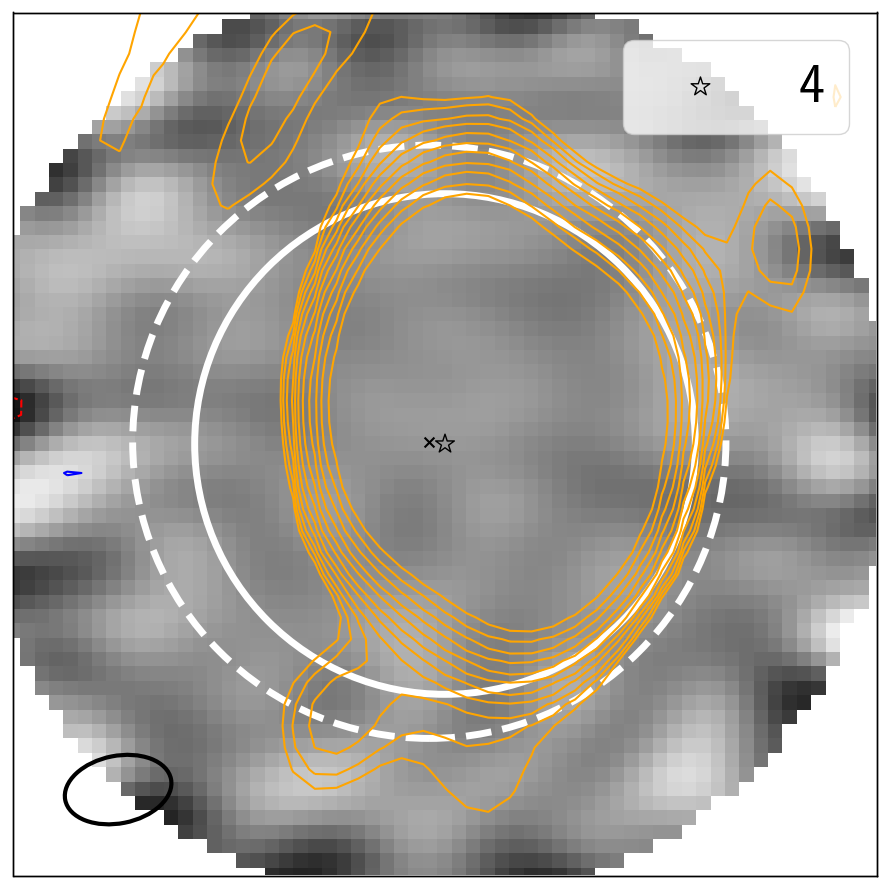}
	\includegraphics[width=0.24\textwidth]{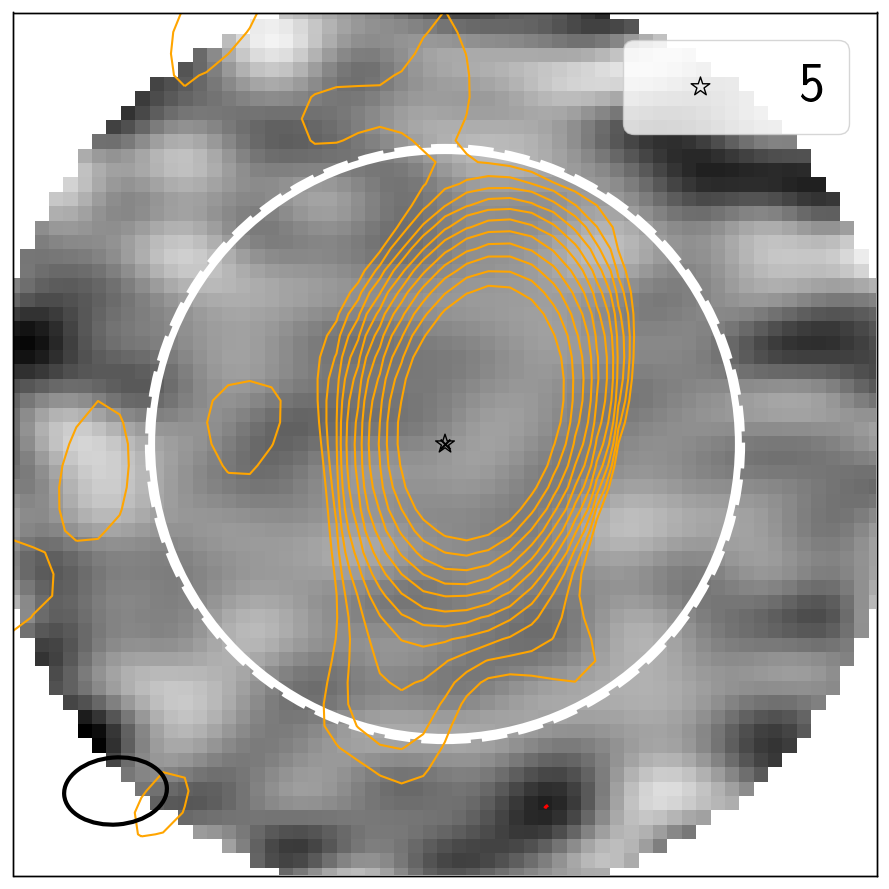}
	\includegraphics[width=0.24\textwidth]{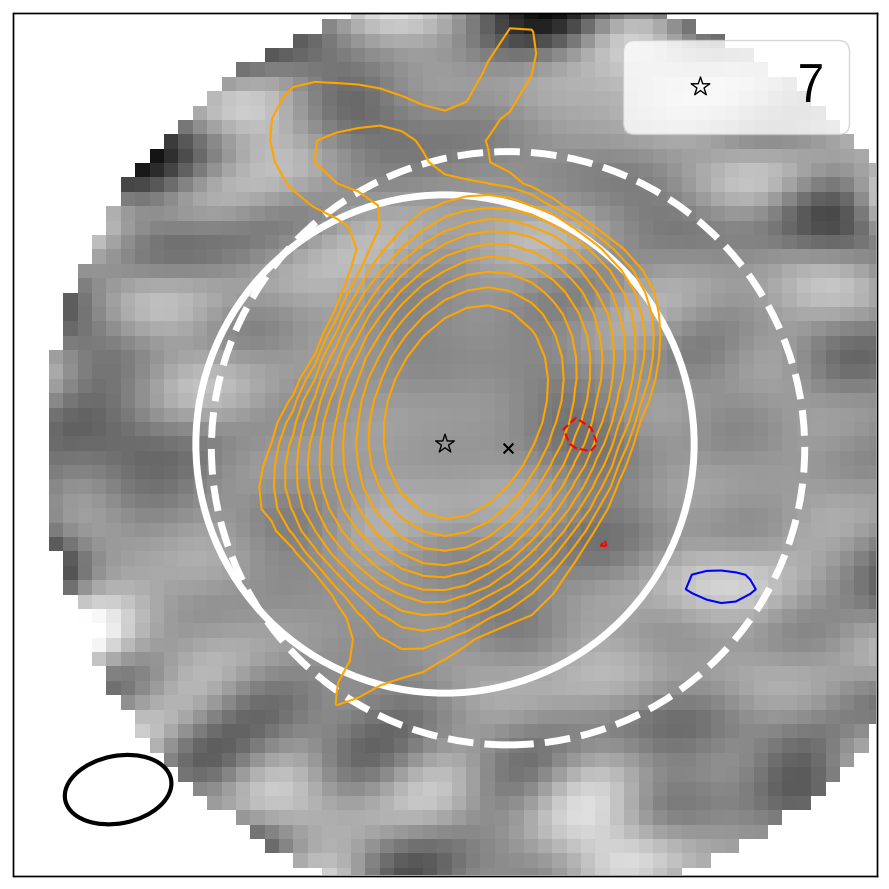}
	\includegraphics[width=0.24\textwidth]{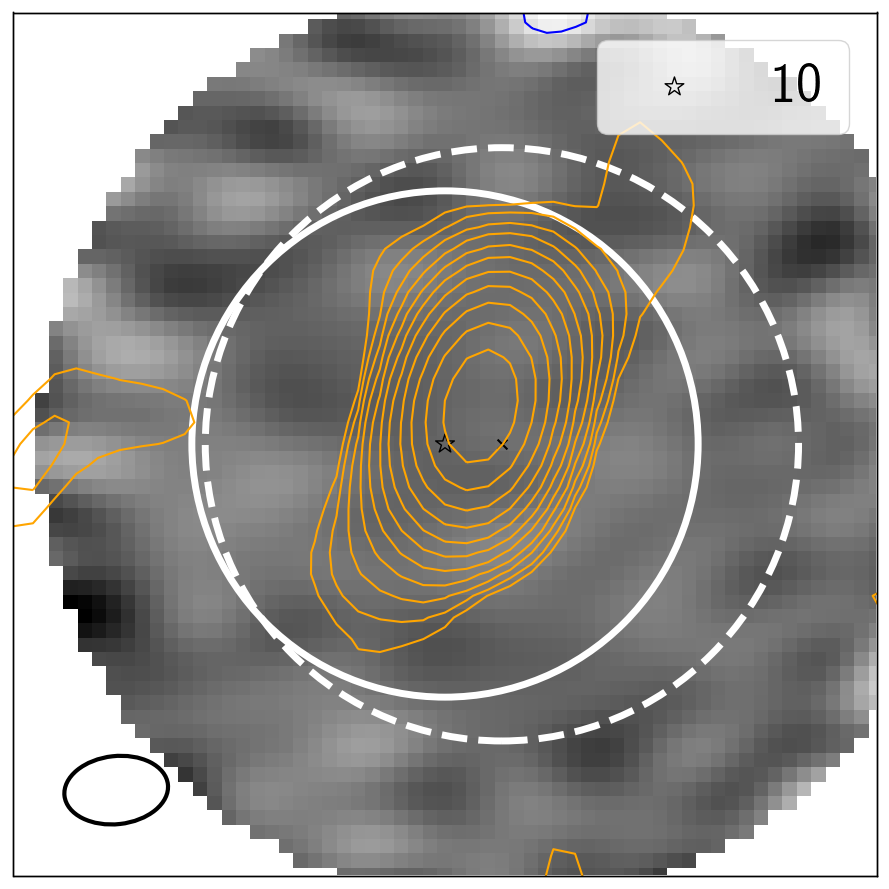}
	\includegraphics[width=0.24\textwidth]{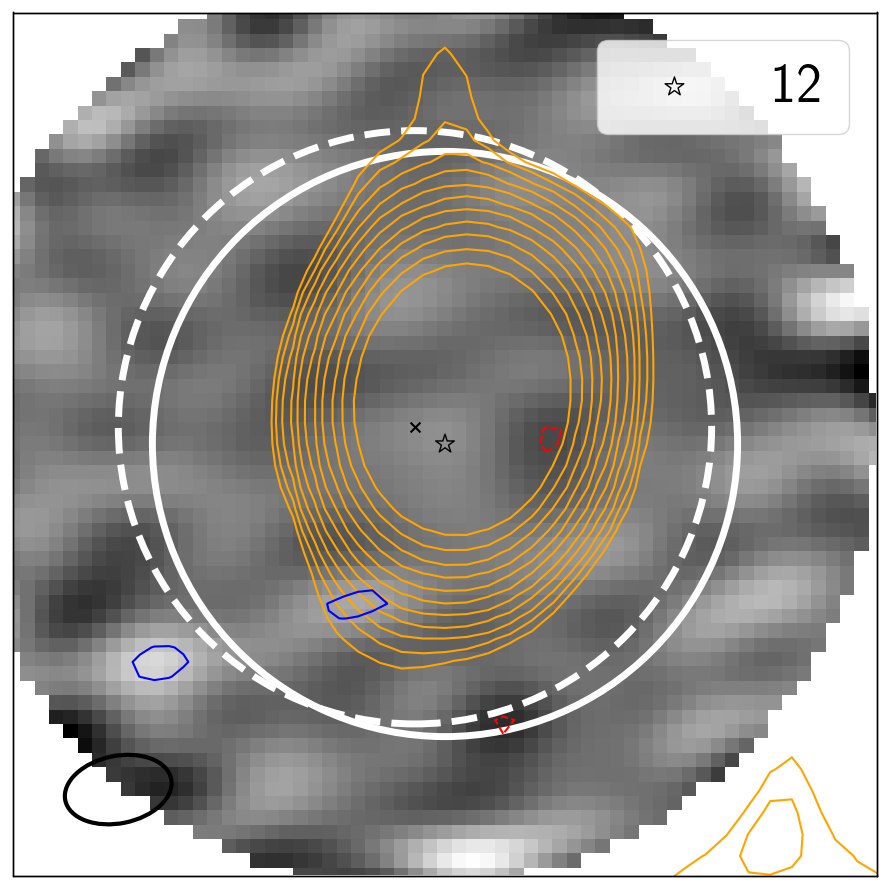}
	\includegraphics[width=0.24\textwidth]{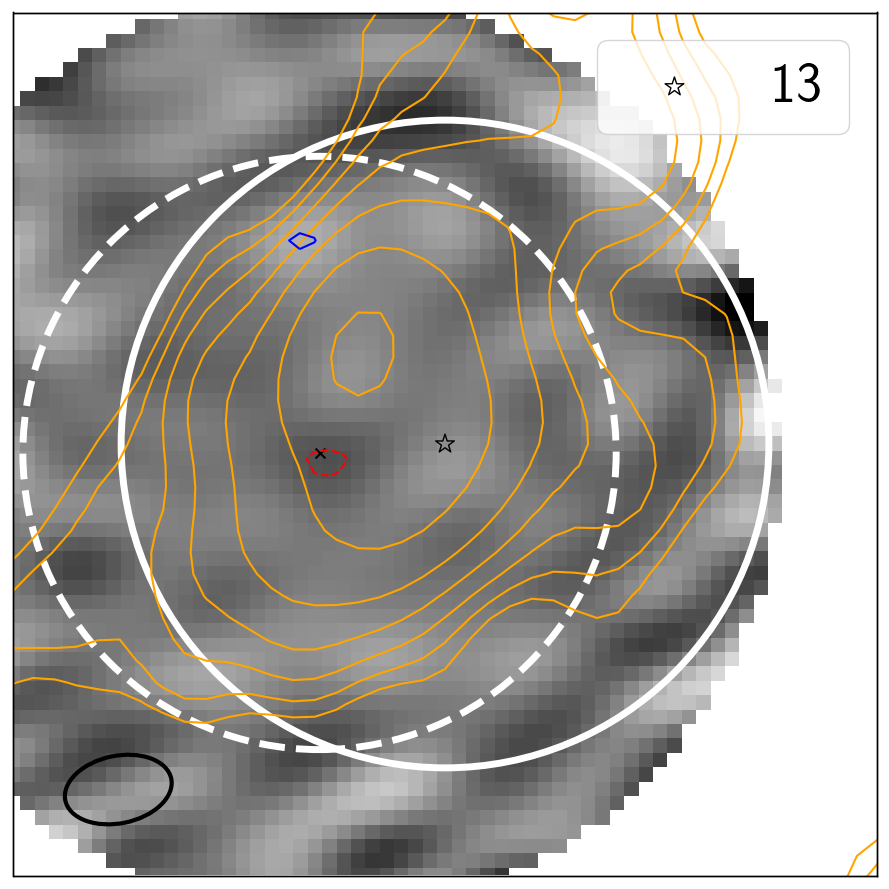}
	\includegraphics[width=0.24\textwidth]{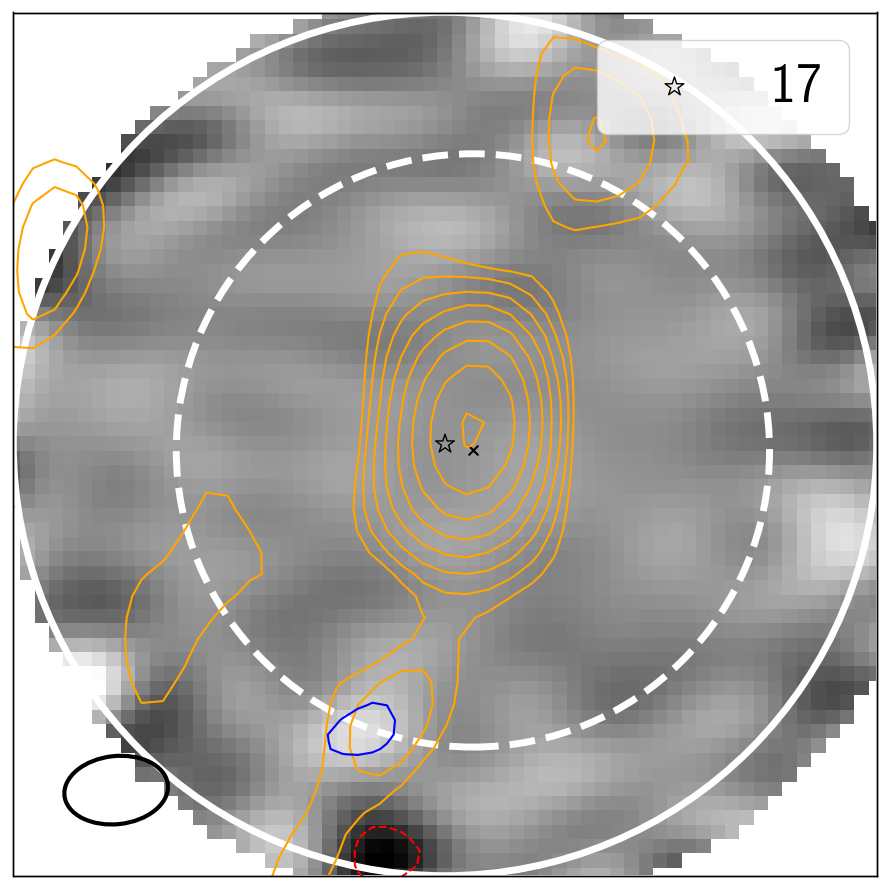}
	\includegraphics[width=0.24\textwidth]{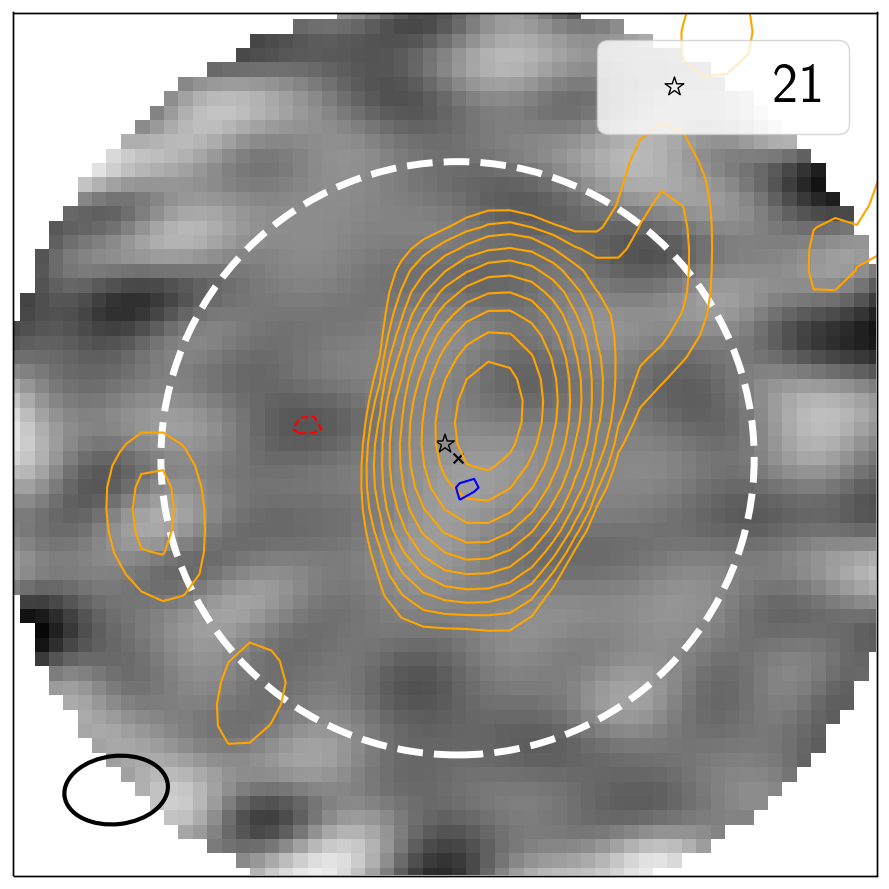}
	\includegraphics[width=0.24\textwidth]{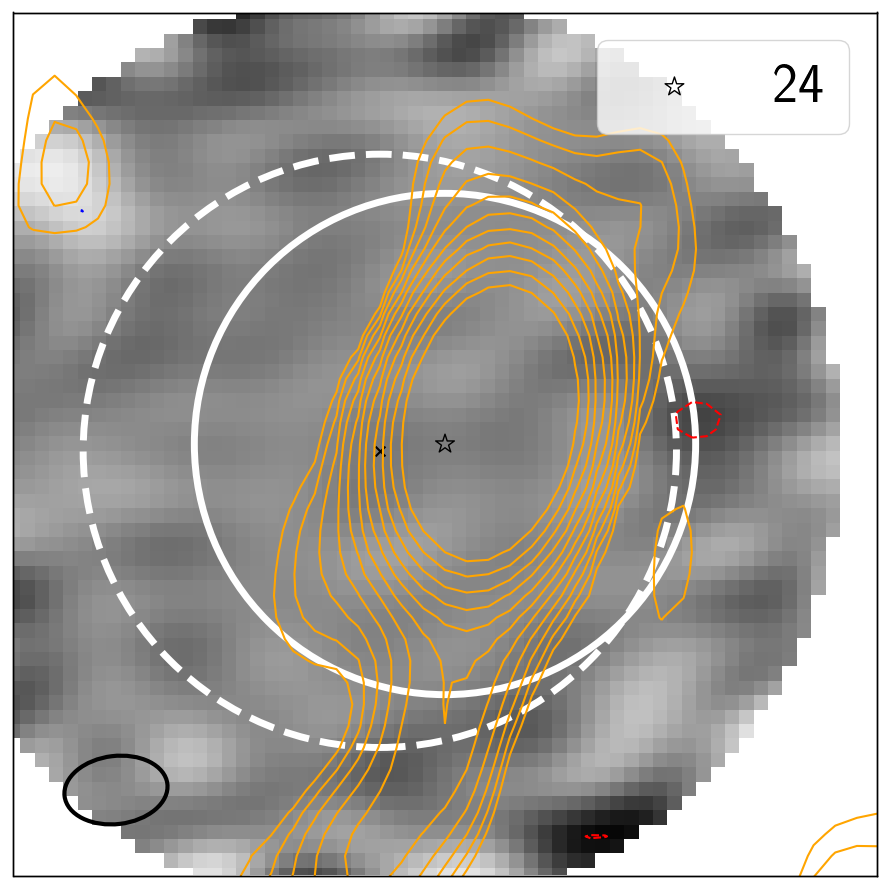}
	\includegraphics[width=0.24\textwidth]{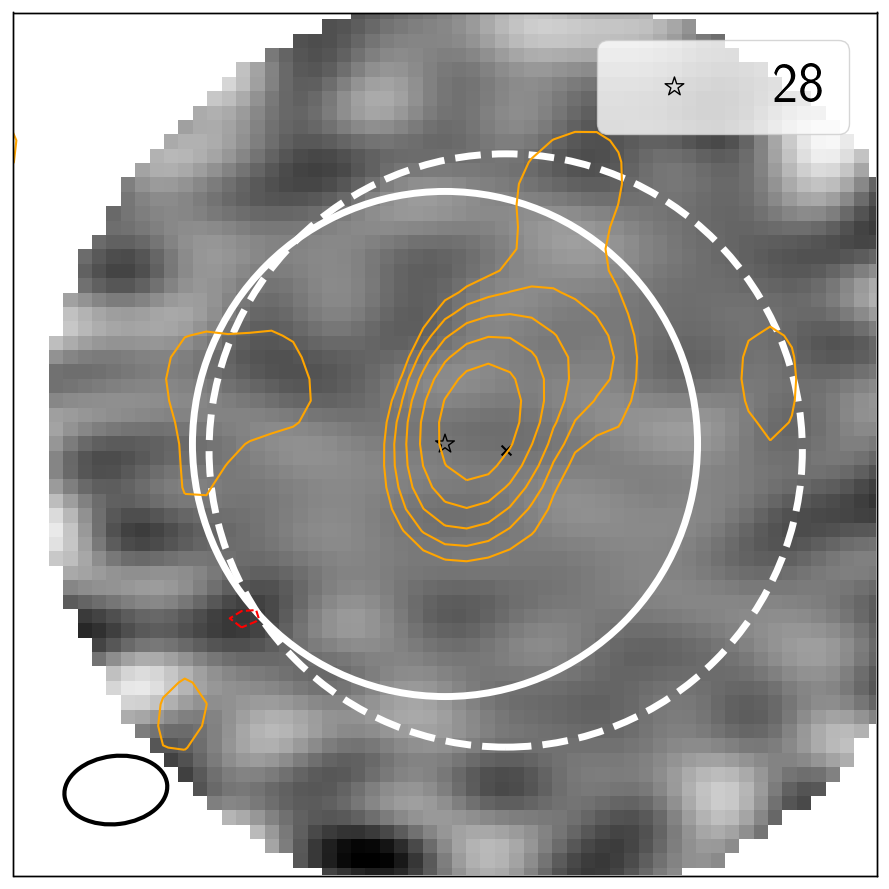}
	\includegraphics[width=0.24\textwidth]{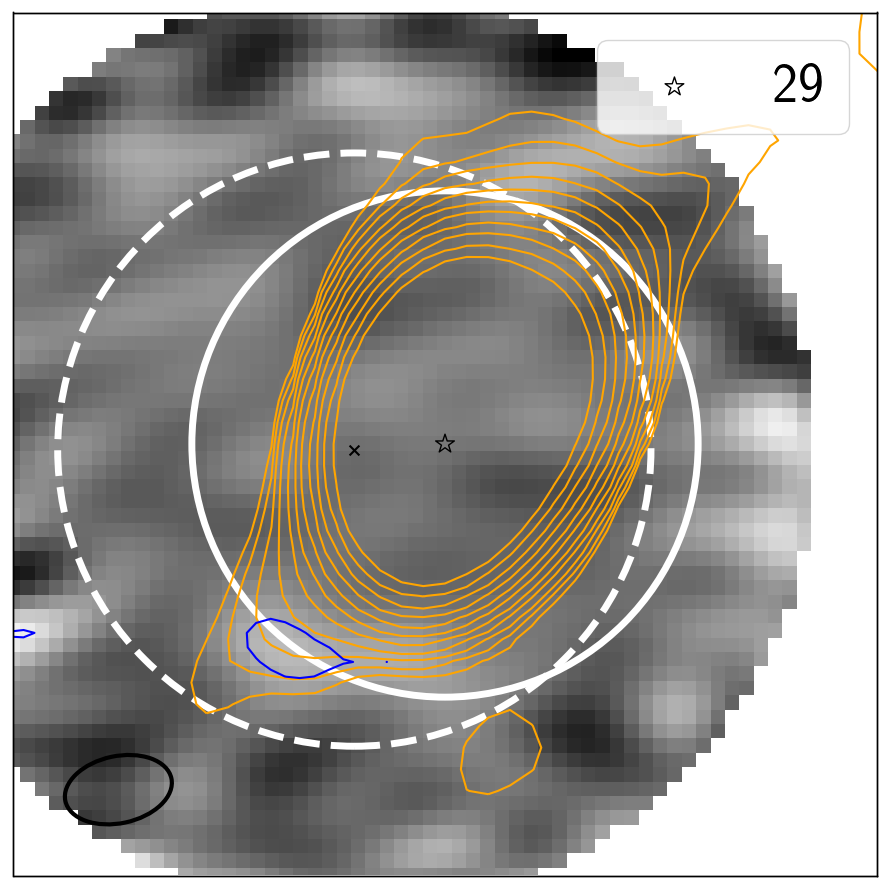}
	\includegraphics[width=0.24\textwidth]{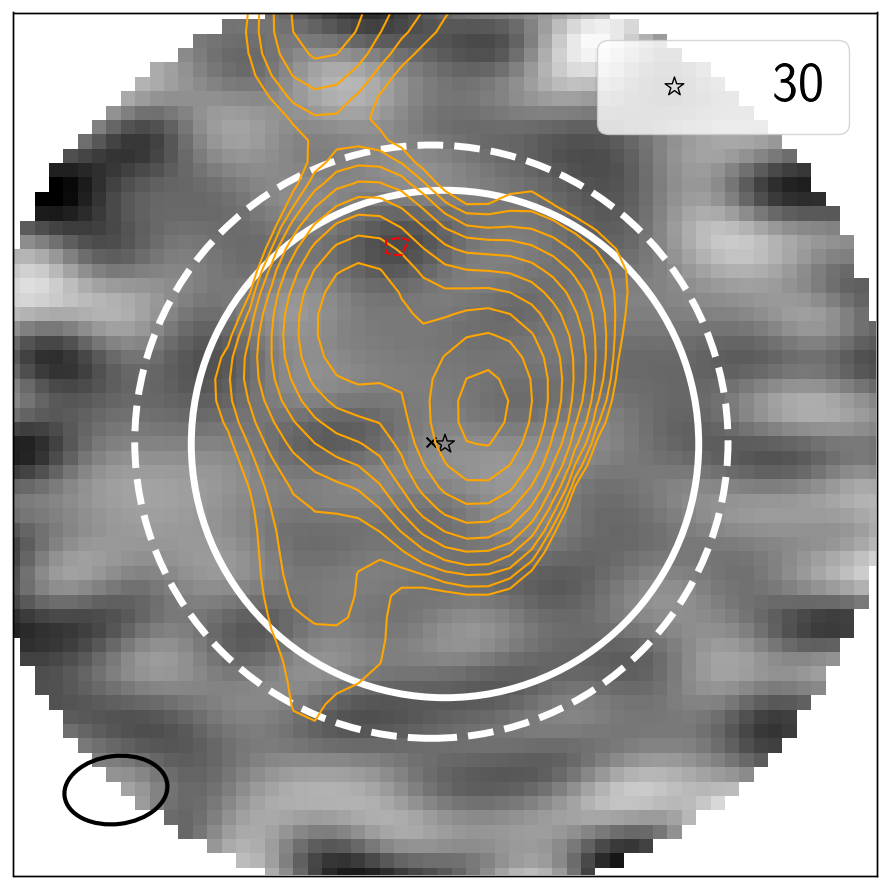}
	\includegraphics[width=0.24\textwidth]{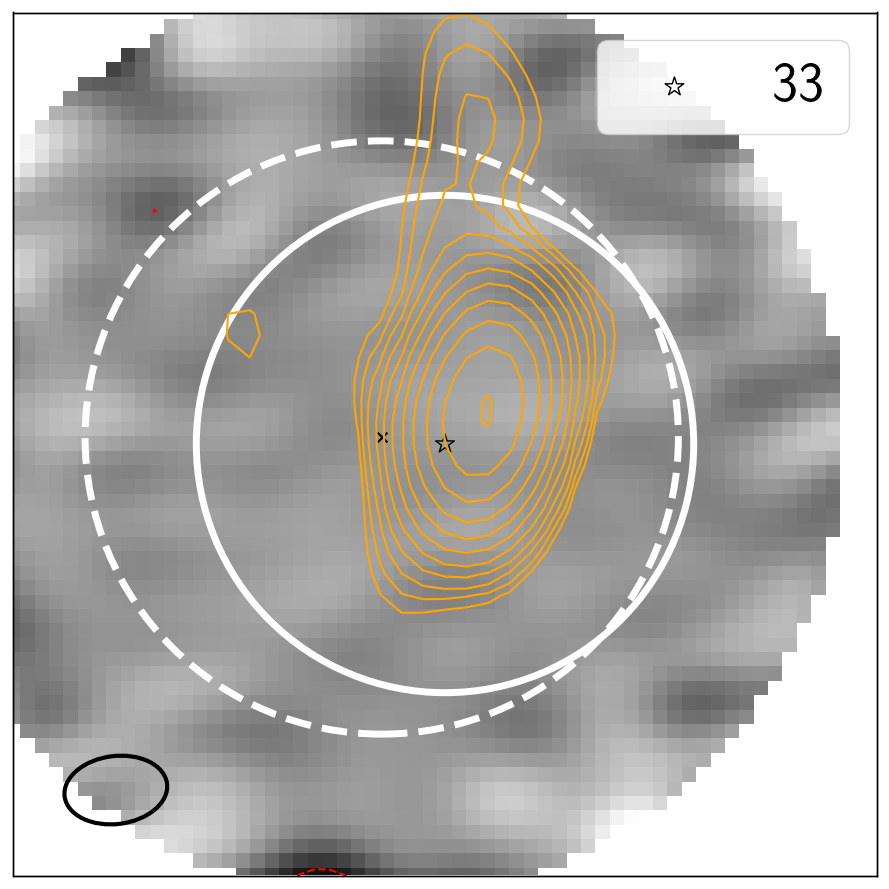}
	\caption{The same as in Figure~\ref{fig:aca1}, but for sources where no significant emission ($|{\rm SNR}|>4.2\sigma$) is detected.}
	\label{fig:aca3}
\end{figure*}

For completeness, in Fig.~\ref{fig:aca3}, we present the 1.3\,mm continuum maps around the sources that do not show any significant detection down to $3\sigma$ ({\sc rms}=0.15\,mJy; see Sec.~\ref{sec:b6flx} for details on detection significance).

% remainder continuum maps

%% file: sections/app-sampdrop.tex
At the time of the initial sample selection, both the photometry reported in M08 and \cite{zinn12}, whose photometry refers to H14, were considered in order to be less affected by possible calibration issues between releases and thus be more complete. However, based on the discussion in Sec.~\ref{sec:radphot} we solely adopt 1.4\,GHz photometry reported by M08 throughout this work. Furthermore, as mentioned in Sec.~\ref{sec:samp}, the IRAC photometry has gone through a fourth data release (DR4), which we adopted throughout this work instead of DR3 that was used for the initial sample selection.
	
\begin{figure*}
\includegraphics[width=0.24\textwidth]{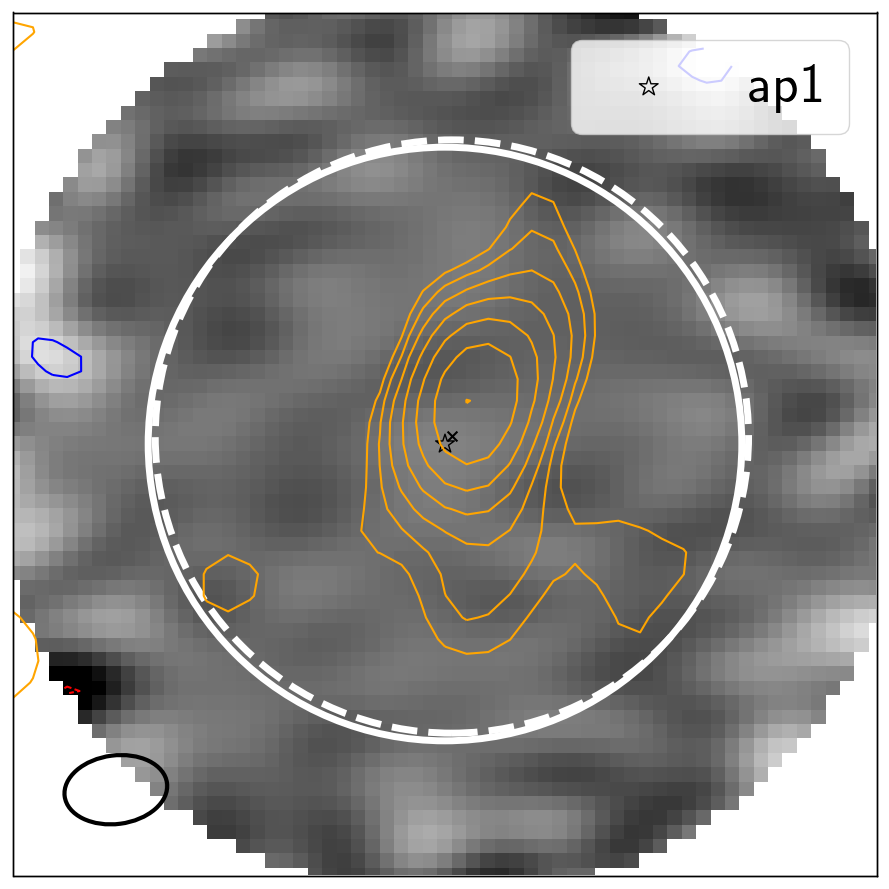}
\includegraphics[width=0.24\textwidth]{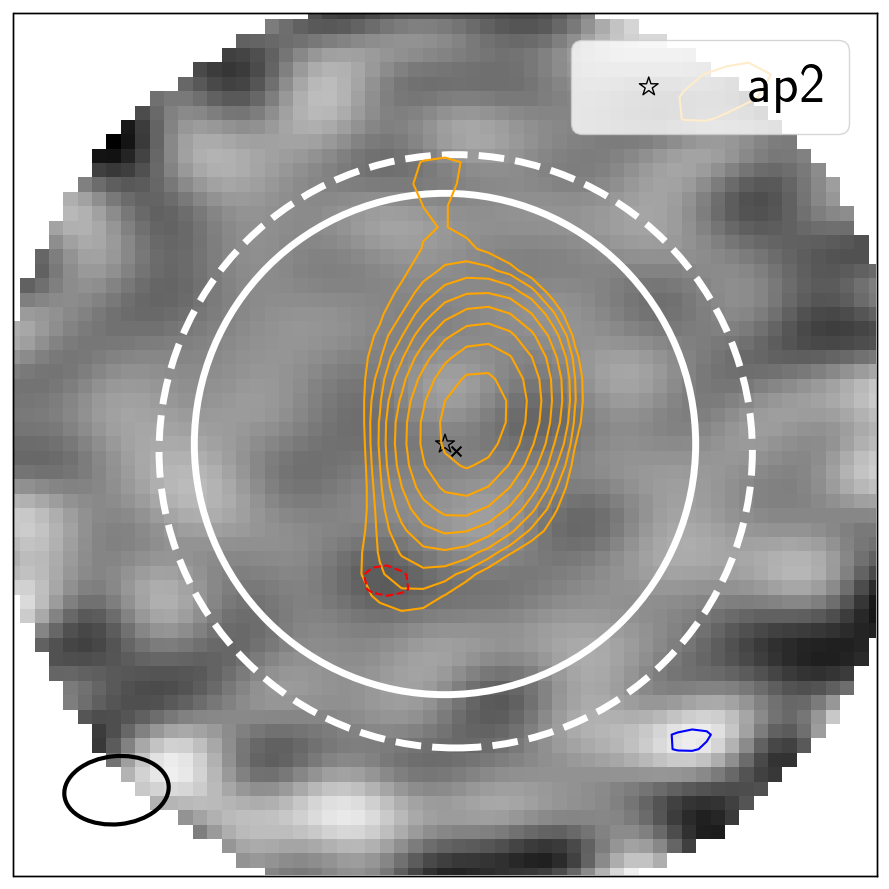}
\includegraphics[width=0.24\textwidth]{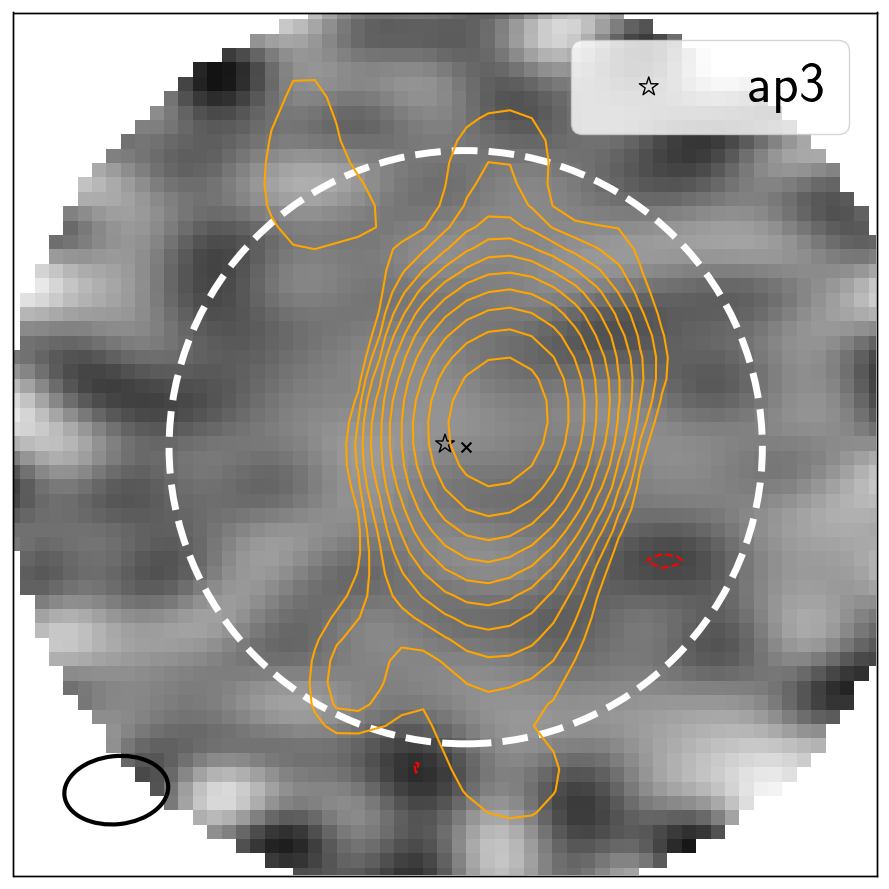}\\
\includegraphics[width=0.24\textwidth]{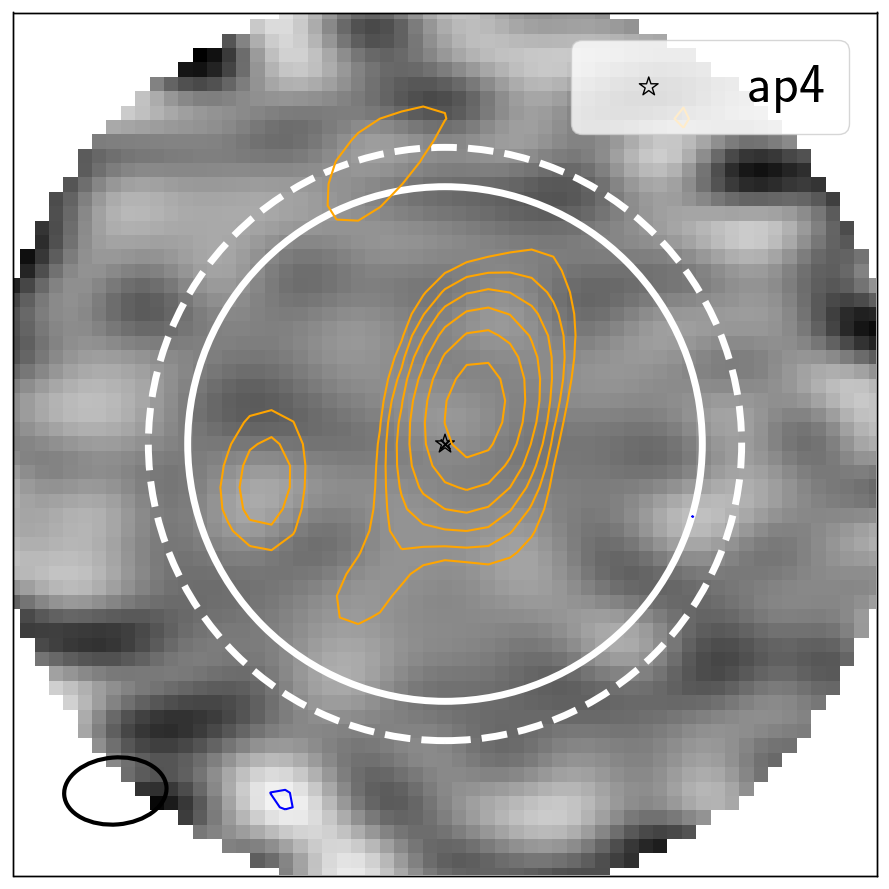}
\includegraphics[width=0.24\textwidth]{figs/acamaps/es1-ap5.png}
\includegraphics[width=0.24\textwidth]{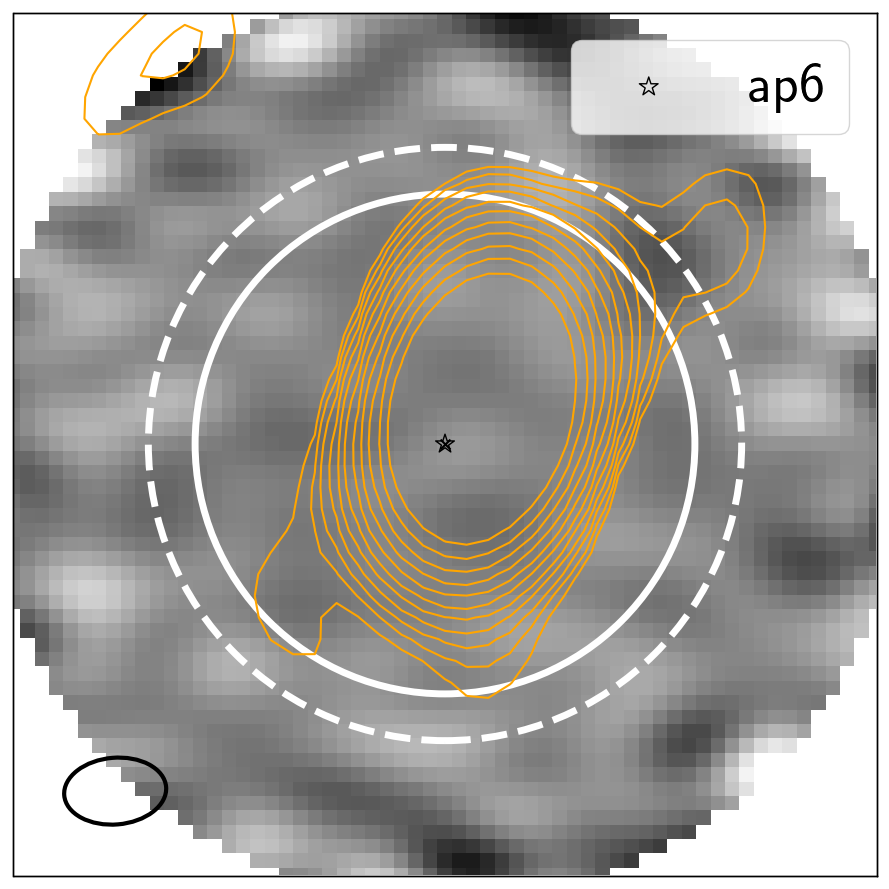}
\caption{The same as in Fig.~\ref{fig:aca1} for the six sources that no longer comply with the selection criteria.}
\label{afig:aca3}
\end{figure*}

As a result, six additional sources (Tab.~\ref{atab:samp}, Fig.~\ref{afig:aca3}) 
 % ap1 - considered USS using Zinn12/Hales14 catalogs
 % ap2 - SEIP/DR3 showed S5.8/S3.6=1.3, but DR4 is 1.2 (prev LRS)
 % ap3 - SEIP/DR3 showed S5.8/S3.6=1.34, but DR4 is 1.1 (prev LRS)
 % ap4 - SEIP/DR3 showed S5.8/S3.6=1.53, but DR4 is 0.95 (USS)
 % ap5 - considered USS using Zinn12/Hales14 catalogs
 % ap6 - SEIP/DR3 showed S5.8/S3.6=0.9, but DR4 is 0.8 (rounded) (IFRS)
 were observed by ACA as part of the initial candidate selection which, based on the above considerations, they no longer comply with any of the original selection criteria described in Sec. \ref{sec:selectioncriteria} and were not included in the main part of the paper. More precisely, for sources ap2, ap3, ap4 and ap6 the DR4 IRAC flux ratios (Tab.~\ref{atab:phot} and \ref{atab:ratios}) are no longer compatible with the MIR criteria (second column in Tab.~\ref{tab:selectioncriteria}). Sources ap1 and ap5, on the other hand, do have 1.4\,GHz fluxes reported for the low-resolution map in \citet[][associated to the DR2 release by H14]{zinn12} that comply with the USS criterion (first row in Tab.~\ref{tab:selectioncriteria}), but do not if the 1.4\,GHz fluxes reported in M08 and F15 are adopted instead. We note that F15 imaging is deeper, hence the fluxes reported here should be less affected by flux boosting with respect to DR1 and DR2, M08 and H14 respectively. Nevertheless, since these six sources have been observed, we report their properties in Tabs.~\ref{atab:phot}, \ref{atab:ratios}, \ref{atab:b6phot}, and \ref{atab:radiosed}
 These tables report the same properties as Tabs. \ref{tab:phot}, \ref{tab:ratios}, \ref{tab:b6phot} and \ref{tab:radiosed}, respectively,  describing the HzRG candidate sample in the main part of the paper.

\begin{table*}
\begin{tabular}{ccrrrrrrcccc}
ID & ID$_{M08}$ & RA$_{rad}$ & Dec$_{rad}$ & RA$_{IR}$ & Dec$_{IR}$ & sep & $z$ & $z_{ref}$ & USS & LRS & IFRS \\
 & & & [deg] & [deg] & [deg] & [deg] & [arcsec] & & & & \\
\hline
ap1 & S50 & 8.22400 & -44.49071 & 8.22418 & -44.49086 & 0.7 & 0.68$_{-0.19}^{+-0.19}$ & DES &  &  &  \\
ap2 & S82 & 9.63380 & -44.42796 & 9.63362 & -44.42791 & 0.5 & 1.95$_{-0.86}^{+4.05}$ & SERVS &  &  &  \\
ap3 & S230 & 8.67470 & -44.22424 & 8.67556 & -44.22411 & 2.3 & 0.1243 & MVF &  &  &  \\
ap4 & S536 & 9.32458 & -43.84778 & 9.32452 & -43.84757 & 0.8 & 1.0909 & MVF &  &  &  \\
ap5 & S716 & 9.64360 & -43.63541 & 9.64376 & -43.63535 & 0.5 & 0.4532 & M12 &  &  &  \\
ap6 & S827 & 8.56982 & -43.51339 & 8.56971 & -43.51349 & 0.5 & 1.35$_{-0.07}^{+0.84}$ & SERVS &  &  &  \\
\hline
\end{tabular}
\caption{Same as Tab.~\ref{tab:samp} for the six sources that no longer comply with the selection criteria.}
\label{atab:samp}
\end{table*}

\begin{table*}
\begin{tabular}{crrrrrrr}
ID & S$_{3.6}$ & S$_{4.5}$ & S$_{5.8}$ & S$_{2.3}$ & S$_{1.4}$ & S$_{0.8}$ & S$_{0.6}$ \\
 & [$\mu$Jy] & [$\mu$Jy] & [$\mu$Jy]& [mJy] & [mJy] & [mJy] & [mJy] \\
\hline
ap1 & 160$\pm$10 & 448$\pm$2 & 1151$\pm$7 & 0.32$\pm$0.08 & 0.53$\pm$0.09 & \ldots & 1.3$\pm$0.1 \\
ap2 & 31.0$\pm$0.6 & 39.0$\pm$0.8 & 37$\pm$3 & 1.2$\pm$0.2 & 1.3$\pm$0.2 & \ldots & \ldots \\
ap3 & 513$\pm$1 & 444$\pm$2 & 557$\pm$5 & 1.0$\pm$0.1 & 2.2$\pm$0.2 & \ldots & 2.4$\pm$0.2 \\
ap4 & 73.3$\pm$0.7 & 71$\pm$1 & 70$\pm$5 & \ldots & 0.35$\pm$0.07 & \ldots & 1.1$\pm$0.1 \\
ap5 & 216$\pm$1 & 260$\pm$1 & 336$\pm$5 & 0.41$\pm$0.09 & 0.59$\pm$0.08 & \ldots & 1.0$\pm$1.0 \\
ap6 & 33.1$\pm$0.6 & 37.5$\pm$0.6 & 26$\pm$4 & 4.4$\pm$0.5 & 6.9$\pm$0.8 & 12$\pm$2 & 14$\pm$1 \\
\end{tabular}
\caption{Same as Tab.~\ref{tab:phot} for the six sources that no longer comply with the selection criteria.}
\label{atab:phot}
\end{table*}

\begin{table*}
\begin{tabular}{crrrrrrrrr}
ID & $\alpha^{2.3}_{1.4}$ & $\alpha^{2.3}_{0.8}$ & $\alpha^{2.3}_{0.6}$ & $\alpha^{1.4}_{0.8}$ & $\alpha^{1.4}_{0.6}$ & $\alpha^{0.8}_{0.6}$ & S$_{5.8}$/S$_{3.6}$ & S$_{1.4}$/S$_{3.6}$ & S$_{1.4}$/S$_{4.5}$\\
\hline
ap1 & -1.0$\pm$0.6 & \ldots & -1.1$\pm$0.2 & \ldots & -1.1$\pm$0.2 & \ldots & 7.2$\pm$0.5 & 3.3$\pm$0.6 & 1.2$\pm$0.2 \\
ap2 & -0.0$\pm$0.4 & \ldots & \ldots & \ldots & \ldots & \ldots & 1.2$\pm$1.0 & 41$\pm$5 & 32$\pm$4 \\
ap3 & -1.7$\pm$0.4 & \ldots & -0.7$\pm$0.1 & \ldots & -0.1$\pm$0.2 & \ldots & 1.09$\pm$0.10 & 4.2$\pm$0.5 & 4.9$\pm$0.6 \\
ap4 & \ldots & \ldots & \ldots & \ldots & -1.4$\pm$0.3 & \ldots & 0.95$\pm$0.07 & 4.8$\pm$0.9 & 5$\pm$10 \\
ap5 & -0.7$\pm$0.5 & \ldots & -0.6$\pm$0.2 & \ldots & -0.6$\pm$0.2 & \ldots & 1.56$\pm$0.02 & 2.7$\pm$0.4 & 2.3$\pm$0.3 \\
ap6 & -0.9$\pm$0.3 & -1.0$\pm$0.2 & -0.9$\pm$0.1 & -1.0$\pm$0.3 & -0.8$\pm$0.2 & -0.5$\pm$0.5 & 0.8$\pm$0.1 & 210$\pm$20 & 180$\pm$20 \\
\hline
\end{tabular}
\caption{Same as Tab.~\ref{tab:ratios} for the six sources that no longer comply with the selection criteria.}
\label{atab:ratios}
\end{table*}

\begin{table}
\begin{tabular}{rrrrrr}
ID & S$_p$ & S$_i$ & RA & Dec & Dist \\
& [mJy] & [mJy] & [deg~('')] & [deg~('')] & [''] \\
\hline
\multicolumn{5}{c}{$3\sigma<$~SNR~$<4.2\sigma$}\\
ap5b & 0.7$\pm$0.2 & 0.9$\pm$0.3 & 9.6407~(1) & -43.6374~(0.4) & 11 \\
\hline
\end{tabular}
\caption{Same as Tab.~\ref{tab:b6phot} for one source that no longer comply with the selection criteria.}
\label{atab:b6phot}
\end{table}

\begin{table}
\centering
\begin{tabular}{rrrrr}
ID & S$^{obs}_{233}$ & S$^{fit}_{233}$ & $\alpha$ & Synchr. \\
& [mJy] & [mJy] & & \\
\hline\hline
\multicolumn{4}{c}{without GLEAM and ACA}\\
\hline
ap1 & $<$0.48 & 0.002~(1.9) & -1.1$\pm$0.1 &   \\
ap2 & $<$0.50 & 0.037~(1.4) & (-0.68) &   \\
ap3 & $<$0.49 & 0.08~(1.6) & -0.59$\pm$0.09 &   \\
ap4 & $<$0.47 & 0.010~(1.4) & (-0.68) &   \\
ap5 & $<$0.54 & 0.02~(2.0) & -0.6$\pm$0.1 &   \\
ap6 & $<$0.45 & 0.07~(1.5) & -0.90$\pm$0.08 &   \\
\hline
\end{tabular}
\caption{Same as Tab.~\ref{tab:radiosed} for the six sources that no longer comply with the selection criteria.
}
\label{atab:radiosed}
\end{table}

% neighbour detections

\subsection*{Serendipitous line detection toward ap3}

\begin{figure*}
\includegraphics[width=0.28\textwidth]{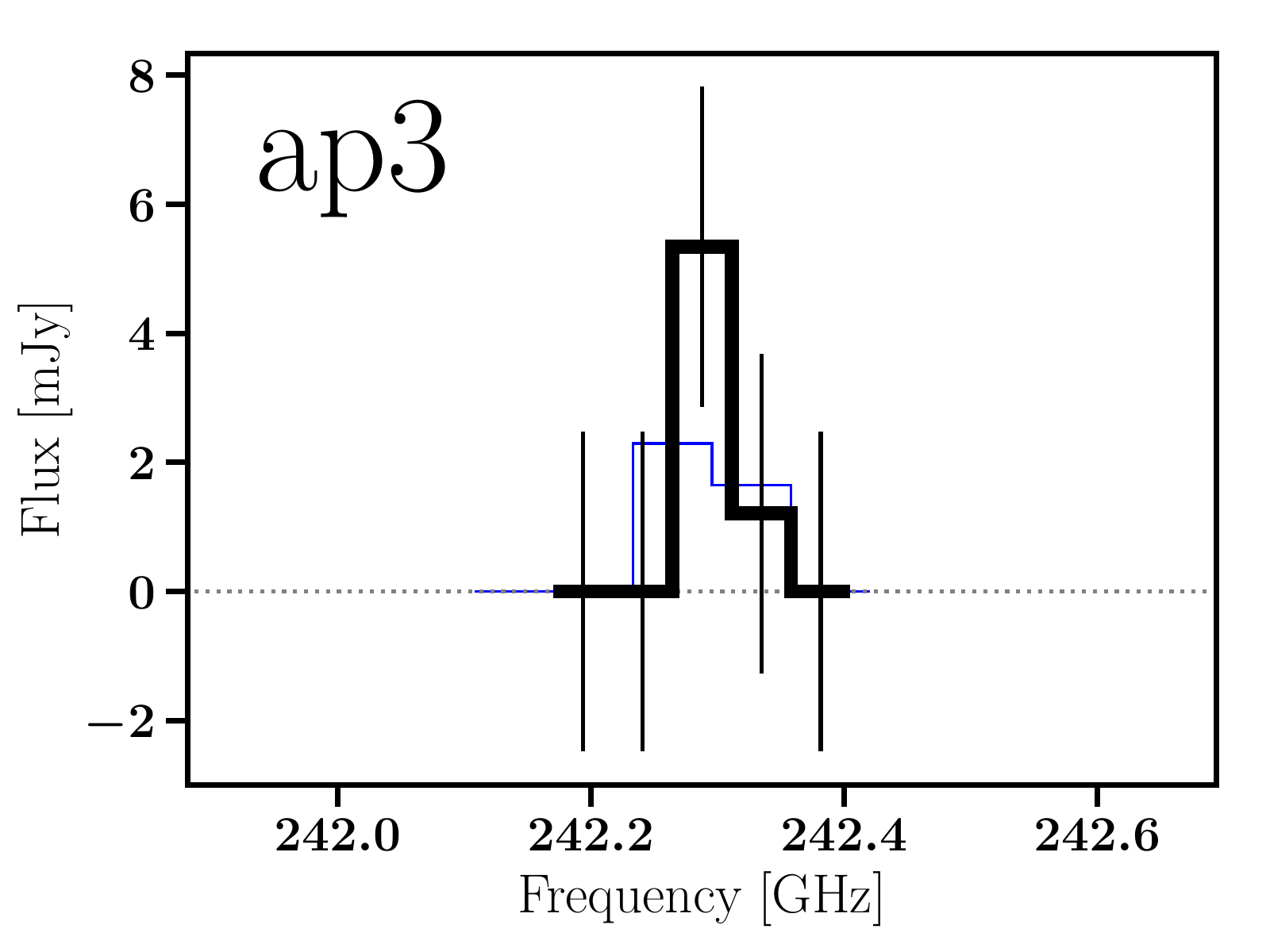}
\includegraphics[width=0.28\textwidth]{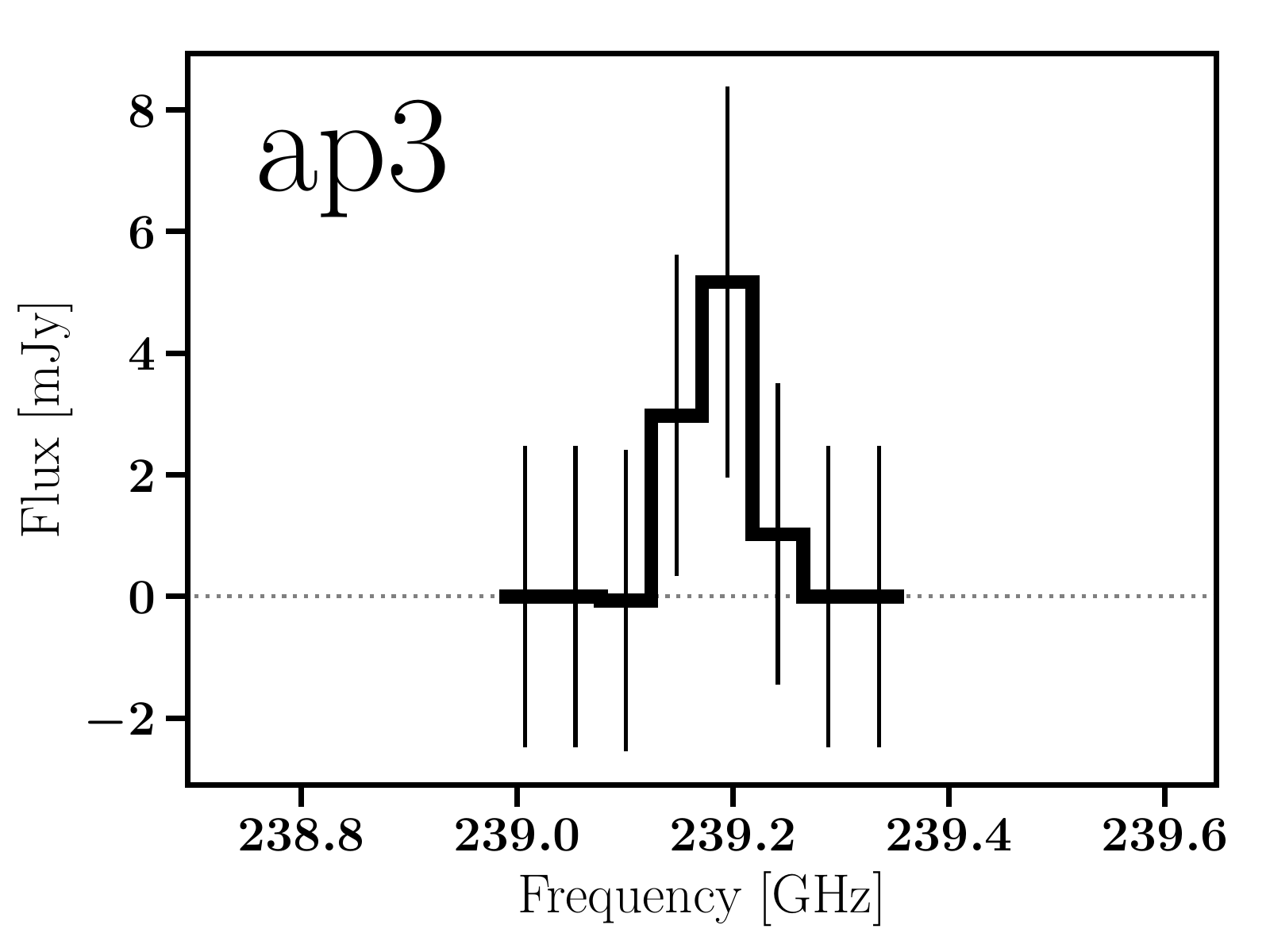}
\includegraphics[width=0.22\textwidth]{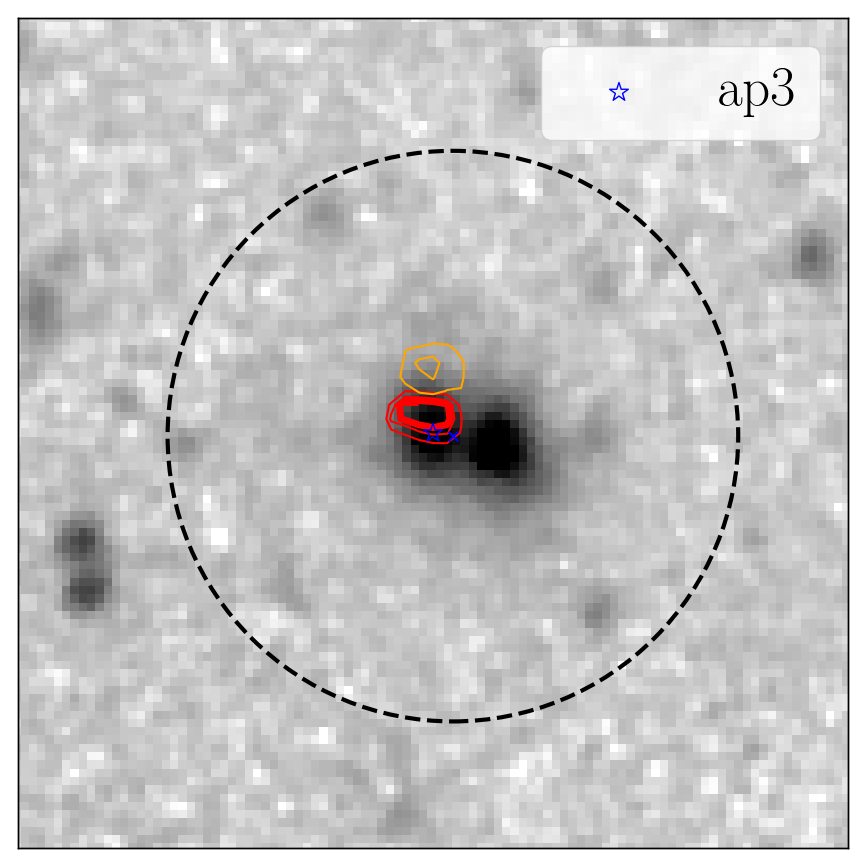}
\caption{Same as in Fig.~\ref{fig:sofia}, but for source ap3. Here, two line features are reported. The spectrum on the left-hand side is that of HNC$_{v=0}$(3-2), whose moment-0 map is displayed with red contours overlaid on the IRAC imaging on the right-hand side panel. The spectrum in the middle is that of HOC$^+_{v=0}$(3-2), whose moment-0 map is displayed with orange contours.}
\label{afig:sofia}
\end{figure*}

The source ap3 is at $z_{\rm spec}=0.12434$, and SoFiA identifies two lines in the 3-channel smoothed cube (Fig.~\ref{afig:sofia}). The more significant one, and equally identified in the 4-channel smoothed cube, 
is found at 242.297$\pm$0.004\,GHz with a velocity-integrated flux of 0.4$\pm$0.1~Jy.km/s and peaking at 5$\pm$2\,mJy. We associate the emission with hydrogen isocyanide, HNC$_{v=0}$(3-2), $\nu_{\rm RF}=271.981~$GHz, $z=0.12251\pm0.00002$). The less significant line-emission is found at 239.184$\pm$0.005\,GHz with a velocity-integrated flux of 0.5$\pm$0.2~Jy.km/s and peaking at 5$\pm$2\,mJy. We associate the emission with isoformyl ion, HOC$^+_{v=0}$(3-2) ($\nu_{\rm RF}=268.451~$GHz, $z=0.12236\pm0.00002$). We note that the light-weighted centroids of the two emissions are spatially offset in the North-South direction by $2.7\pm0.1$\,arcsec. 
The IRAC imaging reveals two distinct sources 5\,arcsec apart in the East-West direction, with the Western one showing a spectrum rising in flux with increasing wavelength revealing the presence of a dusty AGN \citep[e.g.,][]{donley12}. Nevertheless, the HNC$_{v=0}$(3-2) and HOC$^+_{v=0}$(3-2) emissions are associated with the Eastern galaxy. Further characterisation of and discussion about this system is differed to a future work.
	
% specs and moments of SoFiA-identified lines

%% file: sections/app-mastertable.tex
In order to facilitate the access to the information reported in this manuscript, we have assembled a master table with all the information of the primary ACA detection (including the sources presented in Appendix~\ref{app:sampdrop}). The table lists sky coordinates, cross identifications between radio catalogues (including H14 and F15), the reported multi-wavelength photometry, redshifts, as well name under which each source and corresponding data set can be found in the ALMA archive\footnote{https://almascience.eso.org/asax/}. The master-table is available as supplementary material.

%% file: main.bbl
\begin{thebibliography}{99}
    \bibitem[\protect\citeauthoryear{Adam, et al.}{2017}]{adam17} Adam R., et al., 2017, \aap, 598, A115
\bibitem[\protect\citeauthoryear{Adam et al.}{2020}]{Adam2020} Adam R., Goksu H., Leing{\"a}rtner-Goth A., Ettori S., Gnatyk R., Hnatyk B., H{\"u}tten M., et al., 2020,\aap, 644, A70. doi:10.1051/0004-6361/202039091
\bibitem[\protect\citeauthoryear{Afonso et al.}{2005}]{afonso05} Afonso, J., Georgakakis, A., Almeida, C., et al.\ 2005, \apj, 624, 135
\bibitem[\protect\citeauthoryear{Afonso et al.}{2011}]{afonso11} Afonso J. et al.\ 2011, \apj, 743, 122
\bibitem[\protect\citeauthoryear{Alatalo et al.}{2015}]{alatalo15} Alatalo K. et al.\ 2015, \apj, 798, 31
\bibitem[\protect\citeauthoryear{Amarantidis, et al.}{2019}]{amarantidis19} Amarantidis S., et al., 2019, MNRAS, 485, 2694
\bibitem[\protect\citeauthoryear{Anders \& Grevesse}{1989}]{andersGrevesse89} Anders E., Grevesse N., 1989, GeCoA, 53, 197
\bibitem[\protect\citeauthoryear{Archibald et al.}{2001}]{archibald01} Archibald E.~N., Dunlop J.~S., Hughes D.~H., Rawlings S., Eales S.~A., Ivison R.~J.\ 2001, \mnras, 323, 417
\bibitem[\protect\citeauthoryear{Arnouts \& Ilbert}{2011}]{arnouts11} Arnouts S. \& Ilbert O.\ 2011, Astrophysics Source Code Library, record ascl:1108.009
\bibitem[\protect\citeauthoryear{Astropy Collaboration}{2013}]{astropy13} Astropy Collaboration, Robitaille, T. P., Tollerud, E. J., et al., 2013, A\&A, 558, A33
\bibitem[\protect\citeauthoryear{Benson et al.}{2003}]{benson03} Benson, A. J., Bower, R. G., Frenk, C. S., Lacey, C. G., Baugh, C. M., Cole, S.\ 2003, \apj, 599, 38
\bibitem[\protect\citeauthoryear{Bertoldi et al.}{2007}]{bertoldi07} Bertoldi, F., Carilli, C., Aravena, M., et al.\ 2007, \apjs, 172, 132
\bibitem[\protect\citeauthoryear{Bielby et al.}{2010}]{bielby10} Bielby R. et al.\ 2010, \aap, 523, A66
\bibitem[\protect\citeauthoryear{Blumenthal \& Miley}{1979}]{blumenthal79} Blumenthal G. \& Miley G.\ 1973, \aap, 80, 13
\bibitem[\protect\citeauthoryear{Bondi et al.}{2003}] {bondi03} Bondi M. et al.\ 2003, \aap, 403, 857
\bibitem[\protect\citeauthoryear{Bondi et al.}{2007}] {bondi07} Bondi M. et al.\ 2007, \aap, 463, 519
\bibitem[\protect\citeauthoryear{Callingham, et al.}{2017}]{callingham17} Callingham J.~R., et al., 2017, \apj, 836, 174
\bibitem[\protect\citeauthoryear{Carilli \& Walter}{2013}]{carilli13} Carilli C. \& Walter F.\ 2013, \araa, 51, 105
\bibitem[\protect\citeauthoryear{Casey}{2013}]{casey16} Casey C.~M. 2016, \apj, 824, 36
\bibitem[\protect\citeauthoryear{Chapman et al.}{2009}]{chapman09} Chapman S. C., Blain A., Ibata R., Ivison R.~J., Smail I., Morrison G. 2009, \apj, 691, 560
\bibitem[\protect\citeauthoryear{Chluba et al.}{2012}]{Chluba2012} Chluba J., Nagai D., Sazonov S., Nelson K., 2012, \mnras, 426, 510
\bibitem[\protect\citeauthoryear{Cicone et al.}{2014}]{cicone14} Cicone C., Maiolino R., Sturm E., Graci{\'a}-Carpio J., Feruglio C., Neri R., Aalto S., et al., 2014, \aap, 562, A21
\bibitem[\protect\citeauthoryear{Cooke et al.}{2016}]{cooke16} Cooke E. et al. 2016, \apj, 816, 83
\bibitem[\protect\citeauthoryear{Cotton et al.}{2020}]{cotton20} Cotton W.~D., Thorat K., Condon J.~J., Frank B.~S., J{\'o}zsa G.~I.~G., White S.~V., Deane R., et al., 2020, MNRAS, 495, 1271. doi:10.1093/mnras/staa1240
\bibitem[\protect\citeauthoryear{Croton et al.}{2006}]{croton06} Croton D.~J. et al. 2006, \mnras, 365, 11
\bibitem[\protect\citeauthoryear{Dannerbauer et al.}{2014}]{dannerbauer14} Dannerbauer H. et al. 2014, \aap, 570, A55
\bibitem[\protect\citeauthoryear{De Breuck et al.}{2010}]{debreuck10} De Breuck, C., Seymour, N., Stern, D. et al. 2010, \apj, 725, 36
\bibitem[\protect\citeauthoryear{Decarli et al.}{2019}]{decarli19} Decarli R., Walter F., G{\'o}nzalez-L{\'o}pez J., Aravena M., Boogaard L., Carilli C., Cox P., et al., 2019, \apj, 882, 138
\bibitem[\protect\citeauthoryear{DES Collaboration}{2018}]{descol18} DES Collaboration, Abbott T.~M.~C., et al., 2018, ApJS, 239, 18
\bibitem[\protect\citeauthoryear{Donley, et al.}{2012}]{donley12} Donley J.~L., et al., 2012, \apj, 748, 142
\bibitem[\protect\citeauthoryear{Drouart et al.}{2014}]{drouart14} Drouart G. et al. 2014, \aap, 566, A53
\bibitem[\protect\citeauthoryear{Duncan, et al.}{2018}]{duncan18} Duncan K.~J., et al., 2018, \mnras, 473, 2655
\bibitem[\protect\citeauthoryear{Dunlop et al.}{2017}]{dunlop17} Dunlop J.~s. et al. 2017, \mnras, 466, 861
\bibitem[\protect\citeauthoryear{Emonts et al.}{2014}]{emonts14} Emonts B.~H.~C. et al. 2014, \mnras, 438, 2898
\bibitem[\protect\citeauthoryear{Emonts et al.}{2015a}]{emonts15a} Emonts B.~H.~C. et al. 2015, \aap, 584, A99
\bibitem[\protect\citeauthoryear{Emonts et al.}{2015b}]{emonts15b} Emonts B.~H.~C. et al. 2015, \mnras, 451, 1025
%\bibitem[\protect\citeauthoryear{Evans}{2008}]{evans08} Evans, N.~J.\ 2008, Pathways Through an Eclectic Universe, 52
\bibitem[\protect\citeauthoryear{Falkendal et al.}{2019}]{falkendal19} Falkendal T. et al.\ 2019, \aap, 621, 27
\bibitem[\protect\citeauthoryear{Filho et al.}{2011}]{filho11} Filho M.~E., Brinchmann J., Lobo C., Ant{\'o}n S., 2011, \aap, 536, A35. doi:10.1051/0004-6361/201117834
\bibitem[\protect\citeauthoryear{Fixsen}{2009}]{fixsen09} Fixsen D.~J., 2009, \apj, 707, 916
\bibitem[\protect\citeauthoryear{Fowler}{2004}]{fowler04} Fowler J.~W., 2004, SPIE, 5498, 1. doi:10.1117/12.553054
\bibitem[\protect\citeauthoryear{Franco et al.}{2018}]{franco18} Franco M. et al. 2018, \aap, 620, 152
\bibitem[\protect\citeauthoryear{Franzen et al.}{2015}]{franzen15} Franzen, T.~M.~O., Banfield, J.~K., Hales, C.~A., et al.\ 2015, \mnras, 453, 4020
\bibitem[\protect\citeauthoryear{Garc{\'\i}a-Burillo et al.}{2014}]{garciaBurillo14} Garc{\'\i}a-Burillo S., Combes F., Usero A., Aalto S., Krips M., Viti S., Alonso-Herrero A., et al., 2014,\aap, 567, A125. doi:10.1051/0004-6361/201423843
\bibitem[\protect\citeauthoryear{Gardner et al.}{2006}]{gardner06} Gardner, J.~P., Mather, J.~C., Clampin, M., et al.\ 2006, \ssr, 123, 485
\bibitem[\protect\citeauthoryear{Geach et al.}{2017}]{geach17} Geach J.~E. et al. 2017, \mnras, 465, 1789
\bibitem[\protect\citeauthoryear{Gobat et al.}{2013}]{gobat13} Gobat R. et al. 2013, \apj, 776, 9
\bibitem[\protect\citeauthoryear{Gowardhan et al.}{2018}]{gowardhan18} Gowardhan A., Spoon H., Riechers D.~A., Gonz{\'a}lez-Alfonso E., Farrah D., Fischer J., Darling J., et al., 2018, \apj, 859, 35
\bibitem[\protect\citeauthoryear{Griffith \& Stern}{2010}]{griffith10} Griffith, R.~L., \& Stern, D.\ 2010, \aj, 140, 533
\bibitem[\protect\citeauthoryear{Gullberg et al.}{2016}]{gullberg16} Gullberg B. et al. 2016, \aap, 586, 124
\bibitem[\protect\citeauthoryear{Hales et al.}{2014}]{hales14} Hales, C.~A., Norris, R.~P., Gaensler, B.~M., et al.\ 2014, \mnras, 441, 2555
\bibitem[\protect\citeauthoryear{Hand et al.}{2012}]{hand12} Hand, N., Addison, G.~E., Aubourg, E., et al.\ 2012, \prl, 109, 041101
\bibitem[\protect\citeauthoryear{Hatch et al.}{2014}]{hatch14} Hatch N.~A. 2014, \mnras, 445, 280
\bibitem[\protect\citeauthoryear{Heckman \& Best}{2014}]{heckman14} Heckman T.~M. \& Best P.~N. 2014, \araa, 52, 589
\bibitem[\protect\citeauthoryear{Henriques et al.}{2015}]{henriques15} Henriques B.~M.~B., White S.~D.~M., Thomas P.~A., Angulo R., Guo Q., Lemson G., Springel V., Overzier R. 2015, \mnras, 451, 2663
\bibitem[\protect\citeauthoryear{Hickox et al.}{2009}]{hickox09} Hickox, R.~C., Jones, C., Forman, W.~R., et al.\ 2009, \apj, 696, 891
\bibitem[\protect\citeauthoryear{Hilton et al.}{2021}]{Hilton2021} Hilton M., Sif{\'o}n C., Naess S., Madhavacheril M., Oguri M., Rozo E., Rykoff E., et al., 2021, ApJS, 253, 3. doi:10.3847/1538-4365/abd023
\bibitem[\protect\citeauthoryear{H\"onig et al.}{2017}]{honig17} H\"onig, D., Watson, M., Kishimoto, P., et al.\ 2017, \mnras, 464, 1693
\bibitem[\protect\citeauthoryear{Hoyle et al.}{2018}]{hoyle18} Hoyle B., Gruen D., Bernstein G.~M., Rau M.~M., De Vicente J., Hartley W.~G., Gaztanaga E., et al., 2018, \mnras, 478, 592
\bibitem[\protect\citeauthoryear{Humphrey et al.}{2011}]{humphrey11} Humphrey A., Zeballos M., Aretxaga I. et al.\ 2011, \mnras, 418, 74 
\bibitem[\protect\citeauthoryear{Hunter}{2007}]{hunter07} Hunter J.~D., 2007, CSE, 9, 90
\bibitem[\protect\citeauthoryear{Hurley-Walker et al.}{2017}]{hurleyWalker17} Hurley-Walker, N., Callingham, J.~R., Hancock, P.~J., et al.\ 2017, \mnras, 464, 1146
\bibitem[\protect\citeauthoryear{Iguchi et al.}{2009}]{iguchi09} Iguchi S., Morita K.-I., Sugimoto M., Vilar{\'o} B.~V., Saito M., Hasegawa T., Kawabe R., et al., 2009, PASJ, 61, 1. doi:10.1093/pasj/61.1.1
\bibitem[\protect\citeauthoryear{Intema et al.}{2017}]{intema17} Intema H.~T., Jagannathan P., Mooley K.~P., Frail D.~A., 2017, A\&A, 598, A78. doi:10.1051/0004-6361/201628536
\bibitem[\protect\citeauthoryear{Itoh, Kohyama, \& Nozawa}{1998}]{Itoh1998} Itoh N., Kohyama Y., Nozawa S., 1998, \apj, 502, 7
\bibitem[\protect\citeauthoryear{Ivison et al.}{2000}]{ivison00} Ivison R.~J., Dunlop J.~S., Smail I., Dey A., Liu M.~C., Graham J.~R.\ 2000, \apj, 542, 27
\bibitem[\protect\citeauthoryear{Ivison et al.}{2008}]{ivison08} Ivison R.~J., Morrison G.~E., Biggs A.~D., Smail, I., Willner S.~P., Gurwell M.~A., Greve T.~R., Stevens J.~A., Ashby M.~L.~N.\ 2008, \mnras, 390, 1117
\bibitem[\protect\citeauthoryear{Ivison et al.}{2013}]{ivison13} Ivison R.~J. 2013, \apj, 772, 137
\bibitem[\protect\citeauthoryear{Jarvis et al.}{2009}]{jarvis09} Jarvis M.~J., Teimourian H., Simpson C., Smith D.~J.~B., Rawlings S., Bonfield D., 2009, \mnras, 398, L83. doi:10.1111/j.1745-3933.2009.00715.x
\bibitem[\protect\citeauthoryear{Jarvis et al.}{2013}]{jarvis13} Jarvis M.~J., Bonfield D.~G., Bruce V.~A., Geach J.~E., McAlpine K., McLure R.~J., Gonz{\'a}lez-Solares E., et al., 2013, \mnras, 428, 1281
\bibitem[\protect\citeauthoryear{Jones et al.}{2001}]{jones01} Jones, E., Oliphant, T. E., Peterson, P., et al. 2001-, http://www.scipy.org/
\bibitem[\protect\citeauthoryear{Kakkad et al.}{2018}]{kakkad18} Kakkad D., Groves B., Dopita M., Thomas A.~D., Davies R.~L., Mainieri V., Kharb P., et al., 2018, \aap, 618, A6
%\bibitem[\protect\citeauthoryear{Kalfountzou et al.}{2017}]{kalfountzou17} Kalfountzou E., Stevens J.~A., Jarvis M.~J., Hardcastle M.~J., Wilner D., Elvis M., Page M.~J., et al., 2017, MNRAS, 471, 28. doi:10.1093/mnras/stx1333
%https://ui.adsabs.harvard.edu/abs/2017MNRAS.471...28K/abstract
\bibitem[\protect\citeauthoryear{Kelly et al.}{2015}]{kelly15} Kelly G., Viti S., Bayet E., Aladro R., Yates J., 2015, \aap, 578, A70
\bibitem[\protect\citeauthoryear{Kirkpatrick, et al.}{2019}]{kirkpatrick19} Kirkpatrick A., Sharon C., Keller E., Pope A., 2019, \apj, 879, 41
\bibitem[\protect\citeauthoryear{Kormendy \& Ho}{2013}]{kormendyho13} Kormendy J., Ho L.~C., 2013, ARA\&A, 51, 511. doi:10.1146/annurev-astro-082708-101811
\bibitem[\protect\citeauthoryear{Lacy et al.}{2007}]{lacy07} Lacy M., Petric A.~O., Sajina A., Canalizo G., Storrie-Lombardi L.~J., Armus L., Fadda D., Marleau F.~R. 2007, \aj, 133, 186 
\bibitem[\protect\citeauthoryear{Lacy et al.}{2019}]{lacy19} Lacy M., Mason B., Sarazin C., Chatterjee S., Nyland K., Kimball A., Rocha G., Rowe B., Surace J.\ 2019, \mnras, 483, L22
\bibitem[\protect\citeauthoryear{Lacy et al.}{2021}]{lacy21} Lacy M.,  Surace J.~A.,  Farrah D.,  Nyland K.,  Afonso J.,  Brandt W.~N.,  Clements D.~L.,  Lagos C.~D.~P.,  Maraston C.,  Pforr J. et al., 2021, \mnras, 501, 892
\bibitem[\protect\citeauthoryear{Lada, Evans, \& Falgarone}{1997}]{lada97} Lada E.~A., Evans N.~J., Falgarone E., 1997, \apj, 488, 286
\bibitem[\protect\citeauthoryear{Lada, Lombardi, \& Alves}{2009}]{lada09} Lada C.~J., Lombardi M., Alves J.~F., 2009, \apj, 703, 52
\bibitem[\protect\citeauthoryear{Lada, Lombardi, \& Alves}{2010}]{lada10} Lada C.~J., Lombardi M., Alves J.~F., 2010, \apj, 724, 687
\bibitem[\protect\citeauthoryear{Leroy et al.}{2015}]{leroy15} Leroy A.~K. et al. 2015, \apj, 801, 25
\bibitem[\protect\citeauthoryear{Lilly}{1988}]{lilly88} Lilly S.~J. 1988, \apj, 333, 161
\bibitem[\protect\citeauthoryear{Lonsdale et al.}{2003}]{lonsdale03} Lonsdale C.~J., Smith H.~E., Rowan-Robinson M., Surace J., Shupe D., Xu C., Oliver S., et al., 2003, PASP, 115, 897
\bibitem[\protect\citeauthoryear{Lynden-Bell}{1969}]{lynden-bell69} Lynden-Bell, D. 1969, \nat, 223, 690
\bibitem[\protect\citeauthoryear{Madau \& Dickinson}{2014}]{madau14} Madau P. \& Dickinson M. 2014, \araa, 52, 415
\bibitem[\protect\citeauthoryear{Magnelli et al.}{2019}]{magnelli19} Magnelli, B., Karim, A., Staguhn, J., et al.\ 2019, \apj, 877, 45
\bibitem[\protect\citeauthoryear{Mao et al.}{2012}]{mao12} Mao M.~Y. et al.\ 2012, \mnras, 426, 3334
\bibitem[\protect\citeauthoryear{Marchesini et al.}{2009}]{marchesini09} Marchesini D., van Dokkum P. G., F\"orster Schreiber N.~M., Franx M., Labb\'e I., Wuyts S. 2009, \apj, 701, 1765
\bibitem[\protect\citeauthoryear{Mart{\'\i}-Vidal et al.}{2014}]{martividal14} Mart{\'\i}-Vidal, I., Vlemmings, W.~H.~T., Muller, S., et al.\ 2014, \aap, 563, A136
\bibitem[\protect\citeauthoryear{Mauduit et al.}{2012}]{mauduit12} Mauduit J.~C. et al.\ 2012, PASP, 124, 714
\bibitem[\protect\citeauthoryear{McMullin et al.}{2007}]{mcmullin07} McMullin J.~P., Waters B., Schiebel D., Young W., Golap K., 2007, ASPC, 376, 127
\bibitem[\protect\citeauthoryear{Messias et al.}{2012}]{messias12} Messias H., Afonso J., Salvato M., Mobasher B., Hopkins A.~M. 2012, \apj, 754, 120
\bibitem[\protect\citeauthoryear{Messias et al.}{2014}]{messias14} Messias H., Dye S., Nagar N., Orellana G., Bussmann R.~S., Calanog J., Dannerbauer H., et al., 2014, \aap, 568, A92
\bibitem[\protect\citeauthoryear{Middelberg et al.}{2008}]{middelberg08} Middelberg et al. 2008, \aj, 135, 1276
\bibitem[\protect\citeauthoryear{Miley \& De Breuck}{2008}]{miley08} Miley G. \& De Breuck C. 2008, \araa, 15, 67
\bibitem[\protect\citeauthoryear{Mills}{1981}]{mills81} Mills B.~Y., 1981, PASAu, 4, 156. doi:10.1017/S1323358000016222
\bibitem[\protect\citeauthoryear{Mroczkowski}{2009}]{mroczkowski09} Mroczkowski T., 2009, PhDT
\bibitem[\protect\citeauthoryear{Mroczkowski et al.}{2019}]{mroczkowski19} Mroczkowski, T., et al.\ 2019, SSR, 215, 17
\bibitem[\protect\citeauthoryear{Naess et al.}{2020}]{Naess2020} Naess S., Aiola S., Austermann J.~E., Battaglia N., Beall J.~A., Becker D.~T., Bond R.~J., et al., 2020, JCAP, 2020, 046. doi:10.1088/1475-7516/2020/12/046
\bibitem[\protect\citeauthoryear{Nakagawa et al.}{2014}]{nakagawa14} Nakagawa T., Shibai H., Onaka T., Matsuhara H., Kaneda H., Kawakatsu Y., Roelfsema P. 2014, Space Telescopes and Instrumentation 2014: Optical, Infrared, and Millimeter Wave, 91431I, doi:10.1117/12.2055947
%\bibitem[\protect\citeauthoryear{Norris}{1998}]{norris88} Norris R., 1988, S&T, 76, 615
\bibitem[\protect\citeauthoryear{Norris et al.}{2006}]{norris06} Norris R.~P. et al. 2006, \aj, 132, 2409
\bibitem[\protect\citeauthoryear{Norris et al.}{2011}]{norris11} Norris R.~P., Afonso J., Cava A., Farrah D., Huynh M.~T., Ivison R.~J., Jarvis M., et al., 2011, ApJ, 736, 55. doi:10.1088/0004-637X/736/1/55
\bibitem[\protect\citeauthoryear{Oliver et al.}{2000}]{oliver00} Oliver S. et al. 2000, \mnras, 316, 749
\bibitem[\protect\citeauthoryear{Oteo et al.}{2016}]{oteo16} Oteo I. et al. 2016, \apj, 827, 34
\bibitem[\protect\citeauthoryear{Padovani}{2016}]{padovani16} Padovani P. 2016, \araa, 24, 13
\bibitem[\protect\citeauthoryear{Papadopoulos et al.}{2000}]{papadopoulos00} Papadopoulos P.~P., R\"ottgering H.~J.~A., van der Werf P.~P., Guilloteau, S., Omont A., van Breugel W.~J.~M., Tilanus R.~P.~J. 2000, \apj, 528, 626
\bibitem[\protect\citeauthoryear{Perez \& Granger}{2007}]{perez07} Perez F., Granger B.~E., 2007, CSE, 9, 21
\bibitem[\protect\citeauthoryear{Perna et al.}{2017}]{perna17} Perna M., Lanzuisi G., Brusa M., Cresci G., Mignoli M., 2017, A\&A, 606, A96
\bibitem[\protect\citeauthoryear{Pforr et al.}{2019}]{pforr19} Pforr J., Vaccari M., Lacy M., Maraston C., Nyland K., Marchetti L., Thomas D.\ 2019, \mnras, 483, 3168
\bibitem[\protect\citeauthoryear{Pilbratt et al.}{2010}]{pilbratt10} Pilbratt G.~L., Riedinger J.~R., Passvogel T., Crone G., Doyle D., Gageur U., Heras A.~M., et al., 2010, A\&A, 518, L1. doi:10.1051/0004-6361/201014759
\bibitem[\protect\citeauthoryear{Pitchford et al.}{2016}]{pitchford16} Pitchford L.~K., Hatziminaoglou E., Feltre A., Farrah D., Clarke C., Harris K.~A., Hurley P., et al., 2016, MNRAS, 462, 4067. doi:10.1093/mnras/stw1840
\bibitem[\protect\citeauthoryear{Polletta, et al.}{2007}]{polletta07} Polletta M., et al., 2007, ApJ, 663, 81
\bibitem[\protect\citeauthoryear{Pope et al.}{2006}]{pope06} Pope A. et al. 2006, \mnras, 370, 1185
\bibitem[\protect\citeauthoryear{Pratt et al.}{2019}]{Pratt2019} Pratt G.~W., Arnaud M., Biviano A., Eckert D., Ettori S., Nagai D., Okabe N., et al., 2019, SSRv, 215, 25
\bibitem[\protect\citeauthoryear{Randall et al.}{2012}]{randall12} Randall, K.~E., Hopkins, A.~M., Norris, R.~P., et al.\ 2012, \mnras, 421, 1644
\bibitem[\protect\citeauthoryear{Rees}{1982}]{rees82} Rees, M.~J.\ 1982, The Galactic Center, 166
\bibitem[\protect\citeauthoryear{Rees}{1984}]{rees84} Rees, M. J. 1984, \araa, 22, 471
\bibitem[\protect\citeauthoryear{Rephaeli}{1995}]{rephaeli95} Rephaeli Y., 1995, ARA\&A, 33, 541. doi:10.1146/annurev.aa.33.090195.002545
\bibitem[\protect\citeauthoryear{Reuland et al.}{2004}]{reuland04} Reuland M., R\"ottgering H., van Breugel W., De Breuck C.\ 2004, \mnras, 353, 377
\bibitem[\protect\citeauthoryear{Rigby et al.}{2014}]{rigby14} Ribgy E.~E. et al. 2014, \mnras, 437, 1882
\bibitem[\protect\citeauthoryear{Robitaille \& Bressert}{2012}]{robitaille12} Robitaille T., Bressert E., 2012, ascl.soft
\bibitem[\protect\citeauthoryear{Rowan-Robinson et al.}{2013}]{rowanrobinson13} Rowan-Robinson M., Gonzalez-Solares E., Vaccari M., Marchetti L.\ 2013, \mnras, 428, 1958
\bibitem[\protect\citeauthoryear{Rupke \& Veilleux}{2011}]{rupkeVeilleux11} Rupke D.~S.~N., Veilleux S., 2011, ApJL, 729, L27
\bibitem[\protect\citeauthoryear{Salvato et al.}{2009}]{salvato09} Salvato M. et al.\ 2009, \apj, 690, 1250
\bibitem[\protect\citeauthoryear{Sayers, et al.}{2013}]{sayers13} Sayers J., et al., 2013, ApJ, 778, 52
\bibitem[\protect\citeauthoryear{Sayers, et al.}{2019}]{sayers19} Sayers J., et al., 2019, ApJ, 880, 45
\bibitem[\protect\citeauthoryear{Sazonov \& Sunyaev}{1998}]{sazonov98} Sazonov S.~Y., Sunyaev R.~A., 1998, ApJ, 508, 1. doi:10.1086/306406
\bibitem[\protect\citeauthoryear{Schinnerer et al.}{2010}]{schinnerer10} Schinnerer E. et al. 2010, \apjs, 188, 384
\bibitem[\protect\citeauthoryear{Scott et al.}{2008}]{scott08} Scott, K.~S., Austermann, J.~E., Perera, T.~A., et al.\ 2008, \mnras, 385, 2225
\bibitem[\protect\citeauthoryear{Scoville et al.}{2016a}]{scoville16a} Scoville N., Sheth K., Aussel H., Vanden Bout P., Capak P., Bongiorno A., Casey C.~M., et al., 2016, ApJ, 820, 83. doi:10.3847/0004-637X/820/2/83
\bibitem[\protect\citeauthoryear{Scoville et al.}{2016b}]{scoville16b} Scoville N., Sheth K., Aussel H., Vanden Bout P., Capak P., Bongiorno A., Casey C.~M., et al., 2016, ApJ, 824, 63. doi:10.3847/0004-637X/824/1/63
\bibitem[\protect\citeauthoryear{Serra et al.}{2015}]{serra15} Serra P.,  Westmeier T.,  Giese N. et al.\ 2015, \mnras, 448, 1922
\bibitem[\protect\citeauthoryear{Seymour et al.}{2007}]{seymour07} Seymour N. et al. 2007, \apjs, 171, 353
\bibitem[\protect\citeauthoryear{Shakura \& Sunyaev}{1973}]{shakura73} Shakura N.~I., Sunyaev R.~A., 1973, A\&A, 500, 33
\bibitem[\protect\citeauthoryear{Silva et al.}{2015}]{silva15} Silva A., Sajina A., Lonsdale C., Lacy M. 2015, \apj, 806L, 25
\bibitem[\protect\citeauthoryear{Simpson et al.}{2006}]{simpson06} Simpson C., Martínez-Sansigre A., Rawlings S., Ivison R.~J., Akiyama M., Sekiguchi K., Takata T., Ueda Y., Watson M. 2006, \mnras, 372, 741
\bibitem[\protect\citeauthoryear{Smolci\'c et al.}{2014}]{smolcic14} Smolci\'c V. et al. 2014, \mnras, 443, 2590
\bibitem[\protect\citeauthoryear{Staguhn et al.}{2014}]{staguhn14} Staguhn, J.~G., Kov{\'a}cs, A., Arendt, R.~G., et al.\ 2014, \apj, 790, 77
\bibitem[\protect\citeauthoryear{Stevens et al.}{2003}]{stevens03} Stevens J.~A., Ivison R.~J., Dunlop J.~S., Smail I.~R., Percival W.~J., Hughes D.~ H., R\"ottgering H.~J.~A., van Breugel W.~J.~M., Reuland M.\ 2003, \nat, 425, 264
\bibitem[\protect\citeauthoryear{Sturm et al.}{2011}]{sturm11} Sturm E., Gonz{\'a}lez-Alfonso E., Veilleux S., Fischer J., Graci{\'a}-Carpio J., Hailey-Dunsheath S., Contursi A., et al., 2011, ApJL, 733, L16
\bibitem[\protect\citeauthoryear{Sunyaev \& Zeldovich}{1972}]{sunyaev72} Sunyaev, R.~A., Zeldovich, Y.~B., 1972, Comments Astrophys. Space Phys., 4, 173;
\bibitem[\protect\citeauthoryear{Sunyaev \& Zeldovich}{1980}]{sunyaev80} Sunyaev R.~A., Zeldovich I.~B., 1980, \mnras, 190, 413
\bibitem[\protect\citeauthoryear{Swarup}{1990}]{swarup90} Swarup G., 1990, IJRSP, 19, 493
\bibitem[\protect\citeauthoryear{Swinyard et al.}{2009}]{swinyard09} Swinyard B. et al. 2009, ExA, 23, 193
\bibitem[\protect\citeauthoryear{Tasse et al.}{2006}]{tasse06} Tasse C. et al. 2006, \aap, 456, 791
\bibitem[\protect\citeauthoryear{Tasse et al.}{2007}]{tasse07} Tasse C., R\"ottgering H.~J.~A., Best P.~N., Cohen A.~S., Pierre M., Wilman R. 2007, \aap, 471, 1105
\bibitem[\protect\citeauthoryear{Taylor}{2005}]{taylor05} Taylor M.~B., 2005, ASPC, 347, 29
\bibitem[\protect\citeauthoryear{Tielens et al.}{1979}]{tielens79} Tielens A.~G.~G.~M., Miley G.~K., Willis A.~G.\ 1979, \aas, 35, 156 
\bibitem[\protect\citeauthoryear{Umehata et al.}{2017}]{umehata17} Umehata H. et al. 2017, \apj, 835, 98
\bibitem[\protect\citeauthoryear{Vaccari}{2015}]{vaccari15} Vaccari M., 2015, fers.conf, 27
\bibitem[\protect\citeauthoryear{van der Walt, Colbert, \& Varoquaux}{2011}]{vanderwalt11} van der Walt S., Colbert S.~C., Varoquaux G., 2011, CSE, 13, 22
\bibitem[\protect\citeauthoryear{Viti et al.}{2014}]{viti14} Viti S., Garc{\'\i}a-Burillo S., Fuente A., Hunt L.~K., Usero A., Henkel C., Eckart A., et al., 2014, A\&A, 570, A28. doi:10.1051/0004-6361/201424116
\bibitem[\protect\citeauthoryear{Wayth et al.}{2015}]{wayth15} Wayth, R.~B., Lenc, E., Bell, M.~E., et al.\ 2015, \pasa, 32, e025
\bibitem[\protect\citeauthoryear{Werner et al.}{2004}]{werner04} Werner M.~W., Roellig T.~L., Low F.~J., Rieke G.~H., Rieke M., Hoffmann W.~F., Young E., et al., 2004, ApJS, 154, 1. doi:10.1086/422992
\bibitem[\protect\citeauthoryear{Wilson et al.}{2011}]{wilson11} Wilson W.~E., Ferris R.~H., Axtens P., Brown A., Davis E., Hampson G., Leach M., et al., 2011, MNRAS, 416, 832. doi:10.1111/j.1365-2966.2011.19054.x
\bibitem[\protect\citeauthoryear{Wright et al.}{2010}]{wright10} Wright E.~L., Eisenhardt P.~R.~M., Mainzer A.~K., Ressler M.~E., Cutri R.~M., Jarrett T., Kirkpatrick J.~D., et al., 2010, AJ, 140, 1868. doi:10.1088/0004-6256/140/6/1868
\bibitem[\protect\citeauthoryear{Wu et al.}{2010}]{wu10} Wu, J., Evans, N.~J., Shirley, Y.~L., et al.\ 2010, \apjs, 188, 313
\bibitem[\protect\citeauthoryear{Wu et al.}{2015}]{wu15} Wu et al. 2015, \nat, 518, 512
\bibitem[\protect\citeauthoryear{Zinn et al.}{2012}]{zinn12} Zinn, P.-C., Middelberg, E., Norris, R.~P., et al.\ 2012, \aap, 544, A38
	\end{thebibliography}
